\documentclass[12pt,a4paper]{rg-hivon-thesis-en-arxiv}
\usepackage{graphicx}
\usepackage{rotating}
\usepackage[normalem]{ulem}
\usepackage{hyperref}
\usepackage[acronym,toc]{glossaries}
\makeglossaries
\newacronym{mimo}{}{multiple-input multiple output}
\newacronym{dft}{DFT}{density functional theory}
\newacronym{dag}{DAG}{directed acyclic graph}
\newacronym{gpu}{GPU}{graphics processing unit}
\newacronym{gui}{GUI}{graphical user interface}
\newacronym{neb}{NEB}{nudged elastic band}
\newacronym{idpp}{IDPP}{image dependent pair potential}
\newacronym{cineb}{CI-NEB}{climbing-image nudged elastic band}
\newacronym{lbfgs}{L-BFGS}{limited-memory Broyden-Fletcher-Goldfarb-Shanno}
\newacronym{cg}{CG}{conjugate gradient}
\newacronym{scg}{SCG}{scaled conjugate gradient}
\newacronym{gpd}{GPDimer}{Gaussian process regression dimer}
\newacronym{rpc}{RPC}{remote proceedure call}
\newacronym{mep}{MEP}{minimum energy path}
\newacronym{otgp}{OTGP}{optimal transport Gaussian process framework}
\newacronym{otgpd}{OTGPD}{optimal transport Gaussian process dimer}
\newacronym{mmf}{MMF}{minimum mode following}
\newacronym{mlip}{MLIP}{machine learned interatomic potential}
\newacronym{kmc}{KMC}{kinetic Monte Carlo}
\newacronym{akmc}{AKMC}{adaptive kinetic Monte Carlo}
\newacronym{hpc}{HPC}{high performance computing}
\newacronym{roneb}{MMF-NEB}{minimum mode following nudged elastic band}
\newacronym{md}{MD}{molecular dynamics}
\newacronym{hf}{HF}{Hartree-Fock}
\newacronym{htst}{H-TST}{harmonic transition state theory}
\newacronym{tst}{TST}{transition state theory}
\newacronym{scf}{SCF}{self consistent field}
\newacronym{ks}{KS}{Kohn-Sham}
\newacronym{wbo}{WBO}{Wiberg bond order}
\newacronym{mvn}{MVN}{multivariate normal}
\newacronym{rmsd}{RMSD}{root-mean-square displacement}
\newacronym{emd}{EMD}{earth mover's distance}
\newacronym{psd}{PSD}{positive semi-definite}
\newacronym{spd}{SPD}{symmetric positive definite}
\newacronym{mll}{MLL}{marginal log-likelihood}
\newacronym{fps}{FPS}{Farthest point sampling}
\newacronym{rkhs}{RKHS}{reproducing-kernel Hilbert space}
\newacronym{ira}{IRA}{iterative rotations and assignments}
\newacronym{se}{SE}{squared exponential}
\newacronym{fem}{FEM}{finite element method}
\newglossaryentry{sp}{name=saddlepoint,description={{Point of zero force, with a single negative eigenvalue of the Hessian}}}
\newglossaryentry{mgp}{name=marginal likelihood,description={{Evidence; probability of observed data under a model, integrating over parameters}}}
\newglossaryentry{extrema}{name=extrema,description={{Minima, maxima, saddle points}}}
\newglossaryentry{hessian}{name=Hessian,description={{Second derivative matrix of a function, used to analyze curvature and stationary points}}}
\newglossaryentry{eigenvec}{name=eigenvector,description={{Vector indicating a principal direction in a matrix, often used in optimization}}}
\newglossaryentry{eigenval}{name=eigenvalue,description={{Scalar associated with an eigenvector; quantifies curvature along that direction}}}
\newglossaryentry{glmm}{name=Generalized Linear Mixed Model,description={{Statistical model incorporating both fixed and random effects}}}
\newglossaryentry{pes}{name=Potential Energy Surface,description={{Multidimensional surface representing energy as a function of geometry}}}
\newglossaryentry{gpr}{name=Gaussian Process Regression,description={{Non-parametric Bayesian approach for regression and function approximation}}}
\newglossaryentry{gp}{name=Gaussian Process,description={{A collection of random variables, any finite number of which have a joint multivariate normal (Gaussian) distribution; defines a distribution over functions.}}}
\newglossaryentry{kernel}{name=kernel function,description={{Function defining similarity between data points in GPR; determines covariance structure}}}
\newglossaryentry{cholesky}{name=Cholesky decomposition,description={{Matrix decomposition into lower and upper triangular matrices, used for solving systems}}}
\newglossaryentry{pruning}{name=data pruning,description={{Technique to reduce dataset size by removing less informative or distant data points}}}
\newglossaryentry{posterior}{name=posterior distribution,description={{Probability distribution representing updated beliefs after observing data}}}
\newglossaryentry{prior}{name=prior distribution,description={{Initial probability distribution before observing data}}}
\newglossaryentry{armijo}{name=Armijo line search,description={{Method for step size selection in optimization to ensure sufficient decrease}}}
\newglossaryentry{hyperparam}{name=hyperparameters,description={{Parameters governing model behavior but not directly optimized during training}}}
\newglossaryentry{rbf}{name=Radial Basis Function,description={{Common kernel function in GPR, based on distance between points}}}
\newglossaryentry{gram-schmidt}{name=Gram-Schmidt process,description={{Procedure for orthonormalizing a set of vectors}}}
\newglossaryentry{variance}{name=variance,description={{Measure of spread in a dataset or uncertainty in a model prediction}}}
\newglossaryentry{likelihood}{name=likelihood,description={{Probability of observed data given specific model parameters}}}
\newglossaryentry{credint}{name=credible interval,description={{Bayesian equivalent of a confidence interval; range of parameter values with given probability}}}
\newglossaryentry{randint}{name=random intercept,description={{Model term allowing each group (e.g., a chemical system) to have its own baseline value}}}
\newglossaryentry{fixedeffect}{name=fixed effect,description={{Model term estimating the effect of predictors assumed to be constant across all groups}}}
\newglossaryentry{robust}{name=robust,description={{Describes an estimator or method that maintains performance even when assumptions are violated}}}
\newglossaryentry{homosced}{name=homoscedasticity,description={{Statistical property where the variance of residuals remains constant across all observations}}}
\newglossaryentry{pseudorep}{name=pseudoreplication,description={{Error in statistical inference caused by treating non-independent observations as independent}}}
\newglossaryentry{stderr}{name=standard error,description={{Estimate of the variability of a sample statistic, often used to quantify uncertainty in parameter estimates}}}
\newglossaryentry{nonindep}{name=non-independence,description={{Condition where observations are correlated or related, violating the assumption of independent samples}}}
\newglossaryentry{deflatederr}{name=deflated standard error,description={{Standard error estimate that is artificially small, typically due to ignoring data correlations or non-independence. Leads to overconfident statistical conclusions.}}}
\thesisauthor{Rohit Goswami}
\thesistitle{Efficient Exploration of Chemical Kinetics}
\thesissubtitle{Development and application of tractable Gaussian Process Models}
\thesismonth{October}
\thesisyear{2025}
\thesisdegree{Philosophiae Doctor}
\thesisdegreeletters{Ph.D.}
\thesisdocument{Dissertation}
\thesisfield{Physical Chemistry}
\thesisschool{School of Engineering and Natural Sciences}
\thesisfaculty{Physical Sciences}
\thesisaddress{Dunhagi 5}
\thesispostalcode{107}
\thesistelephone{525-4000}
\thesisprinter{XXPrinter}
\thesissupervisionbody{Ph.D. Committee}
\thesiscommittee{Hannes Jónsson\\Birgir Hrafnkelsson\\Morris Riedel\\Egill Skúlason\\Thomas Bligaard}
\thesisevaluator{Opponents}
\thesisopponents{Sigurður I. Erlingsson\\ Normand Mousseau}
\thesiskeywords{Chemical Reaction Dynamics, Gaussian Process Regression, Computational Representation, Statistical Machine Learning, High-Performance Simulation}
\thesisORCID{0000-0002-2393-8056}
\thesisISBN{978-9935-9826-5-0}
\thesislicense{CC-BY-4.0}%

\date{\today}
\title{}
\hypersetup{
 pdfauthor={Rohit Goswami (HaoZeke)},
 pdftitle={},
 pdfkeywords={},
 pdfsubject={},
 pdfcreator={Emacs 30.2 (Org mode 9.7.39 + HaoZeke)}, 
 pdflang={English}}
\begin{document}

\makecoverpage
\maketitlepage
\thesisabstract{
Spatio-temporal control of chemical systems to tune relative rates of competing reactions has been the goal of chemistry since early alchemy. Today, the estimation of the products and rates of chemical reactions as well as the stability of chemicals and materials are fundamental tasks for the chemical industry. Despite leaps in mathematical modeling, with insightful representations of electronic structure to describe many body quantum systems, and inspite of exascale computing resources, efficient methods for determining reaction rates in large scale simulations has remained out of reach. Direct simulation of atomic dynamics is limited by short timescale and small length scale. Recently, there has been rapid advance in the generation of machine learned potential functions, but they require large data sets as input and are not practical when the task is to quickly screen thousands of chemicals or materials to identify optimal candidates for technological applications. They have, furthermore, been limited so far to regions of stable configurations of the atoms and are not reliable for the transition state regions which are needed for estimating reaction rates.  Attempts to explore reaction networks in an automated manner at sufficient accuracy suffer from the large computational cost of the electronic structure calculations. Simplifying approximations for rate calculations recognise that reactions represent slow processes on the time scale of atomic vibrations and thermal equilibration, and make use of statistical approximations for chemical rate calculations. In the simplest approximation, the harmonic approximation to transition state theory, they boil down to finding first order saddle points on the energy surface describing how the system's energy depends on the position of the atoms. Even so, the computational effort in saddle point searches is prohibitively large in many cases especially when the energy and atomic forces are obtained from electronic structure calculations. Surrogate model based acceleration of saddle point searches have been described as promising for almost a decade now, but in practical terms have remained crippled by large computational overhead and numerical instabilities that negate the advantage in wall time.

This dissertation presents a solution based on a holistic approach that co-designs the physical representation, statistical model, and systems architecture. This philosophy is embodied in the Optimal Transport Gaussian Process (OT-GP) framework, which uses a physics-aware representation based on optimal transport metrics to create a compact and chemically relevant surrogate of the potential energy surface. This defines a statistically robust approach and uses targeted sampling to reduce the computational effort. Alongside rewrites for the EON software for long timescale simulations, we present a reinforcement-learning approach for the minimum-mode following method when final state is not known and nudged elastic band method when both initial and final state are specified. Collectively, these advances establish a representation-first, service-oriented paradigm for chemical kinetics simulations. The success of this paradigm is demonstrated through large-scale benchmarks where the framework shows state of the art performance characteristics, validated with Bayesian hierarchical models. By delivering a framework for high performance open-source tooling, this work transforms a long-held theoretical promise into a practical engine for exploring chemical kinetics.
}
\thesisutdrattur{Stjórnun efnakerfa í rúmi og tíma til að hafa áhrif á samverkandi efnahvörf hefur verið markmið efnafræðinnar allt frá dögum gullgerðarlistarinnar. Í dag er mat á afurðum og hraða efnahvarfa, ásamt mati á stöðugleika efna og efniviða, grundvallarverkefni í efnaiðnaði. Þrátt fyrir stökk í stærðfræðilegri líkanagerð, með nákvæmum lýsingum á rafeindaskipan til að lýsa fjöleinda skammtafræðikerfum, og þrátt fyrir aðgengi að stórauknu reikniafli (exascale), vantar enn skilvirkar aðferðir til að ákvarða hvarfhraða í stórum hermunum. Bein hermun á gangverki atóma takmarkast af stuttum tímaskala og litlum lengdarkvarða. Nýlega hefur orðið hröð framþróun í gerð véllærðra mættisfalla (machine learned potential functions), en þær krefjast stórra gagnagrunna sem inntaks og eru ekki hagnýtar þegar verkefnið er að skima hratt í gegnum þúsundir efna eða efniviða til að finna bestu kandídatana fyrir tæknilega nýtingu. Þær hafa ennfremur hingað til takmarkast við svæði þar sem atómin eru í stöðugri uppröðun og eru ekki áreiðanlegar fyrir hvarfástönd (transition state regions) sem ákvarða að miklu leiti hvarfhraðann. Tilraunir til að kanna hvarfanet á sjálfvirkan hátt með nægilegri nákvæmni fela í sér of háan kostnað við reikninga á rafeindaskipan. Einfaldandi nálganir fyrir hraðaútreikninga gera ráð fyrir því að efnahvörf séu hægir ferlar miðað við titring atómanna svo að varmalegt jafnvægi náist og nýta því tölfræðilegar nálganir fyrir útreikninga á hvarfhraða. Í einföldustu nálguninni, kjörsveifilsnálgun (harmonic approximation) við virkjunarástandskenninguna (transition state theory), snúast þær um að finna fyrsta stigs söðulpunkta á orkuyfirborðinu sem lýsir því hvernig orka kerfisins er háð staðsetningu atómanna. Jafnvel þá er reikniþörfin við leit að söðulpunktum of mikil í mörgum tilfellum, sérstaklega þegar orka og atómkraftar eru fengnir úr reikningum á rafeindaskipaninni. Hröðun á söðulpunktaleit byggð á staðgengilslíkönum (surrogate models) hefur verið lýst sem vænlegri í nærri áratug, en hefur í reynd verið hömluð af mikilli yfirbyggingu og tölulegum óstöðugleika sem gera að engu ávinninginn í rauntíma.

Þessi ritgerð kynnir lausn sem byggir á heildrænni nálgun á þessu verkefni sem samþættir hönnun á eðlisfræðilegri framsetningu, tölfræðilegu líkani og kerfisarkitektúr. Þessi hugmyndafræði birtist í Optimal Transport Gaussian Process (OT-GP) umgjörðinni, sem notar eðlisfræðilega meðvitaða (physics-aware) framsetningu byggða á mælikvörðum fyrir bestun flutnings (optimal transport) til að búa til þjappaðan og efnafræðilega viðeigandi staðgengil fyrir stöðuorkuyfirborðið. Þetta skilgreinir tölfræðilega trausta nálgun og notar markvissa sýnatöku til að draga úr reikniþörfinni. Samhliða endurskrifun á EON hugbúnaðinum fyrir hermun á löngum tímaskala, er sett fram styrktarnámsnálgun (reinforcement-learning) fyrir lágháttarfylgni (minimum mode following) aðferðina þegar lokaástand er ekki tiltekið og hnikateygjubands (nudged elastic band) aðferðina þegar bæði upphafs- og lokaástand eru tilgreind. Samanlagt marka þessar framfarir nýja hugmyndafræði fyrir hermun á efnahvörfum sem byggir á framsetningunni fyrst (representation-first) og er þjónustumiðuð (service-oriented). Árangur þessarar aðferðafræði er sýndur með stórum viðmiðunarprófunum sem sýna góða frammistöðu, greinda með líkönum Bayes. Með því að þróa aðferð fyrir afkastamikil opinn-hugbúnaðar (open-source) verkfæri, umbreytir þessi vinna gömlu fræðilegu loforði í hagnýta tæki til að kanna gang og hraða efnahvarfa.}
\thesisdedication{TRUE}{Dedication\\\mbox{} \\
To anyone who still wants to write a monograph, and to my family and pets, for waiting longer for my own.
}
\thesispreface{TRUE}{
This document serves as a monograph that complements the research which forms
my doctorate. By design, all text and figures within this thesis are original 
and do not appear in the associated papers.

This work is therefore not a standalone repository of those publications but is
specifically intended to be read in conjunction with them. The chapters herein
provide a guiding narrative, expanded theoretical context, software design, and
novel analysis that link, support, and build upon the findings presented in my
published articles. The reader is encouraged to consult the primary papers for
the detailed methodologies and core results, using this thesis as a companion
for deeper integration and supplementary insight.

\par

I have always known I would like to be an academician. My preferred field of study has been a bit fluid though. For academia as a whole, I've been waiting since I was six pontificating on my father's manuscript. My mom has always fanned the flames of academic curiosity, as my sister races ahead with brilliant deductions. For computers, never trusting to learn as a trade what I enjoy-I started messing with machines in high school. Computational chemistry was a bit of a late interest, mostly when I realized I thought I knew better. I have worn many hats over the years, software engineer, editor, reviewer, author. This is a short demi-monograph, but it has truly been a lifetime in the making.}
\thesislists

\thesislistofpapers{TRUE}{%
\subsection*{Publications directly related to the thesis}

\mypaper{Paper I} {Čertík, Ondřej, Pask, John E., Fernando, Isuru, \textbf{Rohit Goswami}, Sukumar, N., Collins, Lee. A., Manzini, Gianmarco, and Vackář,
Jiří, 2023, High-Order Finite Element Method for Atomic Structure Calculations. \textit{Computer Physics Communications}, Vol. 315, pp. 109051. Accessed at \url{https://dx.doi.org/10.1016/j.cpc.2023.109051}. Rohit finalized the code, reproducibly generated figures and revised the article.}

\mypaper{Paper II} {\textbf{Rohit Goswami}, Maxim Masterov, Satish Kamath, Alejandro Pena-Torres, and Hannes Jónsson, 2025, Efficient Implementation of Gaussian Process Regression Accelerated Saddle Point Searches with Application to Molecular Reactions. \textit{Journal Chemical Theory and Computation}, Vol. 1, Issue 2, pp. 1--10. Accessed at \url{https://dx.doi.org/10.1021/acs.jctc.5c00866}. Rohit and Maxim led the development of the C++ code. Rohit performed calculations, created figures and wrote the article.}

\mypaper{Paper III} {\textbf{Rohit Goswami}, 2025, Bayesian Hierarchical Models for Quantitative Estimates for Performance Metrics Applied to Saddle Search Algorithms. \textit{AIP Advances}, Vol. 15, Issue 8. Accessed at \url{https://dx.doi.org/10.1063/5.0283639}. Rohit developed and validated the models described, wrote the paper, and made the figures.}

\mypaper{Paper IV} {\textbf{Rohit Goswami}, Hannes Jónsson, 2025, Adaptive Pruning for Increased Robustness and Reduced Computational Overhead in Gaussian Process Accelerated Saddle Point Searches. \textit{ChemPhysChem}. Accessed at \url{https://doi.org/10.1002/cphc.202500730}. Rohit developed the OTGPD algorithm, ran simulations, made figures, and wrote the article.}

\subsection*{Other publications submitted during the doctorate}
\mypaper{Paper A} {Laurence Kedward, Balint Aradi, Ondrej Certik, Milan Curcic, Sebastian Ehlert, Philipp Engel, \textbf{Rohit Goswami}, Michael Hirsch, Asdrubal Lozada-Blanco, Vincent Magnin, Arjen Markus, Emanuele Pagone, Ivan Pribec, Brad Richardson, Harris Snyder, John Urban, and Jeremie Vandenplas, 2022, The State of Fortran. \textit{Computing in Science \& Engineering}. Accessed at \url{https://dx.doi.org/10.1109/MCSE.2022.3159862}.}

\mypaper{Paper B} {Moritz Sallermann, Amrita Goswami, Alejandro Peña-Torres, and \textbf{Rohit Goswami}, 2025, Flowy: High Performance Probabilistic Lava Emplacement Prediction. \textit{Computer Physics Communications}, Vol. 315, pp. 109745. Accessed at \url{https://dx.doi.org/10.1016/j.cpc.2025.109745}}

\mypaper{Paper C} {\textbf{Rohit Goswami}, Ashwini Kumar Rawat, Sonaly Goswami, and Debabrata Goswami, 2025, Compositional Analysis of Fragrance Accords Using Femtosecond Thermal Lens Spectroscopy. \textit{Chemistry – an Asian Journal}, e00521. Accessed at \url{https://dx.doi.org/10.1002/asia.202500521}.}

\subsection*{Publications submitted before the doctorate}
\mypaper{Paper 1} {Prerna, \textbf{Rohit Goswami}, Atanu K. Metya, S. V. Shevkunov, and Jayant K. Singh, 2019, Study of Ice Nucleation on Silver Iodide Surface with Defects. \textit{Molecular Physics}, pp. 1–13. Accessed at \url{https://dx.doi.org/10.1080/00268976.2019.1657599}.}

\mypaper{Paper 2} {\textbf{Rohit Goswami}, Amrita Goswami, and Jayant Kumar Singh, 2020, d-SEAMS: Deferred Structural Elucidation Analysis for Molecular Simulations. \textit{Journal of Chemical Information and Modeling}. Accessed at \url{https://dx.doi.org/10.1021/acs.jcim.0c00031}.}

\mypaper{Paper 3} {Ligesh Theeyancheri, Subhasish Chaki, Nairhita Samanta, \textbf{Rohit Goswami}, Raghunath Chelakkot, and Rajarshi Chakrabarti, 2020, Translational and Rotational Dynamics of a Self-Propelled Janus Probe in Crowded Environments. \textit{Soft Matter}. Accessed at \url{https://dx.doi.org/10.1039/D0SM00339E}.}

}
\glsaddall \printglossaries
\thesisacknowledgements{TRUE}{My work has spanned a couple of fields, and for that, I'm grateful to have been given an opportunity to be inspired by so many.

My family comes first. The PhD has been a bit of a white whale for me—the bare minimum requirement I demanded of myself. I know this ambition must have been grating to them, as I am the youngest and have long been looking up to their accomplishments. Their unwavering support has sustained this dream since I was in pre-school.

I'd also like to specifically thank Prof. Rajarshi Chakrabarti at IIT Bombay. He was instrumental early on, introducing me to the fascinating world of soft matter systems when I was a first-year undergraduate. My sister Dr. Amrita Goswami, has been a colleague and inspiration, both when we were under Prof. Singh and after.

By day, of course, I can think of no one who embodies my degree more than my advisor, Prof. Hannes Jónsson. Since we met at IISc so many years ago, I've learned a lot from him. He has been an inspiration whose support provided the diving board I needed to go off into statistics, tensor methods, and software development.

Dr. Vilhjálmur Ásgeirsson stands out as someone who not only got Hannes' attention back to my tentative email but who has also been a staunch friend.

My degree spanned the COVID-19 pandemic, but thankfully, for me, this only opened the doors to more science and more people who had time outside the usual bubble of real life. Dr. Ondřej Čertík was my first collaborator and confidant, helping steer me through both corporate shenanigans and academic endeavors. He introduced me to the delight that is modern Fortran and compilers, and for that, I'm ever grateful.

All my co-authors are special; especially family, Dr. Moritz, Dr. Amrita, Sonaly, Ruhila, and Prof. Debabrata. Ondrej, Maxim and Satish in particular, spent long hours with me everyday for discussing the nuances of Fortran and MATLAB respectively. Amongst my near or to-be collaborators come Dr. Andreas Vishart, Prof. Laurent Béland, and Dr. Miha Gunde. My in-laws Nagaselvamani and Sankaranarayan, though non-collaborators have provided immense warmth. My non-human collaborators, Arí, Yoda, Jude, Crystee, TuiTui, Thor, Loki, and all the garden cats are ever driving towards being more curious about everything. Dr. Suryakant and Dr. Ashwini have both been great sources of support for the family and me.

I haven't been very social, but my labmates managed to find a way to connect, especially Ellie, Ivan, Alejandro, Oskar, Olafur, Andre, Liam, and everyone else. At the University, I was honored to be able to teach for Prof. Helmut Neukirchen and support Prof. Steinn Guðmundsson. My graduate committee as a whole, Prof. Birgir Hrafnkelsson, Prof. Morris Riedel, Prof. Hannes Jónsson, Prof. Egill Skúlason and Prof. Thomas Bligaard greatly enriched my time at the university. A special mention to my cactus, Ami, named recently, but who has been by my side since my first apartment in Alfheimar, and led to having many more plants.

Early on, when I was shopping for my grant, I was introduced to Prof. Birgir, who has been invaluable and a pillar of knowledge, statistical and otherwise. I met Prof. Morris Riedel at the first Icelandic HPC meeting and have been attending ever since. He has always given great advice, not to mention providing support for me when I was between funds.

Funding has been a long thread for me. I briefly received income from Quansight Labs for working on foundational open-source projects, an association which was initially interesting but eventually tiresome. I was lucky to have a chance to work under Prof. Gianluca Levi on software enhancements for GPAW within the context of plane wave calculations, leaving even fewer aspects of chemistry untouched.

A few years ago, I was lucky enough to have Dr. Miha Gunde sit behind me in the lab. I've had the pleasure of being able to count on him at all hours of the day or night to respond and discuss any and every aspect related to kinetic Monte Carlo and molecular systems.

It was around the same time that I met several others who provided support, financially and academically, within the context of adaptive kinetic Monte Carlo. In particular, Prof. Laurent Béland has supported me for months while I visited his lab in spite of great administrative hurdles, all the while working actively with me on several interesting projects. Prof. Normand Mousseau, whom I also met around the same time, has provided inputs and insight on much of my recent work, and I am very grateful. He has done a stellar job helping shape my thesis through his role as my opponent as well.

I owe a great debt to Prof. Michele Ceriotti at EPFL. Prof. Ceriotti, whom I first met at several conferences, offered unbridled support despite the baggage of my extended degree timeline. He, along with Dr. Guillaume Fraux, exhibited remarkable patience and commitment while I worked on this thesis. Dr. Fraux also took time to provide suggestions for improvement.

In my conviction that my work must be read widely, I solicited many external reviews from people who earned my respect. I have been humbled by the positive responses, and am even more grateful to those who provided constructive criticism along with approbation. In particular, Prof. Martin Gruebele furnished historical context and insightful academic points. I relished the opportunity to reconnect collaborators and wellwishers including Dr. John Pask, Prof. Baron Peters, Prof. Zubin Jacbin and Prof. Pedro Costa amongst many others.

I'd also like to take the opportunity to thank every other person I've met with or interacted with, with special note for my landlady Guðrun, who gave my family and I a lovely house. Finally, I thank my wife, Ruhi again, for making us feel home wherever we are, and for lending me her brilliant perceptually uniform, colorblind friendly colorscheme ("ruhi") used for most of the figures.
}
\thesisbody
\section{Introduction}
\label{sec:thesis:introduction}
\epigraph{If you wish to make an apple pie from scratch, you must first invent the universe.}{Carl Sagan}

The central pursuit of chemistry is the rational control and transformation of matter. Progress in this domain hinges on the strategic use of abstraction, specifically in how we choose to represent chemical systems. This evolution of representation is profound, moving from the empirical models of alchemy to the rigorous mathematical frameworks of today. The advent of formalisms like second quantization, for instance, provided a language to systematically treat many-body quantum effects, fundamentally changing our ability to model molecular interactions.

Words, too, function as representations. They imperfectly ferry an idea from my mind to yours and, in doing so, enact a subtle form of control. A leaky abstraction, like a rumor retold, degrades signal and invites failure. We therefore cultivate models that compress without distorting, that guide computation without surrendering physics.

By creating simplified yet powerful models that capture essential phenomena, we make intractable problems computationally solvable. Implementing these representations computationally introduces a physical cost that is often overlooked. The ultimate goal of this pursuit is not merely descriptive understanding, but predictive control \cite{goswamiControlChemicalDynamics1994} with direct applications in materials science, pharmacology, and industry \cite{goswamiCrystalNucleationChallenges2021}. Achieving this requires immense computational power, and the efficiency of these computations is constrained by fundamental thermodynamics. Echoing the principles behind Maxwell's Demon \cite{denbighPrinciplesChemicalEquilibrium1997}, Landauer's principle \cite{landauerInformationPhysical1991} dictates a minimum energy cost for erasing information: the ``cost of forgetting.'' An ``efficient Demon'' would avoid accessing the entire distributed dataset simultaneously, thereby eliminating the need to continually churn space in memory. From this perspective, modern high-performance computing—specifically distributed networks—functions as a strategy to manage this information burden, even though the total energy expenditure often remains orders of magnitude higher in practice. This trade-off succeeds by reducing the time-to-solution and facilitating work with datasets that exceed the storage capacity of a single node. Thus, by partitioning a large problem across many computational units, we reduce the information load—the specific memory the core must retain and process. While not offsetting the total energy cost of distributed computing, this distribution mitigates the local difficulty of information processing. Practically, this approach enables explorations of reactive systems on experimentally relevant time and length scales, accelerating the pace of discovery.
\subsection{Chemistry for computers: Space, Time and Temperature}
\label{sec:orgd0ba196}
Computational chemistry relies on a foundational spatial representation. We begin by defining a high-dimensional space where each point corresponds to a unique configuration of atomic nuclei. A scalar potential energy associates with each point to generate a landscape known as the \gls{pes}. The exploration of this landscape—finding stable minima, identifying transition saddle points, and defining reaction paths—manifests fundamentally as a problem in mechanics and optimization. This framework remains atemporal and zero-temperature; a distance metric defines ``closeness,'' and geometric paths measure ``progress'' rather than the passage of time.

\begin{figure}[htbp]
\centering
\includegraphics[width=.9\linewidth]{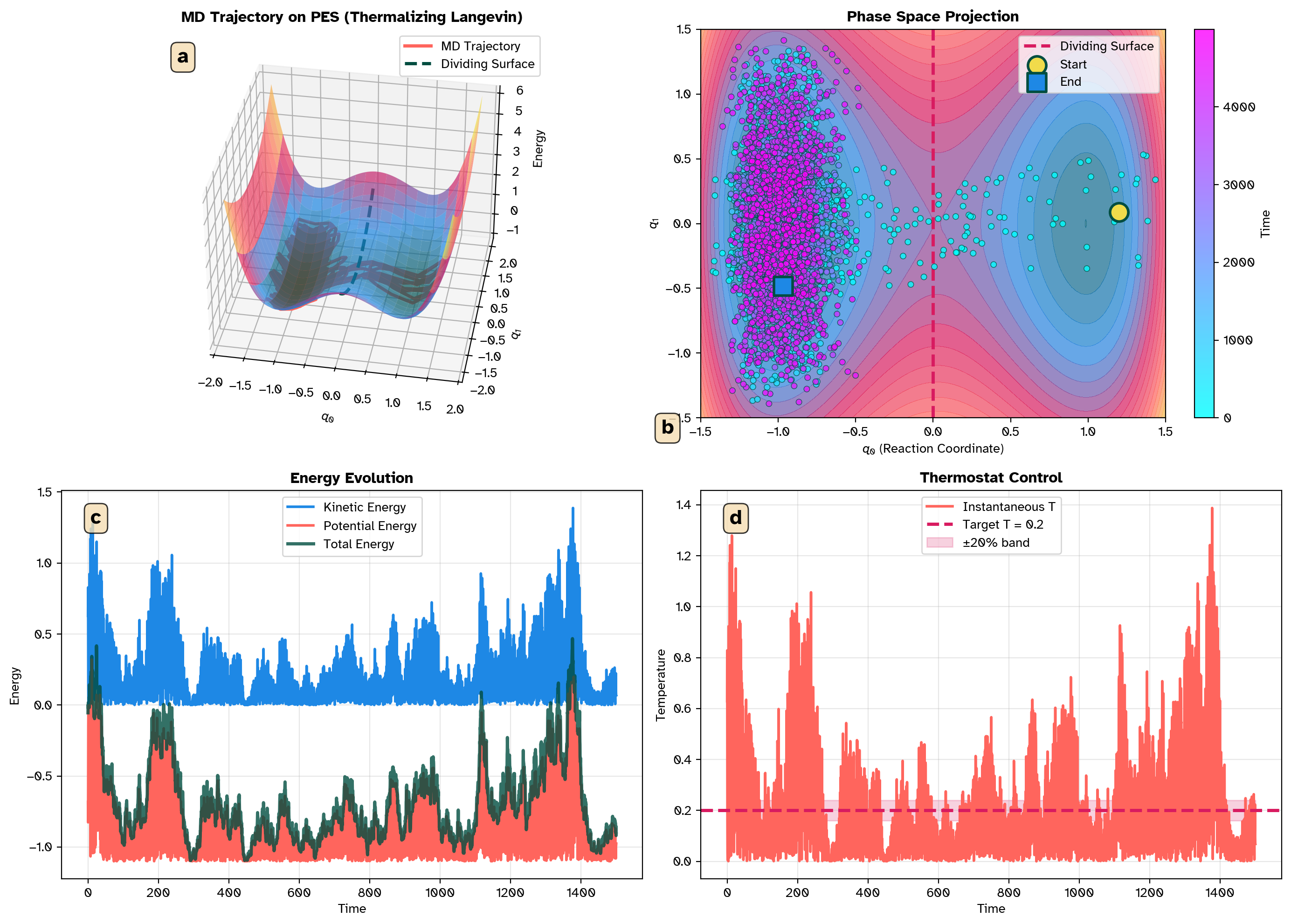}
\caption{\label{fig:mdsurf}\textbf{\textbf{(a)}} A simulated Molecular Dynamics trajectory (red line) on a double-well Potential Energy Surface, shown in 3D with the dividing surface (black dashed line) separating reactant and product regions. \textbf{\textbf{(b)}} Top-down projection of the same trajectory onto the reaction coordinate \(q_0\) and bath coordinate \(q_1\), with energy contours shown as background. The trajectory begins at the reactant minimum (yellow circle) and eventually reaches the product basin (cyan circle), with multiple recrossings of the dividing surface. \textbf{\textbf{(c)}} Energy evolution during the simulation, showing kinetic, potential, and total energy. \textbf{\textbf{(d)}} Temperature control by the Langevin thermostat, maintaining the target temperature around \(T = 0.2\) K within a ±20\% band.}
\end{figure}

Upon this static, spatial landscape, we superimpose representations of dynamics that introduce time and temperature. The most direct of these, \gls{md} traces a time-resolved trajectory of the system according to Newton's laws. Figure \ref{fig:mdsurf} \cite{frenkelUnderstandingMolecularSimulation2001} illustrates this concept. A variety of processes, from phase transitions in fixed-topology force fields to reactive potentials with bond formation and breaking may be understood in this context. Here, temperature arises naturally from the kinetic energy of the particles.

From numerical stability considerations \gls{md} requires a time step on the order of one to two femtoseconds (\(10^{-15}\) s), to capture the bond vibrations as the motion of interest. Consequently, achieving simulation times on the order of milliseconds necessitates billions (\(10^{12}\)) of steps. This computational expense often prohibits the observation of diffusion effects and point defects that occur on macro-scales.

\begin{figure}[htbp]
\centering
\includegraphics[width=0.5\linewidth]{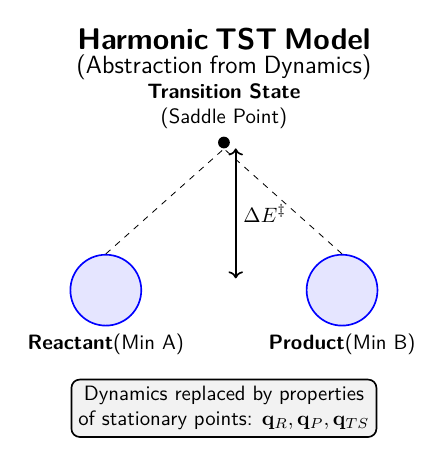}
\caption{\label{fig:htst}The conceptual abstraction for Harmonic Transition State Theory (H-TST) simplifies the complex dynamics. This model replaces the entire trajectory with an analysis of the energetic and vibrational properties of three critical stationary points: the reactant minimum, the product minimum, and the transition state saddle point connecting them.}
\end{figure}

To address these temporal constraints, we move to a representation that is thermodynamically equivalent in the long-time limit but computationally distinct. This approach abstracts the explicit integration of femtosecond vibrations by leveraging statistical mechanics to connect key spatial features of the landscape—the reactant minima and the transition state saddle points—to a macroscopic rate, which has units of inverse time. Temperature enters not through kinetic energy, but through the partition functions that describe the probability of occupying these critical states (Figure \ref{fig:htst}) \cite{petersReactionRateTheory2017}.

These form the class of \Gls{kmc} methods which exploit the fact that the long-term dynamics of many chemical systems may be modeled as a series of jumps from state to state. Instead of following the trajectory through every vibrational period, \gls{kmc} treats state-to-state transitions directly. By assuming that the first escape time from a state follows an exponential distribution, \gls{kmc} allows access to much longer time scales—typically seconds and beyond. In standard implementations, the algorithm selects a transition path from a catalog of rates, typically calculated with \gls{htst}, updates the time, and places the system in a new state. 

In practice, \gls{akmc} \cite{pedersenDistributedImplementationAdaptive2010,henkelmanLongTimeScale2001,el-mellouhiKineticActivationrelaxationTechnique2008} discovers relevant transitions on the fly, builds local catalogs of events, and samples waiting times from exponential clocks. This effectively rewrites dynamics in the language of statistically weighted hops. From a thermodynamic perspective, this is not an approximation but a change in basis: provided the rate constants are accurate, the master equation describes the exact same probabilistic evolution of the system as the ensemble of MD trajectories, but efficiently skips the non-reactive vibrational basins. We shift therefore from covering every possible state, to the study of rare events.

Thus, the challenge manifests as two-fold: first, to efficiently map the high-dimensional spatial landscape, and second, to employ either direct (\gls{md}) or statistical (\gls{tst}, \gls{kmc}) methods to model the temporal evolution of a system across it. This dissertation focuses on developing efficient representations for both aspects. We recognize that predictive control—steering a reaction towards a desired product—requires mastery over both the spatial representation of what exists as possible and the temporal representation of what becomes probable at a given temperature.

\begin{figure}[htbp]
\centering
\includegraphics[width=\textwidth]{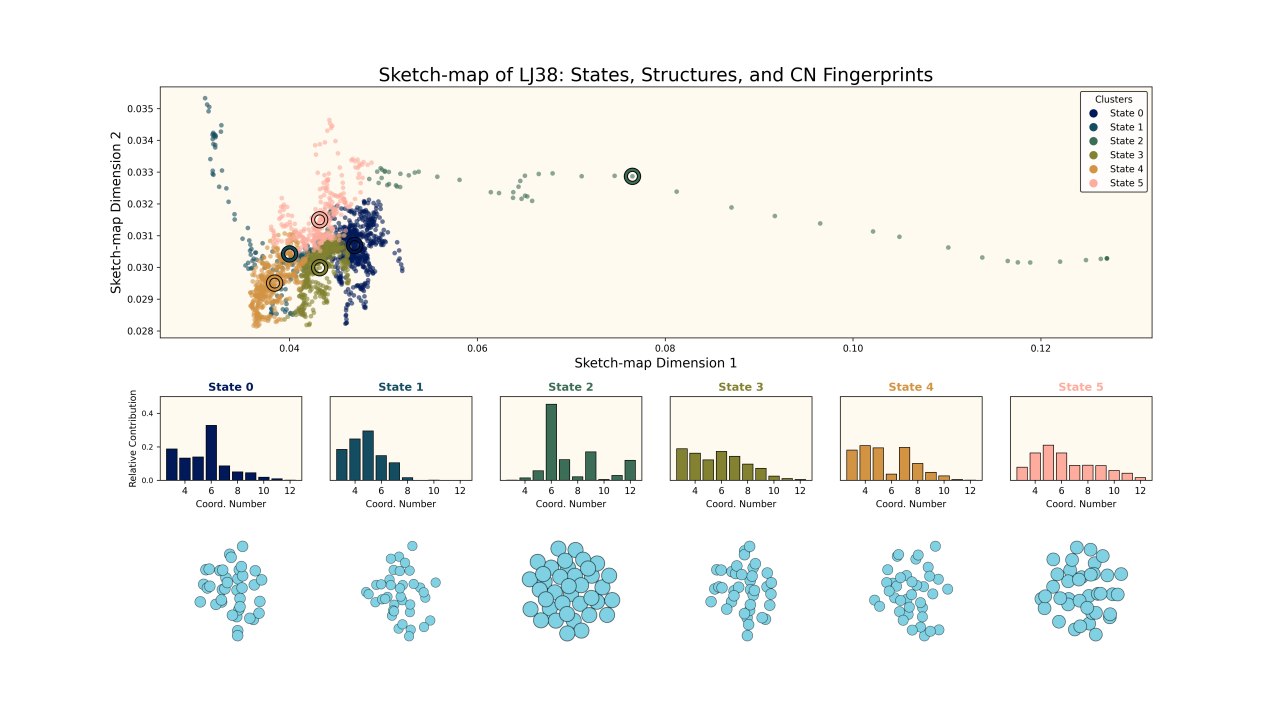}
\caption{\label{fig:sketchy38}The complex structural landscape of the Lennard-Jones 38-atom (LJ38) cluster is visualized using a sketch-map projection. The data is sourced from a molecular dynamics simulation thermostatted at 80.0 K, modeling an Argon cluster. Each point in the trajectory was characterized by a high-dimensional vector representing its coordination number (CN) histogram, which was then projected onto this 2D map. The resulting visualization clearly separates distinct structural basins. An unsupervised clustering algorithm identifies six major structural families, shown as colored regions. This approach allows for an intuitive understanding of the system's structural diversity, with the inset panels providing the characteristic CN histogram ``fingerprint'' for each distinct state.}
\end{figure}

We return to space, because intuition begins there. The \gls{pes} derives from the Born–Oppenheimer separation of electronic and nuclear motion \cite{lewarsComputationalChemistry2016}. For anything beyond simple models, the curse of dimensionality \cite{jamesIntroductionStatisticalLearning2013,hastieElementsStatisticalLearning2009} defeats direct visualization. The Lennard-Jones 38-atom cluster (LJ38) offers a classic case \cite{malekDynamicsLennardJonesClusters2000,walesEnergyLandscapesClusters2000}: a rugged landscape with many contending structures. As an interatomic potential, the Lennard-Jones historically has been a stand-in for the study of Argon \cite{rahmanModelingNumericalStudy} which forms the model system in Figure \ref{fig:sketchy38} as well.

To trace a ``meaningful'' path through such a space, we must first create a map. Tools built on principles of dimensionality reduction, unsupervised learning, and the identification of landmark configurations enable projection onto a two dimensional figure \cite{tribelloUsingSketchmapCoordinates2012}. Figure \ref{fig:sketchy38} demonstrates this approach. We must first tame this complexity visually before we can interrogate it physically. Although such maps often aim to preserve metrics \footnote{in the sense that relative Euclidean distances are often preserved}, they lack explicit information regarding thermodynamic and kinetic transformations. Mapping basins in itself, provides no explicit transition paths or rates between them and cannot characterize or categorize the actual reactive events. We dispense with such post-hoc methods; for the remainder of this work, we focus on the means to explore interesting aspects of phase space and trace paths explicitly.

Even without a low dimensional projection, as soon as we characterize the landscape through the \gls{pes} and choose a metric for distance, we can simulate time-dependent behavior. \gls{md} evolves the system's coordinates according to Newton's equations of motion, providing a microscopic view of atomic motion \cite{frenkelUnderstandingMolecularSimulation2001}. The result is a continuous path through configuration space, which mostly samples thermal vibrations within energy wells, with rare stochastic events of barrier crossing, and the characteristic recrossings that occur near the transition state, often called a trajectory. Figure \ref{fig:mdsurf} illustrates such a trajectory on a model double-well potential. Panel (a) shows the three-dimensional landscape with the trajectory (red line) snaking across the surface, while panel (b) projects this motion onto the reaction coordinate plane overlaid with the energy landscape as a colored background. The trajectory begins localized in the reactant basin (yellow circle), undergoes several thermal explorations, and eventually crosses the dividing surface at \(q_0 = 0\) (magenta dashed line) into the product basin (cyan circle). Critically, the path exhibits multiple recrossings of the barrier—the trajectory does not simply traverse from reactant to product, but rather crosses back and forth, reflecting the stochastic nature of barrier passage at finite temperature.

Panels (c) and (d) reveal why \gls{md}, while dynamically rigorous, contains far more information than needed for determining basins. While the temperature and energy fluctuate as expected for a finite-temperature system with a stochastic thermostat, the quantitative accuracy is less important here than the qualitative demonstration of basin residence, barrier crossing, and recrossing behavior. The temperature and energy fluctuate as expected for a finite-temperature system coupled to a heat bath; these fluctuations characterize the canonical ensemble. The direct simulation of time presents a practical barrier.

Most of the computational effort in \gls{md} is expended on thermal basin exploration—the rapid, high-frequency oscillation of atoms within a stable energy well. While this exchange between kinetic and potential energy is physically rigorous and essential for defining the free energy of the state, it contributes little to the analysis of rare events that drive structural evolution.

This observation motivates the further simplification of \gls{htst}. Rather than tracking every atomic coordinate along the entire trajectory, \gls{htst} posits that the reaction rate depends primarily on the properties of a few critical points on the PES. These are the reactant minimum (low-energy starting configuration), the product minimum (low-energy final configuration), and the transition state, often simplified to the highest-energy saddle point connecting them along the minimum free energy path. By focusing on these stationary points rather than the full dynamics, we can derive reaction rates with far fewer calculations.

\gls{htst} approximates the potential near minima and the transition state with quadratic forms, enabling statistical-mechanical evaluation of partition functions and the derivation of a rate constant. Figure \ref{fig:htst} demonstrates how the trajectory of Figure \ref{fig:mdsurf} collapses to three stationary points and their local curvatures. This chain of reasoning from quantum mechanics to a harmonic model underpins how we compute and understand chemical reactivity.

\begin{figure}[htbp]
\centering
\includegraphics[width=.9\linewidth]{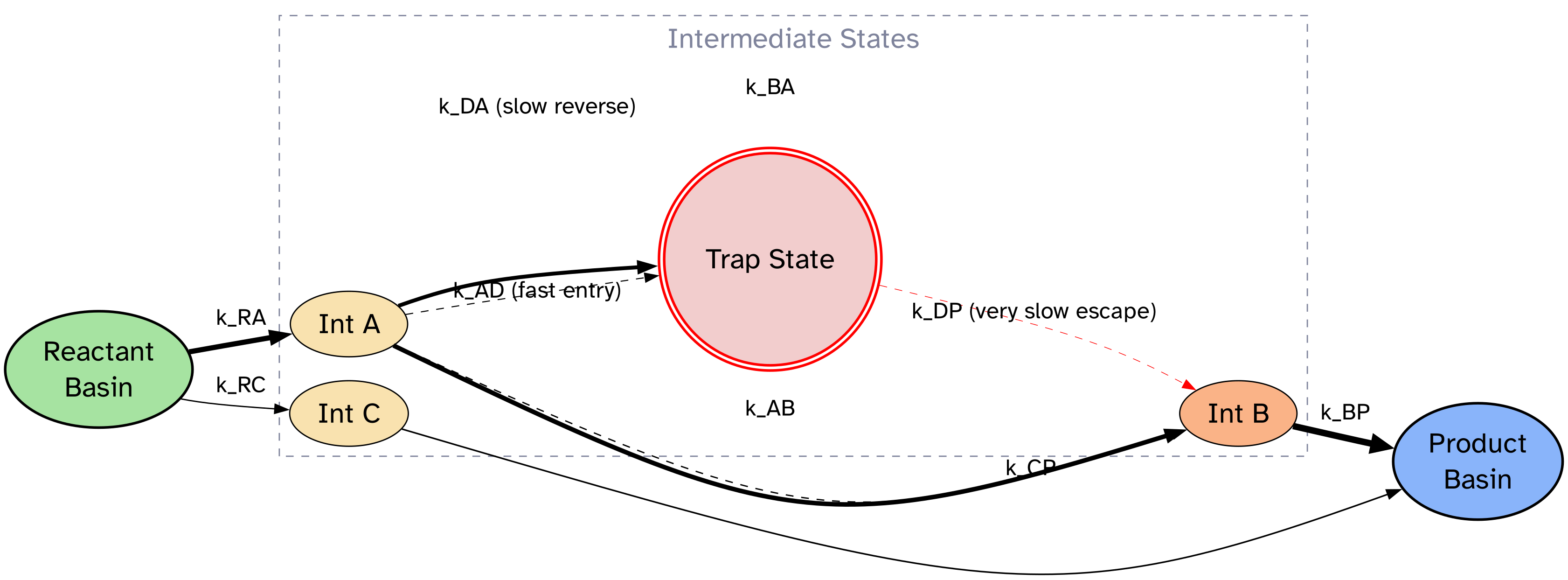}
\caption{\label{fig:akmc_network}Abstraction of a Potential Energy Surface to a Discrete Kinetic Network. The continuous, high-dimensional PES is simplified into a network of states (nodes) and transitions (arrows). Each node represents a stable energy basin. Each arrow's thickness is proportional to its TST-calculated rate constant (\(k\)), visualizing the system's kinetic preferences. The network reveals key dynamical features: a dominant pathway (Reactant \(\to\) Int A \(\to\) Int B \(\to\) Product) with high-flux transitions (thick lines), a slower side-channel (Reactant \(\to\) Int C \(\to\) Product), and a kinetic trap (Trap State). The trap is characterized by a fast entry rate (\(k_{AD}\)) and very slow escape rates (\(k_{DA}\) and \(k_{DP}\)), representing a long-lived metastable state that can dominate the system's evolution. This abstraction allows methods like aKMC to simulate timescales far beyond the reach of direct molecular dynamics.}
\end{figure}

The abstraction from the full dynamics in Figure \ref{fig:mdsurf} to the simple rate model in Figure \ref{fig:htst} forms a compelling progression. However, most systems, even small ones like the LJ38 in Figure \ref{fig:sketchy38} have complex landscapes containing a vast network of states, not just a single reactant–product pair.The notion of ``connectivity'' becomes malleable, as is the notion of ``close'', and many escape paths may exist, with widely varying barriers and time scales. Perhaps the disconnect between kinetics and theromdynamics is most clearly understood from the engineering perspective, where the Haber-Bosch \cite{foglerEssentialsChemicalReaction2011,levenspielChemicalReactionEngineering1999,himmelblauBasicPrinciplesCalculations1989} in practice requires conditions which do not provide maximum yield, but does allow the reaction to proceed in reasonable timeframes. For slowly evolving systems, evaluating \gls{htst} rates for all viable escape paths, allows the continuous \gls{pes} to be recast as a discrete set of states linked by thermally activated transitions (Figure \ref{fig:akmc_network}). The long-term evolution then follows a memoryless, Markovian journey on that network, leading to two distinct simulation methodologies. From the perspective of long-timescale evolution, \gls{md} and \gls{kmc} are thermodynamically equivalent representations; they both aim to sample the equilibrium distribution. The distinction lies in their sampling efficiency for activated processes.

On-lattice \gls{kmc} adopts a predefined lattice and a fixed catalog of allowed processes \cite{andersenPracticalGuideSurface2019,battaileKineticMonteCarlo2008}. Practitioners precompute or fit the energy barriers and rate constants for these processes to various local chemical environments. During a simulation, events are drawn stochastically from this lookup table with probabilities determined by \gls{htst} rates, and the occupation numbers (or site identities) are updated accordingly. The system time advances via an exponential clock: if \(N\) events are available with rates \(k_1, k_2, \ldots, k_N\), the time to the next event is \(\Delta t = -\ln(\xi) / \sum_i k_i\), where \(\xi\) is a uniform random number. This approach delivers exceptional speed for crystalline diffusion and lattice-respecting reactions, yet it omits events that a prior catalog fails to include.

In contrast, off-lattice methods like \gls{akmc} \cite{henkelmanLongTimeScale2001,el-mellouhiKineticActivationrelaxationTechnique2008,trochetOffLatticeKineticMonte2020} lift the fixed lattice and process table constraints and treat states in continuous space. The rate catalog and kinetic network grow on the fly: from the current minimum, the method performs single-ended saddle searches to uncover escape routes, evaluates rates via \gls{htst}, and assembles a local event table. We then proceed as in on-lattice KMC, select an event, advance time with an exponential clock, and resume discovery from the new minimum. This adaptive loop replaces a fixed move set with active discovery and suits defects, amorphous phases, and surface reconstructions without prior assumptions \cite{henkelmanLongTimescaleSimulationsChallenges2018}. The general utility of this method relies on the landscape topology. Systems characterized by high disorder, such as bulk liquids, possess a landscape dense with shallow minima and low barriers. In such regimes, the algorithm tends to trap itself within a superbasin, expending resources to catalog negligible diffusive events rather than describing meaningful phase transitions. Consequently, while clusters and crystalline defects remain tractable, bulk liquids often require different approaches to avoid this combinatorial explosion of states. The cost of discovering pathways in this framework must then be faster than the time taken to traverse the landscape using dynamical methods. The off lattice method is primraily of importance for in principle being free from the bias of having to select events.

We must acknowledge, however, that identifying the saddle point is a necessary but not sufficient condition for describing the full dynamics. Recrossing events, variational effects, and anharmonicity all contribute to the transmission coefficient. Yet, without an efficient method to locate these saddle points in the first place, the higher-order corrections cannot even be attempted.

A significant advantage of this network representation lies in sensitivity analysis; in large reaction networks, the macroscopic evolution often exhibits sensitivity to only a small subset of critical rates, rendering errors in non-rate-limiting steps inconsequential.

From this chain of reasoning then, faster identification of transition states with fewer evaluations of a high-accuracy solver on the \gls{pes} unlocks reaction networks and, with them, practical control over chemical change, feeding into the goal of accurate in-silico control of materials.
\subsection{Motivation}
\label{sec:org0c278a4}
This dissertation concerns itself primarily with the development of computational representations to efficiently model inhospitable regions of chemically interesting phase space.
\subsection{Research objectives and hypotheses}
\label{sec:org8218a3c}
The central thesis of this work is that the efficiency of rare-event sampling is constrained not just by the cost of electronic structure, but by the rigidity of standard software architectures and the fragility of optimization algorithms. To address this towards the in-silico control of materials to accelerate scientific applications, we investigate the following across various length and time scales:

\begin{description}
\item[{Relativistic spurious state evasion}] Can a squared Hamiltonian formulation in a finite element modular Fortran framework provide state of the art stability\footnote{by avoiding spurious states since no variational principle holds for the Dirac case} and numerical performance for single atom all-electron mean-field calculations of both the Schrodinger and Dirac equations?

\item[{Saddle diagnostics}] Can the interpretation of saddle structures be improved beyond the ``eyeball norm'' and baseline methods of calculating bonds based on covalent radii with approximate energy surfaces? Can optimization trajectories towards these points be improved?

\item[{Software designs}] Does the modernization of existing software unlock novel scientific algorithms for reducing the number of samples needed to find a saddle point?

\item[{Internal coordinates}] Can performance gains from complex internal coordinate generating / coordinate drive methods be beaten by better engineered Cartesian coordinate based local surrogate saddle searches?

\item[{Statistical performance modeling}] Average-cost benchmarks and point estimates of performance mask algorithmic reliability despite havnig poor statistical backing due to repeated data. Can more robust Bayesian methods lead to novel insights?

\item[{Wall-time efficiency}] GP acceleration has been known to be theoretically efficient for almost a decade, but no implementation has managed to show high reliability and wall-time gains in time, can this be alleviated by careful analysis of hyperparameters and software?
\end{description}
\subsection{Overview}
\label{sec:org0e9f149}
\begin{figure}[htbp]
\centering
\includegraphics[width=.9\linewidth]{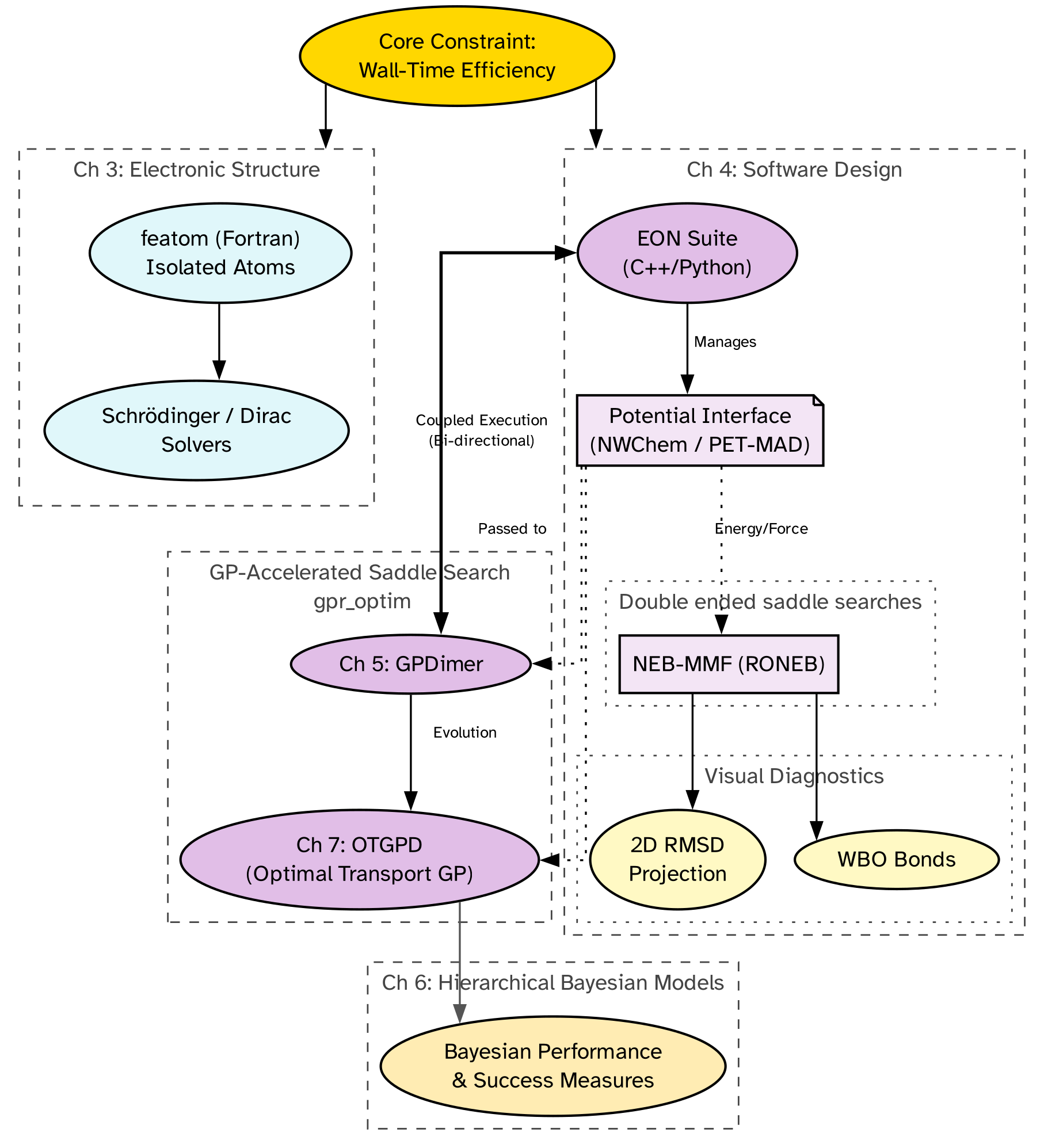}
\caption{\label{fig:thesis_over}The structural connectivity of this dissertation. Wall-time efficiency serves as the primary constraint driving innovation across three domains: relativistic electronic structure (cyan), chemical kinetics software (purple), and statistical diagnostics (amber). EON (Chapter 4) functions as the central orchestrator, managing interfaces to potential energy and force calculators and maintaining a tight, bidirectional coupling with the distinct GP-accelerated package developed for the \gls{gpd} in Chapter 5 and the \gls{otgpd} in 7. Quantification via the Bayesian performance metrics in Chapter 6 and visual diagnostics in Chapter 4 round out the software and methods developed.}
\end{figure}

The introductory Chapter \ref{sec:repasp} outlines basic preliminaries and provides pointers to the wider literature for the three fields under consideration. Across the programming languages and academic domains covered, wall-time efficiency functions as the core constraint, as illustrated in Figure \ref{fig:thesis_over}.

We begin with the fundamental physics in Chapter \ref{sec:fem}, which introduces \texttt{featom}, a high-order \gls{fem} solver for the Dirac and Schroedinger equations in Fortran. Unlike the frozen core approximations common to valence chemistry, \texttt{featom} facilitates heavy element relativistic calculations in the all-electron mean-field approximation. This tool stands distinct from the kinetic methods discussed in the remainder of the thesis.

Chapter \ref{sec:asd:eon} shifts focus to the EON software suite, a hybrid C++ and Python framework designed for long-timescale simulations. We detail the client-server architecture and the evolution of the software stack. Notably, we demonstrate how these scalable designs enable hitherto hard-to-implement workflows, including state-of-the-art nudged elastic band methods and a novel hybrid method, the \gls{roneb}.

These efforts evolve into an implementation of \gls{gp} acceleration for single-ended saddle searches in C++, exploring a large, though historically challenging, benchmark set. We extend the methodology to surface calculations and vet it for use within proximal reaction network explorations. Because exhaustive case studies fail to scale indefinitely, Chapter \ref{sec:brmsgp} explores performance modeling in high-throughput regimes. We employ Bayesian hierarchical methods in R and Stan to highlight computational bottlenecks, before directly tackling the data efficiency of \gls{gp} methods in Chapter \ref{sec:dataeff}.

These efforts culminate in Chapter \ref{sec:otgpd} with the \gls{otgpd}, a framework leveraging optimal transport distance metrics. This approach proves chemically intuitive and transferable across systems, demonstrating state-of-the-art wall-time efficiency. A brief summary and conclusions pave the way for the reader to engage with the primary articles in the appendix \footnote{on ArXiv these are not appended, but are freely available}.
\section{Theory}
\label{sec:repasp}
\epigraph{The derivation can be made to look slightly less juvenile by introducing an obscure notation at this point.}{P. Pechukas \\ Dynamics of Molecular Collisions, Part B}
\subsection{Minimum mode following}
\label{sec:theo:mmf}
\begin{wrapfigure}{l}{0.5\textwidth}
\centering
\includegraphics[width=0.48\textwidth]{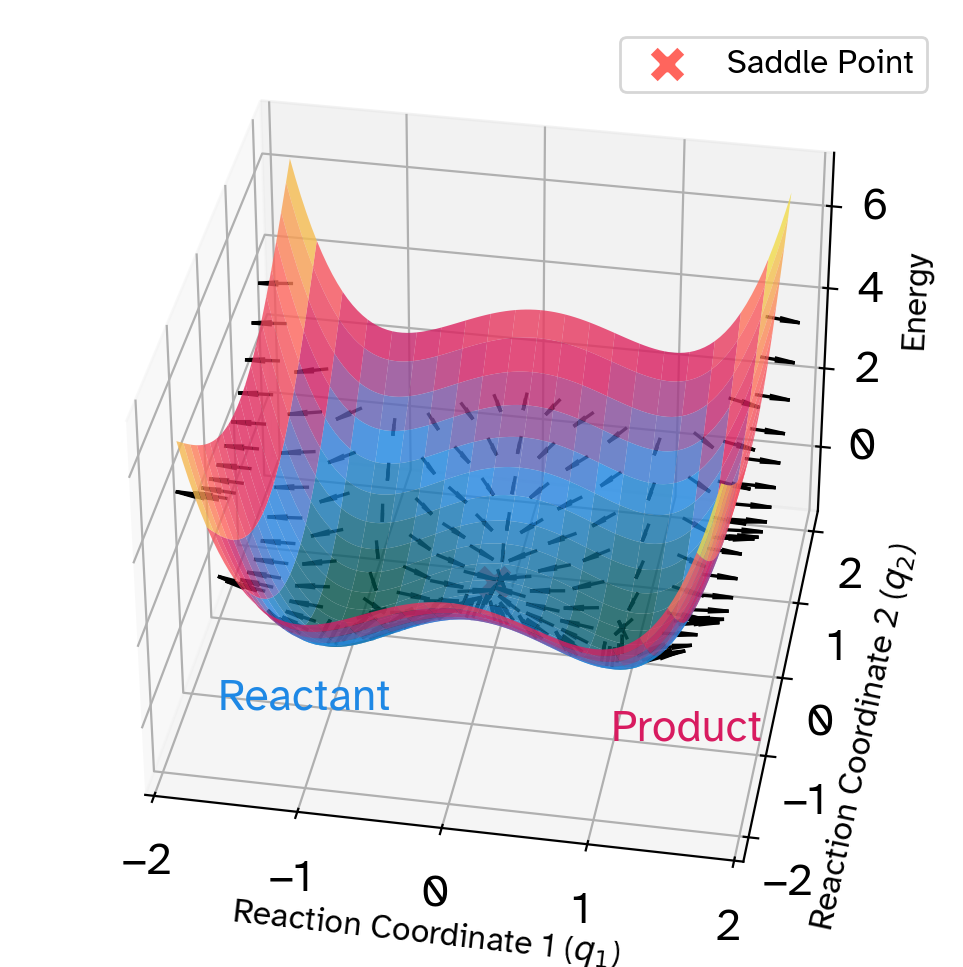}
\caption{\label{fig:dimer_mb}An illustration of the forces from the dimer method for locating a first-order saddle point on a two-dimensional \gls{pes}. The vector field represents the effective dimer force, a transformation of the true potential gradient (\(-\nabla V\)). This modified force guides an optimization uphill along the minimum-energy pathway while minimizing energy in orthogonal directions, enabling an efficient climb to the saddle point (red 'X').}
\end{wrapfigure}

To seek a mountain pass on a \gls{pes}, one need not survey the entire range \cite{bisgardMountainPassesSaddle2015}. The dimer method is a robust algorithm designed to locate first-order saddle points on a \gls{pes} without calculating the full Hessian matrix \cite{henkelmanDimerMethodFinding1999}. It belongs to the class of minimum mode following methods \cite{mousseauTravelingPotentialEnergy1998,munroDefectMigrationCrystalline1999,cerjanFindingTransitionStates1981}, where the search is guided by the eigenvector corresponding to the lowest eigenvalue of the Hessian. The core concept is to apply a pointwise transformation to the force, enabling an efficient climb uphill while simultaneously minimizing the energy in all orthogonal directions, as illustrated in Figure \ref{fig:dimer_mb}. This section recaps the standard formulation, similar to that implemented in software such as EON \cite{chillEONSoftwareLong2014}.

The ``dimer'' itself consists of two replicas (or images) of the system, \(\mathbf{R}_1\) and \(\mathbf{R}_2\), defined by their separation from a central point \(\mathbf{R}\) along a normalized orientation vector \(\hat{\mathbf{N}}\):
\begin{align}
  \mathbf{R}_1 &= \mathbf{R} - \frac{\Delta R}{2} \hat{\mathbf{N}} \\
  \mathbf{R}_2 &= \mathbf{R} + \frac{\Delta R}{2} \hat{\mathbf{N}}
\end{align}

The algorithm proceeds by iteratively alternating between rotation and translation steps.
\subsubsection{Rotational Step}
\label{sec:org23f25ce}
The primary goal of the rotational step involves aligning the dimer orientation vector, \(\hat{\mathbf{N}}\), with the minimum mode at the midpoint \(\mathbf{R}\). To this end, we rotate the dimer to minimize the total energy, \(E = E_1 + E_2\). With this, an effective rotational force, or torque \(\mathbf{F}_{rot}\) can be derived from the forces at the endpoints, \(\mathbf{F}_1\) and \(\mathbf{F}_2\), to estimate the change in local curvature without computing the Hessian. The curvature \(C\) along the dimer axis can be approximated using a finite difference:
\begin{equation}
  C(\hat{\mathbf{N}}) \approx \frac{(\mathbf{F}_2 - \mathbf{F}_1) \cdot \hat{\mathbf{N}}}{\Delta R}
  \label{eq:dimer_curvature_thesis}
\end{equation}

\begin{figure}[htbp]
\centering
\includegraphics[width=0.8\linewidth]{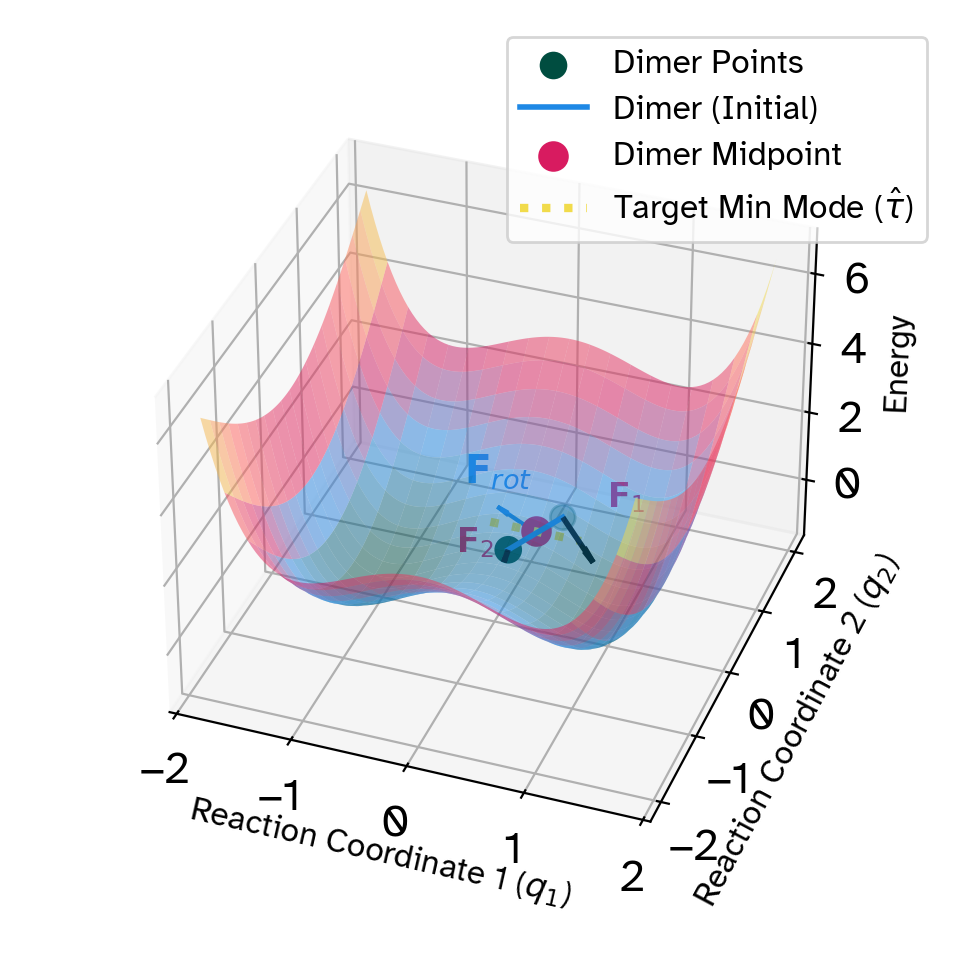}
\caption{\label{fig:dimer_mb_rot}The rotational step of the Dimer Method. The effective rotational force (\(\mathbf{F}_{\text{rot}}\)) is derived from the atomic forces (\(\mathbf{F}_1, \mathbf{F}_2\)) and applies a torque to the misaligned dimer. This torque drives the dimer's orientation to align with the minimum mode (\(\hat{\tau}\)), which is the prerequisite for the translational step.}
\end{figure}

This \(\mathbf{F}_{rot}\) defines the search direction perpendicular to \(\hat{\mathbf{N}}\), depicted in Figure \ref{fig:dimer_mb_rot}. We use an optimization algorithm, such as \gls{cg}, to determine the orientation at which \(\hat{\mathbf{N}}\) is aligned with the minimum mode. Practically, we recognize that the curvature within a rotational plane spanned by the current orientation \(\hat{\mathbf{N}}\) and the normalized torque direction approximates a sinusoidal curve

\begin{equation}
C(\phi) = A + B \cos(2\phi) + D \sin(2\phi)
\end{equation}

By evaluating the gradients at the current position we determine the angle which minimizes the curvature along a given search direction, and the dimer rotates to this orientation. This process occurs without the coordinates of the midpoint of the dimer changing and continues until the predicted rotation angle falls below a fixed convergence threshold. For standard calculations a rather loose tolerance of 5 degrees typically suffices.
\subsubsection{Translational Step}
\label{sec:orgb94c0bc}
Once the dimer is aligned, the translational step moves the midpoint \(\mathbf{R}\) towards the saddle point. This is guided by a modified force, \(\mathbf{F}_{\text{trans}}\), where the component of the true force parallel to the minimum mode is inverted. This transformation effectively turns the saddle point into a local minimum from the perspective of an optimization algorithm. The modified force is given by:
\begin{equation}
  \mathbf{F}_{\text{trans}}(\mathbf{R}) = \mathbf{F}(\mathbf{R}) - 2 (\mathbf{F}(\mathbf{R}) \cdot \hat{\mathbf{N}}) \hat{\mathbf{N}}
  \label{eq:dimer_trans_force_thesis}
\end{equation}

This effective force field, shown in Figure \ref{fig:dimer_mb}, ensures the system moves uphill towards the saddle. An optimizer, commonly the \gls{lbfgs} algorithm (Alg. \ref{alg:lbfgs}), takes a step using this modified force.

\begin{algorithm}
\caption{L-BFGS Two-Loop Recursion}
\label{alg:lbfgs}
\begin{algorithmic}[1]
\State \textbf{Given:} Current gradient $\nabla f_k$, history of $m$ updates $\{s_i, y_i\}_{i=k-m}^{k-1}$, where $s_i = x_{i+1} - x_i$ and $y_i = \nabla f_{i+1} - \nabla f_i$.
\State $q \gets \nabla f_k$
\For{$i = k-1, \dots, k-m$}
    \State $\rho_i \gets 1 / (y_i^T s_i)$
    \State $\alpha_i \gets \rho_i s_i^T q$
    \State $q \gets q - \alpha_i y_i$
\EndFor
\State $H_k^0 \gets \frac{s_{k-1}^T y_{k-1}}{y_{k-1}^T y_{k-1}} I$ \Comment{Initial Hessian approximation}
\State $z \gets H_k^0 q$
\For{$i = k-m, \dots, k-1$}
    \State $\beta \gets \rho_i y_i^T z$
    \State $z \gets z + s_i(\alpha_i - \beta)$
\EndFor
\State \Return Search direction $p_k = -z$
\end{algorithmic}
\end{algorithm}

The cycle of rotation and translation is repeated until the true force at the midpoint \(\mathbf{R}\) falls below a defined convergence threshold, indicating that a saddle point has been successfully located.

\subsection{The Nudged Elastic Band (NEB) Method}
\label{sec:theo:neb}
Minimum mode following methods prove useful for reaction network generation and general kinetic applications. However, when we distinguish specific reactant and product basins, the objective shifts. In such cases, we seek to determine the most probable paths connecting these fixed points in configuration space. A discrete approximation to a trajectory between the reactant and product can be formed through a ``chain of states''. Conceptually, this can be viewed as an extension of the dimer images, and the subsequent string is then connected through springs.

The \gls{neb} method \cite{jonssonNudgedElasticBand1998} is a form of double ended saddle search tecnique, used to determine the \gls{mep} between a known reactant and product state \cite{henkelmanImprovedTangentEstimate2000}. The \gls{mep} on the free energy surface represents the path of highest statistical weight. The method works by creating a discrete representation of the path, known as a ``band,'' which is a series of system configurations (or ``images'') connected by springs.

The total force on each image is a combination of the perpendicular component of the atomic force derived from the \gls{pes} and a fictitious spring force, which are projected to guide the system across the transition state.
\subsubsection{Path Discretization and Initialization}
\label{sec:org715c476}
The continuous reaction path is approximated by \(P+1\) images, denoted \(\left\{\mathbf{R}_0, \mathbf{R}_1, \dots, \mathbf{R}_P\right\}\), where \(\mathbf{R}_0\) and \(\mathbf{R}_P\) are the fixed reactant and product endpoints. The \(P-1\) intermediate images are movable. The initial path is typically generated by a linear interpolation between the endpoints:
\begin{equation}
  \mathbf{R}_i = \mathbf{R}_0 + i \frac{\mathbf{R}_P - \mathbf{R}_0}{P} \quad \text{for } i = 1, \dots, P-1
  \label{eq:neb_linear_interp}
\end{equation}
Alternatively, an initial path can be constructed from a series of provided configuration files, or by using a cheap surrogate potential like the \gls{idpp} \cite{smidstrupImprovedInitialGuess2014} which generates a more physically realistic path by minimizing the energy of a system described by a simple, classical potential. The total potential for an image \(\mathbf{R}_k\) is a sum of pair potentials \(V_{ij}\):

\begin{equation}
  V_{\text{IDPP}}(\mathbf{R}_k) = \sum_{i<j} V_{ij}(\mathbf{R}_k)
  \label{eq:idpp_total}
\end{equation}

Each pair potential is a harmonic spring that is switched on or off by a connectivity function \(C_{ij}\). The spring's equilibrium length, \(d_{ij}^{\text{ref}}\), is interpolated along the path.

\begin{equation}
  V_{ij}(\mathbf{R}_k) = \frac{1}{2} k_{ij} \left( d_{ij}(\mathbf{R}_k) - d_{ij}^{\text{ref}}(k) \right)^2 C_{ij}
  \label{eq:idpp_pair}
\end{equation}

The reference distance for a pair in image \(k\) is a linear interpolation between its distance in the reactant (\(k=0\)) and product (\(k=P\)).
\begin{equation}
  d_{ij}^{\text{ref}}(k) = \left(1 - \frac{k}{P}\right) d_{ij}(\mathbf{R}_0) + \frac{k}{P} d_{ij}(\mathbf{R}_P)
  \label{eq:idpp_ref_dist}
\end{equation}

The connectivity function \(C_{ij}\) ensures that a potential is only applied between atoms that are considered bonded in either the reactant or the product.
\begin{equation}
  C_{ij} =
  \begin{cases}
    1 & \text{if bond }(i,j)\text{ exists in } \mathbf{R}_0 \text{ or } \mathbf{R}_P \\
    0 & \text{otherwise}
  \end{cases}
  \label{eq:idpp_connectivity}
\end{equation}
The primary benefit of the \gls{idpp} method is that it often generates a more chemically reasonable initial path. Because the potential is based on the equilibrium bond connectivity of the endpoints, it avoids the unphysical atomic overlaps that often occur with linear interpolation. However, a significant limitation is that the method is not permutationally invariant, and has no explict handling for mass. The \gls{idpp} relies on a fixed, one-to-one mapping of atomic indices from reactant to product. If two identical atoms are swapped in the product coordinates, the \gls{idpp} method will generate a completely different and likely unphysical path, as it cannot recognize chemical equivalence.
\subsubsection{The NEB Force}
\label{sec:orga503e4e}
The core of the \gls{neb} method is the definition of the force acting on each intermediate image \(i\). This force is constructed to prevent the path from sliding downhill, colloquially called ``corner cutting'', and to maintain equal spacing of the images. It is composed of the perpendicular component of the true force and the parallel component of the spring force:
\begin{equation}
  \mathbf{F}_i^{\text{NEB}} = \mathbf{F}_i^{\perp} + \mathbf{F}_i^{\parallel, \text{spring}}
  \label{eq:neb_total_force}
\end{equation}
Here, \(\mathbf{F}_i^{\perp}\) is the component of the atomic forces on the \gls{pes}, \(\mathbf{F}_i^{\text{true}} = -\nabla V(\mathbf{R}_i)\), perpendicular to the path tangent. \(\mathbf{F}_i^{\parallel, \text{spring}}\) is the component of the spring force parallel to the path tangent.
\subsubsection{Tangent Vector Estimation}
\label{sec:theo:neb:tangent}
A crucial element is the estimation of the local tangent to the path, \(\hat{\tau}_i\), at each image. We include, in EON, several schemes:
\begin{description}
\item[{\textbf{Old Tangent}}] A simple central-difference vector \cite{jonssonNudgedElasticBand1998} \footnote{Implemented as a forward difference in EON \cite{chillEONSoftwareLong2014}}.
\begin{equation}
  \hat{\tau}_i = \text{normalize}(\mathbf{R}_{i+1} - \mathbf{R}_{i-1})
  \label{eq:neb_old_tangent}
\end{equation}
\item[{\textbf{Improved Tangent}}] A more robust method \cite{henkelmanImprovedTangentEstimate2000} that prevents kinks in the path. It selects the tangent based on the local energy landscape:
\begin{equation}
  \hat{\tau}_i =
  \begin{cases}
    \text{normalize}(\mathbf{R}_{i+1} - \mathbf{R}_i) & \text{if } V_{i+1} > V_i > V_{i-1} \\
    \text{normalize}(\mathbf{R}_i - \mathbf{R}_{i-1}) & \text{if } V_{i-1} > V_i > V_{i+1} \\
    \text{weighted average} & \text{otherwise}
  \end{cases}
  \label{eq:neb_improved_tangent}
\end{equation}
At \gls{extrema}, the tangent is a weighted average of the vectors to the neighboring images, giving preference to the vector on the higher energy side.
\end{description}
\subsubsection{Force Components}
\label{sec:theo:neb:forces}
With the tangent \(\hat{\tau}_i\) defined, the force components are calculated as:
\begin{description}
\item[{\textbf{Perpendicular Force}}] This component moves the image to minimize energy perpendicular to the path, relaxing it onto the \gls{mep}.
\begin{equation}
  \mathbf{F}_i^{\perp} = \mathbf{F}_i^{\text{true}} - (\mathbf{F}_i^{\text{true}} \cdot \hat{\tau}_i) \hat{\tau}_i
  \label{eq:neb_force_perp}
\end{equation}
\item[{\textbf{Parallel Spring Force}}] This component adjusts the position of the image along the path to ensure equal spacing.
\begin{equation}
  \mathbf{F}_i^{\parallel, \text{spring}} = k (|\mathbf{R}_{i+1} - \mathbf{R}_i| - |\mathbf{R}_i - \mathbf{R}_{i-1}|) \hat{\tau}_i
  \label{eq:neb_force_spring}
\end{equation}
where \(k\) defines the spring constant.
\end{description}
\subsubsection{Implementation Modalities and Improvements}
\label{sec:theo:neb:modalities}
Variations of the NEB method in EON also include:
\begin{description}
\item[{\Gls{cineb}}] To accurately locate the \gls{sp}, the spring force on the highest energy image (the ``climbing image,'' \(\mathbf{R}_{\text{climb}}\)) is removed, and the parallel component of its true force is inverted. This forces the image to move uphill along the path to converge exactly on the saddle point.
\begin{equation}
  \mathbf{F}_{\text{climb}} = \mathbf{F}_{\text{climb}}^{\text{true}} - 2 (\mathbf{F}_{\text{climb}}^{\text{true}} \cdot \hat{\tau}_{\text{climb}}) \hat{\tau}_{\text{climb}}
  \label{eq:neb_ci}
\end{equation}
\item[{Energy-Weighted Springs}] In a standard \gls{neb} calculation, a uniform spring constant connects all images. A more adaptive approach, known as the energy-weighted spring method \cite{asgeirssonNudgedElasticBand2021}, dynamically adjusts the spring constants along the path. This method applies stronger (stiffer) springs in high-energy regions, typically near the transition state, and weaker springs in lower-energy regions. This procedure effectively concentrates images around the saddle point, improving the resolution of the reaction barrier without necessitating an increase in the total number of images.
\end{description}

The spring constant, \(k_i\), for the segment connecting image \(i-1\) and image \(i\), is determined by a linear interpolation between a defined maximum spring constant, \(k_{\text{max}}\), and a minimum, \(k_{\text{min}}\). The interpolation depends on the energy of that segment.

We define a reference energy, \(E_{\text{ref}}\), as the lower of the two endpoint energies (reactant or product). We also identify the maximum energy found along the current path, \(E_{\text{max}}\). For each spring segment between images \(i-1\) and \(i\), we define an effective energy, \(E_i\), as the higher of the two adjacent image energies:

\begin{equation}
  E_i = \max\left(V(\mathbf{R}_{i-1}), V(\mathbf{R}_i)\right)
\end{equation}

If this effective energy \(E_i\) exceeds the reference energy \(E_{\text{ref}}\), we calculate a dimensionless weighting factor, \(\alpha_i\):

\begin{equation}
  \alpha_i = \frac{E_{\text{max}} - E_i}{E_{\text{max}} - E_{\text{ref}}}
  \label{eq:alpha_weight}
\end{equation}

This factor, \(\alpha_i\), ranges from 0 (when \(E_i = E_{\text{max}}\)) to 1 (when \(E_i = E_{\text{ref}}\)). This factor interpolates the spring constant \(k_i\) between \(k_{\text{max}}\) and \(k_{\text{min}}\). If the segment's energy \(E_i\) does not exceed \(E_{\text{ref}}\), the spring constant defaults to the minimum value, \(k_{\text{min}}\). This leads to:

\begin{equation}
  k_i =
  \begin{cases}
    (1 - \alpha_i)k_{\text{max}} + \alpha_i k_{\text{min}}, & \text{if } E_i > E_{\text{ref}} \\
    k_{\text{min}}, & \text{otherwise}
  \end{cases}
\end{equation}

These dynamically adjusted spring constants are then used to calculate the parallel component of the spring force, \(\mathbf{F}_i^{s, \parallel}\), acting on each image \(i\). This force depends on the tension from the springs on its left (\(k_i\)) and right (\(k_{i+1}\)):

\begin{equation}
  \mathbf{F}_i^{s, \parallel} = \left( k_{i+1} |\mathbf{R}_{i+1} - \mathbf{R}_i| - k_i |\mathbf{R}_i - \mathbf{R}_{i-1}| \right) \hat{\boldsymbol{\tau}}_i
\end{equation}

Here, \(\hat{\boldsymbol{\tau}}_i\) represents the normalized tangent vector at image \(i\). This formulation ensures that the net effect of the springs pulls images toward the saddle point, refining the path's most critical region. It should be noted that the implicit assumption here is of a single maxima along the path, and the formulation as implemented and presented does not have special handling for spanning multiple basins. Numerical jitter may be used to handle degeneracy from the denominator in Eq. \ref{eq:alpha_weight}.
\subsubsection{Optimization and Path Analysis}
\label{sec:theo:neb:optpath}
The set of movable images is relaxed using a standard optimization algorithm (e.g., \gls{lbfgs}), which iteratively updates the image positions based on the calculated \gls{neb} forces until a convergence criterion is met. In EON the most common criteria is on the largest force component on any atom.

After convergence, a Hermite polynomial interpolation between the final image energies and forces estimates the location and height of the energy barrier, providing a more refined value than the energy of the highest image alone.
\subsection{Acceleration strategies}
\label{sec:org794c198}
The core methodology described thus-far has formed an integral part of computational chemistry calculations for several decades. In order to obtain results on larger systems or longer time-scales, these must be accelerated. Within this thesis we consider two primary approaches to this acceleration:
\begin{description}
\item[{Software design}] As an engineering endeavor, generating efficient machine code leads to faster solutions and thus unlocks larger systems. Under this broad remit falls parallelization strategies, caching, handling large data, better use of specialized hardware like \gls{gpu} computations etc.
\item[{Algorithmic improvements}] The total computational time depends on the algorithms employed, which in turn demand energy and atomic force evaluations from high-level theory. Consequently, algorithms that minimize the number of calculations required to achieve a specific solution accuracy constitute the primary metric for optimization. Utilizing cheaper approximate surfaces for portions of the algorithm represents a significant focus of current research.
\end{description}

An ideal surrogate demands ease of computation, autonomy from manual intervention, and ``transparent'' progression without the manual construction of training sets. We therefore mandate that the method:
\begin{description}
\item[{Learns on the fly}] Efficiently learning from low amounts of often correlated data.
\item[{Encode physics}] Providing, at minimum, a functional form that aids analysis.
\item[{Trains quickly}] Enabling the efficient incorporation of new points into the model.
\end{description}

We explicitly define the scope: strictly avoiding the goal of matching the underlying potential energy surface globally. These surrogates serve only to guide algorithms toward specific points in the solution space; thus, absolute errors in energy and forces hold little value compared to local gradient accuracy. Similarly, we exclude general energy surface construction, the basis of machine learning interatomic potentials. We therefore focus on regression techniques \cite{jamesIntroductionStatisticalLearning2013}. Computing atomic descriptors are based on cartesian coordinates \cite{willattAtomdensityRepresentationsMachine2019} requires both user intervention for parameter selection, and increases the computational cost of on-the-fly learning. Consequently, an ideal surrogate for our consideration operates directly on raw Cartesian coordinates. We briefly evaluate the candidates:
\begin{description}
\item[{Linear regression}] The oldest fitting technique, which may be competitive with enough data over a small region of space, but only with descriptors to account for non-linearities, which are often incomplete by construction \cite{pozdnyakovIncompletenessAtomicStructure2020}.
\item[{Neural networks}] Deep neural networks \cite{goodfellowDeepLearning2016} interpolate arbitrary systems well and can encode specific qualities through filters (e.g., convolutional architectures). However, they tend to overfit in the low-data regime \cite{murphyProbabilisticMachineLearning2023}, and require careful selection of both data and long training times for the hyperparameters. The resulting models have no closed form analytical solution and rely on being able to backpropagate gradients \cite{berlandAutomaticDifferentiation,paszkeAutomaticDifferentiationPyTorch}, which also forms a computational bottleneck.
\item[{Gaussian processes}] Often understood as limiting cases of spline models \cite{calderKrigingVariogramModels2009}, neural networks \cite{nealPriorsInfiniteNetworks1996,yangWideFeedforwardRecurrent2019}, quadrature rules \cite{hennigProbabilisticNumericsUncertainty2015}, stochastic partial differential equations \cite{sarkkaLinearOperatorsStochastic2011}, finite realizations of these methods form multivariate normal distributions \cite{gramacySurrogatesGaussianProcess2020} which are analytically and computationally compact. The connection to standard statistics enables a rich set of uncertainity quantification, and the ability to constrain the function spaces based on distance measures between data points makes for more interpretable surrogates.
\end{description}

With this in mind, we move on with further contextualizing the \gls{gp} for molecular systems, though we note that descriptor based methods have also been applied for activated searches \cite{sanscartierEvaluatingApproachesOnthefly2023}.
\subsection{Gaussian Process Regression}
\label{sec:theo:gpr}
We begin by positing that the unknown \gls{pes}, a function \(f(\mathbf{x})\), represents a single realization from a \gls{gp}. A \gls{gp} defines a probability distribution over a space of functions. Any finite collection of function values drawn from this process follows a joint \gls{mvn} distribution. It is worth noting that such an assumption can be justified rather rigorously for molecular potential energy surfaces with pronounced global minimum as demonstrated by \cite{madsenApproximateFactorizationMolecular1997} in terms of an asymptotic study of vibrational degrees of freedom, which leads to the potential energy surface on a random coordinate frame manifesting as a sum of many contributions, which in turn through the Central limit theorem \cite{jamesIntroductionStatisticalLearning2013} leads to an \gls{mvn} perspective, equivalent to a perturbative approach for small molecules.

Such theoretical modeling of molecular potential energy surfaces \cite{madsenApproximateFactorizationMolecular1997} rests upon three postulates: the potential constants represent averages over all unitary transformations; the topology possesses a distinct global minimum; and the number of vibrational modes approaches the asymptotic limit. While these constraints, in particular the topological requirement strictly preclude the multiple-minima landscapes inherent to the reactive systems under investigation here, the derivation nevertheless offers robust physical support for approximating the energy surface locally as a multivariate Gaussian.

In practice, one never works with the infinite-dimensional function directly. Instead, we select a finite set of \textbf{M} input configurations, \(\mathbf{X} = \{\mathbf{x}_1, ..., \mathbf{x}_M\}\). The GP specifies that the corresponding vector of function outputs, \(\mathbf{f} = [f(\mathbf{x}_1), ..., f(\mathbf{x}_M)]^T\), constitutes a single draw from an \textbf{M}-dimensional \gls{mvn}. This finite vector becomes our computational representation of the underlying function.

\begin{equation}
\mathbf{o}(\mathbf{x}) = \begin{pmatrix} E(\mathbf{x}) \\ \mathbf{F}(\mathbf{x}) \end{pmatrix} \in \mathbb{R}^{3N+1}
\end{equation}

When we evaluate the \gls{pes} at a set of \textbf{M} distinct configurations,  \(\mathbf{X} = \{\mathbf{x}_1, ..., \mathbf{x}_M\}\), the \gls{gp} framework posits that the collection of all corresponding observation vectors follows a single, large \gls{mvn} distribution:

\begin{equation}
p(\mathbf{o}_1, \mathbf{o}_2, \dots, \mathbf{o}_M) = \mathcal{N}(\mathbf{0}, \mathbf{K})
\end{equation}

The full covariance matrix \(\mathbf{K}\) is built from a kernel function \(k(x_i, x_j)\) that defines the similarity between any two configurations \cite{gramacySurrogatesGaussianProcess2020,rasmussenGaussianProcessesMachine2006}. Because the observation vector \(o\) contains both a scalar (energy) and a vector (forces), the kernel itself produces a block covariance matrix for any two configurations:

\begin{equation}
\mathbf{o}(\mathbf{r}) = \begin{pmatrix} E(\mathbf{r}) \\ \mathbf{F}(\mathbf{r}) \end{pmatrix} = \begin{pmatrix} E(\mathbf{r}) & F_{x_1}(\mathbf{r}) & F_{y_1}(\mathbf{r}) & F_{z_1}(\mathbf{r}) & F_{x_2}(\mathbf{r}) & \ldots & F_{z_N}(\mathbf{r}) \end{pmatrix}^{T} \in \mathbb{R}^{3N+1}
\end{equation}

where:
\begin{itemize}
\item \(E(\mathbf{r})\) is the potential energy at atomic configuration
\(\mathbf{r}\).
\item \(\mathbf{F}(\mathbf{r})\) is the \(3N-\) dimensional vector of atomic forces,
where \(F_{x_i}\), \(F_{y_i}\), and \(F_{z_i}\) are the \(x\), \(y\), and \(z\) components
of the force on atom \(i\), respectively. \(N\) is the number of atoms.
\end{itemize}

Each draw from the \gls{gp} forms a concrete \gls{mvn} fully specified by a mean vector \(\mathbf{m}\) and a covariance matrix \(\mathbf{K}\). We construct these from a mean function \(m(\mathbf{x})\), assumed to be zero throughout this work, and a covariance function or kernel \(k(\mathbf{x}, \mathbf{x}')\) which determines the covariance between the energy and forces at different atomic configurations \(\mathbf{x}\) and \(\mathbf{x}'\). The choice of kernel encodes prior assumptions about the functional form of the \gls{pes}. For systems with strong repulsive forces at short distances, the most commonly used kernel, the infinitely differentiable \gls{se} kernel can be too restrictive \cite{koistinenMinimumModeSaddle2020}, which is only partially resolved by choosing different kernels like the Matern. The stationarity of the \gls{se} kernel, i.e. the assumption of uniform fluctuations across the domain makes it difficult to handle the steep gradients of repulsive walls in Cartesian space. By transforming the input space through an inverse distance metric, we effectively ``precondition'' the energy surface by homogenizing the effective length scale of the landscape, which enables the \gls{gp} to resolve high gradient regions without sacrificing resolution in energetic wells.  We therefore employ a \gls{se} kernel based on an inverse distance metric \cite{koistinenMinimumModeSaddle2020,goswamiEfficientImplementationGaussian2025a} since it provides a strong physical prior, leveling out the sharp increase in repulsive force when pairs of atoms get too close.

For the chosen design sites \(\mathbf{X}\), the components of the mean vector and covariance matrix are:

\begin{equation}
  (\mathbf{m})_i = m(\mathbf{x}_i) \quad \text{and} \quad (\mathbf{K})_{ij} = k(\mathbf{x}_i, \mathbf{x}_j)
\end{equation}

In other words, we model the joint distribution of the energy and forces at multiple configurations \(\{\mathbf{r}_1, \mathbf{r}_2, \dots, \mathbf{r}_M\}\) as a multivariate Gaussian distribution Eq. \ref{eq:mvn_pred} and we now identify
the elements of \(\mathbf{K}\) are given by a kernel function \(k(\mathbf{r}_i, \mathbf{r}_j)\) that measures the similarity between configurations \(\mathbf{r}_i\) and \(\mathbf{r}_j\). This kernel operates on the input space of atomic geometries (\(\mathbb{R}^{N \times 3}\)) and outputs the covariance between the combined energy and force vectors (\(\mathbb{R}^{3N+1}\)) at those geometries.

Including force derivatives improves performance for models with limited samples \cite{solakDerivativeObservationsGaussian2002}. The covariance between the combined energy and force vectors at two different geometries \(\mathbf{r}\) and \(\mathbf{r}'\) is given by:

\begin{equation}
\text{Cov}(\mathbf{o}(\mathbf{r}), \mathbf{o}(\mathbf{r}')) = \begin{pmatrix}
k_{EE}(\mathbf{r}, \mathbf{r}') & \mathbf{k}_{EF}(\mathbf{r}, \mathbf{r}') \\
\mathbf{k}_{FE}(\mathbf{r}, \mathbf{r}') & \mathbf{K}_{FF}(\mathbf{r}, \mathbf{r}')
\end{pmatrix}
\end{equation}
where
\begin{itemize}
\item \(k_{EE}(\mathbf{r}, \mathbf{r}') \in \mathbb{R}\) is the covariance between the energies.
\item \(\mathbf{k}_{EF}(\mathbf{r}, \mathbf{r}') \in \mathbb{R}^{1 \times 3N}\) is the covariance between the energy at \(\mathbf{r}\) and the forces at \(\mathbf{r}'\).
\item \(\mathbf{k}_{FE}(\mathbf{r}, \mathbf{r}') = \mathbf{k}_{EF}(\mathbf{r}', \mathbf{r})^\top \in \mathbb{R}^{3N \times 1}\) is the covariance between the forces at \(\mathbf{r}\) and the energy at \(\mathbf{r}'\).
\item \(\mathbf{K}_{FF}(\mathbf{r}, \mathbf{r}') \in \mathbb{R}^{3N \times 3N}\) is the covariance matrix between the forces.
\end{itemize}

For energy-energy covariance, this kernel takes the form:
\begin{equation}
\label{eq:idist_kernel_ee}
k(\mathbf{x}, \mathbf{x}') = \sigma_c^2 + \sigma_f^2 \exp\left(-\frac{1}{2} \sum_{i} \sum_{\substack{j > i}} \left(\frac{1/r_{ij}(\mathbf{x}) - 1/r_{ij}(\mathbf{x}')}{l_{\phi(i,j)}}\right)^2 \right)
\end{equation}
where \(\sigma_f^2\) is the signal variance, \(\sigma_c^2\) is a constant offset, and \(l_{\phi(i,j)}\) is the characteristic length scale for a specific pair type of atoms \(\phi(i,j)\) \footnote{only pairs are considered, a single atom type, O, for instance will not have an O-O term}.

The force-related blocks of the covariance matrix (\(k_{EF}\), \(k_{FE}\), \(K_{FF}\)) are derived by differentiating the energy-energy kernel with respect to the atomic coordinates, leveraging the relationship \(\mathbf{F} = -\nabla E\). This requires the first and second partial derivatives of the squared distance measure, \(\mathcal{D}_{1/r}^2 = \sum \left( \frac{\Delta(1/r)}{l} \right)^2\):
\begin{equation}
\frac{\partial \mathcal{D}_{1/r}^2(\mathbf{x}, \mathbf{x}')}{\partial x_{i,d}} = \sum_{j \neq i} \left[ \frac{-2(x_{i,d} - x_{j,d})}{l_{\phi(i,j)}^2 r_{ij}^3(\mathbf{x})} \left( \frac{1}{r_{ij}(\mathbf{x})} - \frac{1}{r_{ij}(\mathbf{x}')} \right) \right]
\end{equation}
\begin{equation}
\frac{\partial^2 \mathcal{D}_{1/r}^2(\mathbf{x}, \mathbf{x}')}{\partial x_{i_1,d_1} \partial x'_{i_2,d_2}} =
\begin{cases}
\frac{2(x_{i_1,d_1} - x_{i_2,d_1})(x'_{i_1,d_2} - x'_{i_2,d_2})}{l_{\phi(i_1,i_2)}^2 r_{i_1,i_2}^3(\mathbf{x}) r_{i_1,i_2}^3(\mathbf{x}')}, & \text{if } i_1 \neq i_2 \\
\sum_{j \neq i} \frac{-2(x_{i,d_1} - x_{j,d_1})(x'_{i,d_2} - x'_{j,d_2})}{l_{\phi(i,j)}^2 r_{ij}^3(\mathbf{x}) r_{ij}^3(\mathbf{x}')}, & \text{if } i_1 = i_2 = i
\end{cases}
\end{equation}

The kernel hyperparameters \(\boldsymbol{\theta} = \{\sigma_f^2, \sigma_c^2, l_\psi, \dots\}\) are not known \emph{a-priori}. Most commonly, the hyperparameters derive from the \gls{mgp}. Practically, optimizing takes place in log-space (\(\eta = \log \theta\)) since this provides better numerical scaling through two benefits.

Firstly, it implicitly maintains the strict positivity constraints required by variances and length scales. The mapping \(\theta = \exp(\eta)\) ensures that the optimization algorithm, which typically operates in unconstrained Euclidean space, yields strictly positive physical parameters without requiring barrier functions or Lagrange multipliers.

Secondly, logarithmic scaling preconditions the objective function landscape. In chemical physics applications, hyperparameters frequently vary by orders of magnitude. A gradient step size appropriate for a large signal variance \(\sigma_f^2\) could prove disastrously large for a short-range length scale \(l_\psi\). Working in the log-space equilibrates these sensitivities, as a constant step size in \(\boldsymbol{\eta}\) corresponds to a proportional change in \(\boldsymbol{\theta}\) rather than an absolute one. This effectively balances the gradient contributions across disparate dimensions and accelerates convergence.

From a practical standpoint, the gradient computation for the transformed parameters follows the chain rule:

\begin{equation}
\frac{\partial \mathcal{L}}{\partial \eta_i} = \frac{\partial \mathcal{L}}{\partial \theta_i} \frac{\partial \theta_i}{\partial \eta_i} = \frac{\partial \mathcal{L}}{\partial \theta_i} \theta_i
\end{equation}

This scaling factor \(\theta_i\) naturally dampens the gradient for small parameters and amplifies it for large ones, further stabilizing the descent. Thus the kernel hyperparameters are learned from the training data by maximizing the logarithm of the \gls{mll}:
\begin{equation}
\log p(\mathbf{y} \mid \mathbf{X}, \boldsymbol{\theta}) = -\frac{1}{2} \mathbf{y}^T \mathbf{K}^{-1} \mathbf{y} - \frac{1}{2} \log \det(\mathbf{K}) - \frac{M(3N+1)}{2} \log(2\pi)
\end{equation}
This optimization is performed using gradient-based methods, which require the partial derivatives of the kernel with respect to each hyperparameter. For the length scales \(l_\psi\), this involves the following derivatives of the distance measure:
\begin{equation}
\frac{\partial \mathcal{D}_{1/r}^2(\mathbf{x}, \mathbf{x}')}{\partial l_\psi} = \sum_{\substack{i, j>i \\ \phi(i,j) = \psi}} \frac{-2\left(\frac{1}{r_{ij}(\mathbf{x})} - \frac{1}{r_{ij}(\mathbf{x}')}\right)^2}{l_\psi^3}
\end{equation}
\begin{equation}
\frac{\partial^2 \mathcal{D}_{1/r}^2(\mathbf{x}, \mathbf{x}')}{\partial x_{i,d} \partial l_\psi} = \sum_{\substack{j \neq i \\ \phi(i,j) = \psi}} \left[ \frac{4(x_{i,d} - x_{j,d})}{l_\psi^3 r_{ij}^3(\mathbf{x})} \left( \frac{1}{r_{ij}(\mathbf{x})} - \frac{1}{r_{ij}(\mathbf{x}')} \right) \right]
\end{equation}
\begin{equation}
\frac{\partial^3 \mathcal{D}_{1/r}^2(\mathbf{x}, \mathbf{x}')}{\partial x_{i_1,d_1} \partial x'_{i_2,d_2} \partial l_\psi} =
\begin{cases}
0, & \text{if } i_1 \neq i_2 \wedge \phi(i_1,i_2) \neq \psi \\
-\frac{4(x_{i_1,d_1} - x_{i_2,d_1})(x'_{i_1,d_2} - x'_{i_2,d_2})}{l_\psi^3 r_{i_1,i_2}^3(\mathbf{x}) r_{i_1,i_2}^3(\mathbf{x}')}, & \text{if } i_1 \neq i_2 \wedge \phi(i_1,i_2) = \psi \\
\sum_{\substack{j \neq i \\ \phi(i,j) = \psi}} \frac{4(x_{i,d_1} - x_{j,d_1})(x'_{i,d_2} - x'_{j,d_2})}{l_\psi^3 r_{ij}^3(\mathbf{x}) r_{ij}^3(\mathbf{x}')}, & \text{if } i_1 = i_2 = i
\end{cases}
\end{equation}

Once the model is trained, it can be used to predict the energy and forces at new configurations \(\mathbf{X}_*\). The posterior predictive mean gives the best estimate for the surface at the new locations:
\begin{equation}
\bar{\mathbf{f}}_* = \mathbf{K}_{*y} (\mathbf{K}_{yy} + \sigma_n^2 \mathbf{I})^{-1} \mathbf{y}
\end{equation}
where \(\mathbf{y}\) is the vector of training observations, \(\mathbf{K}_{yy}\) is the covariance of the training data, \(\mathbf{K}_{*y}\) is the covariance between the test and training points, and the \(\sigma_n^2 \mathbf{I}\) term is a regularization or noise term that ensures numerical stability. The posterior predictive covariance quantifies the model uncertainty:
\begin{equation}
\text{cov}(\mathbf{f}_*) = \mathbf{K}_{**} - \mathbf{K}_{*y} (\mathbf{K}_{yy} + \sigma_n^2 \mathbf{I})^{-1} \mathbf{K}_{y*}
\end{equation}
From here, adding atomic features \cite{willattAtomdensityRepresentationsMachine2019,musilPhysicsInspiredStructuralRepresentations2021} leads to the smoothed overlap of atomic positions class of models \cite{caroOptimizingManybodyAtomic2019,bartokGaussianApproximationPotential2010a} and other machine learned interatomic potentials \cite{noeMachineLearningMolecular2020}. Pivoting slightly towards active, or reinforcement learning \cite{suttonReinforcementLearningIntroduction2018}, we utilize the posterior mean to guide the search for stationary points on a series of approximate \gls{pes}. The posterior covariance, while a formal measure of uncertainty, may be unreliable in an iterative refitting scheme (Section \ref{sec:dataeff:varacc}) and we therefore disregard it here. Essentially, because the global hyperparameters are re-optimized with each new data point, a local reduction in variance does not reliably indicate an improvement in the model's true accuracy.
\subsubsection{Scaling in Time and Storage}
\label{sec:scal:timestor}
The computational cost of \gls{gpr} \cite{kochenderferAlgorithmsOptimization} is dominated by the
inversion of the covariance matrix \(\mathbf{K}\), and the determination of the
hyperparameters for conditioning on the data as each new point is acquired. More precisely, the costs of a \gls{gp} involve:

\begin{description}
\item[{Storage}] The covariance matrix \(\mathbf{K}\) carries dimensions \((M(3N+1)) \times (M(3N+1))\), where \(M\) equals the number of training configurations. Storage therefore scales as \(O(M^2 N^2)\).
\item[{Fitting Time}] Inverting \(\mathbf{K}\) requires \(O(M^3 N^3)\). This cubic scaling in both \(M\) and \(N\) renders standard GPR expensive for large systems and datasets.
\item[{Inference time}] Once the kernel components are determined, prediction involves a single kernel vector between the new point and training points, so this scales linearly with the number of training points.
\end{description}

The key advantage of a \gls{gp} approach is that we can constrain the functional form of the posterior, determined by the inverse distance kernel models physical constraints, while the model remains relatively cheap to re-fit and predict with. Physical constraints like smoothness may also be enforced, and data agumentation can encode non-linear prior assumpsions.
\subsection{\gls{gp} as an accelerator}
\label{sec:theo:gpd}
\begin{wrapfigure}{l}{0.5\textwidth}
\centering
\includegraphics[width=0.48\textwidth]{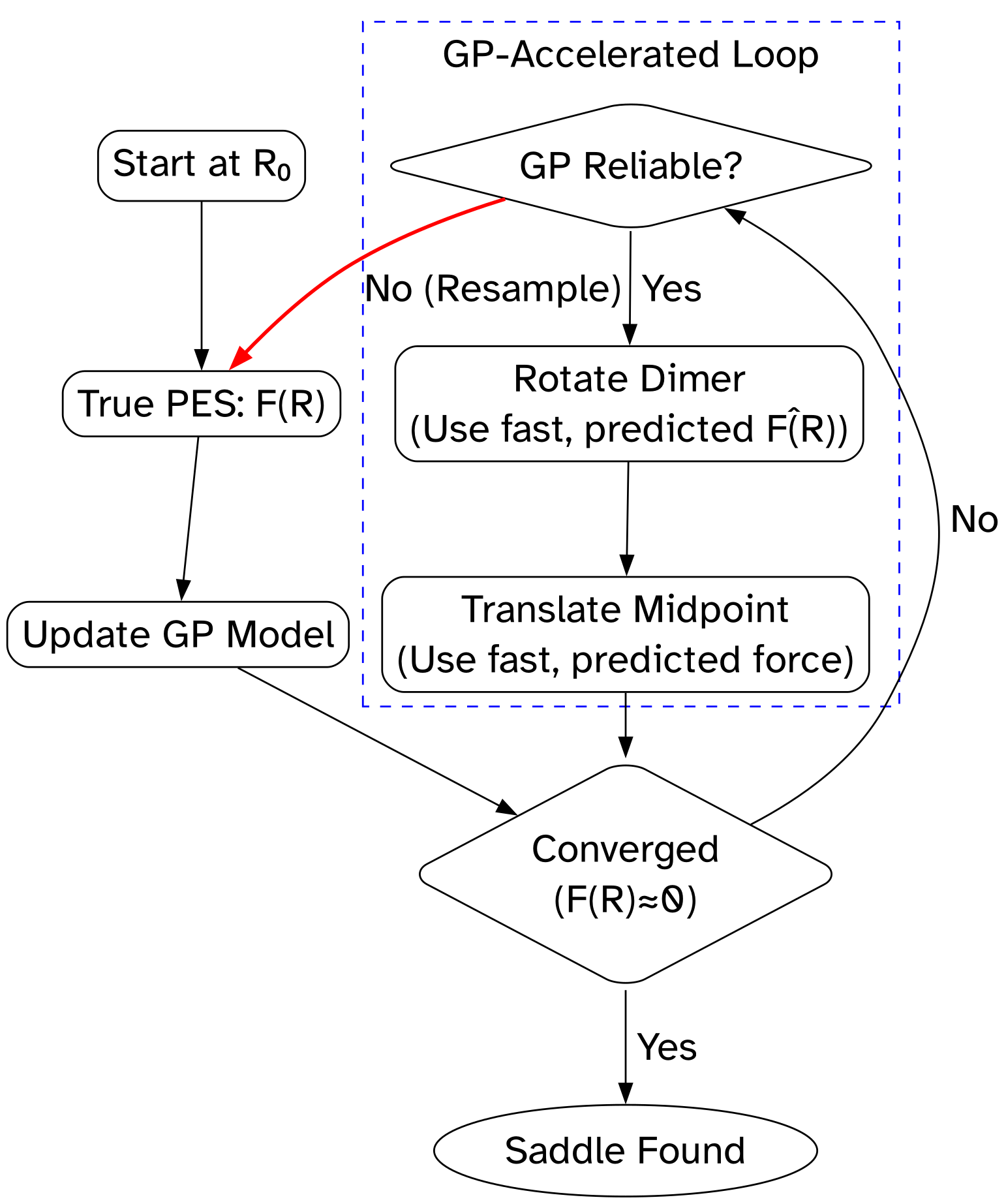}
\caption{\label{fig:gp_dimer_base_thesis}The \gls{gpd} method.}
\end{wrapfigure}

A simplified flowchart of the logic is presented in Figure \ref{fig:gp_dimer_base_thesis}.

The \gls{gpd} method constructs a local surrogate model for the \gls{pes} through a finite, targeted set of samples. The approach functions like a reinforcement learning agent \cite{suttonReinforcementLearningIntroduction2018}. At each step of a search, the model decides whether to trust its current surrogate \gls{pes} or to query the ``true'' \gls{pes} via an expensive quantum chemical calculation. This ``on-the-fly'' model building refines a highly local and task-specific \gls{pes} with a minimal number of data points \footnote{we will demonstrate results typically around 30 samples \cite{goswamiEfficientImplementationGaussian2025a,goswamiAdaptivePruningIncreased2025}}.

As breifly covered before, this strategy differs fundamentally from that of a general-purpose \gls{mlip}. The principal objective of an \gls{mlip}: create a single, global, transferable \gls{pes}, which demands vast datasets that may contain millions of configurations. To achieve transferability across diverse chemical environments, the \gls{mlip} architecture must respect the system's physical symmetries. This requirement demands outputs that remain invariant to translation, rotation, and permutation of identical atoms, often achieved through atom-centered symmetry functions \cite{liAtomcenteredMachinelearningForce2023} or specialized equivariant neural-network layers \cite{musilPhysicsInspiredStructuralRepresentations2021,bigiWignerKernelsBodyordered2024}. Although such models can deliver high energy accuracy, force accuracy generally remains worse and improves more slowly with training-set size, with errors often exceeding 0.1 eV/ \AA{} \cite{zuoPerformanceCostAssessment2020,khanKernelBasedQuantum2023}. Because forces drive the methods in this thesis, most \glspl{mlip} offer limited value here. Foundational models like \cite{mazitovPETMADUniversalInteratomic2025} with easy finetuning support may still prove useful.

By contrast, the \gls{gpd} method pursues a different objective, that is, high fidelity within a localized region of the \gls{pes} tied to a specific process (e.g., a single saddle search). This focused scope obviates large-scale sampling. We develop models that operate effectively on raw Cartesian coordinates, bypassing the heavy data and architectural requirements of global \glspl{mlip}.
\subsection{Conclusions}
\label{sec:theo:conc}
This chapter has delineated the theoretical and algorithmic framework necessary for navigating high-dimensional potential energy surfaces. We detailed the topological search methods based on the number of known minima, specifically the single ended Dimer and double ended Nudged Elastic Band algorithms which serve as the primary vehicles for locating first-order saddle points and minimum energy pathways. While these techniques provide robust convergence properties, in practice, their reliance on expensive \emph{ab-initio} force evaluations necessitates an acceleration strategy capable of operating within the strict data constraints of on-the-fly exploration.

Our analysis identified Gaussian Process Regression as the optimal surrogate for this regime. By combining the asymptotic derivation of local energy landscapes with a physically motivated inverse-distance kernel, we show how one constructs a model that balances flexibility with the regularization required by repulsive atomic cores. Crucially, this approach bypasses the need for the extensive datasets required by global machine learning potentials, favoring instead a rigorous local approximation that respects the underlying physics of the reaction channel. With these mathematical and computational foundations established, we later proceed to their practical implementation and application in the study of complex reactive systems.

However, a surrogate model acts merely as an efficient interpolator; it possesses no intrinsic knowledge of the quantum mechanics. The fidelity of any reaction path search depends ultimately on the accuracy of the underlying electronic structure calculations. Furthermore, while we discuss methods to minimize the number of evaluations, the computational cost per evaluation remains a critical bottleneck, particularly for heavy-element systems where relativistic effects dominate. Therefore, before applying these exploration algorithms to complex reactive systems, we cover some eletronic structure theory for chemical systems and foray into a rigorous and efficient foundation for computing the potential energy surface for single atom systems. The subsequent chapter addresses this challenge, introducing a high-order finite element approach to the self-consistent field problem that balances relativistic accuracy with numerical efficiency.
\section{Electronic structure calculations}
\label{sec:fem}
\epigraph{the chemical difference between silver and gold may mainly be a relativistic effect.}{P. Pyykkö \\ Chemistry – A European Journal}

\begin{quote}
This chapter is based on \fullcite{certikHighorderFiniteElement2023a}
\end{quote}

To explore the \gls{pes} with the methods described previously, we must compute energies and atomic forces for specific nuclear configurations. This task reduces to solving the many-electron Schrödinger equation in a form that scales to useful systems. We proceed by choice of representation: we begin with a mean field to obtain a tractable one-body problem; then we refine the physics or numerics as necessary.

After a brief introduction to the general electronic structure methods for chemical systems, we focus on a state of the art solver, \texttt{featom} which solves for isolated single atom systems, the Dirac and Schrödinger equations.
\subsection{Mean-field quantum chemistry}
\label{sec:fem:mfqc}
The Hartree approximation represents the simplest mean field theory approximation. It assumes that each electron moves in the Coulomb field generated by nuclei and the spherically averaged density of all other electrons \cite{hartreeSelfconsistentFieldExchange1948}. This local, multiplicative field removes two-body integrals and invites efficient solvers. However, it leaves self-interaction uncorrected. The Hartree-Fock (HF) method \cite{hartreeSelfconsistentFieldExchange1935} improves upon this by treating electron correlation through an antisymmetrized wave function characterized by a single Slater determinant to incorporate exact exchange. As such, the non-local Fock operator introduces an integral operator which couples space points and removes self-interaction, though it considers only exchange correlations.

For closed shells, the HF Fock operator reads
\begin{equation}
\hat{F}=\hat{h}+\hat{J}[n]-\hat{K}[\{\phi\}],\qquad
(\hat{K}\psi)(\mathbf r_1)=\sum_j \phi_j(\mathbf r_1)\int \frac{\phi_j^*(\mathbf r_2)\psi(\mathbf r_2)}{|\mathbf r_1-\mathbf r_2|}\,d\mathbf r_2,
\end{equation}
with \(\hat{h}=-\tfrac12\nabla^2+v(\mathbf r)\). In a spherical atom, orbitals separate as \(P_{nl}(r)=rR_{nl}(r)\), and the Hartree term reduces to a 1D Poisson solve for \(V_H(r)\). Exchange remains nonlocal after angular reduction. A multipole expansion yields
\begin{subequations}
\label{eq:hf-exchange-radial-thesis}
\begin{align}
(\hat{K}\,P_a)(r_1) &= \sum_{b\in \mathrm{occ}} \sum_{k=|l_a-l_b|}^{l_a+l_b}
g_k(l_a,l_b)\, Y_k^{(ab)}(r_1)\, P_b(r_1),\\
Y_k^{(ab)}(r) &= \int_0^\infty P_a(r')P_b(r')\, \frac{r_<^k}{r_>^{k+1}} \, dr',\\
g_k(l_a,l_b) &= \frac{4\pi}{2k+1}\,(2l_b+1)
\begin{pmatrix} l_a & k & l_b \\ 0 & 0 & 0 \end{pmatrix}^{\!2}.
\end{align}
\end{subequations}

From this basic formulation, subsequent expansions involving the occupancy of spin-orbitals yield unrestricted, restricted, and restricted open-shell Hartree-Fock methods \cite{bartlettApplicationsPosthartreefockMethods1994}. All of these are solved iteratively through the \gls{scf} approach. The addition of spin-orbit occupancies at this stage leaves the fundamental nature of the mean field approximation unchanged. Since instantaneous interactions typically prove more repulsive than an average interaction, the difference between the exact energy and the \gls{scf} approximation constitutes the ``correlation'' energy. More sophisticated treatments of the two-electron interaction can approximate this energy, leading to ``post-Hartree-Fock'' methods \cite{szaboModernQuantumChemistry1996}. These prescriptions offer mathematical rigor but scale too poorly for frequent computation in large systems. A numerically efficient alternative arrives from a different angle: grounding the theory in the density of electrons rather than exhaustively enumerating their positions.
\subsubsection{Kohn–Sham DFT: exact in principle, orbital constrained by construction}
\label{sec:orgd5e074f}
The Hohenberg–Kohn \cite{hohenbergInhomogeneousElectronGas1964}  theorem formulates the ground-state energy in terms of the electron density. \gls{ks} \gls{dft} builds a noninteracting system of virtual electrons that reproduces the interacting density, which brings the concept of orbitals back to the fore. The \gls{ks} equations employ a single multiplicative potential
\begin{equation}
V_\mathrm{eff}[n](r)=v(r)+V_H[n](r)+v_{xc}[n](r).
\end{equation}
If \(E_{xc}[n]\) were known, we could recover the exact density and energy. In practice we choose approximations \cite{perdewJacobLadderDensity2001a} (LDA/GGA/meta-GGA) and gain correlation corrections at modest cost. This local-potential form dovetails with radial finite elements and enables a fast, stable self-consistent loop.

We find the Hartree potential by solving the Poisson equation, \(\nabla^2V_H({\bf x}) = -4\pi n({\bf x})\), where \(n(r)\) represents the radial electron density constructed from the wave functions. Because the potential depends on the wave functions and the wave functions depend on the potential, these equations require iterative solution until they reach self-consistency.
\subsection{The Physical and Mathematical Problem}
\label{sec:fem:physmath}
We describe the electronic structure of an isolated, spherically symmetric atom at two primary levels of theory, depending on the required rigor: the non-relativistic Schrödinger equation and the relativistic Dirac equation.

In computational terms, the primary entities in this domain are the \texttt{Atom}, the \texttt{ElectronState}, and the \texttt{Potential}.
\begin{itemize}
\item An \texttt{Atom} is the central entity, characterized by its nuclear charge (Z). It is composed of a set of \texttt{ElectronState} s.
\item An \texttt{ElectronState} is defined by its quantum numbers (e.g., \(n\), \(l\), or \(\kappa\)) and is primarily described by a \texttt{WaveFunction} entity and its corresponding energy \texttt{Eigenvalue}.
\item The \texttt{Potential} is an entity that governs the behavior of the \texttt{ElectronState} s.
\end{itemize}

Since the \gls{ks} framework of \gls{dft} maps the complex many-electron problem onto a tractable set of single-particle equations, a cyclic dependence arises: the potential depends on the wave functions (via the electron density) and the wave functions depend on the potential. We must solve the \gls{ks} equations iteratively to a self-consistent solution. The governing mathematical models for the \texttt{WaveFunction} entity thus become the radial Schrödinger and Dirac equations, solved within this self-consistent loop. The discussion thus far does not restrict the number of atoms, however, for the remainder of the chapter we consider single atom systems, though we consider all electrons without approximations typically applied to larger systems \footnote{often treated with non-relativistic methods since core electrons are represented with a pseudo-potential or projected augmented wave \cite{dohnGridBasedProjectorAugmented2017}}.
\subsubsection{The Radial Schrödinger Equation}
\label{sec:org8e7e5b6}
For a spherically symmetric potential \(V(r)\), the wave function separates into radial and angular parts, \(\psi_{nlm}({\bf x})=R_{nl}(r)\,Y_{lm}(\theta, \phi)\). By substituting \(P_{nl}(r) = rR_{nl}(r)\), the problem reduces to solving the one-dimensional radial Schrödinger equation:

\begin{equation}
\label{eq:ch_schroed}
-\frac{1}{2} P_{nl}''(r) + \left(V(r) + \frac{l(l+1)}{2r^2}\right)P_{nl}(r) = E P_{nl}(r)
\end{equation}
where \(l\) is the angular momentum quantum number and \(E\) is the energy eigenvalue. The function \(P_{nl}(r)\) must be normalized such that \(\int_0^\infty P_{nl}^2(r) \, \mathrm{d}r = 1\).
\subsubsection{The Radial Dirac Equation}
\label{sec:orgc53eb54}
For heavy atoms, the appropriate single-particle theory requires a relativistic
treatment. The central-field Dirac equation leads to two coupled first-order
radial equations for the large and small components \((P_{n\kappa},Q_{n\kappa})\)
of a four-component spinor, with the relativistic quantum number \(\kappa\)
encoding \((l,j)\) \cite{grantRelativisticQuantumTheory2007}.

\begin{subequations}
\begin{align}
P_{n\kappa}'(r) &= -{\frac{\kappa}{r}}P_{n\kappa}(r)+\left({\frac{E-V(r)}{c}}+2c\right)Q_{n\kappa}(r), \\
Q_{n\kappa}'(r) &= -\left({E-\frac{V(r)}{c}}\right)P_{n\kappa}(r)+{\frac{\kappa}{r}}Q_{n\kappa}(r),
\end{align}
\end{subequations}
where \(c\) is the speed of light and \(\kappa\) is the relativistic quantum number
that encodes both total and orbital angular momentum.

The Dirac Hamiltonian possesses a spectrum unbounded from below, a feature that
historically plagued basis-set discretizations with ``variational collapse'' and
spurious states
\cite{grantRelativisticQuantumTheory2007,tupitsynSpuriousStatesDirac2008}. Typical
remedies over the past three decades involve shooting methods
\cite{certikDftatomRobustGeneral2013} which require trial solutions for each
eigenfunction and convergence parameteres for the solver grid, complicating the
need for robust and computationally efficient solutions. Basis set methods
\cite{dyallKineticBalanceVariational1990,fischerBsplineGalerkinMethod2009,grantBsplineMethodsRadial2009,almanasrehStabilizedFiniteElement2013}
solve for all states at once through diagonalization, but struggle with the
spurious states of he Dirac Hamiltonian, despite attempts to mitigate these with
changes in basis for large and small components
\cite{dyallKineticBalanceVariational1990,shabaevDualKineticBalance2004,beloyApplicationDualkineticbalanceSets2008,fischerBsplineGalerkinMethod2009,sunComparisonRestrictedUnrestricted2011,jiaoDevelopmentKineticallyAtomically2021},
Hamiltonian modifications
\cite{kutzelniggBasisSetExpansion1984,grantBsplineMethodsRadial2009,almanasrehStabilizedFiniteElement2013,almanasrehFiniteElementMethod2019,fangSolutionDiracEquation2020},
and boundary value constraints
\cite{johnsonFiniteBasisSets1988,sapirsteinUseBasisSplines1996,beloyApplicationDualkineticbalanceSets2008,grantBsplineMethodsRadial2009}.
\subsection{Robust finite element solvers for isolated atoms}
\label{sec:fem:lrepr}
We now consider a methodology which circumvents the spurious states of the Dirac
while providing robust and efficient solutions with a high-order finite element
basis. The \texttt{featom} code employs solves the governing equations using a finite
element basis, implemented in modern modular Fortran. The radial coordinate is
discretized into a mesh of finite elements, and within each element, the
solution is expanded in a basis of high-order Lagrange polynomials defined on
Gauss-Lobatto nodes. This spectral element approach yields exponential
convergence with respect to the polynomial order, providing high accuracy with a
relatively small number of basis functions. The success of this approach,
however, rests on a cascade of intelligent choices in representation at the
mathematical, numerical, and software levels.

The finite element expressions and code derived for the squared Dirac
Hamiltonian \cite{wallmeierUseSquaredDirac1981,kutzelniggBasisSetExpansion1984}
are novel, and these have the same eigenfunctions as the operator, and remains
bounded from below while preserving convergence to the Schrödinger limit
\cite{kutzelniggBasisSetExpansion1984}. Numerical stability arises from the
squared operator providing second derivative terms, while known asymptotic forms
near the origin allow side-stepping solving for large and small components of
the Dirac wavefunction components. We view these concepts and their synthesis
through the lens of varying representations below.
\subsubsection{Layer 1: The Mathematical Representation (Squared Hamiltonian)}
\label{sec:org7a44828}
Direct discretization of the Dirac Hamiltonian operator poses significant
difficulties because its energy spectrum remains unbounded from both above and
below, leading to spurious, unphysical solutions. To circumvent this, we utilize
a different mathematical representation: we solve the eigenvalue problem for the
square of the Dirac Hamiltonian, \((H+\mathbb{I}c^2)^2\). This squared operator
shares eigenfunctions with the original operator, and its eigenvalues relate
simply as the square of the original eigenvalues, \((E+c^2)^2\). Crucially, the
squared operator remains bounded from below. This allows the application of
standard variational techniques, like the \gls{fem}, without generating spurious
states.

A key principle here follows from eliminating the need for kinetic balance and
other such constraints, and can instead proceed with Galerkin discretization
\cite{brennerMathematicalTheoryFinite2008,sastryIntroductoryMethodsNumerical2010}
directly in a polynomial basis, the same for both large and small components.
\subsubsection{Layer 2: The Functional Representation (Asymptotic Correction)}
\label{sec:orgdae90c7}
For Coulombic potentials, the relativistic wave functions for states with
\(\kappa = \pm 1\) exhibit non-polynomial behavior near the origin (\(r \to 0\)),
with derivatives that diverge. This slow convergence poisons standard
polynomial-based approximation schemes. We address this by changing the
functional representation of the solution. Instead of solving for \(P(r)\) and
\(Q(r)\) directly, we solve for modified functions \(\tilde P(r) = P(r)/r^\beta\)
and \(\tilde Q(r) = Q(r)/r^\beta\), where \(\beta = \sqrt{\kappa^2-(Z/c)^2}\) is the
known asymptotic exponent. The new functions \(\tilde P(r)\) and \(\tilde Q(r)\) are
smooth and well-behaved at the origin, allowing for rapid, exponential
convergence in the polynomial basis for all quantum states.
\subsubsection{Layer 3: The Numerical Representation (The Golub-Welsch Algorithm)}
\label{sec:org4bc2ac5}
Beyond the theoretical framework, the choice of numerical representation is
critical for obtaining reliable results
\cite{chapraNumericalMethodsEngineers2015,eppersonIntroductionNumericalMethods2012}.
A pivotal enhancement involved resolving a critical instability in the
Gauss-Jacobi quadrature routine \cite{pressNumericalRecipes3rd2007}, essential
for accurately integrating terms involving the asymptotic correction factor. The
original implementation, based on a direct recurrence relation, was susceptible
to floating-point errors. To correct this, the routine was re-implemented using
the stable Golub-Welsch algorithm, which recasts the problem of finding
quadrature points (\(x_i\)) and weights (\(w_i\)) for integrals of the form
\begin{equation}
\int_{-1}^{1} (1-x)^{\alpha} (1+x)^{\beta} f(x) \mathrm{d}x \approx \sum_{i=1}^{n} w_i f(x_i)
\end{equation}
into a well-conditioned matrix eigenvalue problem. A symmetric tridiagonal
Jacobi matrix, \(\mathbf{J}\), is constructed, and its eigenvalues correspond
precisely to the quadrature nodes \(x_i\):
\begin{equation}
\mathbf{J} \mathbf{v}_i = x_i \mathbf{v}_i
\end{equation}
The corresponding weights \(w_i\) are then calculated from the first components of
the normalized eigenvectors \(\mathbf{v}_i\):
\begin{equation}
w_i = \mu_0 (v_{i,1})^2, \quad \text{where} \quad \mu_0 = 2^{\alpha+\beta+1} \frac{\Gamma(\alpha+1)\Gamma(\beta+1)}{\Gamma(\alpha+\beta+2)}
\end{equation}
This stable numerical representation was essential for guaranteeing the physical
integrity of the simulations.
\subsubsection{Layer 4: The Software Representation (Modern, Maintainable Code)}
\label{sec:orgeeaaa1b}
The final and most concrete layer of representation is the software itself. The
\texttt{featom} library is a modern Fortran 2008 \cite{kedwardStateFortran2022}
implementation with a strong emphasis on modularity, reusability, and the
absence of global state \cite{ousterhoutPhilosophySoftwareDesign2018}. This
design is crucial for enabling its use as a component in larger, more complex
simulation workflows. Interoperability is guaranteed through
backwards-compatible C bindings, allowing the high-performance Fortran core to
be called from other languages like C++ or Python.

This robust code is supported by a professional software engineering
infrastructure and involved introducing the flexible Meson build system
alongside the existing Fortran Package Manager (\texttt{fpm}), establishing a
comprehensive automated test harness, and refining continuous integration (CI)
pipelines. This focus on the software representation ensures correctness through
automated validation, lowers the barrier for collaboration, and guarantees
long-term maintainability and scientific reproducibility.
\subsection{From KS to HF: conceptually simple, practically subtle in spherical FE}
\label{sec:fem:kshf}
Conceptually, Kohn–Sham replaces the nonlocal exchange operator by a
multiplicative \(v_{xc}[n](r)\), which fits perfectly into the radial \texttt{featom}
framework that already solves Schrödinger/Dirac with a local potential.
Practically, three nontrivial points arise:

\begin{description}
\item[{Local vs nonlocal.}] HF exchange is nonlocal; KS uses a local \(v_x[n](r)\).
``Exact exchange'' (EXX) within KS requires solving an optimized effective
potential (OEP) \cite{kriegerSystematicApproximationsOptimized1992} equation
even in spherical symmetry. This adds a numerically involved integral equation
for \(v_x(r)\) to the formulation.

\item[{Orbital-dependent quantities are not a common potential}] Using
\(U_x^{(a)}(r)\) directly as ``the'' primary \(v_x(r)\) breaks the KS structure and
is unstable at nodes. A robust local proxy is the Slater average
\end{description}
\begin{equation}
v_x^\text{Slater}(r)
= \frac{1}{n(r)} \sum_{a\in\text{occ}} \frac{f_a}{2}\, n_a(r)\, U_x^{(a)}(r),
\qquad
n_a(r)=\frac{P_a(r)^2}{4\pi r^2},\quad n(r)=\sum_a f_a n_a(r),
\end{equation}
which is multiplicative and stable in SCF.

\begin{description}
\item[{Partial-wave assembly inside SCF}] Whether building HF (nonlocal) or local approximations, the spherical FE code benefits from the same partial-wave machinery: accurate \(Y_k^{(ab)}(r)\), correct angular algebra, and careful treatment near nodes and \(r\to 0\). Implementations that instead attempt to fold nonlocal exchange into a single multiplicative potential without these steps tend to diverge or collapse the spectrum.
\end{description}

\begin{figure}[h!]
\centering
\includegraphics[width=\linewidth]{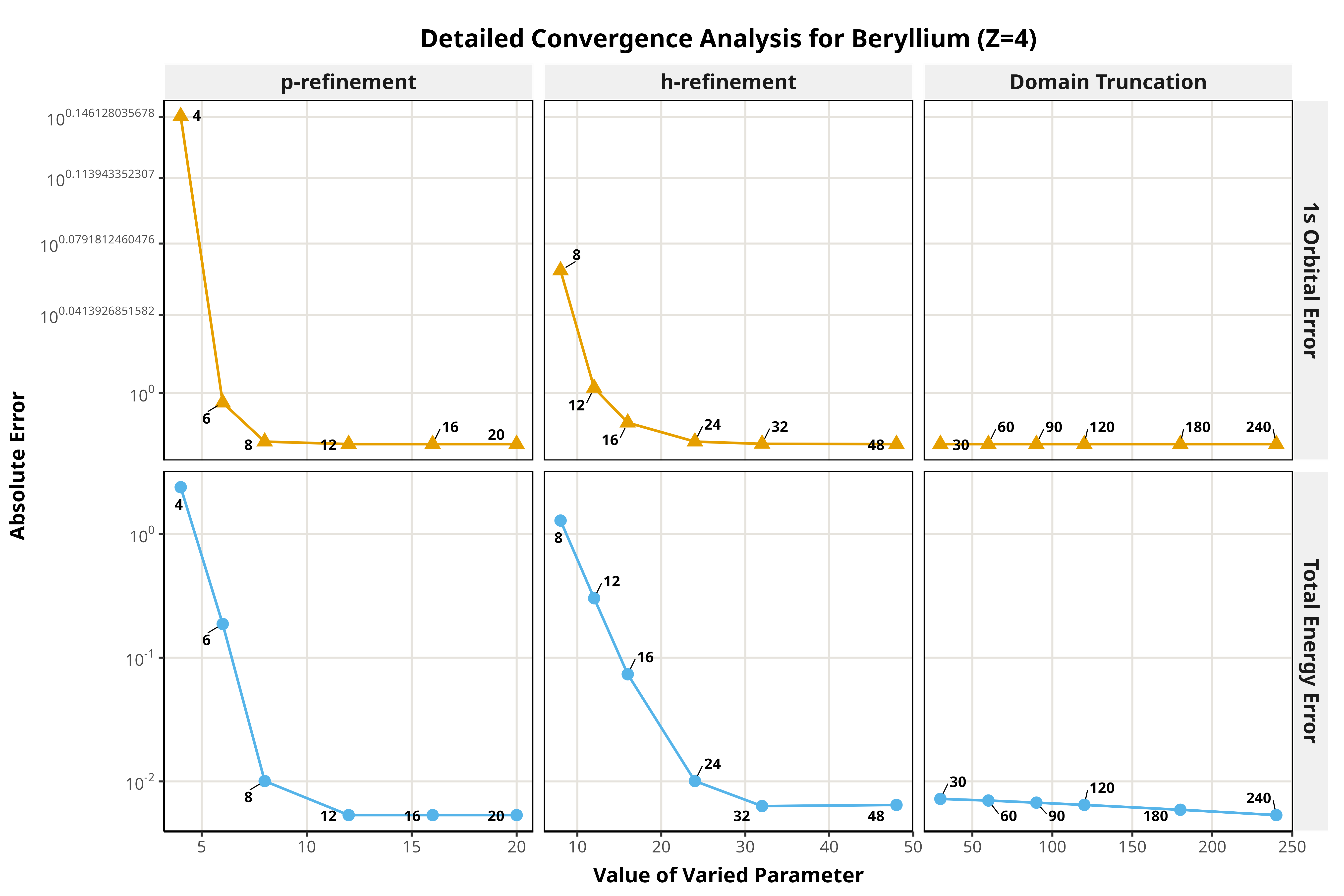}
\caption{Systematic convergence of the radial finite element solver for a Restricted Hartree-Fock calculation on a Beryllium atom (Z=4). The grid validates the two primary modes of convergence. \textbf{Bottom Row (Total Energy)} and \textbf{Top Row (1s Orbital)} show the error for three distinct refinement studies. (\textbf{Left Column}) p-refinement: For a fixed mesh, the error decreases exponentially with increasing polynomial order (p), demonstrating rapid convergence to high accuracy. (\textbf{Middle Column}) h-refinement: For a fixed polynomial order, the error decreases more slowly (algebraically) with the number of elements ($N_e$). (\textbf{Right Column}) Domain Truncation: The solution is stable and well-converged with respect to the domain cutoff ($r_{max}$).}
\label{fig:be_convergence_grid}
\end{figure}

A simple extension towards an HF/KS implementation in the radial FE code follows a ``local-in-the-loop, exact-after'' strategy \footnote{\href{https://github.com/atomic-solvers/featom/pull/24/}{gh-26} to featom}:
\begin{itemize}
\item In the SCF loop we use a multiplicative exchange potential of Slater–LDA form,
\end{itemize}
\begin{equation}
v_x^\text{LDA}(r) = -\left(\frac{3}{\pi}\right)^{\!\!1/3} n(r)^{1/3},
\end{equation}
which is local and stable to iterate together with \(V_H(r)\) from the spherical Poisson solver.
\begin{itemize}
\item After \gls{scf} convergence we compute the exact \gls{hf} exchange energy a posteriori using the multipole machinery in \eqref{eq:hf-exchange-radial-thesis} on the converged orbitals, with numerically stable global-sorting and prefix–suffix accumulations for \(Y_k^{(ab)}(r)\).
\item This yields total energies close to the restricted \gls{hf} benchmarks while preserving the robustness of a local \gls{scf}. For example, for a Beryllium atom (Z=4):
\end{itemize}
\[
E_\text{tot} = -14.57067378~\text{Ha}\quad (\text{RHF ref} -14.57541503~\text{Ha}),
\]
a \(\sim 4.7\) mHa gap consistent with using a local \(v_x\) instead of nonlocal \gls{hf} in the loop. Figure \ref{fig:be_convergence_grid} demonstrates the convergence characterstics. 

\subsection{Performance and accuracy}
\label{sec:fem:perfacc}
\begin{figure}[htbp]
\centering
\includegraphics[width=.9\linewidth]{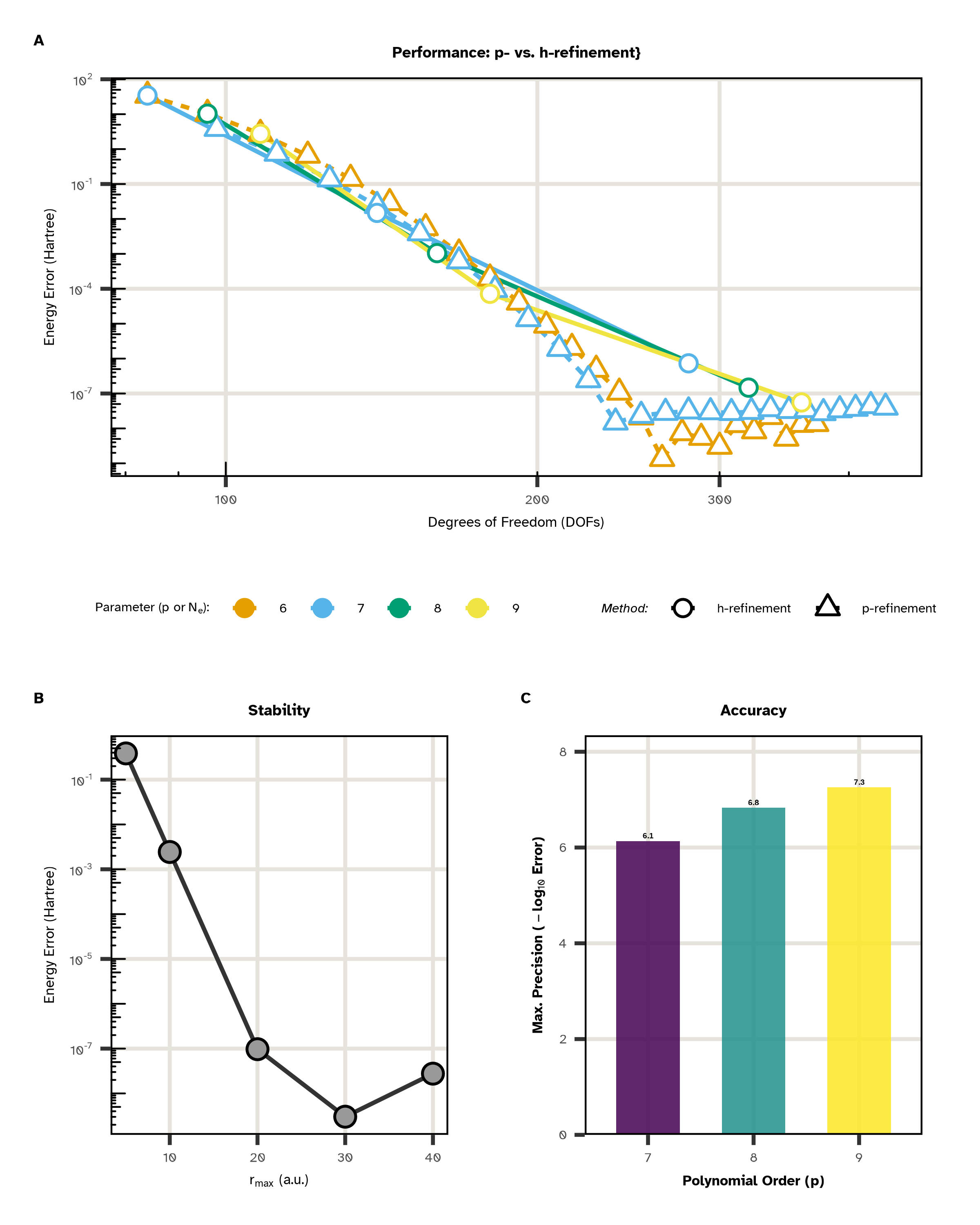}
\caption{\label{fig:featom_dirac_convergence_panel}Systematic convergence and precision of the \texttt{featom} finite element solver for relativistic Dirac–Kohn–Sham calculations of uranium (Z=92). (a) p- vs. h-refinement: Both p-refinement (increasing polynomial order \(p\), colored) and h-refinement (increasing number of elements \(N_e\)) yield systematic error reductions. The plot shows energy error as a function of the total degrees of freedom (DOFs), with shape and color encoding the refinement parameter and method; exponential convergence in \(p\) and algebraic in \(N_e\) are both evident. (b) Domain cutoff stability: The total energy error decreases rapidly as the radial domain boundary \(r_{\max}\) increases and quickly plateaus, demonstrating insensitivity to the outer cutoff. (c) Accuracy: Bar plot of the maximum precision (number of correct digits, -\(\log_{10}\)(error)) reached for each \(p\) value, highlighting the accuracy attainable with moderate \(p\). Collectively, these results establish \texttt{featom} as a robust, high-precision, and reproducible tool for atomic DFT, confirming correct asymptotic error behavior for both p- and h-refinement, as well as stability against domain truncation.}
\end{figure}

Figure \ref{fig:featom_dirac_convergence_panel} demonstrates the systematic,
reproducible convergence of the solver for uranium in three complementary
regimes: domain truncation, \$p\$- and \$h\$-refinement, and the achieved energy
precision.

As designed, the \texttt{featom} code is tailored towards being state-of-the-art as a
\gls{dft} solver for relativistic calculations, which are otherwise treated only
through shooting approaches \cite{certikDftatomRobustGeneral2013}, often in the
context of expensive quantum chemical finite difference based \cite{grasp2k}.
Despite this, for systems with a high number of states, \texttt{featom} outperforms
shooting method state of the art calculators as well. Here, we appeal to the
transitive nature of benchmarks since the state of the art \texttt{dftatom} compares
favorably to Desclaux \cite{desclaux}, \texttt{atompp} \cite{hamann}, \texttt{grasp2k}
\cite{grasp2k}, MCHF \cite{FroeseFischer1991}, \texttt{atompaw}
\cite{tackettProjectorAugmentedWave2001}, \texttt{PEtot} \cite{Jia2013} and Elk
\cite{elk}, we assert superiority by directly comparing against \texttt{dftatom}.
\gls{dft} based formulations are always faster, due to the ability to skip the two
particle exchange integral. For instance, a B-spline based Hartree-Fock solver
\cite{zatsarinnyDBSR_HFBsplineDirac2016}  reports, for a toy 3-electron model of
Uranium (``Lithium like'', \texttt{2s(1)}), a time of 1.53 seconds without the Breit
interaction while we demonstrate a full 29 state Uranium atom:

$$1s^2 2s^2 2p^6 3s^2 3p^6 3d^{10} 4s^2 4p^6 4d^{10} 4f^{14} 5s^2 5p^6 5d^{10} 5f^3 6s^2 6p^6 6d^1 7s^2.$$

solution in 0.36 seconds. Furthermore, we compare the wall time against the
state-of-the-art shooting-method code \texttt{dftatom}. For this DFT calculation of
uranium converged to an accuracy of \(10^{-6}\) Hartree, \texttt{featom} shows a
significant speedup for non-relativistic calculations and competitive
performance for relativistic ones, validating the effectiveness of the chosen
representations. 

\begin{table}[htbp]
\caption{\label{tbl:wtimedftrel}Timings for a DFT calculation of a uranium atom on an Apple M-1 Max processor.}
\centering
\begin{tabular}{lll}
Solver & \texttt{featom} & \texttt{dftatom}\\
\hline
Schrodinger & 28 ms & 166 ms\\
Dirac & 360 ms & 276 ms\\
\end{tabular}
\end{table}

The benchmark results in Table \ref{tbl:wtimedftrel} reflect the distinct scaling
characteristics of the \gls{fem} used in \texttt{featom} versus the shooting method used
in \texttt{dftatom}.

For the Schrödinger equation, \texttt{featom} is significantly faster (28 ms vs 166 ms)
because it solves for all eigenvalues in a given angular momentum channel
simultaneously via a single diagonalization of a small matrix (\(N_b \times
N_b\)). In contrast, \texttt{dftatom} must perform iterative radial integrations for
each individual electronic state, which accumulates a higher computational cost
for high-Z atoms with many electrons.

For the Dirac equation, the performance shifts. In the FEM framework (\texttt{featom}),
the basis set size effectively doubles to represent both the large and small
spinor components, resulting in matrices of size \(2N_b \times 2N_b\). Since the
cost of the dense eigensolver scales cubically with matrix size
(\(\mathcal{O}(M^3)\)), the computational effort increases by a factor of roughly
8 (\(2^3\)), alongside increased matrix assembly costs. In contrast, the shooting
method (\texttt{dftatom}) only sees a linear increase in cost associated with
integrating two coupled first-order ODEs rather than one second-order ODE.
Consequently, while \texttt{featom} slows down by a factor of \(\sim 12\) (28 ms to 360
ms) due to the cubic scaling penalty, \texttt{dftatom} slows down by only \(\sim 1.6\)
(166 ms to 276 ms), making it slightly faster in the fully relativistic case for
this specific benchmark. Despite the associated overhead, the optimized software
ensures \texttt{featom} remains competitive in absolute time compared to \texttt{dftatom} and
outstrips other solvers.
\subsection{Conclusions}
\label{sec:fem:conclusions}
This chapter has established the theoretical foundations—specifically the
mean-field approximations of Hartree-Fock and Kohn-Sham density functional
theory—requisite for interpreting electronic structure calculations throughout
this work. While the specific numerical innovations presented here address the
challenge of relativistic atomic precision, the broader investigations into
reactive systems presented in the remainder of this thesis rely on these
fundamental mean-field principles. The calculations for molecular systems in
this thesis are at Hartree-Fock level, with spin-unrestricted formalisms for
radical species and restricted closed-shell formalisms for stable intermediates.

Within this theoretical framework, we presented a robust finite element
formulation for the solution of the radial Schrödinger, Dirac, and Kohn-Sham
equations for isolated atoms. The success of the \texttt{featom} solver exemplifies the
central thesis that overcoming computational barriers in physics requires a
holistic approach to representation. By moving from the Dirac Hamiltonian to its
squared operator, we eliminated the spectral pollution of spurious states that
has historically plagued basis-set methods. By transforming the dependent
variables to account for asymptotic behavior near the nucleus, we resolved the
representational conflict between singular Coulombic potentials and smooth
polynomial bases, recovering exponential convergence. The practical outcome of
these choices is a solver that occupies a unique niche in the current software
ecosystem. As demonstrated by the benchmarks, \texttt{featom} provides sub-second wall
times for full relativistic DFT calculations of heavy elements like uranium
(\(Z=92\)), outperforming state-of-the-art shooting methods in the relativistic
regime while avoiding the complexity of kinetic balance constraints required by
B-spline approaches.

This efficiency is not merely a convenience but should translate eventually
towards generating better basis sets for high-fidelity reference data needed to
train Gaussian Process surrogates discussed in the previous chapter.

Finally, the implementation of \texttt{featom} underscores the critical importance of
software accessibility and reproducibility in computational science. For
instance, while reviewing the landscape of relativistic solvers, we attempted to
evaluate the \texttt{BERTHA} \cite{belpassiBERTHAImplementationFourcomponent2020} code
but were unable to locate source code, finding only Python wrappers dependent on
opaque, closed-source binaries\footnote{to say nothing of ``closed source'' code
results}. In contrast, \texttt{featom} is provided as a modern, modular, open-source
library. This ensures that the distinct layers of representation—mathematical,
numerical, and algorithmic—remain transparent, verifiable, and adaptable.

However, the converse is also true, not every ``conceptually simple'' mapping
(e.g. Hartree-Fock \(\to\) ``a potential'') respects the representation. Where the
physics demands an operator (HF exchange), we either keep the operator or solve
a re-representation problem, e.g. through an optimized effective potential
\cite{kriegerSystematicApproximationsOptimized1992,nagyAlternativeDerivationKriegerLiiafrate1997,staroverovEffectiveLocalPotentials2006}.

We will return to this concept in later chapters; when we formulate efficient
reaction-path searches and have \Gls{gp} accelerators succeed because their
internal representations (forces, curvatures, kernels) are chosen to make the
numerics stable and the computation scalable.

This serves to clarify that the ``correctness'' of a scientific result is
inextricably linked to the design of the software that produces it. Physical
rigor cannot be separated from software architecture. In the subsequent chapter,
we will expand this scope from the design of a single solver to the design of
the broader computational frameworks required for complex chemical exploration.
\section{Aspects of software design}
\label{sec:asp}
\epigraph{Pray, Mr. Babbage, if you put into the machine wrong figures, will the right answers come out?}{Member of the House of Commons \\ asked of Charles Babbage}

Computational science confronts a fundamental representational problem: physical
laws, typically expressed as continuous differential equations, require
evaluation on digital hardware that operates with discrete logic and
finite-precision arithmetic
\cite{goldbergWhatEveryComputer1991,overtonNumericalComputingIEEE2001}. The
evaluation of a function \(f(x)\) on a computer therefore necessitates its
approximation by a discrete counterpart \(\hat{f}(\mathbf{x})\) that maps a finite
vector of inputs to a finite vector of outputs. This transition from the
continuous domain to a discrete, floating-point representation introduces
unavoidable errors, including truncation error from the discretization scheme
and rounding error from the limitations of floating-point number representation
\cite{sauerNumericalAnalysis2018,sastryIntroductoryMethodsNumerical2010,eppersonIntroductionNumericalMethods2012,gottschlingDiscoveringModernIntensive2021}.
The central challenge of scientific software engineering lies not in eliminating
these errors, an impossible task, but in designing computational structures that
control them and guarantee the physical fidelity of the final result.

The preceding chapter on relativistic atomic calculations provided a concrete
example of a successful strategy for managing this challenge. The accuracy and
stability of the \texttt{featom} solver originate from a deliberate, multi-layered
cascade of representations, each chosen to mitigate a specific class of error.
Briefly, this involved a mathematically stable squared Hamiltonian to handle the
unbounded Dirac spectrum, a functionally smooth set of corrected wave functions
to accelerate polynomial convergence, a numerically robust algorithm to
guarantee accurate quadrature, and finally, a modular
\cite{ousterhoutPhilosophySoftwareDesign2018} software implementation to ensure
correctness and maintainability.

This chapter dissects the principles of such software redesign and the novel
scientific capabilities enabled by such an undertaking. We examine how conscious
architectural choices directly impact the quality and reliability of scientific
outcomes, focusing on the implementation of novel scientific algorithms, such as
the hybrid \gls{roneb} method, which were made possible only after a fundamental
re-engineering of the software's state management and potential interfaces; the
choice between geometric and electronic-structure representations for defining a
chemical bond; the interpretation of double ended saddle point data; the
representation of a complex scientific protocol as a formal \gls{dag} using a
workflow engine; along with future directions. In each case, the software design
reflects a conscious strategy to build powerful and reliable computational
models upon the discrete and finite foundation of the computer.
\subsection{Bonding analysis}
\label{sec:asp:bondy}
To unambiguously distinguish covalently bonded molecular fragments from transient non-covalent contacts, a robust analysis of the system's bonding network is essential. This can be approached from two distinct perspectives, namely a heuristic geometric definition or a more rigorous definition based on the system's electronic structure. We implement in \texttt{rgpycrumbs} both methods, with a \texttt{pyvista} backend \footnote{inspired by \texttt{solvis}} allowing for a flexible and chemically aware analysis of molecular connectivity.
\subsubsection{Geometric method: Covalent cutoff}
\label{sec:asp:geom}
The simplest and most computationally efficient method for defining a bonding network is based on geometry. In this approach, we define that a bond exists between two atoms, \(i\) and \(j\), based on their interatomic distance, \(d_{ij}\). Specicially, the bond exists when atoms are closer to each other than a scaled sum of their tabulated covalent radii, \(r_i\) and \(r_j\). This relationship is governed by the inequality:

\begin{equation}
  d_{ij} < M \cdot (r_i + r_j)
  \label{eq:geom_bond}
\end{equation}

Here, \(M\) is a dimensionless scaling multiplier \footnote{typically between \(1.1\) and \(1.3\)} used to adjust the strictness of the criterion. While this method is extremely fast, and widely available due to being the method used by the ASE \cite{larsenAtomicSimulationEnvironment2017} \gls{gui}; a purely geometric measure for molecules is a significant drawback, as the base unit of calculations are centered on electrons (Chapter \ref{sec:fem}). Such measures lack ``chemical intuition'' and can fail in sterically crowded environments where non-bonded atoms are forced into close proximity, leading to the false identification of covalent bonds (as illustrated in Figure \ref{fig:wbo_reaction_D004} B and C.
\subsubsection{Electronic density: \glsdesc{wbo}}
\label{sec:asp:bondy:wbo}
A more physically meaningful approach defines connectivity based on the electronic structure of the system, specifically; from the density matrix obtained in a quantum chemical calculation. A simple form of this is \gls{wbo}, which represents the electron density shared between two atoms, \(A\) and \(B\), by the sum of squares of the density matrix elements corresponding to the atomic orbitals on each atom \cite{wibergApplicationPoplesantrysegalCNDO1968}. We define:

\begin{equation}
  \text{WBO}_{AB} = \sum_{\mu \in A} \sum_{\nu \in B} (P_{\mu\nu})^2
  \label{eq:wbo}
\end{equation}

where
\begin{itemize}
\item \(WBO_{AB}\) is the Wiberg Bond Order between atom A and atom B.
\item \(\sum_{\mu \in A}\) sums over all atomic orbitals \(\mu\) on atom A.
\item \(\sum_{\nu \in B}\) sums over all atomic orbitals \(\nu\) on atom B.
\item \(P_{\mu\nu}\) is an element of the density matrix.
\end{itemize}

The density matrix element \(P_{\mu\nu}\) for a closed-shell system is calculated from the molecular orbital coefficients (\(C\)) of the occupied molecular orbitals (\(i\)):
\begin{equation}
  P_{\mu\nu} = 2 \sum_{i}^{\text{occupied}} C_{\mu i} C_{\nu i}
  \label{eq:density_matrix}
\end{equation}

The \gls{wbo} correlates well with the intuitive chemical concept of single, double, and triple bonds. A bond between atoms \(i\) and \(j\) is defined to exist only if their calculated bond order, \(\text{WBO}_{ij}\), exceeds a predefined threshold, \(T_{\text{bond}}\):

\begin{equation}
  \text{WBO}_{ij} > T_{\text{bond}}
  \label{eq:wbo_bond}
\end{equation}

\begin{figure}[htbp]
\centering
\includegraphics[width=.9\linewidth]{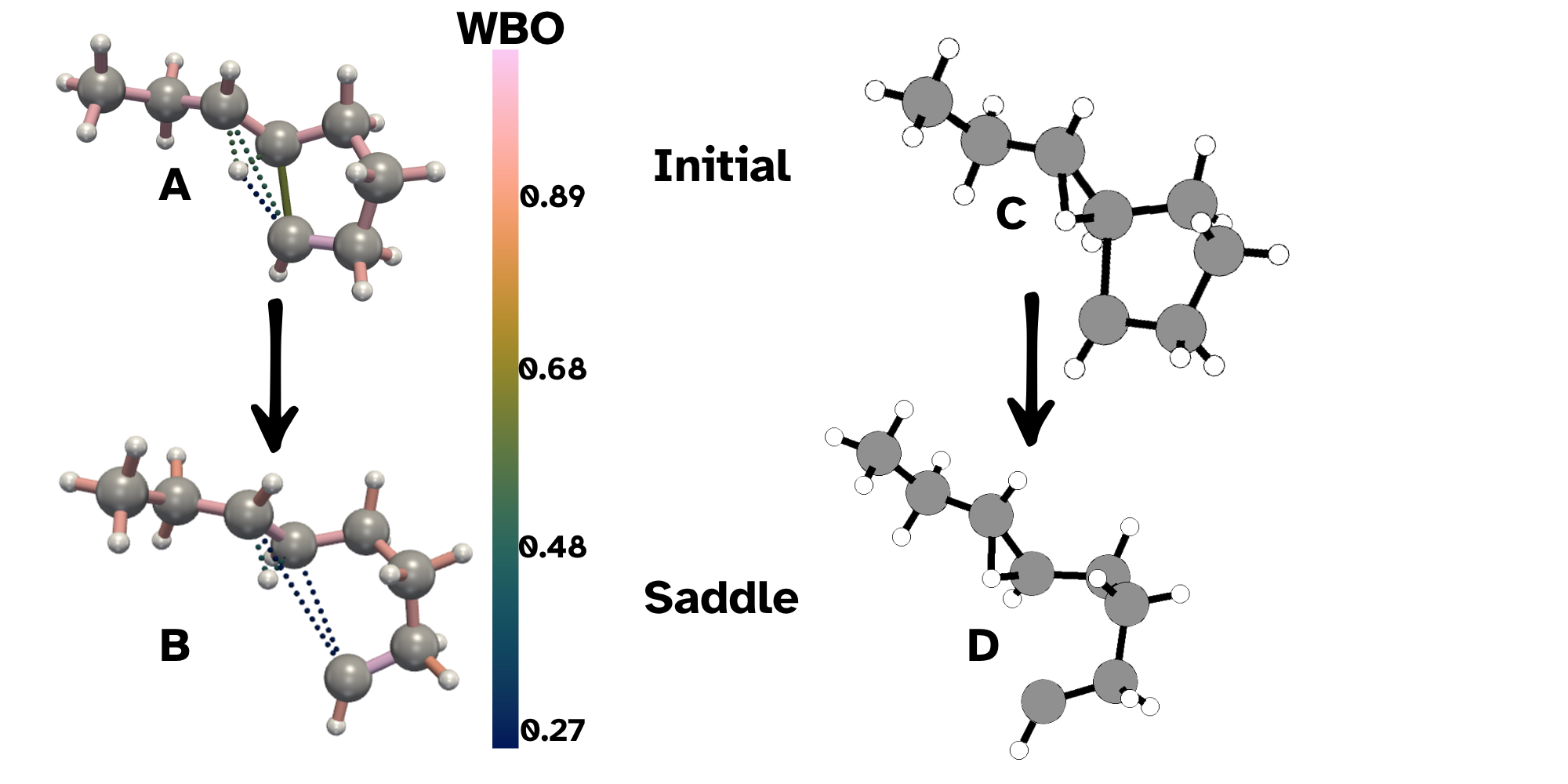}
\caption{\label{fig:wbo_reaction_D004}Wiberg Bond Order (WBO) analysis of a radical hydrogen transfer reaction (doublet system D004) from an initial reactant complex to the saddle point. Panels (a) and (b) visualize the system with interatomic connections colored by their WBO, where bonds are above 0.5, revealing the subtle electronic changes during the reaction: the weak C-C bond with a WBO of \textasciitilde{}0.5 in the initial state (a) is broken (b). In contrast, the standard geometric stick representation from ASE in panels (c) and (d) shows a nonsensical three center bond involving hydrogen, which is geometrically close but not actively bonded.}
\end{figure}

Such a measure can be significantly more robust than the geometric approach as
it hinges on the calculation of the actual chemical interactions. It can
reliably distinguish between genuine covalent bonds, which have significant
shared electron density (typically WBO > 0.7), and close non-covalent contacts,
which exhibit negligible bond orders. As shown in Figure
\ref{fig:wbo_reaction_D004}, the WBO between sterically close but non-bonded atoms
evaluates to a near zero value, correctly identifying them as belonging to
separate molecular fragments. The primary trade-off for this increased accuracy
is the higher computational cost associated with performing the underlying
electronic structure calculation, which is largely alleviated by using GFN2-xTB
semi-empirical calculation \cite{bannwarthGFN2xTBanAccurateBroadly2019} \footnote{this is still not quick enough for extended systems however.}.

\subsection{RMSD projections for path visualization}
\label{sec:asd:pviz}
Methods like the \gls{neb} form chains-of-states pathways to connect configurations on the \gls{pes}, and visualzation of these is provided within \texttt{rgpycrumbs} \footnote{A pure Python library for snippets, here \url{https://github.com/HaoZeke/rgpycrumbs/}}. Within EON \cite{chillEONSoftwareLong2014} profiles of the \gls{extrema} are written out in quadruplets of the terms of the image number, the energy difference relative to the ``reactant'', ``path'' reaction coordinate and parallel force. We define the ``path'' reaction coordinate (\(s\)), to be the cumulative Cartesian distance between successive images \(\mathbf{R}_i\) along the path, or

\begin{equation}
  s_i = \sum_{j=1}^{i} \left\| \mathbf{R}_j - \mathbf{R}_{j-1} \right\|_2
  \label{eq:path_coord}
\end{equation}

where \(\mathbf{R}_j\) are atomic positions for the \(j\)-th image and \(s_0 =
0\). The energy difference against the path coordinate for the converged path is
the most common visualization with insets indicating the climbing image and
end-points. To create a smooth curve from the discrete images, we use a Cubic
Hermite Spline
\cite{eppersonIntroductionNumericalMethods2012,virtanenSciPyFundamentalAlgorithms2020},
as is used internally in EON as well. Unlike a standard spline, this constructs
a piecewise cubic polynomial \(H(s)\) that matches the energy \(E_i\) at each image
\(i\) but also the projected derivative \(E'_i\) which we define to be the negative
of the force component parallel to the path tangent \(\hat{\tau}_i\): 
\begin{equation}
  \left. \frac{dE}{ds} \right|_{s_i} = -\mathbf{F}(\mathbf{R}_i) \cdot \hat{\tau}_i = -F_{\parallel, i}
  \label{eq:hermite_deriv}
\end{equation}
where \(\hat{\tau}_i\) is the normalized tangent for image \(i\). Hermite
interpolation ensures that both the energy and the slope are matched at each
discrete point, yielding a consistent and smooth profile as used internally,
that preserves barrier heights and avoids artifacts. In this thesis, we use the
augmented form of this visualization with the history of the path optimization
\cite{asgeirssonNudgedElasticBand2021,bigiMetatensorMetatomicFoundational2025}, as
shown later, in Figure \ref{fig:combneb}.

For higher-dimensional visualization, we project the \gls{neb} path onto the plane
defined by \gls{rmsd} from reactant and \gls{rmsd} from product:

\begin{equation}
  (x_i, y_i) = \left( d_{\mathrm{RMSD}}(\mathbf{R}_i, \mathbf{R}_\mathrm{reactant}),
                   d_{\mathrm{RMSD}}(\mathbf{R}_i, \mathbf{R}_\mathrm{product}) \right)
  \label{eq:rmsd_pair}
\end{equation}

To resolve atom mapping and orientation ambiguities, particularly in symmetric
systems, we use the \gls{ira} Fortran routine from Python
\cite{gundeDevelopmentIRAShape2021}. This finds the optimal atom permutation,
rotation, and translation to minimize \gls{rmsd}:

\begin{equation}
  d_{\mathrm{RMSD}}(\mathbf{A}, \mathbf{B}) =
    \min_{\mathbf{R},\,\mathbf{P}} \sqrt{ \frac{1}{N} \sum_{j=1}^{N}
    \left\| \mathbf{R}\mathbf{a}_{\mathbf{P}(j)} - \mathbf{b}_j \right\|^2 }
  \label{eq:ira_rmsd}
\end{equation}

where \(\mathbf{P}\) is the atom permutation and \(\mathbf{R}\) is the rotation
matrix.

To visualize the local structure of the potential energy surface, we interpolate
the scattered energies onto the \gls{rmsd} plane. This is achieved using a \gls{rbf}
interpolator with a Thin Plate Spline kernel
\cite{wahbaSplineModelsObservational1990}, as implemented in SciPy
\cite{virtanenSciPyFundamentalAlgorithms2020}.

\begin{equation}
  f(\mathbf{x}) = \sum_{i=1}^{N} w_i \phi(||\mathbf{x} - \mathbf{x}_i||) + P(\mathbf{x})
\end{equation}

where \(\phi(r) = r^2 \log(r)\) is the thin plate spline radial function and
\(P(\mathbf{x})\) is a low-degree polynomial term included to ensure solvability.
The weights \(w_i\) are determined by solving a linear system that interpolates
the energy values \(E_i\) at the sampled coordinates \((x_i, y_i)\). A smoothing
parameter \(\lambda\) (set to 0.009 in this work) is applied to the diagonal of
the interpolation matrix to handle noise in the optimization data and prevent
overfitting to high-energy artifacts. This produces the smooth, physically
continuous contour maps used to visualize the topography of the energy
landscape, shown in section \ref{sec:asd:eon:case}, Figure \ref{fig:viz_analysis}.

The combination of Hermite-spline profile interpolation and two-dimensional
landscape projection provides mechanistic insights into both the energetic
barriers and the geometric progression of the reaction. The 1D profile
quantifies how forces and energy change along the path, while the 2D landscape
exposes the multidimensional structure of the transition region. Together, these
methods allow us to visualize not only the minimum energy pathway but also the
broader context of atomic rearrangements and surface topography that govern
chemical transformations.
\subsection{EON}
\label{sec:asd:eon}
Most of the calculations in this thesis go through EON. Rather than exessively
modifying the SVN copy, a new release was drafted, v2.8.0 \footnote{accompanying
documentation : \url{https://eondocs.org}}. All the methods presented in this thesis
are either the direct result of, or stemmed from the landmark modernization of
the EON client code. This multi-year development effort, spawning millions of
lines of code and documentation, overhauled the entire framework to be more
powerful, flexible, and robust, transforming it from a legacy tool into a modern
scientific platform. The effort focused on several key areas of software
engineering.

First, the core C++ backend was modernized to the C++17 standard, and adopting
modern STL libraries like \texttt{<filesystem>} for cross-platform compatibility.
Second, the entire build process was migrated from legacy Makefiles to the Meson
build system, and a comprehensive continuous integration (CI) pipeline was
established to test automatically across Linux, Windows, and macOS. This
professionalized infrastructure guarantees correctness, portability, and
long-term maintainability.

Crucially, this architectural refactoring enabled a fundamental shift in the
software's capabilities. The redesigned state management and potential
interfaces made it possible, for the first time, to instantiate and control
multiple, different potential energy surface evaluators within a single
simulation. This unlocked a vast expansion of interoperability, with new
interfaces to a dozen external quantum chemistry and machine learning codes
(NWChem \cite{apraNWChemPresentFuture2020}, ORCA
\cite{neeseORCAQuantumChemistry2020}, XTB
\cite{bannwarthGFN2xTBanAccurateBroadly2019}, ASE
\cite{larsenAtomicSimulationEnvironment2017}, PET-MAD
\cite{bigiMetatensorMetatomicFoundational2025,mazitovPETMADUniversalInteratomic2025}).
More importantly, this architectural flexibility provided a platform for
inventing arbitrarily novel and efficient hybrid algorithms. The Hybrid
\gls{roneb} method, detailed next, is a direct product of this new design, as it
leverages the ability to combine different optimizers and energy-weighted spring
forces within a single, cohesive calculation—a capability that was previously
impossible.
\subsubsection{Eliminating I/O Bottlenecks with a Client-Server Architecture}
\label{sec:orgf5ed595}
A primary performance bottleneck in complex simulation workflows is the reliance
on file-based I/O to communicate with external potential energy surface
calculators. This traditional approach—repeatedly writing input files, executing
an external program as a separate process, and parsing text output files—suffers
from immense overhead from disk access and process creation, rendering many
computationally demanding methods infeasible, especially those which use large
wavefunctions.

To overcome this limitation, a significant part of the EON refactoring involved
implementing a high-performance, in-memory communication layer based on the i-PI
\cite{kapilIPI20Universal2019} client-server protocol. In this modern
architecture, EON acts as a persistent server that orchestrates the simulation,
while a quantum chemistry code like NWChem runs as a long-lived client. The
communication of atomic coordinates and the resulting energies and forces occurs
directly through low-latency TCP/IP or high-performance UNIX domain sockets.
This transforms the external potential from a slow, ``black box'' command-line
tool into a responsive, integrated library.

This effort required not only developing the server architecture within EON
\footnote{\href{https://github.com/TheochemUI/eOn/pull/244}{gh-244} to EON} but also contributing directly to the NWChem codebase to
improve its capabilities as a client. A key pull request, which was merged into
the official NWChem repository \footnote{\href{https://github.com/nwchemgit/nwchem/pull/1145}{gh-1145} to NWChem} , enhanced its socket
client with a robust polling and retry mechanism. This modification allows the
NWChem client to wait patiently for the EON server to become available, a
critical feature for ensuring stable, loosely-coupled communication between the
two persistent programs.

The performance impact of this architectural shift is dramatic, as shown in the
benchmark timings for a 16-step minimization.

\begin{table}[htbp]
\caption{Performance comparison of communication methods for an identical 16-step minimization task. Wall times were measured on a ThinkPad X1.}
\centering
\begin{tabular}{llr}
Communication Method & Wall Time & Speedup vs. File-based\\
\hline
File-based (ASE Wrapper) & 78 s & 1.0x\\
TCP/IP Socket & 47 s & 1.7x\\
UNIX Domain Socket & 40 s & 2.0x\\
UNIX Socket (Patched NWChem) & 17 s & 4.6x\\
\end{tabular}
\end{table}

The socket-based communication layer provides a \(2\times\) speedup out of the
box, and a remarkable \(4.6\times\) speedup when combined with a fully optimized
NWChem build. This architectural change from a file-based to a socket-based
representation of the potential energy surface is another enabler for the wall
time efficient methods explored in this thesis.
\subsubsection{Hybrid Climbing Image NEB with Minimum Mode Following (CI-NEB-MMF)}
\label{sec:asp:eon:roneb}
While the standard Climbing Image NEB (CI-NEB) method is effective at converging
to a saddle point, the final stages of relaxation for the climbing image can be
slow, particularly on flat potential energy surfaces. To accelerate this final
convergence, a hybrid approach has been implemented that integrates a dedicated
minimum-mode following (MMF) saddle search directly into the NEB optimization
cycle. This method can be seen as a two-stage refinement strategy within each
NEB iteration once the path is sufficiently relaxed \footnote{\href{https://github.com/TheochemUI/eOn/pull/230}{gh-230} to EON}.

The core idea is to use the robust path-finding capability of NEB to bring the
climbing image close to the saddle point, and then switch to a more aggressive
and efficient local saddle search algorithm for a few steps to rapidly refine
the climbing image's position.

The methodology is controlled by several key parameters: a boolean switch to
enable the feature (\texttt{nebciWithMMF}), a force threshold for activation
(\texttt{nebciMMFAfter}), and the number of MMF steps to perform per NEB iteration
(\texttt{nebciMMFnSteps}).

The force applied during the local refinement phase is the standard MMF force,
which inverts the true force component along the lowest-energy mode
(approximated by the NEB tangent \(\hat{\tau}_{\text{climb}}\)):
\begin{equation}
  \mathbf{F}_{\text{MMF}}(\mathbf{R}_{\text{climb}}) = \mathbf{F}(\mathbf{R}_{\text{climb}}) - 2 (\mathbf{F}(\mathbf{R}_{\text{climb}}) \cdot \hat{\tau}_{\text{climb}}) \hat{\tau}_{\text{climb}}
  \label{eq:neb_mmf_force}
\end{equation}

This force is identical to the one used in the standard CI-NEB (Equation \ref{eq:neb_ci}), but its application within a dedicated saddle search optimizer (such as one based on the Dimer method) allows for more efficient convergence on the saddle point. The overall process is outlined in Algorithm \ref{alg:neb_mmf}.

\begin{algorithm}
\caption{Hybrid CI-NEB with Minimum Mode Following (CI-NEB-MMF)}
\label{alg:neb_mmf}
\begin{algorithmic}[1]
\State Initialize NEB path $\{\mathbf{R}_0, \dots, \mathbf{R}_P\}$
\While{not converged}
    \State Find highest energy image, $\mathbf{R}_{\text{climb}}$
    \State Calculate tangents $\hat{\tau}_i$ for all images
    \State Calculate NEB forces $\mathbf{F}_i^{\text{NEB}}$ for all non-climbing images using Eq. \ref{eq:neb_total_force}
    \State Calculate force for climbing image $\mathbf{F}_{\text{climb}}$ using Eq. \ref{eq:neb_ci}

    \State $F_{\text{max}} \gets \max_i(|\mathbf{F}_i^{\text{NEB}}|)$

    \If{$F_{\text{max}} < F_{\text{MMF\_threshold}}$ \textbf{and} MMF enabled}
        \State \Comment{Switch to local MMF refinement for the climbing image}
        \State Create a temporary MMF optimizer for $\mathbf{R}_{\text{climb}}$
        \For{$k=1$ to $N_{\text{MMF\_steps}}$}
            \State Update $\mathbf{R}_{\text{climb}}$ using a step from the MMF optimizer with $\mathbf{F}_{\text{MMF}}$ (Eq. \ref{eq:neb_mmf_force})
        \EndFor
        \State Update the full path's forces after MMF refinement
        \State Take a global optimization step on all images with their respective forces
    \Else
        \State \Comment{Perform standard NEB optimization step}
        \State Take a global optimization step on all images
    \EndIf
\EndWhile
\end{algorithmic}
\end{algorithm}
\subsubsection{Case Study: Isomerization of Ethylene Oxide to Acetaldehyde}
\label{sec:asd:eon:case}
\begin{figure}
\centering
\begin{subfigure}[b]{0.49\textwidth}
   \centering
   \includegraphics[width=\textwidth]{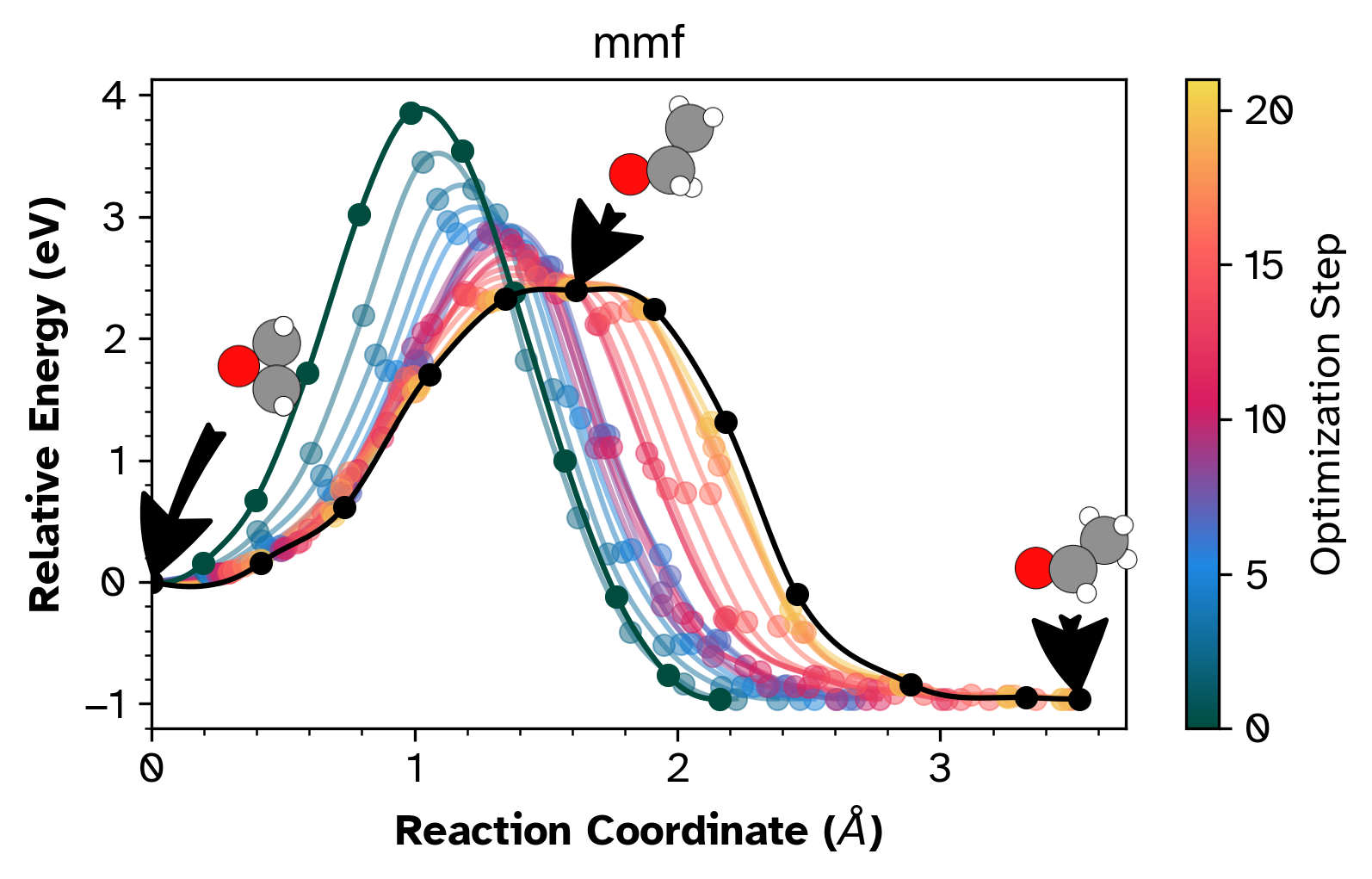}
   \caption{Hybrid CI-NEB-MMF}
   \label{fig:roneb}
\end{subfigure}
\hfill
\begin{subfigure}[b]{0.49\textwidth}
   \centering
   \includegraphics[width=\textwidth]{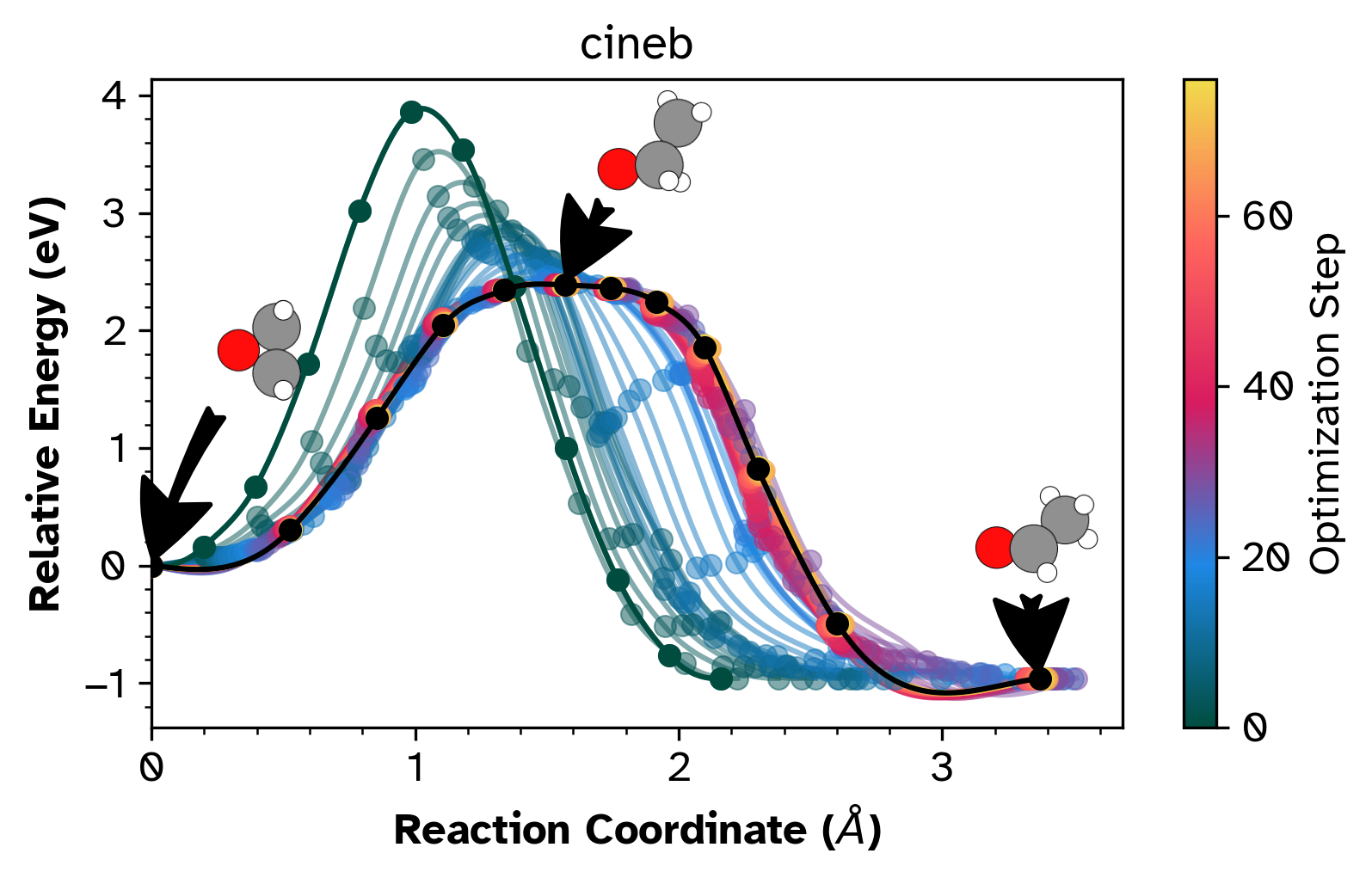}
   \caption{Standard CI-NEB}
   \label{fig:cineb}
\end{subfigure}
\caption{Comparison of the optimization process for the ethylene oxide to acetaldehyde isomerization using (a) the hybrid CI-NEB-MMF and (b) the standard CI-NEB methods. Each colored line represents the reaction path at a specific point in the optimization, progressing towards the final, converged path (based on the climbing image) shown in black. The reaction coordinate on the x-axis is defined as the cumulative Cartesian distance (\AA) between successive images along the path. While both methods find the identical transition state estimate, the color bars highlight the significantly greater efficiency of the hybrid method, which converges in approximately 20 steps, whereas the standard method requires over 70 steps.}
\label{fig:combneb}
\end{figure}

To demonstrate the effectiveness and computational efficiency of the hybrid
CI-NEB-MMF method, it was applied to the isomerization reaction of ethylene
oxide to acetaldehyde, the results of which are shown in Figure
\ref{fig:combneb}.

The reaction involves the rearrangement of ethylene oxide, a three-membered
cyclic ether (epoxide), into its more stable isomer, acetaldehyde. The primary
thermodynamic driving force for this exothermic reaction is the release of
significant ring strain present in the ethylene oxide molecule. The C-C-O bond
angles in the epoxide ring are constrained to approximately \(60\deg\), a severe
deviation from the ideal \(109.5\deg\) for \$sp\textsuperscript{3}\$-hybridized atoms. This strain
makes ethylene oxide a high-energy, reactive species. The rearrangement allows
the ring to open, forming the more stable carbonyl and methyl groups of
acetaldehyde and releasing the stored strain energy.

\begin{wrapfigure}{l}{0.5\textwidth}
\centering
\includegraphics[width=0.48\textwidth]{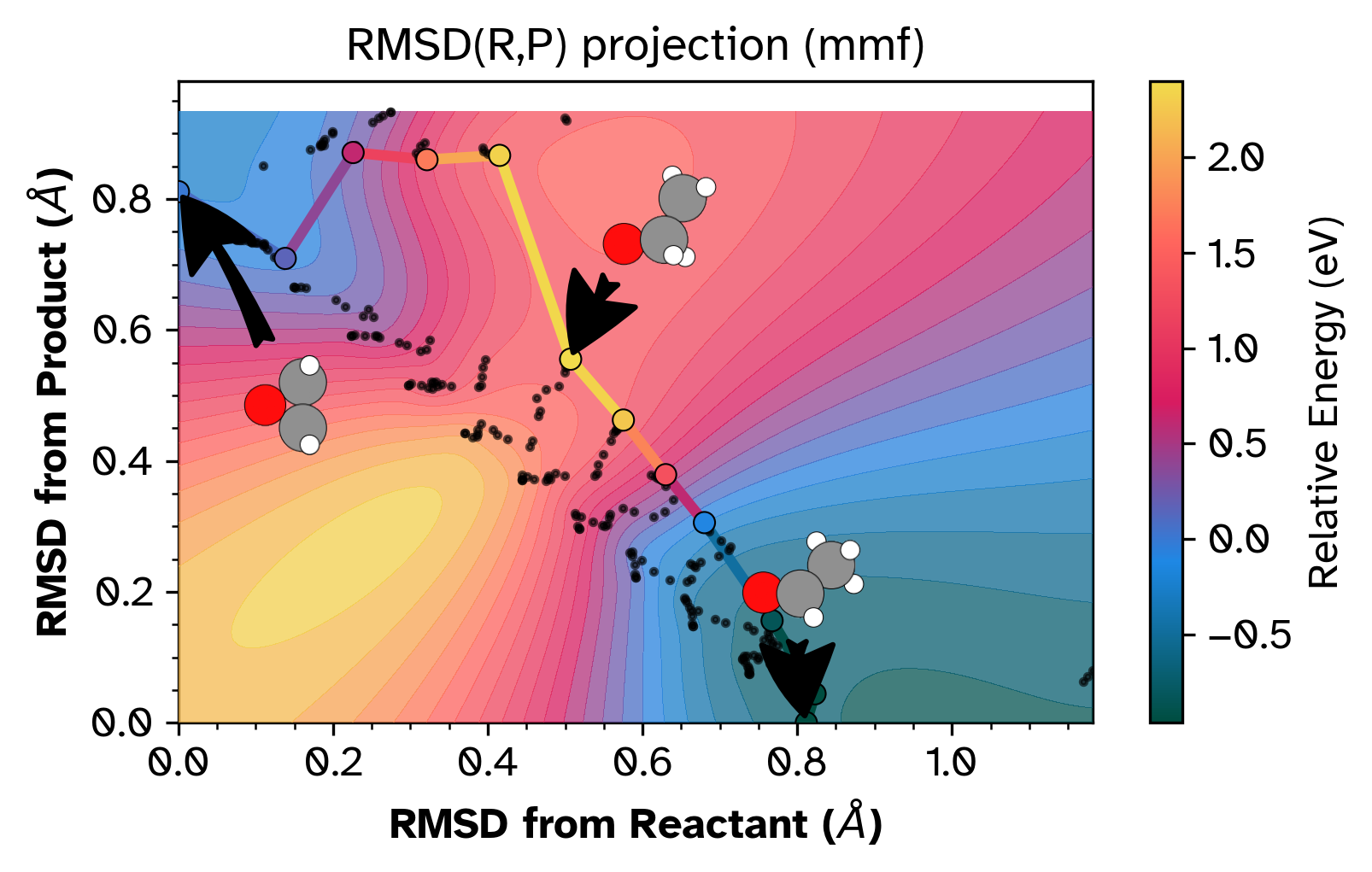}
\caption{\label{fig:viz_analysis}2D landscape projection of the converged hybrid CI-NEB-MMF path for the ethoxy acetal system. The trajectory is plotted on a coordinate system of \gls{rmsd} from the reactant vs. \gls{rmsd} from the product. Black points represent the discrete sampling history during optimization. The interpolated energy contours (color scale in eV) reveal the topography of the potential energy surface, highlighting the minimum energy path connecting the metastable reactant to the stable product via a high-energy transition state.}
\end{wrapfigure}

While the 1D projection in Figure \ref{fig:combneb} is essential for comparing
optimization efficiency and visualizing the energy barrier, it fundamentally
compresses the multidimensional pathway into a single coordinate. We can
visualize the geometric progression of the reaction and the topographic context
of the \gls{pes} with the landscape projection method developed in section
\ref{sec:asd:pviz}, shown in Figure \ref{fig:viz_analysis}. This projects the
converged path onto a coordinate system defined by the \gls{rmsd} of the
endpoints. The trajectory traces the channel from the reactant basin (top-left,
\(RMSD_P \approx 0.8 \text{ \AA}\)) to the product basin (bottom-right, \(RMSD_R
\approx 0.8 \text{ \AA}\)).

The path is overlaid on an energy surface interpolated from the discrete image
data (scattered black points). This ``top-down'' view provides clear mechanistic
insights detailed in \textcite{goswamiTwodimensionalRMSDProjections2025}. The
reactant valley (blue, top-left) represents the metastable basin corresponding
to the strained ethylene oxide molecule in the reactant state, at a relative
energy of approximately 0.0 eV. The path follows a ridge-line to cross the
transition state—the highest energy point on the path, indicated in bright
yellow/orange—located at an RMSD of \(\approx 0.50 \text{ \AA}\) from the reactant
and \(\approx 0.55 \text{ \AA}\) from the product. This saddle point corresponds
to a barrier exceeding 2.0 eV. Following the transition, the path descends into
the product valley (dark teal, bottom-right), which forms a broad, deep basin
with a relative energy below -0.5 eV. This visual confirmation of the product's
stability relative to the reactant aligns with the thermodynamic release of ring
strain expected in the ring-open acetaldehyde species.

The reaction was modeled using the PET-MAD
\cite{mazitovPETMADUniversalInteratomic2025} machine learning potential (v1.1.0)
through the novel Metatomic interface
\cite{bigiMetatensorMetatomicFoundational2025} implemented in EON. The entire
process, from initial path generation using the \gls{idpp}
\cite{smidstrupImprovedInitialGuess2014} method to the final \gls{neb}
calculations, was automated using a Snakemake workflow
\cite{molderSustainableDataAnalysis2021}. Both the standard and hybrid NEB
calculations started from identical, pre-minimized endpoints and an
IDPP-generated \cite{smidstrupImprovedInitialGuess2014} initial path from ASE
\cite{larsenAtomicSimulationEnvironment2017}. An \gls{lbfgs} optimizer was used to
relax the path until the maximum force on any image fell below the convergence
criterion of 0.01 eV/\AA{}.

In the standard CI-NEB calculation, the climbing image was activated once the
maximum force fell below 0.5 eV/\AA{}. In the hybrid CI-NEB-MMF calculation, the
climbing image was activated earlier at 1.5 eV/\AA{}, while the local minimum-mode
following refinement was triggered only after the force reached 0.5 eV/\AA{}. Up
to 20 \gls{mmf} steps were applied to the climbing image during each subsequent
NEB iteration.

\begin{table}[htbp]
\caption{\label{tbl:roneb}Performance comparison for the standard and hybrid NEB methods.}
\centering
\begin{tabular}{lrr}
Method & NEB Steps & PES Calls\\
\hline
Standard CI-NEB & 77 & 1564\\
Hybrid CI-NEB-MMF & 22 & 582\\
\end{tabular}
\end{table}

The hybrid approach required approximately 63\% fewer \gls{pes} calls and converged
in 71\% fewer NEB steps. This efficiency gain stems from the \gls{mmf} method's
ability to rapidly converge the climbing image to the saddle point once the
\gls{neb} path is in the correct region, avoiding the slow relaxation
characteristic of standard \gls{cineb} on flat or nearly flat potential energy
surfaces.
\subsection{Workflow engines}
\label{sec:asd:ba}
\begin{wrapfigure}{l}{0.5\textwidth}
\centering
\includegraphics[width=0.48\textwidth]{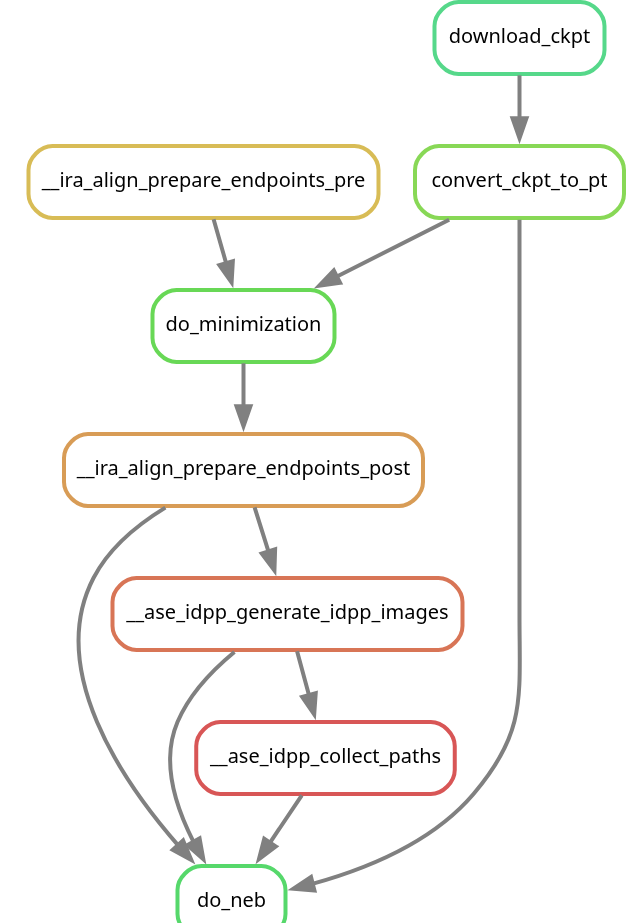}
\caption{\label{fig:neb_workflow}\gls{dag} ensures critical pre-processing steps, e.g. endpoint minimization and initial path generation, are systematically executed before the main \gls{neb}.}
\end{wrapfigure}

Modern computational science relies on complex, multi-step procedures that can be difficult to manage, reproduce, and scale. Workflow engines are software tools designed to address this challenge by providing a framework to define, execute, and automate these computational pipelines. For this work, the Snakemake workflow management system was used. Snakemake utilizes a Python-based, human-readable syntax to define a series of rules in a file known as a \texttt{Snakefile}. These rules, along with their specified input and output dependencies, implicitly form a \gls{dag}, which Snakemake automatically resolves to determine the correct order of execution for all required tasks.

A key advantage of Snakemake in a research environment is its seamless integration with \gls{hpc} resources. It abstracts the underlying job scheduler (e.g., Slurm, PBS, SGE), allowing the same workflow definition to be executed on a local machine for testing or scaled up to a large cluster for production runs. This portability ensures that the computational environment can be easily adapted without altering the scientific logic of the workflow itself.

Another critical function of a workflow engine in a scientific context is the ability to programmatically encode best practices and enforce reproducibility. Many computational methods, such as the \gls{neb} technique, require a specific sequence of preparatory and execution steps for reliable results. Manually performing this sequence can be tedious and prone to human error, such as forgetting a critical step or using mismatched model versions. By defining the entire protocol as a series of dependent rules, Snakemake transforms a manual checklist into a robust, automated, and self-documenting scientific component.

For example, a best-practice \gls{neb} workflow, as illustrated in Figure \ref{fig:neb_workflow}, involves several distinct stages: first, ensuring the reactant and product structures are fully minimized and consistently aligned; second, generating a sensible initial path between these mapped endpoints; and only then, executing the main NEB optimization. Additionally, the workflow enforces reproducibility by automatically fetching and converting the specific version of the machine learning potential required for the simulation. By encoding this logic in a \texttt{Snakefile}, one can guarantee that the initial path is never generated with unrelaxed or unaligned endpoints, thus preventing erroneous calculations and embedding expert knowledge directly into the computational tool. This approach ensures that every calculation is performed consistently and correctly, forming the foundation of truly reproducible research.
\subsection{Towards maximal concurrency}
\label{sec:asd:concurrency}
The architectural modernizations detailed in the preceding sections represent crucial steps away from a traditional, monolithic software design paradigm and towards a more flexible and powerful future. The dominant model in High-Performance Computing has historically relied on large, statically linked executables communicating via the Message Passing Interface (MPI). While effective for tightly-coupled, homogeneous tasks, this approach has significant drawbacks: component libraries must be chosen at compile time, leading to massive, inflexible binaries; the static allocation of resources can be inefficient; and the model is fragile, as a fault in any single component can terminate the entire multi-node calculation.

\begin{figure}[htbp]
\centering
\includegraphics[width=\textwidth]{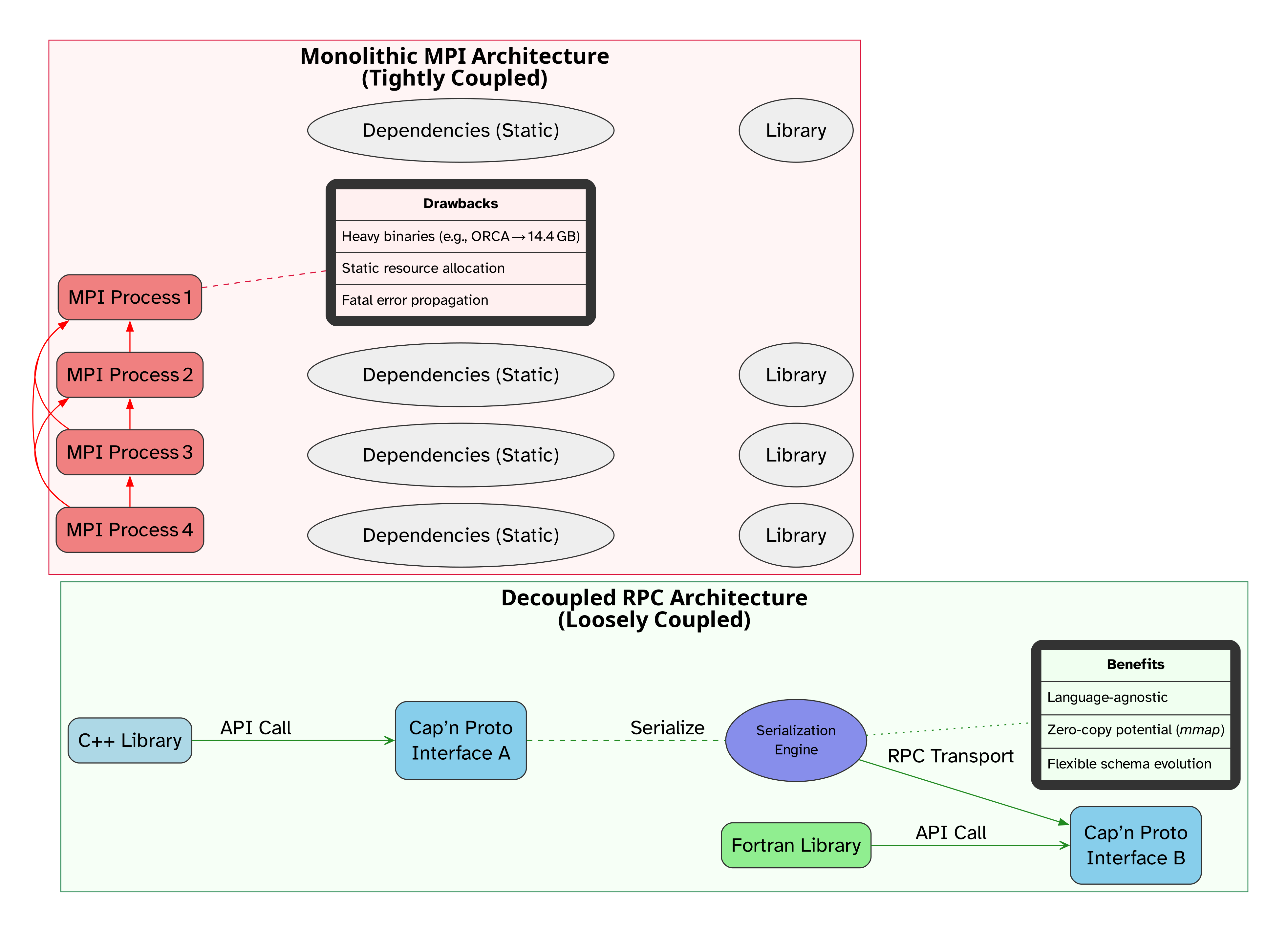}
\caption{\label{fig:rpc}A schematic comparison of two software architecture paradigms in scientific computing. (Left) The traditional monolithic model, based on MPI, statically links all dependencies into large, identical processes. This tight coupling results in heavy binaries, static resource allocation, and system-wide fragility where an error in one process can be fatal to the entire calculation. (Right) The modern decoupled model separates components into independent services (e.g., C++ and Fortran libraries) that communicate through a well-defined, language-agnostic \gls{rpc} interface. This loose coupling enables interoperability, modularity, and flexibility, allowing components to be developed and deployed independently.}
\end{figure}

Legacy scientific codes, often developed over decades, typically feature tightly-coupled components that communicate through global state. A canonical example of this design is the Runtime Database (RTDB) in NWChem. The RTDB functions as a centralized, string-keyed, key-value store—a clever design for its time to decouple modules from the input file, but one that comes at the cost of type safety and creates a strong dependency on a single, shared resource that complicates external interoperability. This monolithic model hinders rapid prototyping, the integration of new tools, and the creation of flexible, polyglot workflows, conceptually shown in Figure \ref{fig:rpc}.

A more robust and flexible paradigm, inspired by modern distributed systems, recasts scientific workflows as a collection of smaller, decoupled services that communicate ``on the wire.'', e.g. through ZeroMQ. Realizing this vision, however, presents two orthogonal philosophies for achieving high-performance interoperability.

The first philosophy relies on a common shared library and a standardized C Application Binary Interface (ABI). The Metatensor library exemplifies this approach \cite{bigiMetatensorMetatomicFoundational2025}. It defines a language-agnostic C ABI and a strict in-memory data layout for its core tensor structures. This allows a library written in Rust to operate on the exact same memory buffer created by a Python script without any serialization or data copying, achieving true zero-copy performance. The trade-off is a strong dependency at the build and link stages: all components must compile and link against the same version of the shared library, which can lead to a cascade of bindings (e.g., from C++ to a C-API, then to Python) to achieve interoperability.

Part of this work demonstrates an initial step in this direction with the client-server architecture for potential energy calculations, and parameters within the \gls{gp} and EON. The logical extension of this concept is a system of interchangeable, polyglot libraries communicating through \gls{rpc}. In this model, a Python-based workflow engine could orchestrate a simulation by sending requests to a high-performance Fortran optimizer, which in turn queries a potential energy surface provided by a C++ machine learning library, with each component running concurrently on the most appropriate hardware.

\begin{listing}[htbp]
\begin{minted}[]{capnp}
@0xbd1f89fa17369103;

struct ForceInput {
  natm   @0 :Int32;
  pos    @1 :List(Float64);
  atmnrs @2 :List(Int32);
  box    @3 :List(Float64);
}

struct PotentialResult {
  energy @0: Float64;
  forces @1: List(Float64);
}

interface Potential {
  calculate @0 (fip :ForceInput) -> (result :PotentialResult);
}
\end{minted}
\caption{A Cap'n Proto schema defines a strict, language-agnostic contract for \gls{rpc}, based on \texttt{potlib}}
\end{listing}

This schema acts as an unambiguous contract that completely decouples the client and server. The Cap'n Proto compiler auto-generates the necessary code, enabling a C++ server to communicate seamlessly with a Python client, even across a network. This design provides strong type safety, a stark contrast to the RTDB's untyped lookups, and remarkable flexibility. Because the schema can evolve, a client can ignore new fields it does not understand, allowing for independent updates without downtime. Furthermore, for co-located processes, the serialized message can be memory-mapped (\texttt{mmap}) \cite{vahaliaUNIXInternalsNew1996}, providing a path to zero-copy communication without the rigid dependency of a shared library.

Both philosophies work towards the same grander vision: a ``BLAS for computational science.'' Just as BLAS \cite{vandegeijnBLASBasicLinear2011} standardized low-level linear algebra, a future ecosystem could be built upon standardized high-level interfaces for tasks like geometry optimization or kinetic Monte Carlo. The choice of implementation, a tightly-coupled C-ABI for maximum on-node performance, or a loosely-coupled RPC for maximum flexibility and distributability, or both, would become a design decision rather than a fundamental limitation. This architecture represents the future of scientific software: a federated system of specialized, best-in-class tools, seamlessly interoperable, enabling maximal concurrency and accelerating the pace of discovery.

With these concepts and pre-emptively developed tools, we can return to the problem of discovering sadde points.
\subsection{Conclusions}
\label{sec:asp:conc}
The algorithmic and architectural developments presented in this chapter serve a specific, unifying purpose: to extend the timescale and complexity of accessible chemical transformations. While standard computational packages offer some solutions for routine equilibrium calculations, the exploration of rare events on high-dimensional potential energy surfaces imposes strict requirements on efficiency, stability, and reproducibility that monolithic software architectures cannot meet.

We have established that the ``technical'' details of implementation are inseparable from the validity of the physical model. The distinction between geometric and electronic criteria for bonding (Section \ref{sec:asp:bondy:wbo}) enables interpretations of saddle geometries. Similarly, the projection and interpolation methods developed for path visualization (Section \ref{sec:asd:pviz}, \cite{goswamiTwodimensionalRMSDProjections2025}) provide the necessary topological verification that the computed minimum energy pathways connect the intended basins of attraction, a non-trivial confirmation in complex molecular rearrangements.

Furthermore, the restructuring of the EON framework demonstrates that algorithmic efficiency determines scientific feasibility. We demonstrate this through the hybrid CI-NEB-MMF optimizer (Section \ref{sec:asp:eon:roneb}), which reduces the computational cost of double ended saddle point searches by an order of magnitude \footnote{submission in progress to Frontiers chemistry.}. Thus, modularization in code allows for the determination of transition states with machine learned interatomic potentials that are otherwise untreatable by standard \gls{cineb} methods due to pathologies of the landscape or the cost of the electronic structure theory. This concept leads to the more wide ranging concept of a distributed micro-service ecosystem for interopble code, a fom of ``chemical BLAS (Basic linear algebra subprograms \cite{vandegeijnBLASBasicLinear2011})'', leading to a new epoch of standardization.

Finally, the encoding of these procedures into a formal Directed Acyclic Graph via the Snakemake workflow engine (Section \ref{sec:asd:ba}) moves the methodology beyond manual execution. By strictly defining the dependencies between minimization, alignment, and path optimization, we ensure that the complex simulation protocols are robust against operator error and computationally reproducible.

This infrastructure forms the necessary foundation for the work in the subsequent chapter \footnote{Chronologically developed alongside all chapters.}. The application of \gls{gpr} to saddle point searches with the \gls{gpd} algorithm requires an underlying engine capable of rapid, asynchronous querying of the potential and robust error recovery. We demonstrated how this software environment minimizes overhead and maximizes modularity while providing diagnostic. We can now move towards addressing the theoretical challenge of constructing data-efficient surrogate models for on-the-fly exploration of the energy landscape, demonstrating state of the art saddle search methods using local \gls{gp} acceleration in a production setting.
\section{Efficient Gaussian Process Regression}
\label{sec:gpjctc}
\epigraph{With four parameters I can fit an elephant, and with five I can make him wiggle his trunk.}{John von Neumann}

\begin{quote}
This chapter is based on \fullcite{goswamiEfficientImplementationGaussian2025a}
\end{quote}

\begin{figure}[htbp]
\centering
\includegraphics[height=0.9\textheight]{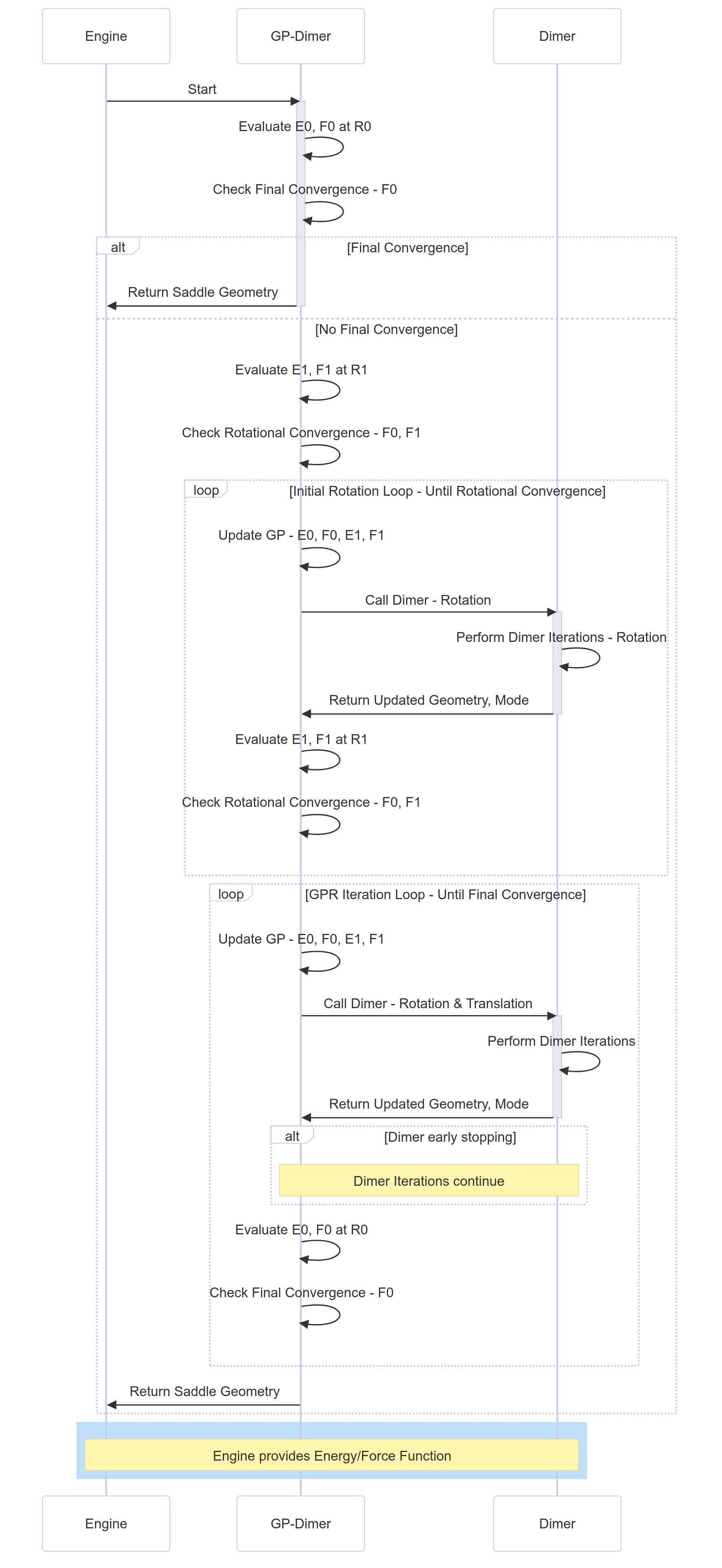}
\caption{\label{fig:gp_dimer_alg}The \gls{gpd} method as an entity-relation diagram showing connections to EON and the Dimer method.}
\end{figure}

In modern computational chemistry, the discovery process is often a fragmented and manual workflow. A researcher might use one high-performance engine to calculate energies (e.g., ORCA \cite{neeseORCAQuantumChemistry2020}, Psi4 \cite{turneyPsi4OpensourceInitio2012}, NWChem \cite{apraNWChemPresentFuture2020}), export the results to a text file, import that data into a scripting environment (e.g., MATLAB, Python) for analysis, and finally use a specialized library (e.g., GPStuff \cite{vanhataloGPstuffBayesianModeling}) for machine learning. This process, while functional for a single system, is untenable at scale. By establishing a clear, internal representation of the energy and force engines, and decoupling algorithms from specific data implementations, we can create a framework that replaces the brittle manual workflow with a robust and scalable platform as discussed in the previous chapter.

This chapter presents the \texttt{gpr\_optim} as a concrete first step towards solving this workflow problem for scientific discovery. The core contribution of this work is afterall, not only to achieve state of the art \footnote{always a moving target} performance in terms of the number of calculations, but also to provide an architectural blueprint that demonstrates how to move away from monolithic, single-purpose applications towards an interoperable ecosystem.
\subsection{Design}
\label{sec:gpjctc:design}
At its highest level, a simulation is a stateful process. The GPR model maintains the state of the learned \gls{pes} in terms of the internal state, which includes training data, hyperparameters, matrix decompositions; and the Dimer method maintains the state of the geometric search, i.e. the dimer position, orientation, optimization history. To orchestrate this at a per-instance scale, EON is used, as it provides generics for potentials, and the parallelism across systems is handled by Snakemake. Thus the overall framework captures this scientific endeavor with an object-oriented design.

The algorithm itself proceeds as described in Alg. \ref{alg:gpr}.

\begin{algorithm}
\caption{GPR Prediction (Energy and Gradient)}
\label{alg:gpr}
\begin{algorithmic}[1]
\State \textbf{Given:} Training data $\{\mathbf{X}, \mathbf{y}\}$, new configuration $\mathbf{x}^*$, covariance function $k(\mathbf{x}_i, \mathbf{x}_j; \theta)$ with hyperparameters $\theta$.
\State \textbf{Training Phase (executed once per hyperparameter update):}
\State Construct the training covariance matrix $K$ where $K_{ij} = k(\mathbf{x}_i, \mathbf{x}_j; \theta)$.
\State Add observation noise: $K_y = K + \sigma_n^2 I$.
\State Perform Cholesky decomposition: $L = \text{chol}(K_y)$.
\State Solve for $\alpha = L^T \backslash (L \backslash \mathbf{y})$. \Comment{This is the weight vector}
\State
\State \textbf{Prediction Phase:}
\State Compute the vector of covariances between the new point and training points: $\mathbf{k}^* = [k(\mathbf{x}^*, \mathbf{x}_1), \dots, k(\mathbf{x}^*, \mathbf{x}_n)]^T$.
\State Predict mean energy: $\bar{E}^* = (\mathbf{k}^*)^T \alpha$.
\State Predict mean gradient: $\bar{\mathbf{g}}^* = (\nabla_{\mathbf{x}^*}\mathbf{k}^*)^T \alpha$.
\State \Return Predicted energy $\bar{E}^*$ and gradient $\bar{\mathbf{g}}^*$.
\end{algorithmic}
\end{algorithm}

Hyperparameters are optimized with the \gls{scg} \cite{mollerScaledConjugateGradient1993} detailed in Alg. \ref{alg:scg-concise}. Since the energy and forces are modeled as a single output vector, the \gls{scg} is required for stability, though the code also implements a per-iteration fixed factor scaling for other optimizers like ADAM \cite{kingmaAdamMethodStochastic2017}. We find that the \gls{scg} is more efficient as implemented.

\begin{algorithm}
\caption{Scaled Conjugate Gradient (SCG)}
\label{alg:scg-concise}
\begin{algorithmic}[1]
\Require Initial weights $\mathbf{w}_0$, data $\mathbf{x}$, target $\mathbf{y}$, loss $f$, gradient $\nabla f$
\State Initialize search direction $\mathbf{p}_0 = -\nabla f(\mathbf{w}_0)$ and scaling factor $\lambda = 1$
\For{optimization iterations}
    \State Compute gradient difference $\mathbf{r}_k = \nabla f(\mathbf{w}_k) - \nabla f(\mathbf{w}_{k-1})$
    \State Compute curvature approximation from gradient differences
    \State Scale search direction and evaluate function and gradient at a trial point
    \State Compute step size $\alpha$ using the scaled curvature
    \State Update weights: $\mathbf{w}_{k+1} \leftarrow \mathbf{w}_k + \alpha \mathbf{p}_k$
    \State Check for non-finite function values and adjust step size if needed
    \State Check convergence criteria
    \State Adapt scaling factor $\lambda$
    \State Update search direction using Conjugate Gradients (or restart periodically)
\EndFor
\State \Return best weights found
\end{algorithmic}
\end{algorithm}

Finally the dimer itself is translated and rotated through the \gls{lbfgs}, which is indpendently implemented within the codebase. The scaling of the \gls{gp} methodology in this chapter proceeds as described in Section \ref{sec:scal:timestor}, that is, exhibiting \(O(M^2 N^2)\) complexity for storage and \(O(M^3 N^3)\) for training dominated by the inversion of the covariance matrix, whereas the inference cost scales linearly with the size of the training set \(M\).
\subsection{Surface systems}
\label{sec:gpjctc:surfaces}
\begin{wrapfigure}{l}{0.5\textwidth}
\centering
\includegraphics[height=0.2\textheight]{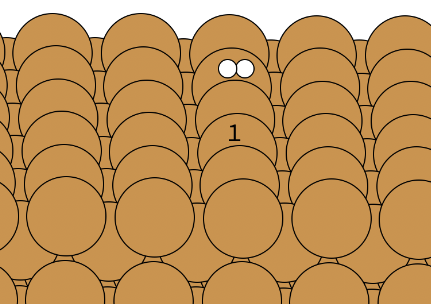}
\caption{\label{fig:cuh2}Hydrogen molecule dissociating on a copper slab. Including the two nearest copper atoms to any moving hydrogen atom in the active set proved to further accelerate convergence, achieving a runtime of just 4.3 seconds.}
\end{wrapfigure}

To handle extended systems, a finite cutoff is taken for determining pairs. Active pairs are updated on each new iteration. For molecular systems, the cutoff is arbitrarily large. The copper hydrogen dissociation in Figure \ref{fig:cuh2} has been the unit test for the development, and the MATLAB results of 230.6 seconds dropped reliably to 12.9 seconds \footnote{From \texttt{gpr\_optim} at commit \texttt{a4d1fdaa0ed943d0c9e8b2931db12a4148be0ba4}}.

Radial cutoffs are generally used \cite{koistinenNudgedElasticBand2019} to define the local environment. These can fail in complex systems \cite{goswamiDSEAMSDeferredStructural2020}. Structural defects, interfaces, or thermal fluctuations cause the number of atoms within a fixed radius to vary, creating an incompatibility with machine learning models that require a fixed-size input vector.

To resolve this, the radial cutoff has been replaced with a nvel nearest-neighbor selection scheme\footnote{From HaoZeke/gpr\textsubscript{optim} at a837ec75f537c551d058e7a170ec880031879f52}. This approach guarantees a fixed number of atoms are chosen to represent the local environment, ensuring a consistent number of degrees of freedom for the model's input. The environment is constructed by selecting a constant number of atoms that are closest to any of the primary moving atoms (e.g., the dissociating molecule or a hopping adatom).

The number of user defined parameters remains the same, swapping a radial cutoff for the number of neighbors to consider. This directly controls the scaling of the method, though sensitivity analyses show that simply increasing the number of atoms does not necessarily improve model accuracy. Future studies will involve using hyperparameter searches \cite{watanabeTreeStructuredParzenEstimator2023} with surrogates on more diverse systems.

The results here highlight the manner in which the nearest-neighbor cutoff effectively decouples the computational cost from the total system size, solving the scalability issue inherent to GPR without succumbing to the instability of radial cutoffs. While the remainder of this thesis focuses on molecular benchmarks, this validation ensures the architecture remains applicable to extended systems for future applications.
\subsection{Gas phase molecular benchmark}
\label{sec:datadrege}
Throughout this thesis, a data set of small organic molecules curated by
\textcite{hermesSellaOpenSourceAutomationFriendly2022} has been used.  This
consists of 500 initial configurations of small gas-phase organic molecules,
ranging from 7 to 25 atoms.

In all cases, the \gls{pes} is evaluated at the \gls{hf} level of theory with the
3-21G basis set, using the NWChem software package
\cite{apraNWChemPresentFuture2020}.  The calculations employ a spin-restricted
formalism for singlet states and a spin-unrestricted formalism for doublet
states. The \gls{scf} convergence threshold is set to \(10^{-8}\) Hartree. A saddle
point search is considered converged when the maximum per-atom norm of the
atomic forces falls below 0.01 eV/\AA{}.

We highlight that the criteria for inclusion in the dataset is that \texttt{sella}
converges for each system. Automated tests are carried out to ensure a single
negative eigenvalue is present at the saddle configuration. However, due to the
inherent selection bias, success is not considered to be a factor in assessing
performance in the work introducing the benchmark, since, by construction,
\texttt{sella} failures are excluded \cite{hermesSellaOpenSourceAutomationFriendly2022}.
This circular dependency limits in some sense the ability to compare the results
across algorithms.

\begin{figure}[htbp]
\centering
\includegraphics[height=0.3\textheight]{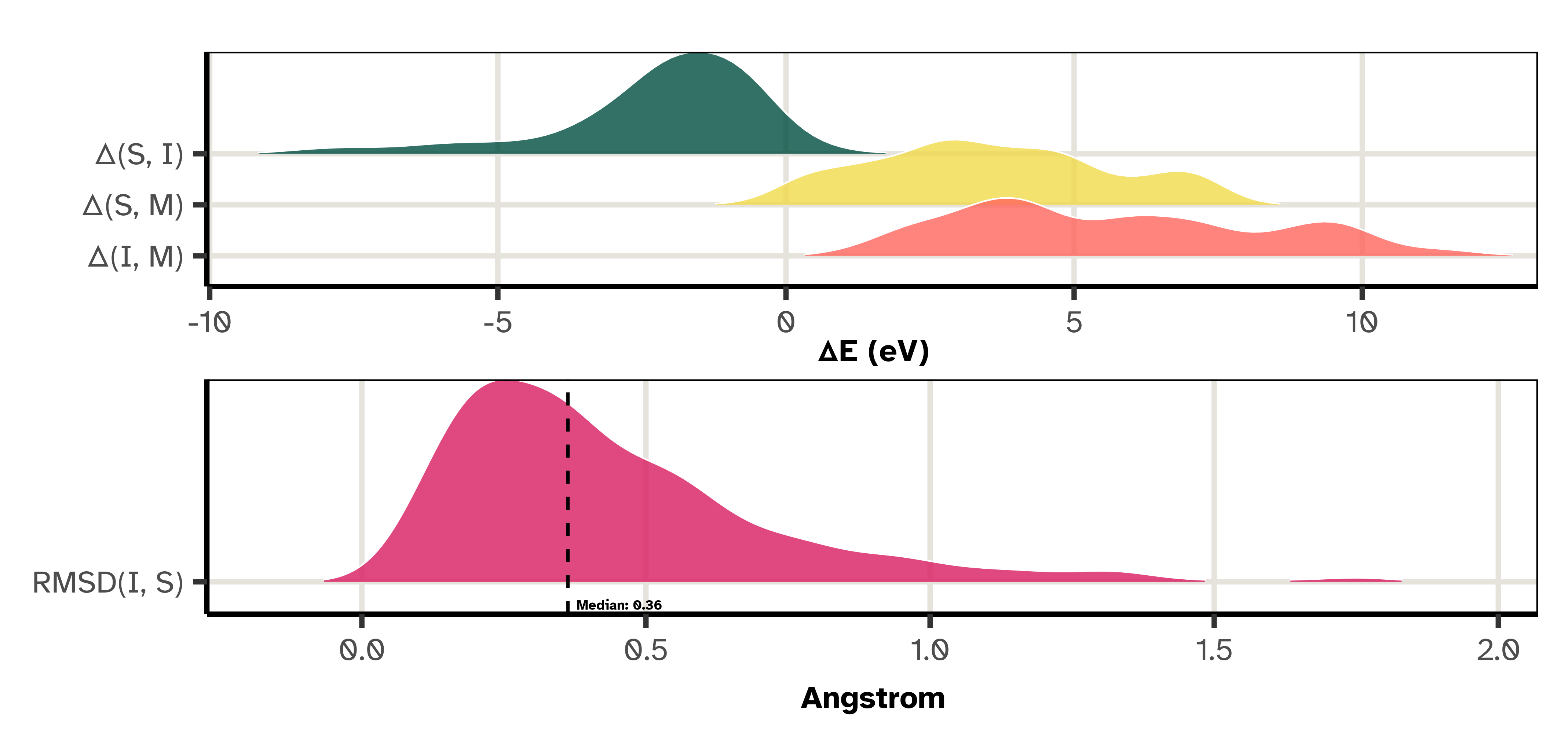}
\caption{\label{fig:sella_propdist}Distribution of Energetic and Structural Properties for the Sella Transition State Dataset. \textbf{(Top)} Probability density of energy differences: \(\Delta\)(S, I) between the final saddle and initial geometry, \(\Delta\)(S, M) between the final saddle and the minimized initial geometry, and \(\Delta\)(I, M) between the initial geometry and its minimized form. The distributions, particularly for \(\Delta\)(S, I), are extremely broad, spanning nearly 20 eV. This range far exceeds expected energy barriers at reasonable temperatures and pressures for the small organic molecules in this dataset, suggesting many initial geometries are highly unstable and unrepresentative of approximate saddle points. \textbf{(Bottom)} Probability density of the root-mean-square deviation (RMSD) between the initial (I) and final saddle (S) structures. The distribution is highly skewed, with a median of 0.36 \AA{} and a significant tail extending to large structural deviations. This, coupled with the wide energy distributions, indicates that the optimization process often involves large, chemically questionable geometric changes rather than the refinement of a reasonable guess. Plotted from data published in \cite{goswamiEfficientImplementationGaussian2025a}.}
\end{figure}

It is also important to note that \gls{mmf} methods are often ``finishing'' methods,
after a reasonable guess to the saddle has been generated. To this end, many
single ended searches in practice are rarely over an eV or two away in energy
from the nearest minima or point of interest. We show in Figure
\ref{fig:sella_propdist} that the distribution of energy differences \((\Delta E)\)
is excessively broad, ranging over a span of nearly 20 eV.

The bottom panel in Figure \ref{fig:sella_propdist} displays the \gls{rmsd}
deviation between the initial (I) and final saddle (S) structures. The \gls{rmsd}
distribution peaks sharply, with a median near 0.36 \AA{}, yet a non-negligible
tail stretches toward large structural deviations. Thus, while most starting
guesses lie structurally close to their target saddle, a notable fraction
requires substantial geometric rearrangement.

A direct inspection of the energy distributions reveals significant energetic
instability in many initial guesses, a result that contrasts with the low median
RMSD. The \(\Delta(\text{S,I}) = E_S - E_I\) distribution (top panel, green) does
not center near 0 eV. Instead, its principal mode lies near \(-2.5\) eV and a broad
tail extends down to about \(-10\) eV. This pattern indicates that initial
geometries often occupy much higher electronic energy than their converged
saddle points.Comparing \(\Delta(\text{I,M})\) (red) and \(\Delta(\text{S,M})\)
(yellow) further quantifies this instability. The \(\Delta(\text{I,M})\) ridge
shifts to more positive energy relative to \(\Delta(\text{S,M})\). This
displacement simply reflects the \(E_I - E_S\) energy difference, reproducing the
finding from the \(\Delta(\text{S,I})\) plot and confirming that initial points
frequently have higher energies than the corresponding saddles.

This disconnect between structural proximity and energetic instability
introduces a substantive complication. A starting geometry that sits
energetically far from the saddle region may not lie in the quadratic basin that
leads to that saddle despite having proximal configurations in space (w.r.t the
\gls{rmsd}); instead, the high potential energy can place the start in a region of
the potential energy surface that grants access to multiple competing downhill
pathways. A local finishing method that follows the local gradient therefore
cannot guarantee convergence toward the structurally proximal saddle. The
elevated starting energy can allow the optimization to descend along an
alternative reaction coordinate and converge to an unrelated minimum or a
different stationary point. The tail in the RMSD(I,S) distribution corroborates
this concern, as many cases require large structural changes before convergence.
In sum, the dataset displays a pronounced disconnect between structural and
energetic closeness. Many initial geometries function as highly unstable
configurations that happen to lie near a saddle in Cartesian space but fail to
occupy the correct energetic basin, which in turn makes them unreliable
``reasonable initial structures''.
\subsection{Performance characteristics}
\label{sec:gpjctc:perfacc}
A comparative analysis of the \gls{gpd} and Sella methods was conducted on a
subset of 345 systems for which both optimizers converged to the same saddle
point, defined as having an energy difference of less than 0.01 eV. Although
this quantification does not rigorously rule out that the geometries of the
saddle point configurations are comparable, the distributional analysis and
general conclusions do not change. The primary metric for comparison is the
number of \gls{hf} calculations required to reach convergence. On average, the
performance of the two methods is comparable. The \gls{gpd} method required a
median of 29 \gls{hf} calculations, while Sella required a median of 31. Notably,
the \gls{gpd} achieves this efficiency using Cartesian coordinates
\cite{goswamiEfficientImplementationGaussian2025a}, whereas Sella employs
internal coordinates, which are generally considered more suitable for the
varied stiffness of molecular degrees of freedom
\cite{bakerLocationTransitionStates1996,pengUsingRedundantInternal1996,denzelGaussianProcessRegression2018a,hermesSellaOpenSourceAutomationFriendly2022}.
The \gls{gpd} proved more efficient in 57\% of the cases, on average reducing the
computational cost by 8 \gls{hf} calculations relative to Sella.

\begin{figure}[htbp]
\centering
\includegraphics[height=0.3\textheight]{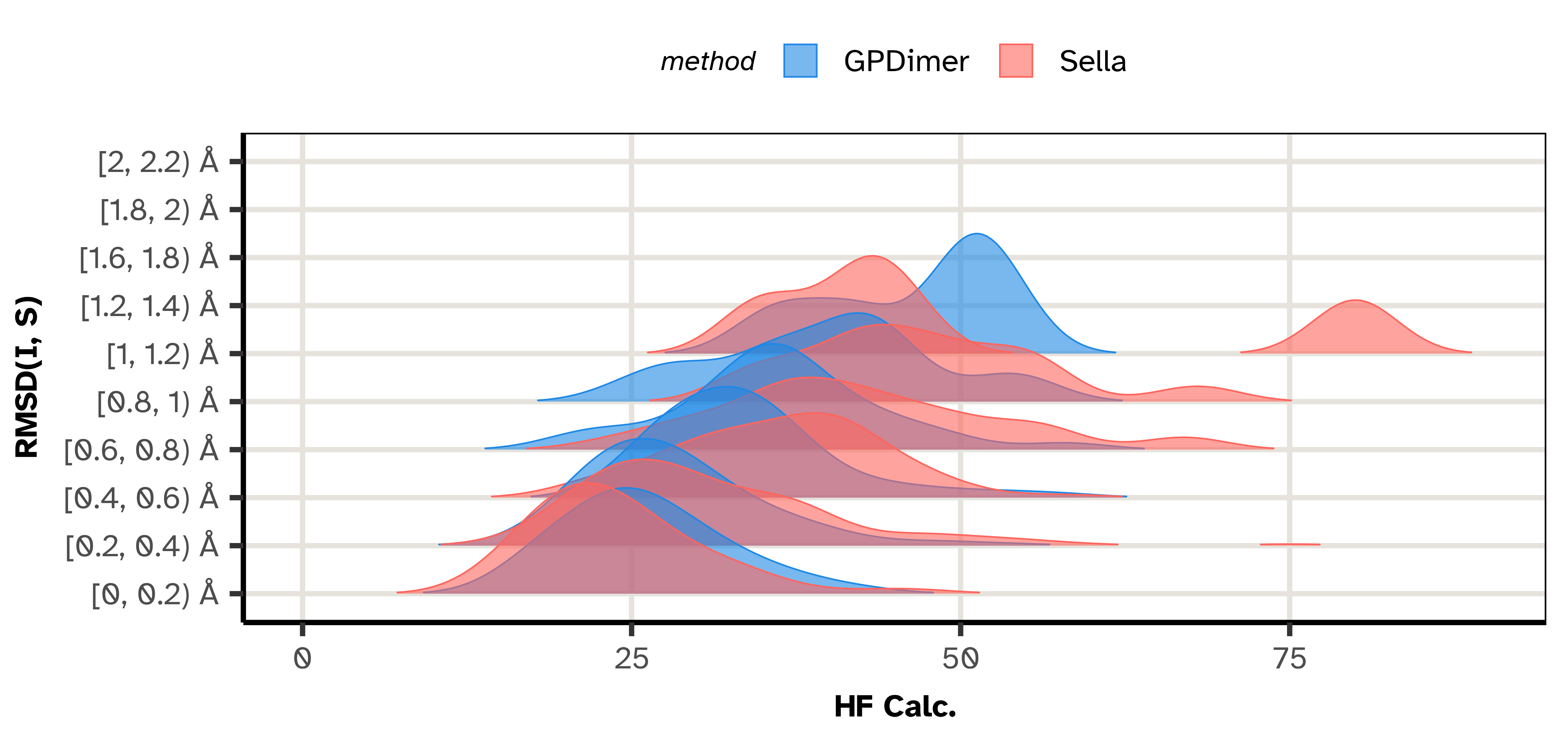}
\caption{\label{fig:sella_gpd_perf}Ridgeline plot showing the distribution of Hartree-Fock (HF) calculation counts required for convergence for the GPDimer (blue) and Sella (red) methods. The data fall into bins according to the root-mean-square deviation (RMSD) between the initial and final saddle geometries. At higher RMSD values (> 0.6 \AA{}), GPDimer shows clear efficiency. The methods perform comparably in the intermediate RMSD range. At high RMSD values, GPDimer again holds a performance advantage. The visualization confirms that algorithm efficiency depends strongly on the quality of the initial guess. Plotted from data published in \cite{goswamiEfficientImplementationGaussian2025a}.}
\end{figure}

Aggregate statistics alone obscure a key relationship. Figure
\ref{fig:sella_gpd_perf} reveals how computational cost depends on the quality of
the initial guess. For initial structures close to the final saddle point,
possessing a \gls{rmsd} below 0.6 \AA{}, the \gls{gpd} often requires slightly fewer
\gls{hf} calculations, as the initial surfaces constructed are not optimal. Its
efficiency advantage increases as the initial structural deviation increases. In
the intermediate range, between 0.6 \AA{} and 1.2 \AA{} RMSD, the methods perform
almost identically. For the most difficult cases, with initial displacements
greater than 1.2 \AA{}, the \gls{gpd} again demonstrates superior efficiency.

The use of internal coordinates in Sella presents challenges for specific
molecular geometries. In systems that approach a near-linear arrangement of
three or more atoms, Sella introduces algorithmic ``ghost atoms'' to avoid
coordinate singularities. In the 10 systems where this occurred, Sella required
an average of 47.1 \gls{hf} calculations, representing an 18.8\% increase over its
typical performance of 39.6 \gls{hf} calculations (median of 31 from previous
analysis). The GPR-dimer, which operates entirely in Cartesian coordinates,
showed no such penalty and needed an average of 35 \gls{hf} calculations on the
same subset, consistent with its overall performance of 34.5 calculations. These
results confirm that the Sella internal coordinate framework introduces
additional computational overhead when handling challenging geometries, and is
explored further in Sec. \ref{sec:otgpd}.
\subsection{Cataloging saddles}
\label{sec:gpjctc:sadcat}
As a diagnostic, the \gls{neb} (described in Section \ref{sec:theo:neb} and Section
\ref{sec:asp:eon:roneb}) can be used to determine the quality of the saddles found
by \texttt{sella} and the \gls{gpd}. We examine the first system in the benchmark, \texttt{singlet
000}, a 16-atom acyclic ether (\(\mathrm{C}_5\mathrm{OH}_{10}\)), for which both \texttt{sella} and
\gls{gpd} converge to distinct saddle points.

The \gls{gpd} identifies a hydrogen transfer saddle point with a \gls{rmsd} of just
0.2 \AA{} after only 23 HF calculations, while Sella locates a saddle
corresponding to a methyl group rotation which is lower in energy by 0.4 eV, but
the configuration is significantly further away (\gls{rmsd} of 0.6 \AA{}).

\begin{figure}[htbp]
\centering
\includegraphics[width=.9\linewidth]{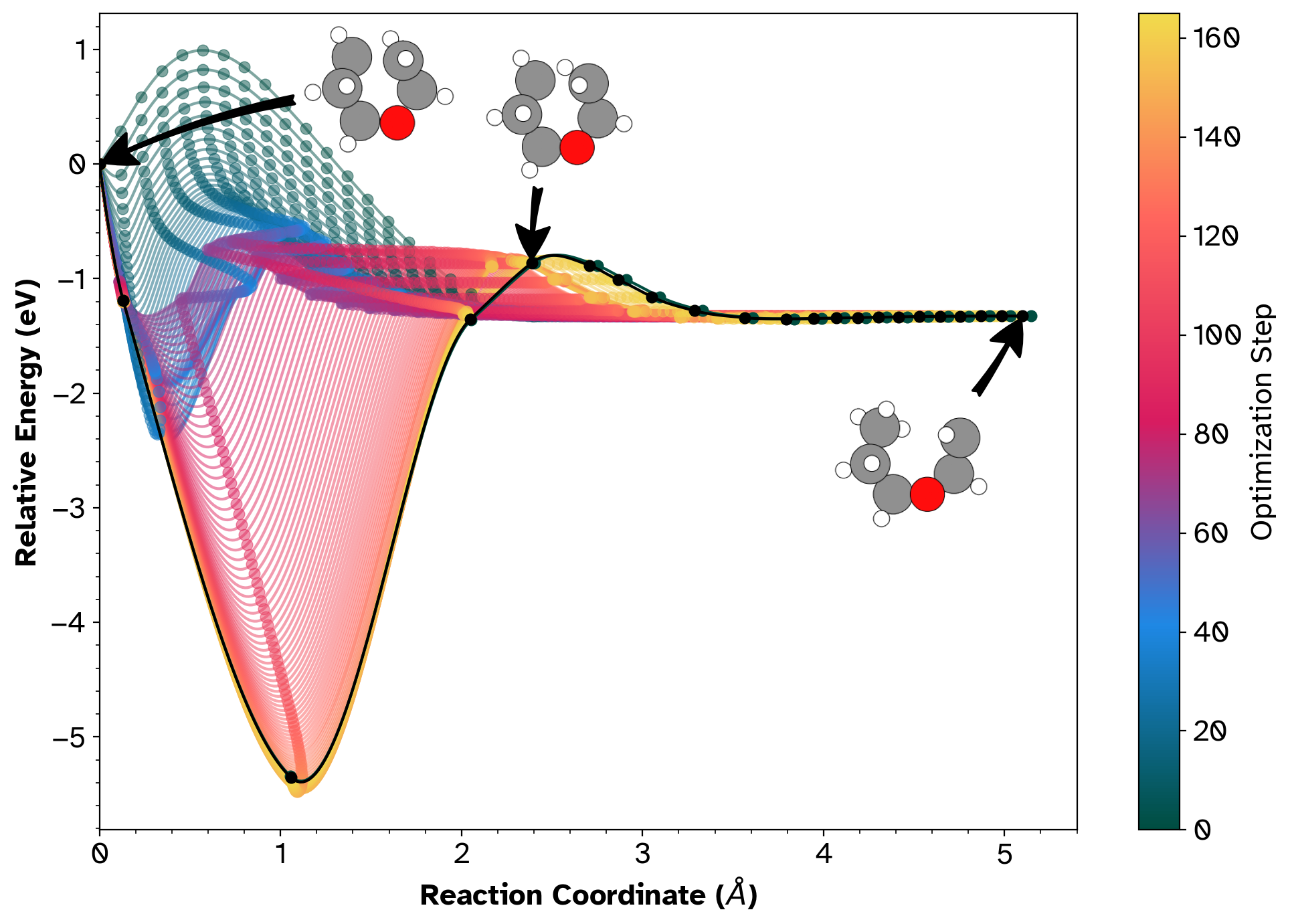}
\caption{\label{fig:s000_rcpath}The optimization history of an NEB connecting the initial reactant (left, \(s=0\)) to the Sella-located saddle point (right, \(s \approx 5.2 \text{ \AA}\)). The pathway reveals that the initial structure is not metastable; it relaxes barrierlessly into a deep intermediate minimum (\(s \approx 1.1 \text{ \AA}\), \(E \approx -5.3 \text{ eV}\)). To reach the Sella saddle, the system must subsequently climb out of this minimum, crossing a distinct transition state (peak at \(s \approx 2.4 \text{ \AA}\)) before proceeding to the distal Sella endpoint.}
\end{figure}

Since energy alone cannot be considered sufficient for cataloging saddles
\cite{poberznikPARTnPluginImplementation2024} we explore the connectivity
between the initial geometry and the Sella saddle, we employ the \gls{roneb}
protocol (Section \ref{sec:asp:eon:roneb}), leveraging the fact that \gls{neb} bands
can be formed between arbitrary points. The resulting optimization history,
plotted against the 1D path coordinate (defined in Section \ref{sec:asd:pviz}, Eq.
\ref{eq:path_coord}), is shown in Figure \ref{fig:s000_rcpath}. The ``reactant''
configuration is the initial configuration (at \(s=0\), \(E=0 \text{ eV}\)), and the
``product'' is the Sella-located saddle (at \(s \approx 5.2 \text{ \AA}\), \(E
\approx -1.3 \text{ eV}\)).

\begin{figure}[htbp]
\centering
\includegraphics[width=0.8\textwidth]{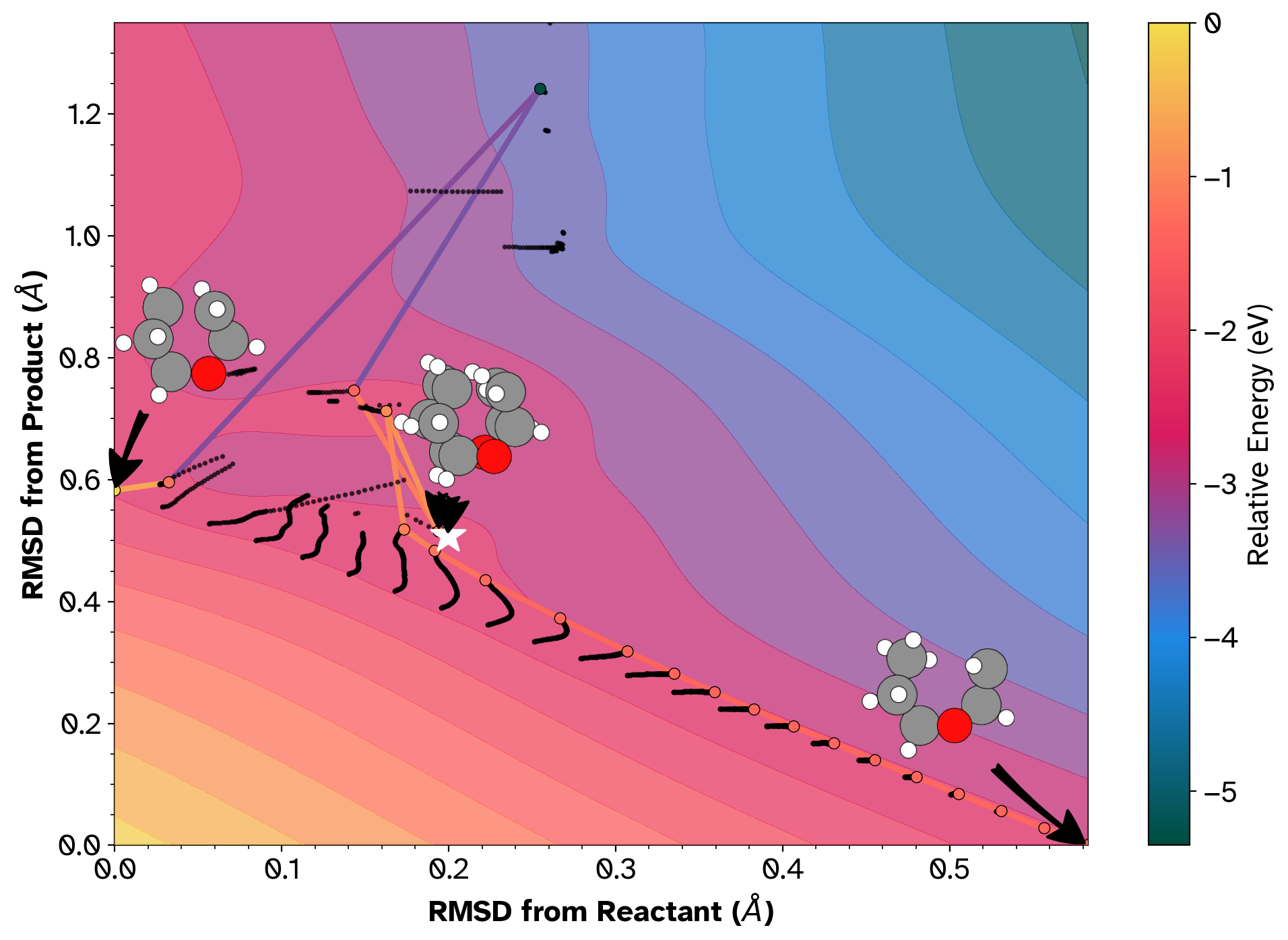}
\caption{\label{fig:s000_surface}A 2D landscape projection of the NEB optimization trajectory connecting the initial reactant (top-left) to the deep intermediate minimum (bottom-right). The white star explicitly marks the GPDimer saddle, which coincides exactly with the true transition state on this path, separating the deep intermediate minimum from the subsequent product channel. Sella's trajectory (red/pink path endpoints) overshoots this state to find a lower-energy, but distal, saddle point.}
\end{figure}

The 1D reaction coordinate profile (Figure \ref{fig:s000_rcpath}) shows that the
``initial'' configuration (\(s=0\)) is highly unstable and relaxes barrierlessly
into a deep intermediate minimum (\(E \approx -5.3 \text{ eV}\)). From this
localized basin, the system must cross a significant barrier (the peak at \(s
\approx 2.4 \text{ \AA}\)) to access the region containing the Sella saddle.The
critical finding is revealed when we project these states onto the 2D landscape
\cite{goswamiTwodimensionalRMSDProjections2025} (Figure
\ref{fig:s000_surface}).  The saddle point found by \gls{gpd} (marked by the white
star) coincides exactly with the barrier peak at \(s \approx 2.4 \text{ \AA}\).
This indicates that the Cartesian-based \gls{gpd} successfully identified the
proximal barrier governing the immediate relaxation of the system, whereas the
internal coordinate search bypassed this feature in favor of a distal stationary
point.  For automated kinetic network discovery (\gls{akmc}) as implemented in
EON, identifying the immediate connectivity is paramount; skipping proximal
barriers in favor of lower-energy distal states leads to disconnected reaction
networks and missing kinetic links.  Thus, GPDimer's ability to localize the
geometrically nearest saddle ensures the correct step-wise mapping of the
potential energy surface.  The result also underscores the necessity of
characterizing the local gradient manifold when sampling from high-energy
initial configurations.

\subsection{Conclusions}
\label{sec:gpjctc:conc}
This chapter presented \texttt{gpr\_optim}, an implementation of Gaussian Process
Regression designed to accelerate minimum mode following methods. By coupling
the dimer method with a local surrogate model, we achieved an order-of-magnitude
reduction in the number of electronic structure calculations required to locate
transition states compared to standard dimer methods.

A critical finding of this work is that the GPR-Dimer, operating in simple
Cartesian coordinates, achieves computational parity with—and often
outperforms—state-of-the-art methods utilizing complex internal coordinates
(Sella). This is a significant architectural advantage. For surface science and
catalysis, defining non-redundant internal coordinates is fraught with
difficulties due to coordinate singularities and the need for ``ghost atoms.'' The
results on the molecular benchmark demonstrate that the learning capability of
the GPR can effectively compensate for the lack of coordinate sophistication.
The model implicitly learns the local Hessian structure that internal
coordinates attempt to explicitly encode.

Furthermore, the detailed connectivity analysis using NEB diagnostics revealed
that ``convergence'' is an insufficient metric for success. In complex landscapes,
algorithms may converge to physically valid but chemically irrelevant saddle
points. The GPR-Dimer showed a robust tendency to locate the proximal saddle
point governing the immediate exit from the reactant basin, whereas aggressive
optimization in internal coordinates occasionally led to landscape traversal and
unrelated isomerizations.

Ultimately, this framework provides a robust, coordinate-system-agnostic engine
for saddle point searches. While the cubic scaling of exact GPR remains a
theoretical bottleneck, the implementation of local active sets (as demonstrated
with the CuH\(_2\) unit test) ensures the method remains viable for extended
systems.

Validation of results in a chemical context with efficient tools such as the
\gls{roneb} must be undertaken, as common point measures of success (single
negative eigenvalue, convergence, ``barrier'' from initial) may be misleading as
shown in the case study. In the twenty-first century, visual analysis is
infeasible, even for the 500 systems described in this thesis (Section
\ref{sec:datadrege}). Typical line plot comparisons and standard errors are
dangerously oversimplified for the physical chemistry context, and we address
this in the next chapter with Bayesian hierarchical models applied to modalities
of dimer searches.
\section{Dimer rotations and Hierarchical Bayesian models}
\label{sec:brmsgp}
\epigraph{In God we trust; all others must bring data.}{W. Edwards Deming}

\begin{quote}
This chapter is based on \fullcite{goswamiBayesianHierarchicalModels2025a}  with applications in \fullcite{goswamiAdaptivePruningIncreased2025b}
\end{quote}
\subsection{Revisiting dimer rotations}
\label{sec:brmsgp:dimrot}
A discriminating factor among minimum mode following methods, and indeed, one of
the core principles of the dimer algorithm, is to form an estimate of the
minimum mode without calculating the Hessian explicitly. The dimer, as described
in Section \ref{sec:theo:mmf}, uses an explicit rotation phase where the
coordinates of the image at the midpoint remain unchanged. This procedure,
designed to find the direction of lowest curvature, has evolved through several
key conceptual improvements.

The original formulation involved a single finite-difference step to estimate
the force gradient, followed by a single rotation to align the dimer
\cite{henkelmanDimerMethodFinding1999}. For more complex potential energy
surfaces, this strategy failed, and evolved into an iterative rotation process
\cite{olsenComparisonMethodsFinding2004}. This transforms the rotation phase
into a nested optimization problem: before each translation step, the dimer
orientation is iteratively rotated until it converges upon the minimum curvature
mode. The choice of numerical algorithm to perform this search profoundly
impacts the method's efficiency and reliability.

The simplest iterative approach restricts the search to a sequence of
two-dimensional plane. At each rotational step \(k\), this plane is defined by the
current dimer orientation, \(\hat{\mathbf{N}}_k\), and the normalized rotational
force, \(\hat{\mathbf{\Theta}}_k = \mathbf{F}^{\perp}_k
/ |\mathbf{F}^{\perp}_k|\). A new trial orientation, \(\hat{\mathbf{N}}(\phi)\), is
generated by a rotation within this plane:
\begin{equation}
  \hat{\mathbf{N}}(\phi) = \hat{\mathbf{N}}_k \cos\phi + \hat{\mathbf{\Theta}}_k \sin\phi
\end{equation}
The energy \(V_D(\phi)\) is then minimized as a one-dimensional function of the
angle \(\phi\). This method, however, can converge slowly.

Though \cite{kastnerSuperlinearlyConvergingDimer2008} demonstrated results on the
utility of the \gls{lbfgs} for translations, it has also been applied for
rotations, a quirk which has cast a long shadow filtering into many different
implementations including EON \cite{chillEONSoftwareLong2014}. Recall that, as
shown in Alg. \ref{alg:lbfgs}, \gls{lbfgs} is a quasi-newton method which constructs
a low-rank approximation of the inverse Hessian, \(\mathbf{B}_k \approx
\mathbf{H}^{-1}\), to determine a search direction. This approximation, however,
provides no guarantee that the resulting search direction will converge
specifically to the lowest eigenmode, \(\mathbf{v}_1\). The algorithm can be
deflected by or become trapped in subspaces corresponding to higher-energy
eigenmodes, particularly if their eigenvalues lie close to \(\lambda_1\). This can
cause the optimizer to fail in its primary task of identifying the true softest
mode \cite{lengEfficientSoftestMode2013}.

In contrast, methods based on the \gls{cg} algorithm align more naturally with the
mathematical structure of the problem. The \gls{cg} method and its variants, such
as the Lanczos algorithm, function as powerful iterative eigensolvers
specifically designed to find the extremal eigenpairs of a large symmetric
matrix. The process constructs a sequence of \(\mathbf{H}\) orthogonal search
directions that systematically and stably isolates the lowest eigenvector.
Furthermore, in the context of the dimer's outer loop of geometry steps, the
converged eigenvector from the previous step provides a high-quality initial
guess for the current rotation, a feature that the \gls{cg} method naturally
exploits to accelerate convergence. The theoretical argument therefore favors
\gls{cg} as the more mathematically appropriate and robust tool for the dimer
rotation phase.

While theoretical arguments favor the stability of the conjugate-gradient
approach, the practical performance of these optimizers can depend heavily on
the specific chemical system and implementation details. To move beyond these
arguments and rigorously quantify the performance trade-offs, we adopted a
Bayesian statistical framework
\cite{gelmanBayesianDataAnalysis2014,gelmanBayesianWorkflow2020,gabryVisualizationBayesianWorkflow2018,mcelreathStatisticalRethinkingBayesian2020}.

Traditional benchmarks often neglect system-to-system variability and lack
robust uncertainty quantification, making it difficult to draw reliable
conclusions. Performance metrics such as the number of \gls{pes} calls (positive
and increasing counts), total computation time (positive, skewed), and
convergence (binary) frequently violate the assumptions of normality and
\gls{homosced} inherent in standard linear models. Furthermore, benchmark designs
typically involve repeated measures, where multiple algorithmic variants are
tested on the same set of chemical systems. Failing to account for the resulting
\gls{nonindep} non-independence of observations can lead to \gls{pseudorep},
\gls{deflatederr}, and invalid statistical inferences.

Bluntly, any model for interpretation of success data that allows values other
than true or false, any model for ``efficiency'' in terms of the number of
calculations that permits numbers below zero, or any model for time that does
not remain strictly positive throughout cannot be used to estimate metrics
reliably. Implicit Gaussian distributions used for standard deviation or errors
on \gls{pes} calls, time, or success metrics are therefore mostly meaningless.

To address these methodological shortcomings, we apply a Bayesian \gls{glmm}
framework using \texttt{brms} \cite{burknerBrmsPackageBayesian2017}. This approach is
explicitly designed to handle such complexities, since it allows for the
selection of appropriate response distributions and link functions for each type
of metric, ensuring that the model's assumptions align with the data's
underlying properties. We explicitly include random effects, specifically,
glsp:randint for each chemical system to correctly model the hierarchical data
structure, partitioning the variance between system-specific effects and the
glsp:fixedeffect of the algorithmic variants. This provides a principled method
for obtaining \gls{robust} estimates and valid glsp:credint for the parameters of
interest.

\begin{figure}[htbp]
\centering
\includegraphics[height=0.4\textheight]{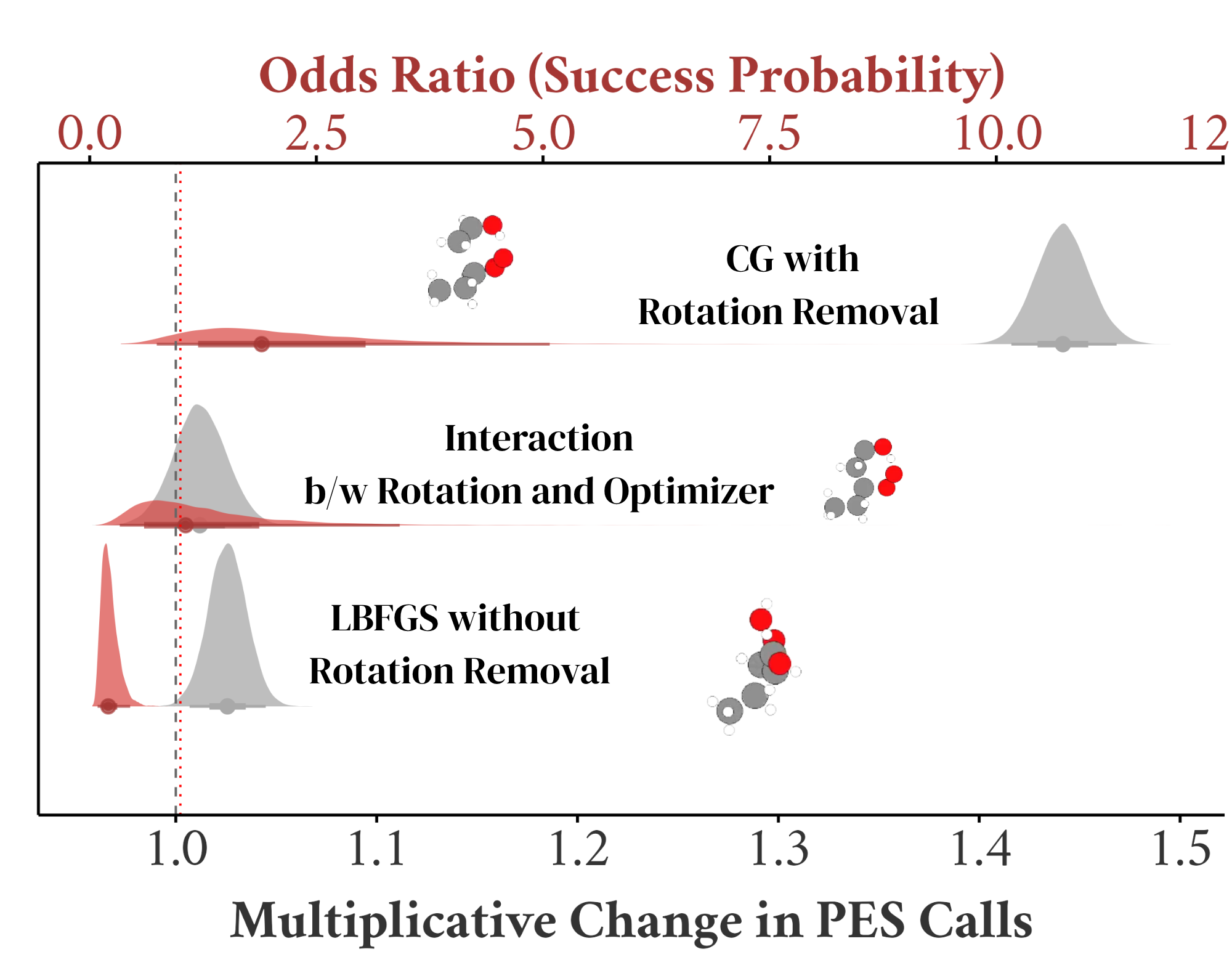}
\caption{\label{fig:brms_idrot_toc}Posterior distributions from generalized linear mixed-effects models showing the effects of algorithmic choices on computational cost (multiplicative change in PES calls, grey) and convergence probability (odds ratio, red). The analysis compares variants to a baseline of the Conjugate Gradient (CG) optimizer with rotation removal disabled. A value of 1.0 (dashed line) indicates no change relative to this baseline. \textbf{(Top)} Effect of Rotation Removal (for CG): Enabling rotation removal substantially increases the required PES calls by a factor of \textasciitilde{}1.44 (95\% CrI: [1.42, 1.47]) but has no statistically credible effect on the odds of success (the distribution overlaps 1.0). \textbf{(Middle)} Interaction Effect: The interaction between the optimizer and rotation removal settings is negligible for both cost and success probability, with distributions centered at 1.0. \textbf{(Bottom)} Effect of Optimizer (without Rotation Removal): Using the L-BFGS optimizer instead of CG results in a small but credible increase in computational cost (factor of \textasciitilde{}1.03) and a significant reduction in the odds of a successful convergence (OR \textasciitilde{}0.2, 95\% CrI: [0.09, 0.45]). Plotted from data published in \cite{goswamiBayesianHierarchicalModels2025a}.}
\end{figure}

We present here the results from our full interaction models, which
simultaneously estimate the main effects of the rotational optimizer (\gls{cg} vs
\gls{lbfgs}), the use of quaternion based external rotation removal
\cite{melanderEffectAdsorptionMagnetic2019} as implemented in EON
\cite{chillEONSoftwareLong2014}, and their interaction term. The findings are
summarized visually in Figure \ref{fig:brms_idrot_toc}. This comprehensive
analysis provides the most statistically powerful view of how these algorithmic
choices jointly influence computational cost and convergence success.

\subsection{Bayesian hierarchical model results}
\label{sec:brmsgp:bhmres}
\subsubsection{Computational Effort}
\label{sec:brmsgp:bhmres:ceff}
It is common during benchmark calculations to use the number of samples from a
energy and force calculator be used as a proxy variable for efficiency, since
most studies are done with an eye towards regimes where single point
calculations take a long time.

We modeled the number of calls for the \gls{pes} using a negative binomial
distribution, as specified in Eq. \ref{eq:pes_full_interaction}. This choice is
motivated by the physical constraints of the metric: \gls{pes} calls are discrete,
non-negative integers. Standard linear models assume normality, which allows for
non-integers and negative values. Furthermore, algorithmic performance data
typically exhibits overdispersion, where the variance exceeds the mean—due to the
heterogeneous nature of chemical landscapes. The Negative Binomial distribution
explicitly models this overdispersion, preventing the underestimation of
standard errors common in Poisson or Gaussian approximations.

\begin{equation} \label{eq:pes_full_interaction}
\begin{aligned}
\text{PESCalls}_{ij} &\sim \text{NegativeBinomial}(\mu_{ij}, \phi) \\
\log(\mu_{ij}) &= \beta_0 + \beta_{1} \text{DR}_{i(j))} + \beta_{2} \text{RR}_{i(j))} \\
&\quad + \beta_{3} (\text{DR}_{i(j))} \times \text{RR}_{i(j))}) + u_j
\end{aligned}
\end{equation}

Our analysis of this model revealed two key findings. First, enabling rotation
removal (\(\beta_2\)) incurred a substantial computational penalty, increasing the
median number of PES calls by a factor of 1.44 (95\% CrI: [1.42, 1.47]). Second,
using the L-BFGS optimizer (\(\beta_1\)) resulted in a small but statistically
credible increase in cost by a factor of 1.03 (95\% CrI: [1.01, 1.05]) compared
to CG. The interaction term (\(\beta_3\)) was not credibly different from zero,
indicating the effects were largely additive. The model also quantified
significant system-to-system variability in baseline computational cost
(\(\sigma_u \approx 0.63\)).
\subsubsection{Wall time estimates}
\label{sec:brmsgp:bhmres:cwall}
While the number of PES calls serves as the standard theoretical proxy for
computational effort, the ultimate metric for a practitioner is the total
wall-clock time elapsed. To ensure our conclusions were not merely an artifact
of our chosen proxy, and to quantify the real-world time costs, we performed a
parallel analysis on the total time for each calculation.

Wall time is a strictly positive, continuous variable. The distribution of
runtimes is often right-skewed, featuring a long tail of difficult systems while
the majority converge quickly. A standard Gaussian model would imply a symmetric
distribution, potentially assigning non-zero probability to physically
impossible negative times. We therefore employed a Gamma generalized linear
mixed model, which is naturally defined for \(x > 0\) and flexibly accommodates
the skewness inherent in computational timing data. The model specification is
shown in Eq. \ref{eq:time_full_interaction}.

\begin{equation} \label{eq:time_full_interaction}
\begin{aligned}
\text{TotalTime}_{ij} &\sim \text{Gamma}(\mu_{ij}, \alpha) \\
\log(\mu_{ij}) &= \beta_0 + \beta_{1} \text{DR}_{i(j)} + \beta_{2} \text{RR}_{i(j)} \\
&\quad + \beta_{3} (\text{DR}_{i(j)} \times \text{RR}_{i(j)}) + u_j
\end{aligned}
\end{equation}

The results of this analysis mirrored that of the effort proxy analysis.
Enabling rotation removal (\(\beta_2\)) incurred a substantial time penalty,
increasing the median total time by 43.0\% (95\% CrI: [40.4\%, 45.6\%]). Similarly,
the choice of the L-BFGS optimizer (\(\beta_1\)) led to a small but credible
increase in runtime of 2.6\% (95\% CrI: [0.7\%, 4.5\%]). The interaction term
(\(\beta_3\)) remained negligible, confirming the additive nature of these
effects.

This striking consistency across different metrics and model families provides
strong evidence that the number of electronic structure calculations is the
dominant factor driving total computation time. The performance conclusions
drawn from the PES call analysis are therefore robust and directly translate to
practical runtime considerations.
\subsubsection{Convergence Success}
\label{sec:brmsgp:bhmres:cconv}
We analyzed the probability of a successful search using a Bernoulli logistic
regression model, detailed in Eq. \ref{eq:success_full_interaction}. As
convergence is a binary event (\(y \in \{0,1\}\)), the search either isolates a
saddle point or it does not, the Bernoulli distribution is the canonical
probability model for such dichotomous outcomes.

\begin{equation} \label{eq:success_full_interaction}
\begin{aligned}
\text{Success}_{ij} &\sim \text{Bernoulli}(p_{ij}) \\
\text{logit}(p_{ij}) &= \beta_0 + \beta_{1} \text{DR}_{i(j))} + \beta_{2} \text{RR}_{i(j))} \\
&\quad + \beta_{3} (\text{DR}_{i(j))} \times \text{RR}_{i(j))}) + u_j
\end{aligned}
\end{equation}

This model demonstrated a clear and significant difference in robustness between
the optimizers. Compared to the CG baseline, the L-BFGS optimizer (\(\beta_1\))
was substantially less likely to converge, with an estimated odds ratio (OR) of
0.20 (95\% CrI: [0.09, 0.45]). In contrast, neither the main effect of rotation
removal (\(\beta_2\)) nor the interaction term (\(\beta_3\)) had a statistically
credible impact on the odds of success. The large standard deviation of the
random intercepts (\(\sigma_u \approx 3.6\)) underscored that intrinsic system
properties are a primary determinant of convergence success.

Our full interaction models show that the CG optimizer is both more efficient
and significantly more robust than L-BFGS for this dataset and the current
implementation of the rotation removal
\cite{melanderEffectAdsorptionMagnetic2019} process, while enabling this
implementation of rotation removal increases computational cost without a
corresponding benefit to the success rate. Finally, we discuss how these
algorithmic choices interact with invariance removal procedures within the
context of \gls{gp} models.
\subsection{Revisiting rotation removal}
\label{sec:brmsgp:rotremgood}
Though quaternions may not have been best implemented EON, we return to the
concept of rotations as implemented in
\textcite{goswamiAdaptivePruningIncreased2025b}.

A \gls{gp} model approximates the potential energy surface without inherent
knowledge of the physical invariances of the system. Consequently, a proposed
optimization step may contain spurious components corresponding to the external
degrees of freedom: overall translation and rotation of the entire molecule. The
optimizer actively removes these components from the proposed translation step
vector to ensure that movements occur only along internal coordinates, which
represent genuine changes in molecular geometry.

The procedure first constructs a basis set spanning the space of infinitesimal
rigid-body motions. For a system of \(N\) atoms, this space has six dimensions (or
five for a linear molecule). The procedure generates three basis vectors for
translation, \(\{\mathbf{t}_x, \mathbf{t}_y, \mathbf{t}_z\}\), where each vector
\(\mathbf{t}_k\) represents a unit displacement of all atoms along the Cartesian
axis \(k\).

\begin{equation}
(\mathbf{t}_k)_{3i+k-1} = 1 \quad \forall i \in \{1, ..., N\}
\end{equation}

Next, the procedure generates three basis vectors for rotation, \(\{\mathbf{l}_x,
\mathbf{l}_y, \mathbf{l}_z\}\), derived from the expression for infinitesimal
rotation about the center of mass, \(\mathbf{r}_i' = \mathbf{r}_i -
\mathbf{r}_{\text{com}}\). An infinitesimal rotation of the entire system
corresponds to a displacement \(\delta\mathbf{r}_i = \delta\boldsymbol{\omega}
\times \mathbf{r}_i'\). The rotational basis vectors thus take the form:

\begin{align}
\mathbf{l}_x &= \sum_{i=1}^{N} \hat{\mathbf{e}}_x \times \mathbf{r}_i' \\
\mathbf{l}_y &= \sum_{i=1}^{N} \hat{\mathbf{e}}_y \times \mathbf{r}_i' \\
\mathbf{l}_z &= \sum_{i=1}^{N} \hat{\mathbf{e}}_z \times \mathbf{r}_i'
\end{align}

The algorithm then applies the Gram-Schmidt process to this set of six vectors
to produce an orthonormal basis, \(\{\mathbf{u}_k\}\), that spans the external
degrees of freedom. For any proposed translation step, \(\mathbf{s} \in
\mathbb{R}^{3N}\), the algorithm projects out the external components. The
component of the step corresponding to translation and rotation,
\(\mathbf{s}_{\text{ext}}\), projects onto this basis:

\begin{equation}
\mathbf{s}_{\text{ext}} = \sum_{k} (\mathbf{s} \cdot \mathbf{u}_k) \mathbf{u}_k
\end{equation}

The pure internal step, \(\mathbf{s}_{\text{int}}\), then becomes the original
step minus its external projection:

\begin{equation}
\mathbf{s}_{\text{int}} = \mathbf{s} - \mathbf{s}_{\text{ext}}
\end{equation}

A feedback mechanism enhances the stability of the GP-driven search. The
algorithm computes the magnitude of the removed component,
\(\|\mathbf{s}_{\text{ext}}\|\). If this magnitude exceeds a defined threshold,
\(\theta_{\text{rot}}\), it signals that the GP model likely predicts a large,
unphysical torque on the molecule. In such cases, the procedure discards the
projection and reverts to the original, unprojected step \(\mathbf{s}\).
Subsequent step-size limitation guardrails then typically intercept this large,
physically questionable step, triggering a resampling of the true potential
energy surface to improve the GP model. When the magnitude of the removed
component remains below the threshold, the algorithm accepts the purified
internal step \(\mathbf{s}_{\text{int}}\). This ensures a more precise update to
the molecular geometry, guided only by genuine internal forces which we find to
be more robust in Chapter \ref{sec:otgpd}.

In practice, since energy does not depend on rotations, the threshold tends to
large values. Collectively, these invariance considerations and algorithmic
choices move us closer to the overarching goal: a turnkey, walltime-efficient,
reliable \gls{gp} optimization framework for molecular and extended systems.

In the following section, we shift our focus to data efficiency, discussing
practical strategies for reducing computational cost in \gls{gp} guided
optimizations.
\subsection{Time-to-Solution as a Function of Effort}
\label{sec:brmsgp:spline_time}
While independent analyses of computational effort (PES calls) and wall time
provide valuable baseline metrics, they do not fully capture the dynamic
relationship between the two. In a practical optimization scenario, the ``cost''
of a method is not just how many steps it takes, but how the time-per-step
scales as the optimization trajectory lengthens.

To investigate this, we modeled the total time-to-solution (\(T\)) as a continuous
function of the computational effort (\(N_{\text{calls}}\)), conditioned on the
optimization method. We employed a Bayesian Hierarchical Gamma Spline model (Eq.
\ref{eq:gamma_spline}) to capture the potentially non-linear scaling behaviors
while accounting for the strict positivity and skewness of runtime data.

\begin{equation} \label{eq:gamma_spline}
\begin{aligned}
T_{ij} &\sim \text{Gamma}(\mu_{ij}, \alpha) \\
\log(\mu_{ij}) &= \beta_{0} + \beta_{\text{method}} + f_{\text{method}}(\log(N_{\text{calls}, ij})) + u_j
\end{aligned}
\end{equation}

Here, \(f_{\text{method}}\) represents a thin-plate regression spline smooth term
allowed to vary by method. This approach enables us to separate the fixed
overhead of a method (intercept differences) from its scaling behavior (the
shape of the spline). We also incorporated random intercepts \(u_j\) for each
system to control for the intrinsic expense of different chemical environments.
\subsubsection{Extrapolation and Data Efficiency}
\label{sec:org1bf0576}
A challenge in benchmarking machine-learning enhanced methods (like GPDimer and
OTGPD, introduced in Chapter \ref{sec:otgpd}) against classical baselines (Dimer)
is the disparity in convergence speed. The ML methods often converge with
significantly fewer calls, leaving no observed data in the ``high-effort'' regime
populated by the classical method.

\begin{figure}[htbp]
\centering
\includegraphics[width=.9\linewidth]{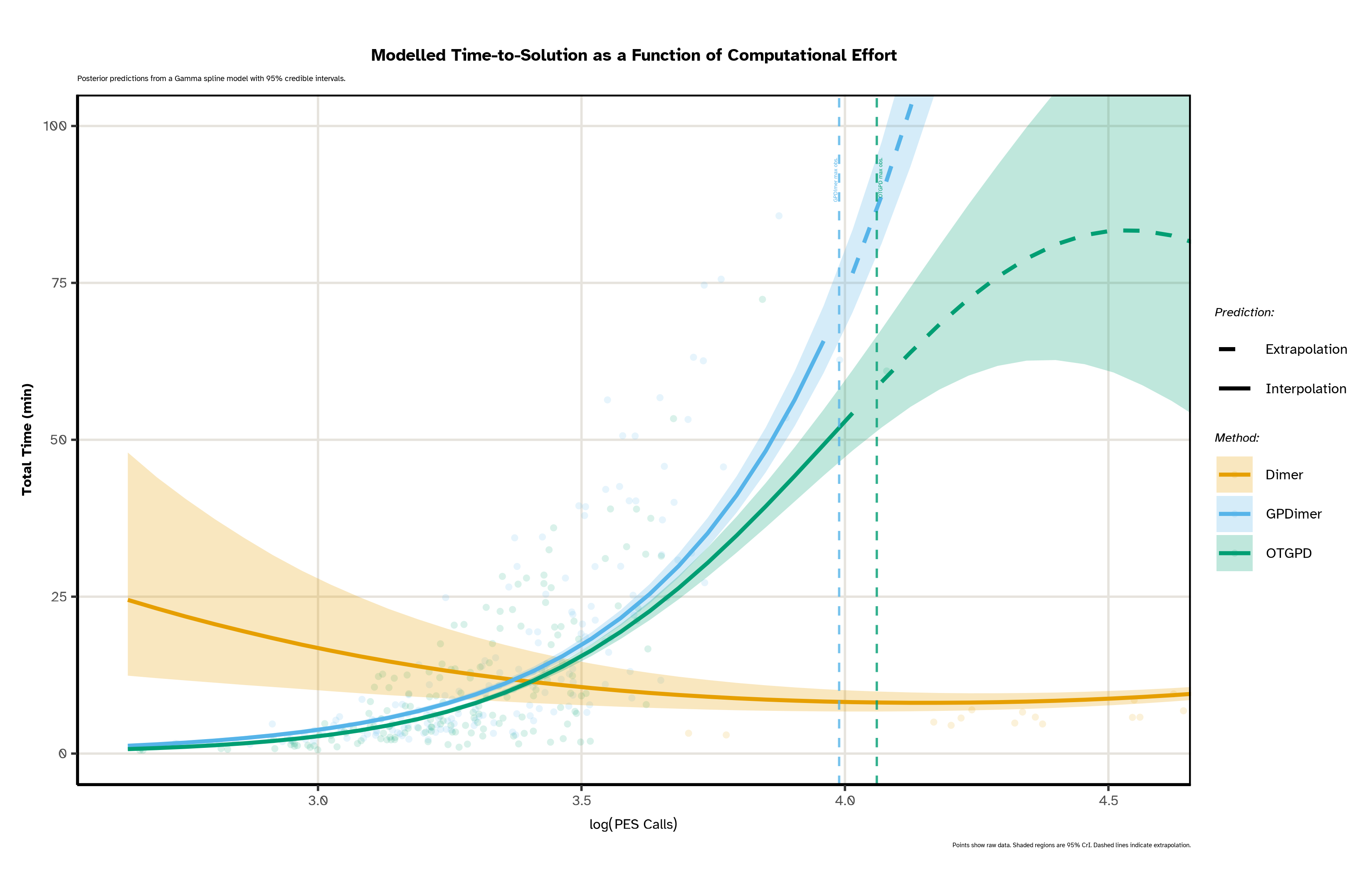}
\caption{\label{fig:spline_time}Modelled Time-to-Solution as a Function of Computational Effort. Posterior predictions from a Gamma spline model. The solid curves indicate the interpolation range where data was observed; dashed curves indicate extrapolation. The vertical dashed lines mark the maximum observed PES calls for the GP-based methods.}
\end{figure}

Figure \ref{fig:spline_time} visualizes this relationship. The solid lines
represent the interpolated predictions within the observed data range for each
method, while the dashed lines indicate extrapolation.

The plot reveals a distinct crossover. For extremely short runs (low PES calls),
the ML-based methods incur a higher time cost due to the overhead of Gaussian
Process regression. However, as the optimization difficulty increases (moving
right on the x-axis), this overhead is amortized. It is also worth noting that
the absolute time per call here is very low, since the Hartree-Fock calculations
considered in this thesis do not take much time per configuration. Systems with
more complex potential energy calculations will perform far better under the GP
as the Dimer methods take an order of magnitude more calls.
\subsubsection{Quantitative Comparison}
\label{sec:orgbbec222}
To rigorously quantify this trade-off, we compared the expected time-to-solution
at distinct regimes of computational effort, the overall results of which are shown in Table \ref{tbl:pesvtime}.

First, we evaluated the methods at a fixed budget of 30 PES calls, a typical
early-phase regime for GP methods. At this point, the classical Dimer method is
slightly faster (11.5 min) compared to GPDimer (12.8 min), reflecting the lower
per-step cost of the classical algorithm. The OTGPD method, however, effectively
mitigates this overhead (11.3 min), performing on par with the classical
baseline even in this low-data regime.

However, comparing methods at a fixed \textbf{x}-value is artificial if the methods
require vastly different amounts of effort to solve the same problem. A more
physically meaningful comparison evaluates each method at its median
convergence effort which approximates the typical ``price'' to solve a system.

\begin{table}[htbp]
\caption{\label{tbl:pesvtime}Results of modelled time to solution as a function of computational effort.}
\centering
\begin{tabular}{lrrl}
\hline
Method & Median PES Calls & Expected Time (min) & 95\% CrI\\
\hline
\textbf{\textbf{OTGPD}} & \textbf{\textbf{28}} & \textbf{\textbf{9.0}} & \textbf{\textbf{{[}8.6, 9.5]}}\\
GPDimer & 30 & 12.8 & {[}12.2, 13.5]\\
Dimer & 254 & 20.5 & {[}19.2, 21.8]\\
\end{tabular}
\end{table}

At their respective operating points, the efficiency gain is stark. The
classical Dimer requires typically 254 calls, resulting in an expected runtime
of 20.5 minutes. The OTGPD method achieves solution with only 28 calls on
average, dropping the total wall time to 9.0 minutes. This represents a factor
of 2.2x speedup in real-world wall time, confirming that the dramatic
reduction in force calls provided by the GP surrogate translates directly to
time savings even for trivial energy surfaces, despite the added computational overhead of the machine learning
model.
\subsection{Conclusions}
\label{sec:brmsgp:conclusions}
This chapter established a rigorous statistical framework for evaluating
computational performance, moving the discourse from anecdotal benchmarking to
hierarchical Bayesian inference. By employing generalized linear mixed models
with appropriate response distributions—Negative Binomial for effort, Gamma for
wall time, and Bernoulli for convergence—we quantified the distinct trade-offs
inherent in algorithmic choices while explicitly accounting for system-specific
variability.

The analysis of the dimer method's rotation phase yielded decisive results. We
demonstrated that the Conjugate Gradient optimizer offers superior robustness
compared to L-BFGS, contradicting default choices in established software
packages. Furthermore, we showed that the specific implementation of rotation
removal within the EON suite incurred a significant computational penalty
without providing a credible improvement in convergence success.

However, the inefficiency of one specific implementation does not negate the
underlying physical principle. As discussed in Section
\ref{sec:brmsgp:rotremgood}, the explicit removal of external invariances
functions as a critical stability mechanism for \gls{gp} surrogates
\cite{goswamiAdaptivePruningIncreased2025b}, preventing unphysical extrapolations
in the active learning loop. This distinction underscores a broader theme in
computational method development: the utility of a physical constraint depends
heavily on its numerical integration within the optimization framework.

Using the same framework, modeling the time-to-solution landscape (Section
\ref{sec:brmsgp:spline_time}) resolves the ``overhead vs. efficiency'' debate for
machine-learning enhanced optimization. While surrogate-based methods like
GPDimer and OTGPD incur a higher computational cost per step for cheap
potentials evident in the crossover observed in the Gamma spline analysis, this
overhead is effectively amortized over the trajectory. Our analysis confirms
that the On-The-Fly Gaussian Process Dimer (OTGPD) method reduces the median
computational effort so drastically (from \textasciitilde{}254 to \textasciitilde{}28 calls) that it yields a
net reduction in total wall time of approximately 56\% compared to the classical
baseline. This result statistically validates the core premise of this thesis:
that data efficiency translates directly to wall-clock acceleration, provided
the surrogate model is strictly stabilized.

With a validated statistical methodology to measure success and a refined
understanding of how to handle physical invariances, we now address the primary
bottleneck in machine-learning-guided optimization, that of total wall time. The
following chapter applies these insights to the challenge of data efficiency,
developing strategies to scale Gaussian Process surrogates to chemically
relevant system sizes.
\section{Data efficiency for Gaussian Processes}
\label{sec:dataeff}
\epigraph{I have only made this letter longer because I have not had the time to make it shorter.}{Blaise Pascal \\ The Provincial Letters}

\begin{quote}
This chapter is partially based on \fullcite{goswamiAdaptivePruningIncreased2025}.
\end{quote}

\glsdesc{gp} (Section \ref{sec:scal:timestor}) surrogates promise analytic posteriors and gradients, yet their naive scaling, \(O(M^3 N^3)\) in time and \(O(M^2 N^2)\) in memory for \(M\) geometries and \(N\) atoms, blocks routine use. This chapter develops a narrative from lossless algebraic reorganizations that preserve fidelity, to lossy approximations that trade information for wall-time. Along the way, we confront pathologies that emerge in active learning loops and establish safeguards that let the later chapters stand on firm computational ground.
\subsection{Quicker inversions through reshaping}
\label{sec:dataeff:invreshape}
In this work, the optimization routine circumvents the explicit inversion of the full kernel matrix. Instead, the log \gls{mll} and its gradient with respect to kernel hyperparameters are evaluated using efficient matrix-vector products and decompositions. This is realized by assembling a rectangular matrix \(\mathbf{R}\) of size \(M(3N+1) \times (3N+1)\), where rows correspond to training targets and columns to specific observables.

Reordering or reshaping a matrix is algebraically a no-op regarding fundamental operations such as inversion and decomposition \cite{gentleMatrixAlgebraTheory2007}. Any permutation or block reorganization of a matrix \(\mathbf{K}\) can be written as
\[
\mathbf{K}' = \mathbf{P} \mathbf{K} \mathbf{P}^T
\]

where \(\mathbf{P}\) is an orthogonal permutation matrix (\(\mathbf{P}^T = \mathbf{P}^{-1}\)). For any vector \(\mathbf{y}\),
\[
\mathbf{K}\mathbf{y} = \mathbf{0} \iff \mathbf{K}'(\mathbf{P}\mathbf{y}) = \mathbf{0}
\]

and
\[
\mathbf{K}^{-1} = \mathbf{P}^T (\mathbf{K}')^{-1} \mathbf{P}
\]

Thus, the solution to \(\mathbf{K}\mathbf{a} = \mathbf{y}\) is unchanged under permutation:
\[
\mathbf{a} = \mathbf{K}^{-1} \mathbf{y} = \mathbf{P}^T (\mathbf{K}')^{-1} \mathbf{P} \mathbf{y}
\]
and the determinant and decomposition (e.g., Cholesky) are similarly invariant up to permutation:
\[
\det \mathbf{K} = \det \mathbf{K}'\,, \quad \text{and} \quad \mathbf{K} = \mathbf{L}\mathbf{L}^T \implies \mathbf{K}' = (\mathbf{P}\mathbf{L})(\mathbf{P}\mathbf{L})^T
\]

Consider a ``reshaped'' or block-organized version of \(\mathbf{K}\), e.g., storing targets grouped by configuration or observable. The Cholesky decomposition (used in \gls{scg} and \gls{gp} marginal likelihood) and matrix-vector solves are unaffected:
\[
\mathbf{K} = \mathbf{L}\mathbf{L}^T \implies \mathbf{K}' = (\mathbf{P}\mathbf{L})(\mathbf{P}\mathbf{L})^T
\]
and
\[
\mathbf{K}^{-1}\mathbf{y} = \mathbf{P}^T (\mathbf{K}')^{-1} (\mathbf{P} \mathbf{y})
\]
This means that, for \gls{scg} optimization, the coefficients obtained from the full kernel or from any reshaped, block, or permuted version are identical (after applying the corresponding permutation to the solution vector).

By exploiting this invariance, practical implementations (including this work and GPstuff \cite{vanhataloGPstuffBayesianModeling}) use block matrices or grouped layouts for efficiency, without loss of mathematical fidelity. The organization is chosen to minimize computational overhead and maximize parallelism, but the GP predictions, marginal likelihood, and \gls{scg} updates remain unchanged.

Thus, the storage and computational cost of this approach scale as \(\mathcal{O}(MN^2)\), which is linear in the number of training geometries and quadratic in the number of atoms. This reduction is achieved without any loss of information or accuracy, as all targets are still included in the optimization; the difference is purely the result of efficient organization and evaluation of the required matrix operations. Scaling from a software design perspective is known to unlock linear scaling in computational chemistry \cite{nakataLargeScaleLinear2020}.

For example, for a small system with \(N=5\) atoms and \(M=1\) configuration, both the block and full kernel have only \(256\) elements (\(10^{2.4}\)). However, at \(M=75\) configurations, the full kernel would require storage for \(1.44 \times 10^6\) elements (\(10^{6.2}\)), while the block matrix requires only \(19,\!200\) elements (\(10^{4.3}\)). For a larger molecule (\(N=18\)), the difference is even more dramatic: at \(M=75\), the full kernel has \(1.7 \times 10^7\) elements (\(10^{7.2}\)), while the block matrix contains just \(226,\!875\) elements (\(10^{5.4}\)).

\begin{figure}[htbp]
\centering
\includegraphics[height=0.3\textheight]{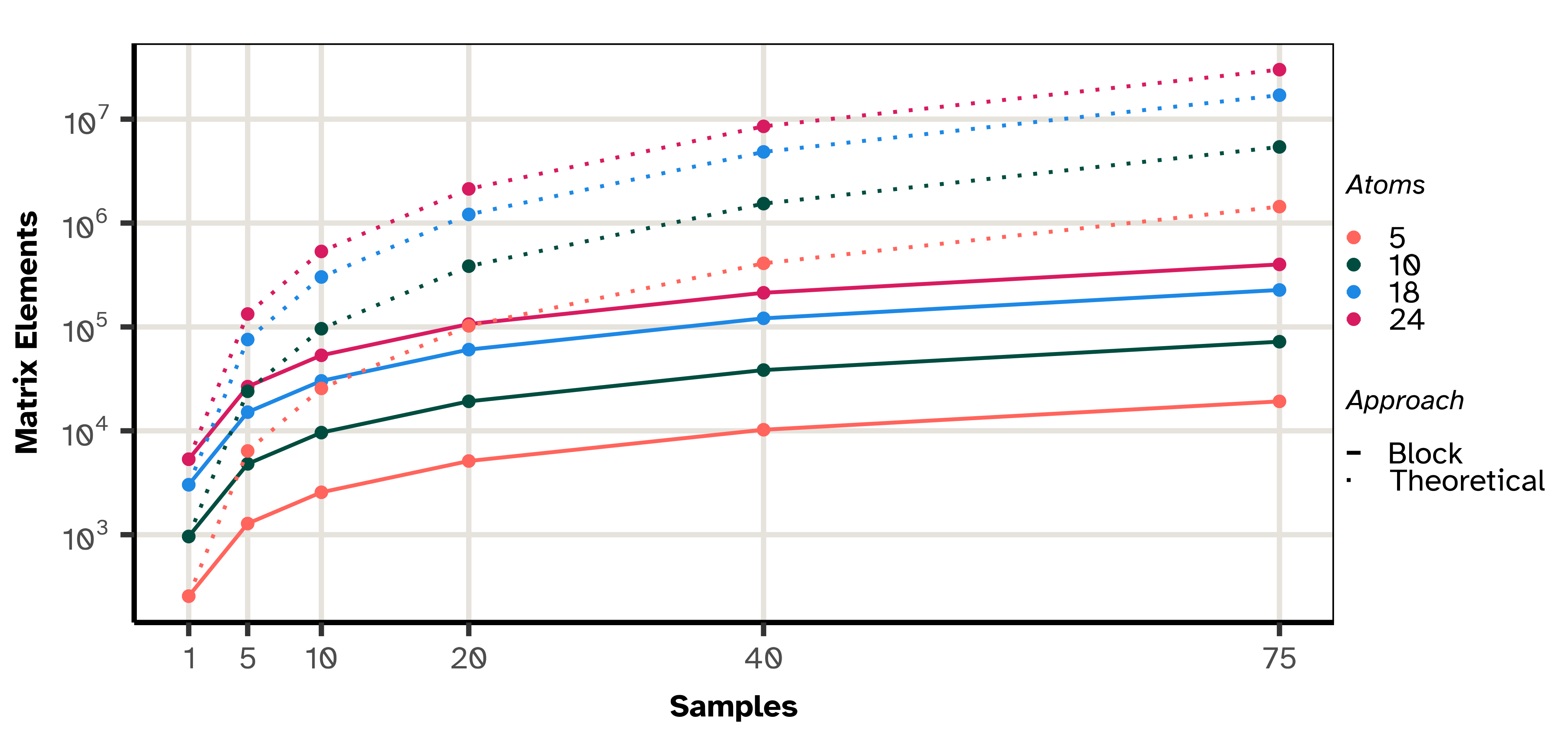}
\caption{\label{fig:gp:scaling:block:vs:full}Kernel matrix element scaling for \gls{gp} training with energy and force data. Solid lines (with points) denote the practical block matrix (\(M(3N+1) \times (3N+1)\)), while dotted lines denote the theoretical full kernel (\((M(3N+1))^2\)). Values are log-scale versus the number of training geometries. For \(N=18\) atoms and \(M=75\) geometries, the block matrix is orders of magnitude smaller than the full kernel formulation.}
\end{figure}

Figure \ref{fig:gp:scaling:block:vs:full} illustrates the practical scaling of the block matrix (solid lines) versus the theoretical full kernel (dashed lines). For a system with \(N=18\) atoms and \(M=75\) geometries, the full kernel requires \(1.7 \times 10^7\) elements (\(10^{7.2}\)), whereas the block matrix requires only \(2.2 \times 10^5\) elements (\(10^{5.4}\)). This efficiency gain enables GPR model training on significantly larger systems.

\subsection{Rank one covariance updates for new data}
\label{sec:dataeff:rankone}
With fixed hyperparameters, one new observation permits an \(O(n^2)\) Cholesky update rather than a full \(O(n^3)\) refactor:
\begin{equation}
\mathbf{L}_{n+1}=
\begin{pmatrix}
\mathbf{L}_n & \mathbf{0}\\
\mathbf{l}^T & l_d
\end{pmatrix},
\quad
\mathbf{L}_n \mathbf{l} = \mathbf{k},
\quad
l_d = \sqrt{k_{dd}-\mathbf{l}^T\mathbf{l}}.
\end{equation}

In the \texttt{gpr\_optim} this approach is not pursued because the formulation of a rank one update procedure for adding the training covariance is overshadowed by the subsequent re-optimization of the hyperparameters, where the entire matrix is inverted repeatedly.

The search for optimal hyperparameters, \(\boldsymbol{\theta}^*\), requires
maximizing the marginal likelihood, but the covariance matrix
\(\mathbf{K}_{\boldsymbol{\theta}}\) is a function of the hyperparameters. Each
candidate set of hyperparameters \(\boldsymbol{\theta}_i\) in the search defines
an entirely new matrix. The incremental update is rendered useless because the
base matrix \(\mathbf{K}_n\) is not fixed; it is constantly being redefined.

Therefore, for a search consisting of \(N_{\text{trials}}\) candidate hyperparameter sets, the total computational cost for the optimization step is dominated by the \(N_{\text{trials}}\) full, \(O(n^3)\) decompositions required to evaluate the marginal likelihood for each candidate. The single, final \(O(n^2)\) update for the chosen \(\boldsymbol{\theta}^*\) is computationally insignificant in comparison.
\subsection{Pruning over data}
\label{sec:dataeff:dataprune}
The block matrix and rank one update formulation so far forms a concrete software representation of a mathematical algebraic form, a design space solution; while providing impressive performance, the fundamental scaling in the context of a high number of points is not addressable in this space.

Approximations to the \gls{gp} \cite{bijlOnlineSparseGaussian2015,rasmussenGaussianProcessesMachine2006,wilsonKernelInterpolationScalable2015} often rely on utilizing a subset of training examples instead of the global update form in Algorithm \ref{alg:gp:no:prune}. 

\begin{algorithm}[H]
\caption{GP-Guided Optimization: No Pruning (Global Model)}
\label{alg:gp:no:prune}
\begin{algorithmic}[1]
\State \textbf{Input:} objective $U(\mathbf{x})$, gradient $\nabla U(\mathbf{x})$, initial $\mathbf{x}_0$, max iterations $T_{\max}$, initial step size $\eta_0$
\State \textbf{Output:} final position $\mathbf{x}_T$, training set $\mathcal{D}_T$

\State Initialize: $\mathcal{D}_1 \leftarrow \{(\mathbf{x}_0, U(\mathbf{x}_0), \nabla U(\mathbf{x}_0))\}$
\State $t \leftarrow 1$

\While{$t \leq T_{\max}$ and not converged}
  \State \Comment{Fit GP on \textbf{all} accumulated data}
  \State Fit GP to $\mathcal{D}_t$ by maximizing marginal likelihood (Eq.~\ref{eq:gp:posterior:mean})
  
  \State \Comment{Predict gradient at current location}
  \State $\hat{\mathbf{g}}_t \leftarrow \nabla_{\mathbf{x}_t} \mu(\mathbf{x}_t)$ via Eq.~\ref{eq:gp:gradient:prediction}
  
  \State \Comment{Line search for step size}
  \State $\mathbf{x}_{t+1}, \eta_t \leftarrow \text{ArmijoLineSearch}(\mathbf{x}_t, \hat{\mathbf{g}}_t, U, \eta_0)$
  
  \State \Comment{Observe objective and gradient at new point}
  \State $u_{t+1} \leftarrow U(\mathbf{x}_{t+1})$, $\nabla u_{t+1} \leftarrow \nabla U(\mathbf{x}_{t+1})$
  
  \State \Comment{Accumulate into training set}
  \State $\mathcal{D}_{t+1} \leftarrow \mathcal{D}_t \cup \{(\mathbf{x}_{t+1}, u_{t+1}, \nabla u_{t+1})\}$
  
  \State $t \leftarrow t + 1$
\EndWhile

\State \Return $\mathbf{x}_t$, $\mathcal{D}_t$
\end{algorithmic}
\end{algorithm}

However, when a data reduction heuristic is applied at each step of an active learning loop (as shown in Algorithm \ref{alg:gp:prune}), it creates a tight coupling between the inference approximation and the data acquisition policy. The chosen approximation affects the posterior, which in turn affects the acquisition function's decision about where to sample next. This new sample then influences the subsequent approximation, creating a feedback loop that can lead to pathological behavior.

\begin{algorithm}[H]
\caption{GP-Guided Optimization: Online Pruning (Local Model)}
\label{alg:gp:prune}
\begin{algorithmic}[1]
\State \textbf{Input:} objective $U(\mathbf{x})$, gradient $\nabla U(\mathbf{x})$, initial $\mathbf{x}_0$, max iterations $T_{\max}$, lengthscale $\ell$, pruning multiplier $\alpha$
\State \textbf{Output:} final position $\mathbf{x}_T$, final training set $\mathcal{D}_T^{\text{pruned}}$

\State Initialize: $\mathcal{D}_1 \leftarrow \{(\mathbf{x}_0, U(\mathbf{x}_0), \nabla U(\mathbf{x}_0))\}$
\State $t \leftarrow 1$
\State $r_p \leftarrow \alpha \ell$ \Comment{Set pruning radius (Eq.~\ref{eq:pruning:radius})}

\While{$t \leq T_{\max}$ and not converged}
  \State \Comment{\textbf{Prune:} retain only nearby observations}
  \State $\mathcal{D}_t^{\text{pruned}} \leftarrow \{ (\mathbf{x}_i, u_i, \nabla u_i) \in \mathcal{D}_t : \|\mathbf{x}_i - \mathbf{x}_t\| \leq r_p \}$ (Eq.~\ref{eq:pruning:mask})
  
  \State \Comment{Fit GP on pruned set only}
  \State Fit GP to $\mathcal{D}_t^{\text{pruned}}$ by maximizing marginal likelihood
  
  \State \Comment{Predict gradient at current location (from limited data)}
  \State $\hat{\mathbf{g}}_t \leftarrow \nabla_{\mathbf{x}_t} \mu(\mathbf{x}_t)$ via Eq.~\ref{eq:gp:gradient:prediction}
  
  \State \Comment{Line search for step size}
  \State $\mathbf{x}_{t+1}, \eta_t \leftarrow \text{ArmijoLineSearch}(\mathbf{x}_t, \hat{\mathbf{g}}_t, U, \eta_0)$
  
  \State \Comment{Observe objective and gradient at new point}
  \State $u_{t+1} \leftarrow U(\mathbf{x}_{t+1})$, $\nabla u_{t+1} \leftarrow \nabla U(\mathbf{x}_{t+1})$
  
  \State \Comment{Add to global history (but pruned set still holds only nearby data)}
  \State $\mathcal{D}_{t+1} \leftarrow \mathcal{D}_t \cup \{(\mathbf{x}_{t+1}, u_{t+1}, \nabla u_{t+1})\}$
  
  \State $t \leftarrow t + 1$
\EndWhile

\State \Return $\mathbf{x}_t$, $\mathcal{D}_t^{\text{pruned}}$
\end{algorithmic}
\end{algorithm}

We demonstrate this pathology with the Rosenbrock potential, defined as:
\begin{equation}
U(\mathbf{x}) = (a - x_1)^2 + b(x_2 - x_1^2)^2
\label{eq:rosenbrock}
\end{equation}
with parameters \(a = 1\) and \(b = 100\). The global minimum lies at \(\mathbf{x}^* = (1, 1)\) with \(U(\mathbf{x}^*) = 0\). 

The gradient is:

\begin{equation}
\nabla U(\mathbf{x}) = \begin{pmatrix}
-2(a - x_1) - 4bx_1(x_2 - x_1^2) \\
2b(x_2 - x_1^2)
\end{pmatrix}
\label{eq:rosenbrock:grad}
\end{equation}

with a starting point at \(\mathbf{x}_0 = (0.0, 1.5)\) chosen outside the valley, and convergence is when the force norm predicted by the \gls{gp} drops below \(1e^{-6}\).

Here, a \gls{gp} optimizer with derivatives is compared against an identical optimizer that employs a naive online pruning rule: at each step, it discards all observations outside a fixed radius from its current position. This can be seen as a crude, state-dependent form of sparsification.

The GP model maintains a joint distribution over function values and their gradients. At each step \(t\), one maintains a training set:

\begin{equation}
\mathcal{D}_t = \left\{ (\mathbf{x}_i, u_i, \nabla u_i) : i = 1, \ldots, n_t \right\}
\label{eq:training:set}
\end{equation}

where \(u_i = U(\mathbf{x}_i)\) is the function value and \(\nabla u_i = \nabla U(\mathbf{x}_i)\) is the gradient vector at each observed location \(\mathbf{x}_i\).

The \gls{gp} prior employs a squared-exponential (RBF) kernel:

\begin{equation}
k(\mathbf{x}, \mathbf{x}') = \sigma_f^2 \exp\left( -\frac{1}{2\ell^2} \|\mathbf{x} - \mathbf{x}'\|^2 \right)
\label{eq:se:kernel}
\end{equation}

with signal variance \(\sigma_f^2\) and lengthscale \(\ell\). The posterior mean prediction at a test point \(\mathbf{x}_*\) is:

\begin{equation}
\mu(\mathbf{x}_*) = \mathbf{k}_*^T \mathbf{K}^{-1} \mathbf{y}
\label{eq:gp:posterior:mean}
\end{equation}

where \(\mathbf{k}_* = [k(\mathbf{x}_*, \mathbf{x}_1), \ldots, k(\mathbf{x}_*, \mathbf{x}_{n_t})]^T\) collects covariances to observed points, and \(\mathbf{K}\) is the full covariance matrix of all observations and their derivatives.

The predicted gradient at \(\mathbf{x}_*\) is obtained by differentiating the posterior mean:

\begin{equation}
\nabla_* \mu(\mathbf{x}_*) = \frac{\partial}{\partial \mathbf{x}_*} \left( \mathbf{k}_*^T \mathbf{K}^{-1} \mathbf{y} \right) = \left( \frac{\partial \mathbf{k}_*}{\partial \mathbf{x}_*} \right)^T \mathbf{K}^{-1} \mathbf{y}
\label{eq:gp:gradient:prediction}
\end{equation}

This predicted gradient \(\hat{\mathbf{g}}_t = \nabla_* \mu(\mathbf{x}_t)\) at the current location \(\mathbf{x}_t\) drives the next step via a simple gradient descent with Armijo line search \cite{pressNumericalRecipes3rd2007}.

Online pruning removes observations deemed ``distant'' from the current location. At iteration \(t\), given the current position \(\mathbf{x}_t\) and the pruning radius \(r_p\), one retains only observations satisfying:

\begin{equation}
\mathcal{D}_t^{\text{pruned}} = \left\{ (\mathbf{x}_i, u_i, \nabla u_i) \in \mathcal{D}_t : \|\mathbf{x}_i - \mathbf{x}_t\| \leq r_p \right\}
\label{eq:pruning:mask}
\end{equation}

The pruning radius we set as a multiple of the lengthscale:

\begin{equation}
r_p = \alpha \ell
\label{eq:pruning:radius}
\end{equation}

where \(\alpha \in (0, 1)\) is a multiplier (e.g., \(\alpha = 0.8\)). This filtered dataset then retrains the \gls{gp} for the next iteration.

The rationale appears sound: observations far from the current location exert minimal influence on the posterior (their kernel weight decays exponentially with distance), so discarding them saves computation without sacrificing local accuracy. However, this reasoning ignores the coupling between data support and inference.

\begin{figure}[htbp]
\centering
\includegraphics[width=1.1\textwidth,height=0.40\textheight]{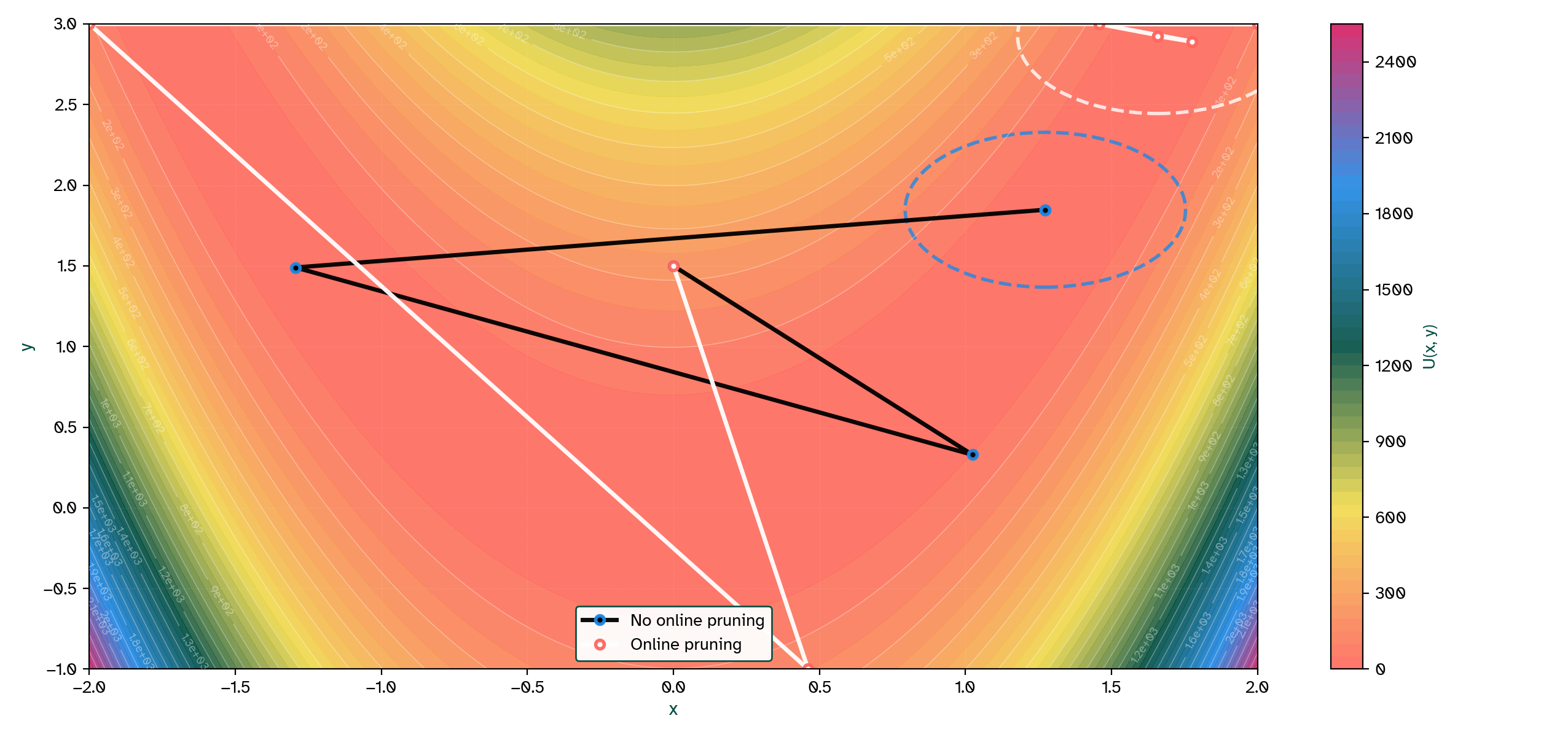}
\caption{\label{fig:pruning:divergence:rosenbrock}Online pruning induces trajectory divergence in GP-guided optimization of the Rosenbrock function (Eq. \ref{eq:rosenbrock}). The landscape is shown with contours. Black path (Algorithm \ref{alg:gp:no:prune}): all observations retained, shown as white circles. White path (Algorithm \ref{alg:gp:prune}): online pruning applied; white circles denote retained observations, black rings mark observations pruned away at the final step (those lying outside radius \(r_p = 0.48\) from the final position). Dashed circles indicate the pruning radius \(r_p\) at each trajectory terminus. The two paths diverge markedly within the first 2–3 steps, demonstrating how the choice to discard distant data fundamentally redirects the optimization dynamics. The unpruned model converges to \(\mathbf{x} \approx (1.27, 1.85)\) in 4 steps, whereas the pruned model takes a longer, misguided path to \(\mathbf{x} \approx (1.66, 2.92)\) in 7 steps.}
\end{figure}

Figure \ref{fig:pruning:divergence:rosenbrock} presents two optimization trajectories guided by \gls{gp} inference with derivative observations on the Rosenbrock function (Eq. \ref{eq:rosenbrock}).  Both trajectories use identical hyperparameters: a lengthscale \(\ell = 1.6\) and signal variance \(\sigma_f=1.1\). The online pruning model uses a multiplier of \(\alpha = 0.3\), yielding a pruning radius of \(r_p = 1.3 \times 0.3 = 0.48\). Each trajectory is ``warm-started'' with 100 randomly selected data points. This initialization is the key mechanism driving the dramatic effect shown. The warm-start points are not sampled uniformly but are instead placed in an asymmetric cloud, predominantly on one side of the starting point \(\mathbf{x}_0\). This biased initial dataset has profoundly different consequences for each optimizer. The global optimizer, following the black path, incorporates all 100 points, and its initial model of the landscape is permanently skewed by this lopsided data distribution. In contrast, the pruned optimizer, following the white path, also observes these 100 points at \(t=0\) but immediately discards the vast majority of them after its first step, as they fall outside its tight pruning radius. Its model is consequently based on only a handful of the closest points. This warm-start strategy ensures the two models begin with fundamentally different ``beliefs'' about the objective function, forcing their optimization paths to diverge from the very beginning.

The pruned trajectory deviates substantially from the unpruned path. At the final iteration, the unpruned model has incorporated all prior observations—a global historical record encoded in the posterior covariance. The pruned model, by contrast, has ``forgotten'' all distant data; its posterior reflects only the local neighborhood bounded by \(r_p\) (Eq. \ref{eq:pruning:radius}). This localization shifts the gradient estimate \(\hat{\mathbf{g}}_t\) (Eq. \ref{eq:gp:gradient:prediction}), which alters the next step direction via line search, which changes the location \(\mathbf{x}_{t+1}\) from which future observations are sampled. The optimization trajectories differ, exploring different data histories as a result of:

\begin{description}
\item[{Loss of global information}] The kernel function \(k(\mathbf{x}, \mathbf{x}')\) (Eq. \ref{eq:se:kernel}) in GP regression assigns non-negligible weight to observations at distances comparable to the lengthscale \(\ell\). For \(\ell = 0.6\) and \(\alpha = 0.3\), the pruning radius is \(r_p = 0.18\)—substantially smaller than the effective support range of the kernel. Discarding observations at distances \(\ell < d \leq r_p\) destroys information about the broader landscape. One sacrifices the ability to maintain a coherent global model in exchange for computational savings that, in practice, amount to microseconds per iteration.

\item[{Discontinuous posterior recomputation}] By construction, the posterior mean (Eq. \ref{eq:gp:posterior:mean}) and gradient prediction (Eq. \ref{eq:gp:gradient:prediction}) depend on the full training set \(\mathcal{D}_t\). Removing observations does not smoothly degrade the posterior—it discontinuously recomputes it on the reduced support \(\mathcal{D}_t^{\text{pruned}}\) (Eq. \ref{eq:pruning:mask}). This recomputation induces jump discontinuities in predicted gradients \(\hat{\mathbf{g}}_t\), leading to erratic step sizes and divergent trajectories. The coupling between the data support set and the inference rule renders the ``optimization'' non-stationary: the effective objective landscape shifts with each pruning event.

\item[{Feedback amplification}] Each pruning event occurs at a new location \(\mathbf{x}_t\). If pruning removes influential historical points that carry information about distant minima or saddle points, the model misdirects the next step to an erroneous \(\mathbf{x}_{t+1}\). From this new position, a fresh set of distant points become candidates for removal (those now outside radius \(r_p\) from \(\mathbf{x}_{t+1}\)). The erroneous step feeds into an erroneous posterior, which guides another erroneous step. Errors compound multiplicatively across iterations.
\end{description}

The coupling between data support (Eq. \ref{eq:pruning:mask}) and inference (Eq. \ref{eq:gp:posterior:mean}, \ref{eq:gp:gradient:prediction}) admits no free lunch.
\subsection{Variance and accuracy}
\label{sec:dataeff:varacc}
Beyond the fundamental problems with pruning data within an active learning loop, the interpetation of variance in a sequential optimization process can be subject to interpretation. To demonstrate this, consider two sampling strategies, with fixed hyperparameters (Algorithm \ref{alg:frozen:theta}) and one where the hyperparameters are optimized at each step, shown in Algorithm \ref{alg:reopt:theta}.

\begin{algorithm}[H]
\caption{Sequential Sampling with Frozen Hyperparameters}
\label{alg:frozen:theta}
\begin{algorithmic}[1]
\State \textbf{Input:} objective $U(\mathbf{x})$, gradient $\nabla U(\mathbf{x})$, frozen $\boldsymbol{\theta}_0 = (\sigma_f^0, \ell^0, \sigma_n^{\text{f},0}, \sigma_n^{\text{d},0})$, sample path $\{\mathbf{x}_1, \ldots, \mathbf{x}_T\}$
\State \textbf{Output:} predictions $\{\mu_t, \sigma_t^2\}_{t=1}^{T}$, RMSE and coverage metrics

\State Initialize: $\mathcal{D}_1 \leftarrow \{(\mathbf{x}_1, U(\mathbf{x}_1), \nabla U(\mathbf{x}_1))\}$

\For{$t = 1$ to $T$}
  \State \Comment{Fit GP with \textbf{fixed} hyperparameters}
  \State Fit GP to $\mathcal{D}_t$ using $\boldsymbol{\theta}_0$ (no optimization)
  
  \State \Comment{Predict on dense probe grid}
  \State $\{\mu(\mathbf{z}), \sigma^2(\mathbf{z})\}_{z \in \text{GRID}} \leftarrow \text{GP.predict}(\text{GRID})$ via Eq.~\ref{eq:gp:posterior:mean}, \ref{eq:gp_variance_posterior}
  
  \State \Comment{Evaluate global and split metrics}
  \State $\text{RMSE}_t \leftarrow \sqrt{\frac{1}{|\text{GRID}|} \sum_{\mathbf{z}} (\mu(\mathbf{z}) - U_{\text{true}}(\mathbf{z}))^2}$
  \State $\{\text{RMSE}_{\text{in},t}, \text{RMSE}_{\text{out},t}\} \leftarrow \text{SplitBySupport}(\mu, U_{\text{true}}, \text{dists}, r_p = 1.5 \ell^0)$
  \State $\text{Cov}_{1\sigma,t} \leftarrow \frac{1}{|\text{GRID}|} \sum_{\mathbf{z}} \mathbb{1}[|\mu(\mathbf{z}) - U_{\text{true}}(\mathbf{z})| \leq \sigma(\mathbf{z})]$
  
  \State \Comment{Accumulate next observation}
  \State $\mathcal{D}_{t+1} \leftarrow \mathcal{D}_t \cup \{(\mathbf{x}_{t+1}, U(\mathbf{x}_{t+1}), \nabla U(\mathbf{x}_{t+1}))\}$
  
\EndFor

\State \Return all metrics for $t = 1, \ldots, T$
\end{algorithmic}
\end{algorithm}

\begin{figure}[htbp]
\centering
\includegraphics[width=1.1\textwidth,height=0.40\textheight]{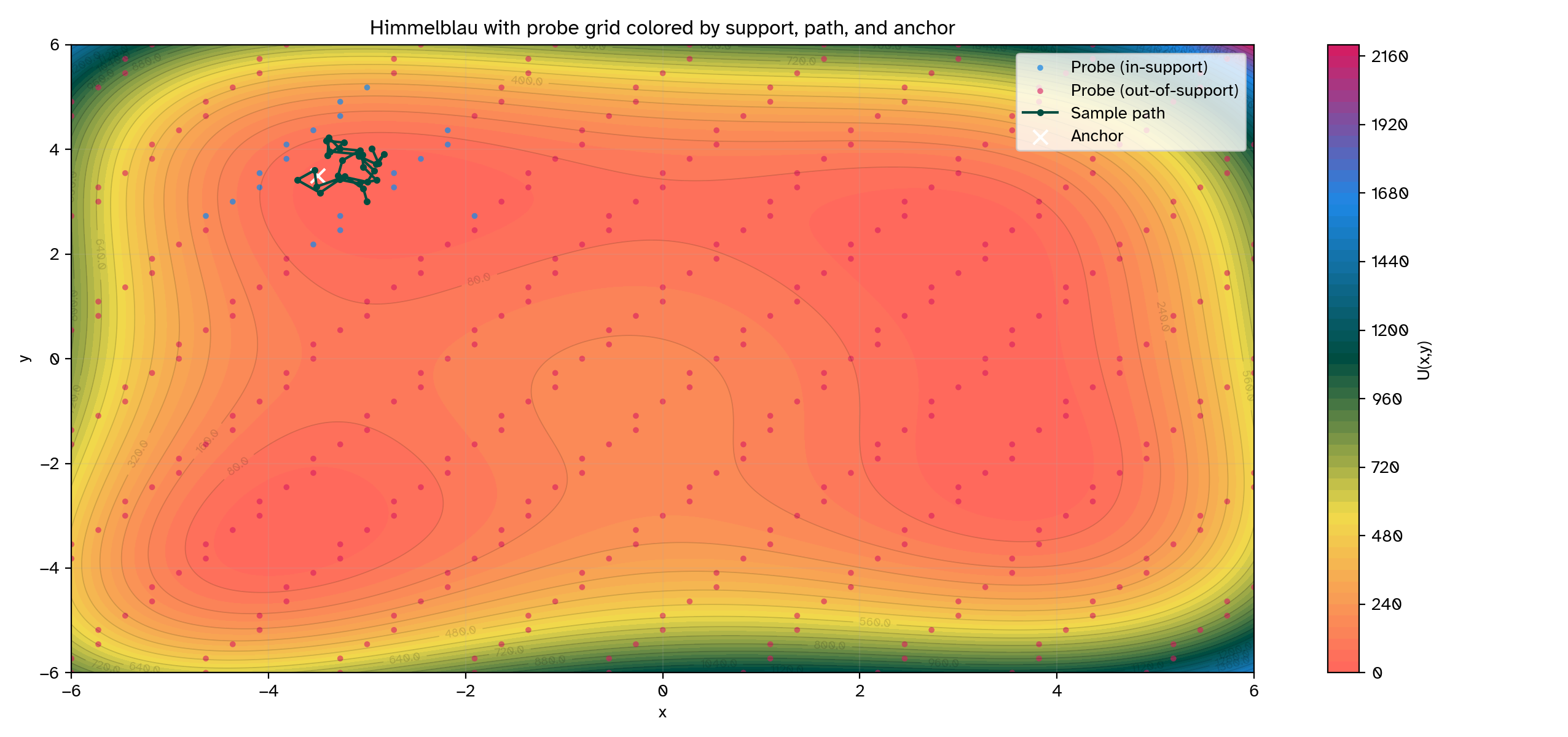}
\caption{\label{fig:hypopt:surface:overview}Himmelblau surface (\(T = 28\) observations). The sample path (teal line) explores the landscape in a local random walk. Probe grid points are colored: blue (in-support, within radius \(r_p = 1.5 \ell_{\text{reopt}}^*\)) and magenta (out-of-support, beyond \(r_p\)). The anchor point (white \(\times\)) is placed within the support region, around 0.1 away from the nearest data point. The visualization reveals that most of the landscape lies out-of-support at any given iteration, a region where the re-optimized model exhibits catastrophic miscalibration.}
\end{figure}

\begin{algorithm}[H]
\caption{Sequential Sampling with Re-Optimized Hyperparameters}
\label{alg:reopt:theta}
\begin{algorithmic}[1]
\State \textbf{Input:} objective $U(\mathbf{x})$, gradient $\nabla U(\mathbf{x})$, initial $\boldsymbol{\theta}_0$, hyperparameter ranges $\Theta$, search trials $N_{\text{trials}}$, sample path $\{\mathbf{x}_1, \ldots, \mathbf{x}_T\}$
\State \textbf{Output:} predictions $\{\mu_t, \sigma_t^2\}_{t=1}^{T}$, RMSE and coverage metrics, optimized $\{\boldsymbol{\theta}_t^*\}_{t=1}^{T}$

\State Initialize: $\mathcal{D}_1 \leftarrow \{(\mathbf{x}_1, U(\mathbf{x}_1), \nabla U(\mathbf{x}_1))\}$, $\boldsymbol{\theta}^* \leftarrow \boldsymbol{\theta}_0$
\State $r_p \leftarrow 1.5 \ell^0$ \Comment{Support radius (will adapt with $\ell_t^*$)}

\For{$t = 1$ to $T$}
  \State \Comment{\textbf{Re-optimize} hyperparameters via random search on marginal likelihood}
  \State $\boldsymbol{\theta}^* \leftarrow \arg\max_{\boldsymbol{\theta} \in \Theta} p(\mathbf{y} | \mathcal{D}_t, \boldsymbol{\theta})$ via Eq.~\ref{eq:mll} (Eq.~\ref{eq:hyperopt:objective})
  \State Extract $\ell_t^* \leftarrow \text{lengthscale}(\boldsymbol{\theta}^*)$; $r_p \leftarrow 1.5 \ell_t^*$ \Comment{Update support radius}
  
  \State \Comment{Fit GP with re-optimized hyperparameters}
  \State Fit GP to $\mathcal{D}_t$ using $\boldsymbol{\theta}^*$
  
  \State \Comment{Predict on dense probe grid}
  \State $\{\mu(\mathbf{z}), \sigma^2(\mathbf{z})\}_{z \in \text{GRID}} \leftarrow \text{GP.predict}(\text{GRID})$ via Eq.~\ref{eq:gp:posterior:mean}, \ref{eq:gp_variance_posterior}
  
  \State \Comment{Evaluate global and split metrics (using re-optimized $r_p$)}
  \State $\text{RMSE}_t \leftarrow \sqrt{\frac{1}{|\text{GRID}|} \sum_{\mathbf{z}} (\mu(\mathbf{z}) - U_{\text{true}}(\mathbf{z}))^2}$
  \State $\{\text{RMSE}_{\text{in},t}, \text{RMSE}_{\text{out},t}\} \leftarrow \text{SplitBySupport}(\mu, U_{\text{true}}, \text{dists}, r_p)$
  \State $\text{Cov}_{1\sigma,t} \leftarrow \frac{1}{|\text{GRID}|} \sum_{\mathbf{z}} \mathbb{1}[|\mu(\mathbf{z}) - U_{\text{true}}(\mathbf{z})| \leq \sigma(\mathbf{z})]$
  
  \State \Comment{Accumulate next observation}
  \State $\mathcal{D}_{t+1} \leftarrow \mathcal{D}_t \cup \{(\mathbf{x}_{t+1}, U(\mathbf{x}_{t+1}), \nabla U(\mathbf{x}_{t+1}))\}$
  
\EndFor

\State \Return all metrics for $t = 1, \ldots, T$; $\{\boldsymbol{\theta}_t^*\}$
\end{algorithmic}
\end{algorithm}

For the Himmelblau function

\begin{equation}
U(\mathbf{x}) = (x_1^2 + x_2 - 11)^2 + (x_1 + x_2^2 - 7)^2
\label{eq:himmelblau}
\end{equation}

with gradient:

\begin{equation}
\nabla U(\mathbf{x}) = \begin{pmatrix}
4x_1(x_1^2 + x_2 - 11) + 2(x_1 + x_2^2 - 7) \\
2(x_1^2 + x_2 - 11) + 4x_2(x_1 + x_2^2 - 7)
\end{pmatrix}
\label{eq:himmelblau:grad}
\end{equation}

. The derivative information is incorporated via cross-covariances between functions and gradients, scaled by a constant factor \(s_d = 10\). This fixed scaling ensures that changes in the marginal likelihood stem from \(\sigma_f\), \(\ell\), and noise parameters alone, not from coupling effects. For this section consider random search over the hyperparameter space:

\begin{equation}
\boldsymbol{\theta}^* = \arg\max_{\boldsymbol{\theta} \in \Theta} p(\mathbf{y} | \mathcal{D}_t, \boldsymbol{\theta})
\label{eq:hyperopt:objective}
\end{equation}

Search ranges are:
\begin{itemize}
\item \(\ell \in [0.25, 3.0]\) (lengthscale)
\item \(\sigma_f \in [0.5, 5.0]\) (signal variance)
\item \(\sigma_n^{\text{f}} \in [10^{-5}, 10^{-1}]\) (function noise)
\item \(\sigma_n^{\text{d}} \in [10^{-5}, 10^{-1}]\) (derivative noise)
\end{itemize}

\begin{figure}[htbp]
\centering
\includegraphics[width=.9\linewidth]{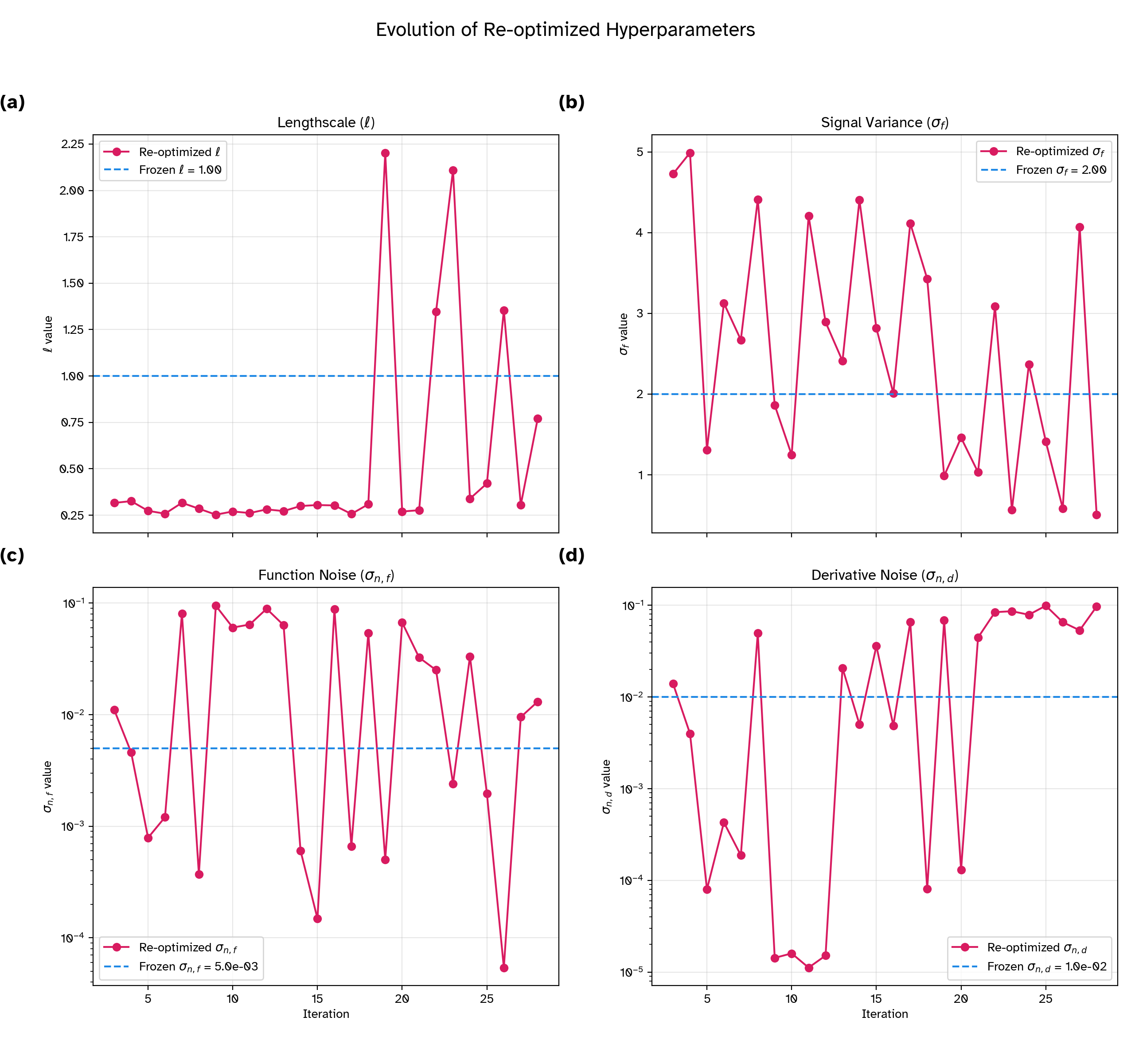}
\caption{\label{fig:hypopt:evol}Hyperparameter re-optimization traces.  The plots show the values of the lengthscale (\(\ell\)), signal variance (\(\sigma_f\)), function noise (\(\sigma_{n,f}\)), and derivative noise (\(\sigma_{n,d}\)) chosen by maximizing the marginal log-likelihood at each step. The dashed blue lines indicate the constant values used by the ``Frozen \(\theta\)'' model. The re-optimized values, particularly for the lengthscale and signal variance, are extremely volatile. They fluctuate dramatically from one iteration to the next, indicating that the MLL optimization landscape is ill-conditioned or has multiple competing maxima, especially when trained on locally clustered data that includes derivatives.}
\end{figure}

We draw \(N_{\text{trials}} = 60\) candidate hyperparameters uniformly in log-space and select the maximum. Figure \ref{fig:hypopt:evol} illustrates the resulting trajectories.

\begin{figure}[htbp]
\centering
\includegraphics[width=.9\linewidth]{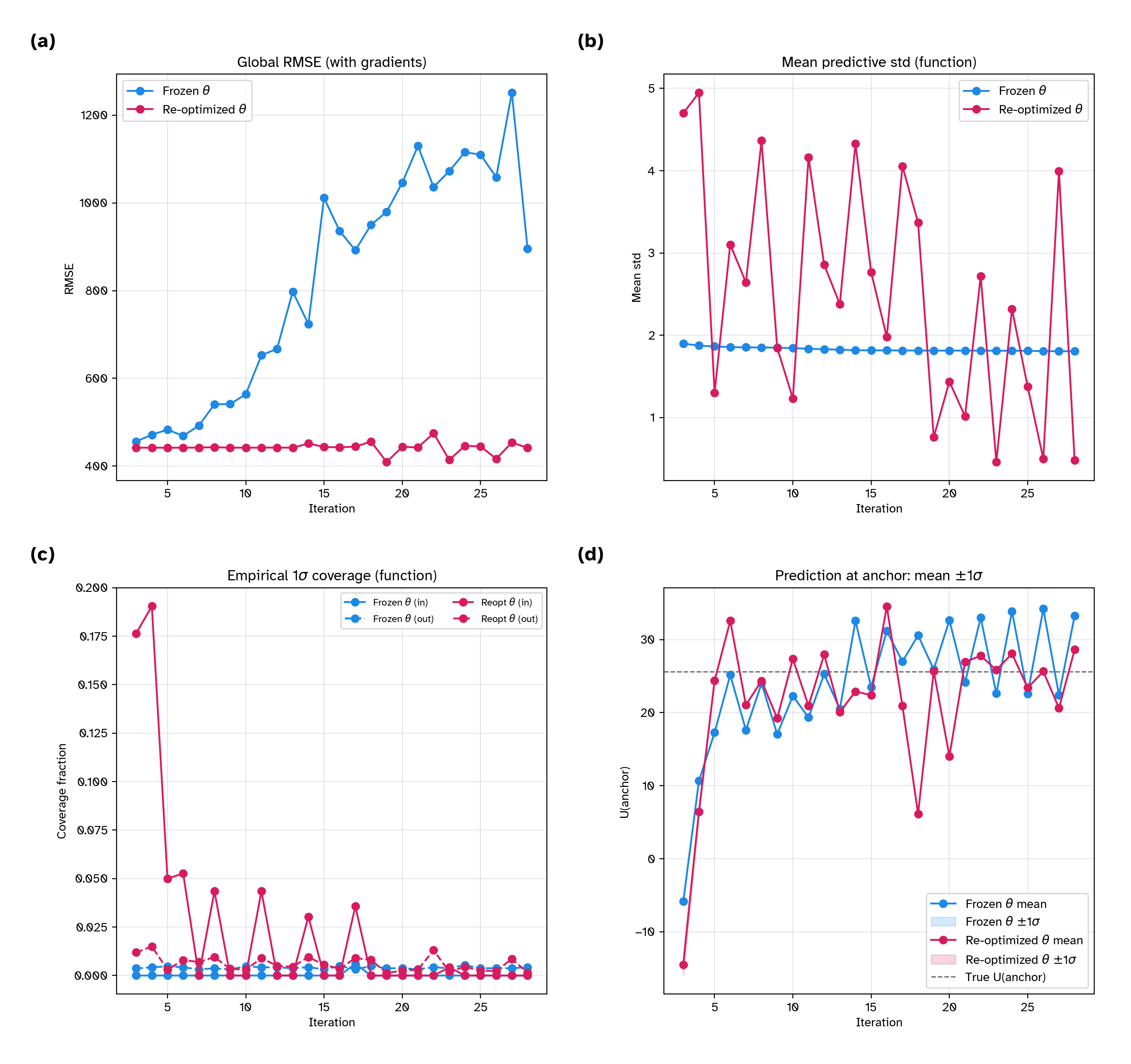}
\caption{\label{fig:hypopt:metrics}Hyperparameter re-optimization effects on accuracy and calibration. (a) Global RMSE: The Root Mean Squared Error over the entire probe grid. The re-optimized model consistently achieves a slightly lower (better) RMSE than the frozen model, suggesting superior global accuracy.(b) Mean Predictive Standard Deviation: The average predictive uncertainty across the grid. The re-optimized model exhibits highly volatile and often significantly larger uncertainty compared to the stable uncertainty of the frozen model.(c) Empirical 1\(\sigma\) Coverage: The fraction of probe points where the true function value falls within the model's predicted \(\pm 1\sigma\) confidence interval. Both models show poor calibration, but the re-optimized model is particularly unreliable for out-of-support points (dashed magenta line), where its coverage fraction is frequently near zero.(d) Prediction at Anchor: The predicted mean and \(\pm 1\sigma\) confidence interval at the anchor point. The frozen model's prediction is stable and converges reasonably close to the true value (dashed black line). In stark contrast, the re-optimized model's prediction can be unstable, with both mean and uncertainty fluctuating with each new data point.}
\end{figure}

Figures \ref{fig:hypopt:surface:overview} and \ref{fig:hypopt:metrics} present the
empirical comparison on the Himmelblau function (Eq. \ref{eq:himmelblau}). \(T =
28\) observations are taken along a random walk initialized at \(\mathbf{x}_0 =
(-3.0, 3.0)\), a point deep in a high-valued basin. A dense probe grid of \(45
\times 45\) points covers the domain \([-6, 6]^2\).

Taken together, these results demonstrate a clear pathology in naively
re-optimizing \gls{gpd} hyperparameters within a sequential learning process,
particularly when using derivative information. While re-optimization can
improve global point-wise accuracy metrics, it does so by sacrificing the
integrity of the model's variance estimates. The resulting model becomes
volatile and severely overconfident, producing miscalibrated uncertainty
predictions that cannot be trusted for decision-making. This reinforces the need
for caution when interpreting predictive variance from models whose
hyperparameters are continually adapted on growing, locally-clustered datasets.
\subsection{Data driven pruning}
\label{sec:dataeff:datadrivenprune}
\begin{figure}[htbp]
\centering
\includegraphics[width=.9\linewidth]{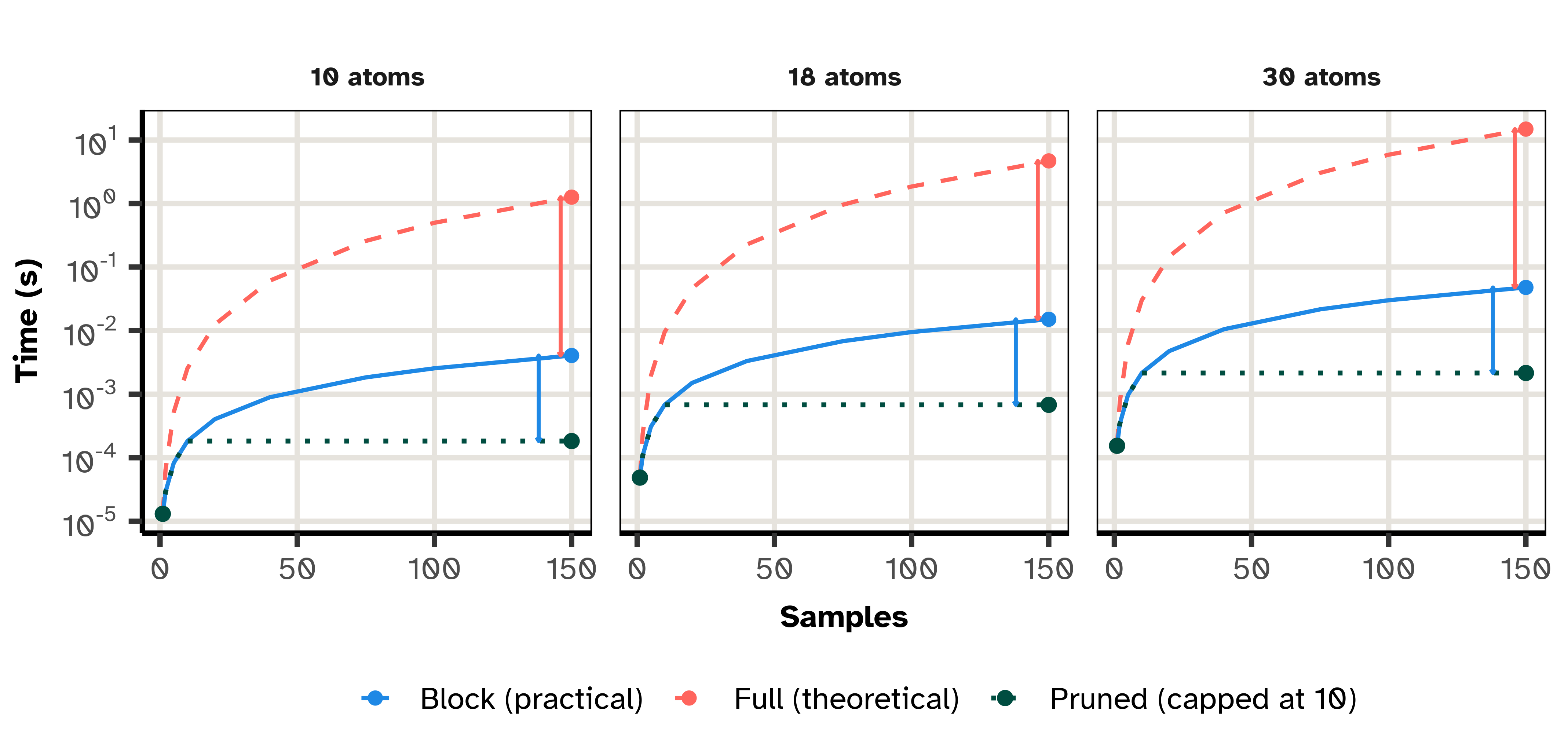}
\caption{\label{fig:gp:scaling:pruned}Computation time scaling with data-driven pruning. Three strategies are compared: full theoretical kernel (dashed red, \((M(3N+1))^2\) elements), practical block matrix (solid blue, \(M(3N+1) \times (3N+1)\) elements), and pruned block matrix (dotted green, capped at 10 configurations, \(10(3N+1) \times (3N+1)\) elements). Time estimates are based on benchmarks from a modern laptop (ThinkPad X1 Carbon 2021; 1538\texttimes{}1538 matrix inversion \textasciitilde{}0.1 s). At 150 samples, pruning would provide consistent \textasciitilde{}22\texttimes{} speedup over block scaling across all molecule sizes, with the benefit growing in absolute time for larger systems.}
\end{figure}

Having established that for active learning settings, standard static data-pruning techniques are not equivalent and may hamper performance, we consider once again the ideal effect of pruning. Figure \ref{fig:gp:scaling:pruned} illustrates this effect across different molecule sizes. At 150 training configurations, the pruned approach achieves a \(22\times\) speedup compared to the full block matrix for all molecule sizes, while the block matrix itself provides a \(300\times\) speedup over the full theoretical kernel. For larger molecules (\(N=30\) atoms), the full kernel computation would require on the order of 15 seconds per optimization step; the block approach reduces this to approximately 0.1 seconds, and pruning further reduces it to approximately 2 milliseconds, corresponding to a combined speedup of over 7000\texttimes{}.

\subsubsection{Hyperparameter trajectories for the GPDimer}
\label{sec:orgba0fe1f}

The hyperparameters in \gls{gpd} runs are seen to stabilize after a modest amount
of data \footnote{though these are not globally optimal like those from \texttt{optuna}
\cite{akibaOptunaNextgenerationHyperparameter2019}}. To this end, we employ a
local, gradient-based \gls{scg} optimizer for the hyperparameters (Alg.
\ref{alg:scg-concise}), warm-started from the converged values of the previous
optimization step. This avoids the cost of a global search and leverages the
fact that the \gls{pes} topology evolves smoothly.

\begin{figure}[htbp]
\centering
\includegraphics[height=0.35\textheight]{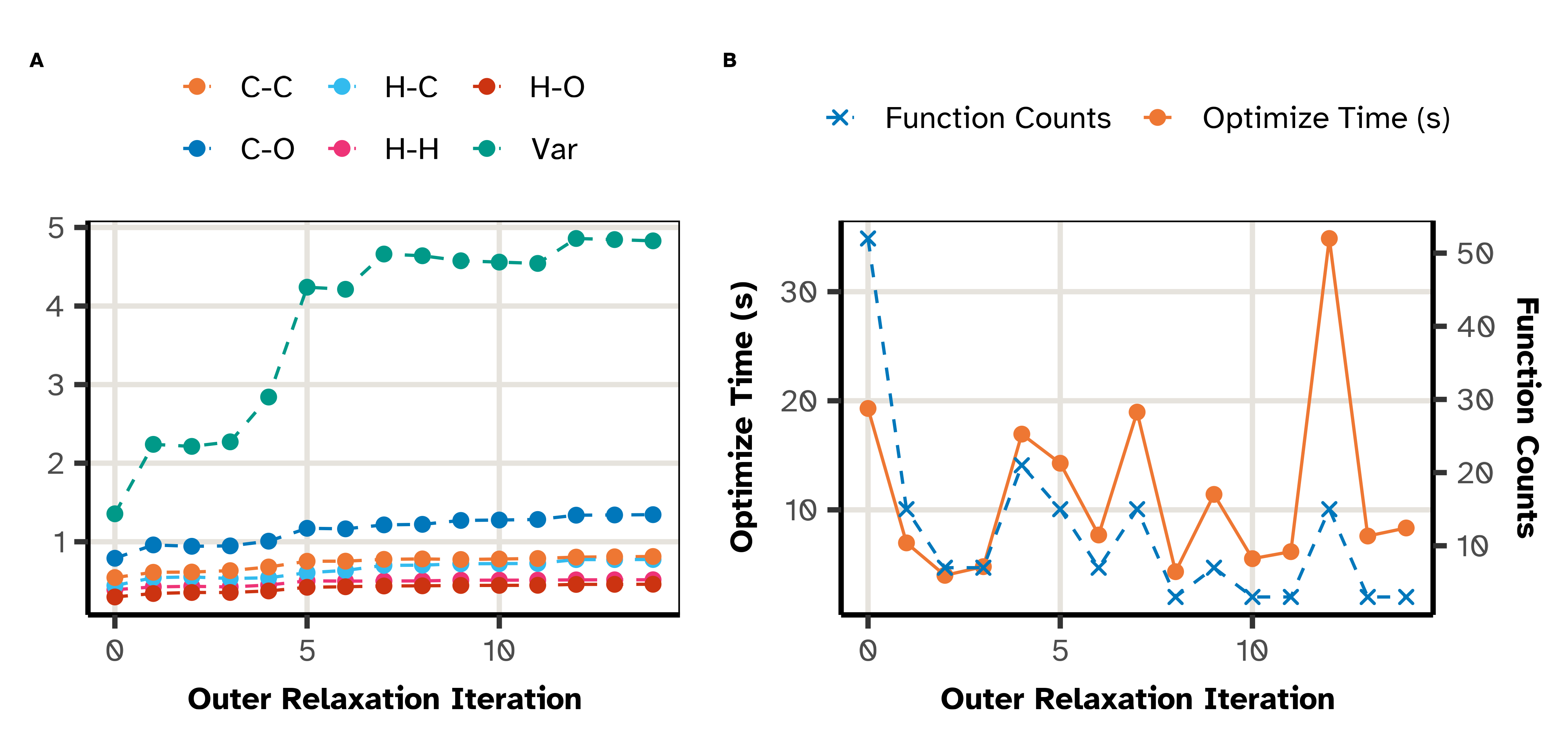}
\caption{\label{fig:gpd:s000_hypot}Hyperparameter and computational cost during \gls{gpd}. \textbf{(A)} Evolution of kernel hyperparameters for the \texttt{S000} show that lengthscales remain stable after an initial adjustment period. The signal variance fluctuates, an artifact of having to fit subsequent points. \textbf{(B)} Computational cost for the hyperparameter optimization at each relaxation loop, showing both wall time and the number of function evaluations. As the steps increase, the time taken grows even as the number of function evaluations reduce. Data from \cite{goswamiEfficientImplementationGaussian2025a}.}
\end{figure}

Fig \ref{fig:gpd:s000_hypot} demonstrates this behavior for a representative
system \texttt{S000}, a 16 atom molecule \(\mathrm{C}_5\mathrm{OH}_{10}\) which starts
from an acyclic ether with a separation of 2.2 \r{A} between the carbon
endpoints. Most of hyperparameters stabilize rapidly, which suggests that the
local maximum on the likelihood surface is a function of a small subset of data.
However, the signal variance (\(\sigma_f^2\)) exhibits fluctuations, hinting at a
potential source of instability in the model. As the geometry changes, the
optimizer may struggle to fit new, challenging data points, causing the variance
to oscillate as an artifact of the dynamic dataset. Furthermore, the cost of
optimization does not decrease monotonically; the time per step can grow even as
the number of function evaluations falls, reflecting the increasing cost of
matrix operations on the growing dataset.

A large signal variance causes the model to lose its physical meaning, behaving
as a pure mathematical interpolator that can guide the search into unphysical
regions (e.g., overlapping atoms), leading to calculation failure. To counteract
this, we introduce an adaptive barrier for the signal variance, and an
oscillation detection heuristic.
\subsubsection{Adaptive Barrier for Signal Variance}
\label{sec:orgc95bf1c}
To directly prevent the pathological growth of the signal variance, we augment the marginal log-likelihood (MLL) objective function with a logarithmic barrier:
\begin{equation}
\mathcal{L}_{\text{eff}}(\boldsymbol{\theta})
= \underbrace{\log p\bigl(\mathbf{y}\mid\mathcal{S},\boldsymbol{\theta}\bigr)}_{\text{MLL}}
  \;-\;
  \mu\,\log\bigl(\lambda_{\text{max}}-\log\sigma_{f}^{2}\bigr)
\label{eq:sigma_barrier}
\end{equation}
where \(\lambda_{\text{max}}\) fixes an absolute upper bound for \(\log\sigma_{f}^{2}\), and the barrier strength, \(\mu\), grows linearly with the number of collected data points, \(N\):
\begin{equation}
\mu(N)=\mu_{0}+\alpha N,\qquad \mu(N)\le \mu_{\max}
\label{eq:sigma_linear}
\end{equation}
This schedule allows the model to remain flexible when data is sparse (small \(\mu\)) but enforces an increasingly strict bound as the dataset matures and the model should have settled on a physically reasonable amplitude. This adaptive behavior eliminates pathological variance growth while preserving the surrogate's ability to capture the true curvature of the PES, as seen in Figure \ref{fig:var_explode} (B).
\subsubsection{Hyperparameter Oscillation Detection}
\label{sec:org96912b2}

Re-optimizing hyperparameters on a dynamically changing subset of data can lead to unstable estimates that oscillate between iterations as shown in previous sections. The Hyperparameter Oscillation Detection (HOD) heuristic monitors these fluctuations over a moving window of the last \(W\) steps. We define an oscillation indicator, \(O_{j}(t)\), for each hyperparameter \(\theta_j\) at step \(t\):

\begin{equation}
O_{j}(t)=
\begin{cases}
1 & \text{if }\operatorname{sgn}\bigl[\Delta\theta_{j}(t-1)\bigr]\neq
      \operatorname{sgn}\bigl[\Delta\theta_{j}(t-2)\bigr],\\[4pt]
0 & \text{otherwise}.
\end{cases}
\label{eq:hod_cases}
\end{equation}
where \(\Delta\theta_j(t) = \theta_j(t) - \theta_j(t-1)\). If the total fraction of oscillations, \(f_{\text{osc}}\), across all hyperparameters in the window exceeds a threshold, \(p_{\text{osc}}\), the optimization is flagged as unstable. In response, the algorithm automatically enlarges the subset of data used for fitting, which improves the conditioning of the covariance matrix and typically results in a smoother, more stable MLL surface.
\subsection{Conclusions}
\label{sec:dataeff:conc}
This chapter demonstrated that scaling \gls{gp} methodologies for high-performance chemistry requires a synthesis of algebraic restructuring and careful statistical approximation. We established that matrix reshaping into block-diagonal forms reduces memory complexity to linear scaling without sacrificing mathematical fidelity. However, the wall-time cost of inverting these matrices during iterative hyperparameter optimization remains the primary computational bottleneck.

Our investigation into naive data pruning revealed severe pathologies; coupling the data support set directly to the inference engine creates feedback loops that drive optimization trajectories toward erroneous local minima. Furthermore, aggressive hyperparameter re-optimization on dynamic local subsets sacrifices variance calibration for marginal gains in point-wise accuracy, rendering the model overconfident and unsuitable for autonomous decision-making.

To mitigate these risks while capturing the immense speedups offered by sparse kernels—exceeding \(7000\times\) for larger systems we introduced essential stability mechanisms. The adaptive signal variance barrier and Hyperparameter Oscillation Detection (HOD) prevent the unphysical divergence of the model during the volatile early stages of learning. These safeguards permit the use of reduced datasets for the hyperparameter optimization, yet the specific criterion for which data to retain remains critical. Simple Euclidean metrics fail to capture chemical similarity effectively across diverse landscapes. Consequently, the final component of this framework demands a rigorous, chemically transferable metric for point selection. Chapter \ref{sec:otgpd} addresses this need through the introduction of the \gls{otgpd}, replacing geometric distance with Optimal Transport metrics to guide both pruning and convergence.
\section{Optimal Transport Gaussian Process}
\label{sec:otgpd}
\epigraph{Every method is somewhere between random searches and gradient descent.}{Hannes Jónsson \\ Discussion with Rohit Goswami}

\begin{quote}
This chapter is based on \fullcite{goswamiAdaptivePruningIncreased2025b}
\end{quote}

\begin{figure}[htbp]
\centering
\includegraphics[width=.9\linewidth]{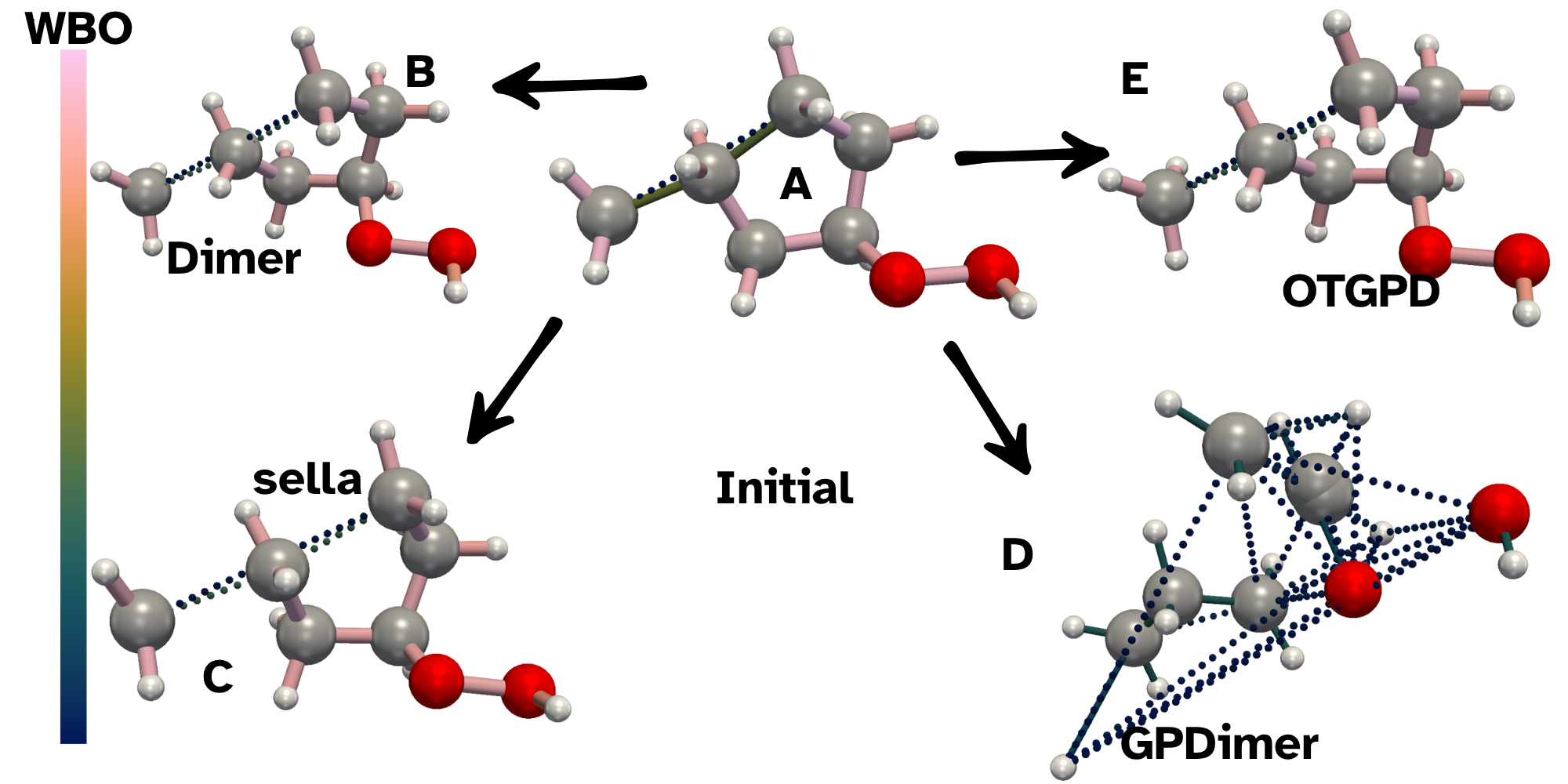}
\caption{\label{fig:wtf:optgd}A comparison of saddle point search trajectories for a ethoxy radical hydrogen abstraction reaction, \texttt{doublet\_150} reaction starting from an initial configuration (A). The standard Dimer method (B), Sella (C), and the OTGPD (E) follow a chemically intuitive path. The previous GPDimer method (D) is guided towards a fractured state, leading to failures in the underlying NWChem calculator.}
\end{figure}

So far, we've demonstrated state of the art performance for the \gls{gpd}, along
with more principled measures of measuring performance, and possible
prescriptions for data efficiency. Proxy-based approaches remain attractive
because they avoid repeated electronic-structure calls, yet several of the
reported trajectories contain chemically implausible features. While the \gls{gp}
is a data driven surrogate the difference between running towards abstract high
dimensional neural networks and the Cartesian representation is the belief that
there is still an interpretation. We note for instance, that the \gls{gpd} on
several systems fails to constrain the generated surrogate surface and thus ends
up exploring pathologically unstable regions of phase space corresponding to
``cold fusion'', shown in Figure \ref{fig:wtf:optgd}. It is relevant to note that
these systems are not atypical in any form, there are several similar hydrogen
abstraction reactions which succeed.

\begin{sidewaysfigure}
\centering
\includegraphics[width=.9\linewidth]{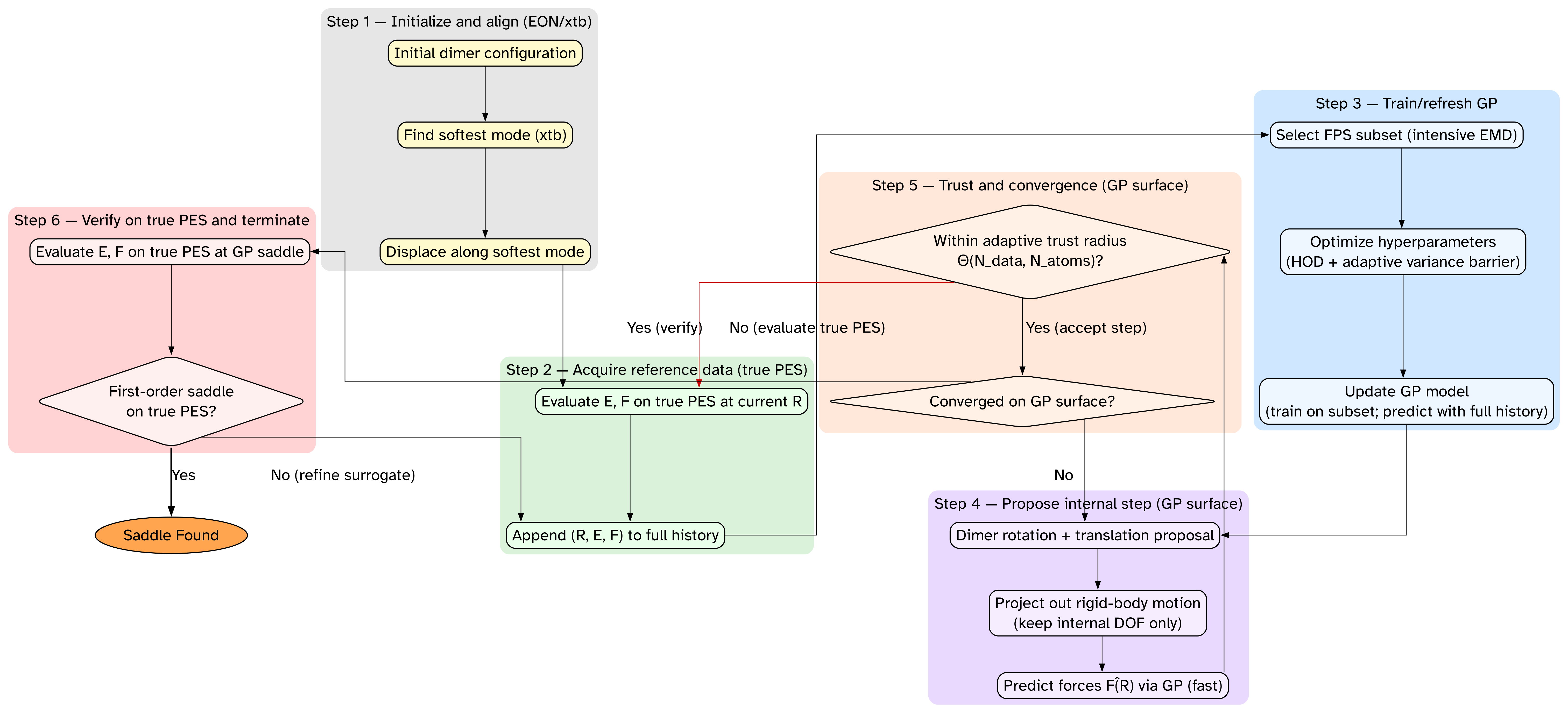}
\caption{\label{fig:otgpdflow}Optimal Transport Gaussian Process Regression framework applied to the dimer method. The algorithm begins with initialization (Step 1, grey) and acquires an initial reference point from the true PES (Step 2, green) to train the GP model (Step 3, blue). An internal, computationally 'cheap' optimization loop (Steps 4 and 5, purple/orange) then searches for a saddle point candidate on the surrogate GP surface. This internal search is governed by an adaptive trust radius (Step 5) to ensure reliability. Calls to the ``expensive'' calculator (Step 2) are only triggered when necessary: either when the optimizer moves outside the trusted region or after the internal optimization on the GP surface has converged. This new data is used to update and refine the GP model (Step 3). Once the entire process converges, the final candidate structure is verified on the true PES (Step 6, red) to confirm it is a valid first-order saddle point before the algorithm terminates.}
\end{sidewaysfigure}

To better understand these failure modes we recall the spline view of a \gls{gp}
\cite{rasmussenGaussianProcessesMachine2004}. With a fixed kernel \(k\) and
observation noise \(\sigma_n^2\), the GP posterior mean is the function that
balances fit and roughness in the \gls{rkhs}:
\begin{equation}
\hat f \;=\; \arg\min_{f\in \mathcal{H}_k} \;\frac{1}{\sigma_n^2}\sum_{i}(y_i - f(x_i))^2 \;+\; \|f\|_{\mathcal{H}_k}^2,
\label{eq:wahba_spline}
\end{equation}
where kernel hyperparameters (length scales, signal variance) control what
counts as ``roughness'' and how it is penalized. In practice, hyperparameters
\(\boldsymbol{\theta}\) are chosen statistically by maximizing the \gls{mll} on the
training set \(\mathcal{S}\),
\begin{equation}
\boldsymbol{\theta}^* = \arg\max_{\boldsymbol{\theta}} \log p(\mathbf{y} | \mathcal{S}, \boldsymbol{\theta}),
\label{eq:mll_theta}
\end{equation}
rather than by minimizing an inaccessible physical discrepancy to the unknown
\gls{pes},
\begin{equation}
\boldsymbol{\theta}_{\text{ideal}} = \arg\min_{\boldsymbol{\theta}} \int |f(x; \boldsymbol{\theta}) - V(x)|^2 \, dx.
\label{eq:ideal_theta}
\end{equation}
The Gaussian Process, defined in terms of finite realizations of multivariate
normal distributions, expressed as a ``function'', or a series of \(x,y\) pairs
reshaped to provide familiarity with 3D matrices; thus suffers greatly from
anthropomorphism of the constituent equations. The physical meaning ascribed to
the hyperparameters are unjustified, and models will only be guaranteed to
interpolate in the noise free regime \footnote{also known to be unstable
\cite{gramacySurrogatesGaussianProcess2020}}. The repeated re-optimization also
doesn't preserve global accuracy as shown earlier. This distinction becomes
decisive in actively learned saddle-point searches, which produce correlated
trajectories of geometries which are atypical from most of the energy surface
geometrically. Under such data, the \gls{mll} surface can be shallow in directions
like the signal variance, encouraging variance blow-up that flattens the mean
and inflates predictive uncertainty, which contributes significantly to the
destabilization of the \gls{gpd}. From a data efficiency perspective, what was
covered in section \ref{sec:dataeff} has even greater significance in terms of
reliability. However, we still require a reasonable measure of distance to
complete algorithm developed in this chapter to make good on wall time
performance, which we summarize in Fig. \ref{fig:otgpdflow}.

\subsection{Intensive \Gls{emd}}
\label{sec:otgpd:iemd}
Before defining the distance measure, we revisit the extant distance metrics
within the methodology outlined in Section \ref{sec:repasp}. We start with the
guardrails on the \gls{gpd} as formulated in Chapter \ref{sec:gpjctc} previous
chapter and in the literature \cite{koistinenMinimumModeSaddle2020}. There are
two, one on the interatomic distances of a given configuration, and one on the
distance from a known point.

An inequality expresses the measure in the literature \cite{koistinenMinimumModeSaddle2020}:
\begin{align}
\frac{2}{3}r_{ij}(\mathbf{x}_{\text{eval}}) &< r_{ij}(\mathbf{x}_{\text{im}}) < \frac{3}{2}r_{ij}(\mathbf{x}_{\text{eval}}) \\
\implies \left| \log \frac{r_{ij}(\mathbf{x}_{\text{im}})}{r_{ij}(\mathbf{x}_{\text{eval}})} \right| &< \log(1.5) \approx 0.405
\end{align}

The core of the problem is that the inverse distance we use in
\ref{eq:idist_kernel_ee} is not invariant to the permutation of identical atoms,
and since kernel's value depends on a direct, index-wise comparison of the
interatomic distance vectors of two configurations. This creates a dependency on
the arbitrary, fixed labels of the atoms, rather than their physical roles.

An easy way to understand this stems from observing symmetric systems. For
instance, consider a proton (indexed k) transferring between two chemically
equivalent sites (m and n). Physically, the initial and final states are
energetically degenerate. However, a fixed-index comparison metric perceives a
significant geometric change, as the distance r(k,m) transitions from short to
long, while r(k,n) simultaneously transitions from long to short. The metric
fails to recognize that the permutation of labels would reconcile the apparent
structural difference.

While the kernel's fitted length-scale hyperparameter may partially average out
this effect, a non-averaged metric for early stopping feels the full impact of
the flaw. The 1D max log distance, by its definition, registers a significant,
non-physical distance for this symmetric swap:

\begin{equation}
D_{\text{1Dmaxlog}}(\mathbf{x}_1, \mathbf{x}_2) = \max_{i,j} \left| \log \frac{r_{ij}(\mathbf{x}_2)}{r_{ij}(\mathbf{x}_1)} \right|
\end{equation}

\begin{figure}[htbp]
\centering
\includegraphics[width=.9\linewidth]{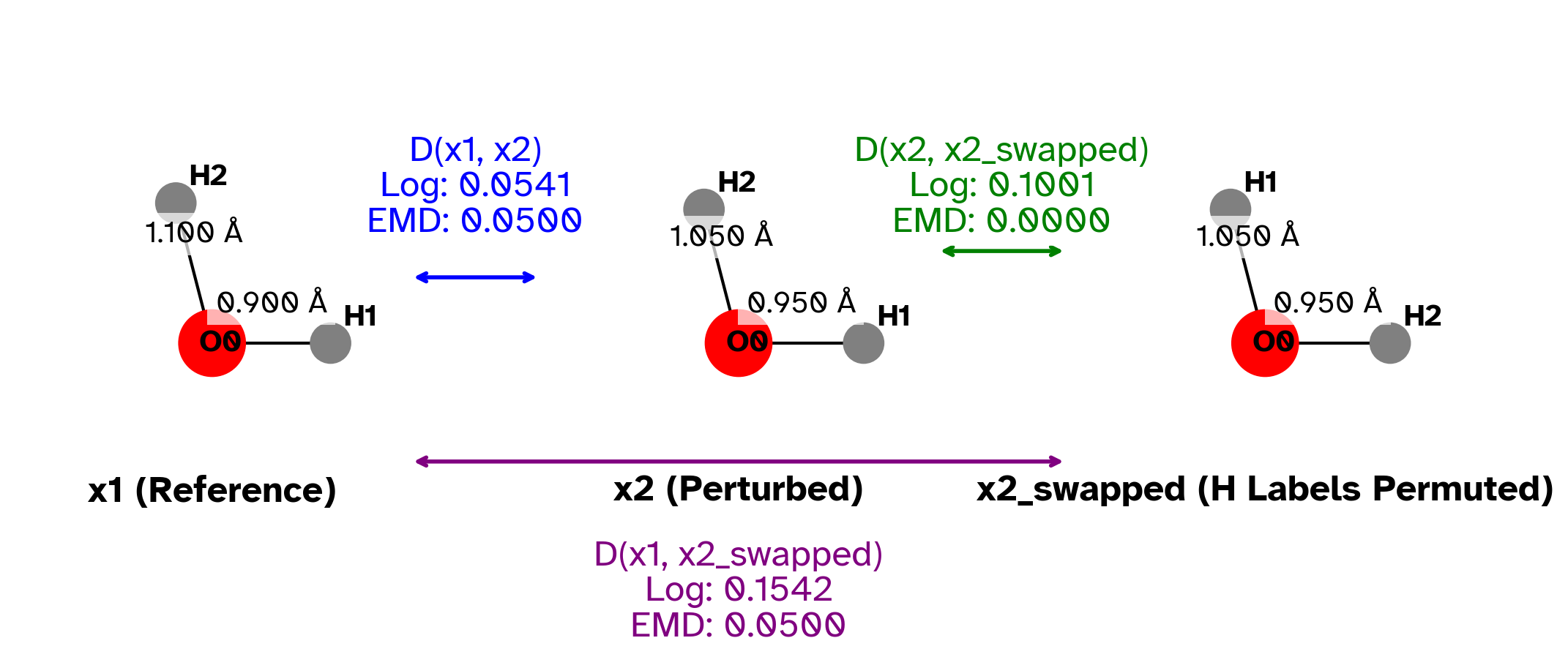}
\caption{\label{fig:suppl:1dmaxEMD}Comparison of the 1D max log distance and the Earth Mover's Distance (EMD) for an asymmetric stretch of a water molecule. While configuration \(x_2\) and \(x_{2,s}\) are physically identical (differing only by the permutation of hydrogen atom labels), the 1D max log metric incorrectly assigns a large distance between them and the reference \(x_1\). In contrast, the EMD correctly identifies them as being equidistant from the reference, demonstrating its permutational invariance.}
\end{figure}

This sensitivity to labeling motivates using the intensive EMD. Figure
\ref{fig:suppl:1dmaxEMD} demonstrates this, by contrasting the behavior of both
metrics for the asymmetric stretching of a water molecule.

So we would preferably have a measure which is invariant to both permutations of
the labels, and not be a property which grows with the system size. Optimal
transport theory \cite{thorpeIntroductionOptimalTransport}, more precisely the
\gls{emd} metric which solves a linear assignment problem, often the Hungarian
\cite{gundeDevelopmentIRAShape2021} to ensure minimal motion required to deform
one discrete distribution into another. However, rather than introducing mass
weighing of graphs, we opt to instead solve the EMD for each atom type, thus we
solve a ``colored'' EMD, divided by the number of atoms to scale intuitively:

\begin{equation}
\bar d_{t}
 = \frac{1}{N_{t}}
   \min_{\pi\in\Pi_{N_{t}}}
   \sum_{k=1}^{N_{t}}
   \bigl\|\mathbf{r}^{(1)}_{k,t}
         - \mathbf{r}^{(2)}_{\pi(k),t}\bigr\|.
\label{eq:emd_type_avgdisp}
\end{equation}
Here, \(N_{t}\) denotes the number of atoms of type \(t\) and \(\Pi_{N_{t}}\)
the set of all permutations of the \(N_{t}\) indices. We then identify the
largest per-type average displacement as the overall distance:
\begin{equation}
D(\mathbf{x}_i,\mathbf{x}_j)=\max_{t}\,\bar d_{t}(\mathbf{x}_i,\mathbf{x}_j).
\label{eq:emd_dist}
\end{equation}
Because each \(\bar d_{t}\) averages over the atoms of a particular element, it
forms an intensive quantity that reflects the collective motion of a specific
chemical group. Adding spectator atoms does not dilute the metric, which makes
it an ideal measure for selecting a chemically diverse subset.
\subsection{Adaptive trust radius}
\label{sec:otgpd:atr}
With this new distance metric, we now re-state the trust region formulation,
with a few additional notes.

While our surrogate model accelerates the search for saddle points, its
reliability is confined to regions of the potential energy surface where it has
been trained. To prevent the algorithm from taking overly ambitious steps into
uncharted territory where the surrogate's predictions may be inaccurate, we
introduce a dynamic ``trust radius.'' This mechanism acts as an intelligent
guardrail, ensuring that any proposed step remains within a zone of confidence
defined by the existing data.

The core of this guardrail is a simple condition. We measure the distance, using
our permutationally-invariant Earth Mover's Distance (EMD), between any new
candidate configuration (\(\mathbf{x}_{\text{cand}}\)) and its nearest neighbor
(\(\mathbf{x}_{\text{nn}}\)) in the current training set. This step is only
accepted if the distance is within an adaptive threshold, \(\Theta\):
\begin{equation}
d_{\text{EMD}}\bigl(\mathbf{x}_{\text{cand}},\mathbf{x}_{\text{nn}}\bigr)\le
\Theta\bigl(N_{\text{data}},N_{\text{atoms}}\bigr)
\label{eq:nn_trust}
\end{equation}

This threshold, \(\Theta\), is not static; it evolves as the surrogate model gathers more information. We designed its functional form to follow an exponential saturation curve, allowing the model to become more adventurous as its knowledge base grows. This ``earned trust'' radius is defined as:
\begin{equation}
\Theta_{\text{earned}}\bigl(N_{\text{data}}\bigr)=
T_{\min }+\Delta T_{\text{explore}}\cdot\Bigl(1-e^{-k\,N_{\text{data}}}\Bigr)
\label{eq:nn_satcurv}
\end{equation}
Here, \(T_{\min}\) provides a minimal safe radius to prevent trivially small steps, while \(\Delta T_{\text{explore}}\) sets the maximum additional exploration distance the algorithm can earn. The rate of this expansion is controlled by \(k\), which is linked to \(N_{\text{half}}\), the number of data points needed for the threshold to reach half of its maximum value.

However, to ensure physical realism, we impose a hard ceiling on this trust radius that prevents it from becoming unphysically large, regardless of the amount of data collected. This ceiling is dependent on the size of the system:
\begin{equation}
\Theta_{\text{phys}}(N_{\text{atoms}})
= \max\Bigl(\,a_{\text{floor}},\; \frac{a_{A}}{\sqrt{N_{\text{atoms}}}}\,\Bigr)
\label{eq:nn_trust_phys}
\end{equation}
with \(a_{\text{floor}}\) a user-defined lower bound and \(a_{A}\) a scaling constant.

The final, operational trust radius is simply the more restrictive of these two bounds: the one the model has earned through data collection and the one imposed by physical constraints.
\begin{equation}
\Theta(N_{\text{data}}, N_{\text{atoms}})
= \min\bigl(\Theta_{\text{earned}}(N_{\text{data}}),\,\Theta_{\text{phys}}(N_{\text{atoms}})\bigr)
\label{eq:nn_trust_exact}
\end{equation}

If a proposed step violates the trust radius in Eq. \ref{eq:nn_trust}, the algorithm intelligently recognizes a gap in its knowledge. It rejects the step and instead evaluates the energy at that very point of failure. This new data point is then added to the training set, and the trust radius is recomputed. This process of targeted data acquisition actively and efficiently improves the surrogate model precisely where it proves to be unreliable, ensuring our search remains both bold in its exploration and grounded in the reality of the potential energy surface. While it is nice to have principled and ultimately relatable distances, the \gls{emd} can be used for much more, in particular, to help with numerical stability.
\subsection{Numerical conditioning for Gaussian Processes}
\label{sec:otgpd:numcond}
We can estimate conditioning of the joint energy–force covariance using Gershgorin's Circle Theorem \cite{vargaGersgorinHisCircles2004}. For a real, symmetric covariance matrix to be \gls{psd}, all its eigenvalues must be non-negative. Gershgorin's theorem provides bounds on these eigenvalues, stating that each eigenvalue must lie within at least one of the intervals (Gershgorin discs) defined by:
\begin{equation}
\label{eq:gershgorin-intervals}
\lambda \in \bigcup_{i=1}^{n} \left[\, K_{ii} - R_i, \; K_{ii} + R_i \,\right], 
\quad R_i := \sum_{j \ne i} |K_{ij}|.
\end{equation}

This yields a lower bound on the smallest eigenvalue:
\begin{equation}
\lambda_{\min}(\mathbf{K}) \ge \min_{i} \left( K_{ii} - R_i \right).
\label{eq:gershgorin-lb}
\end{equation}

For block-structured energy–force kernels, a tighter bound groups terms by configuration \cite{echeverriaBlockDiagonalDominance2018}:
\begin{equation}
\lambda_{\min}(\mathbf{K}) \ge 
\min_{i} \left\{ \lambda_{\min}(\mathbf{K}_{ii}) - \sum_{j \ne i} \|\mathbf{K}_{ij}\|_2 \right\},
\label{eq:block-gershgorin}
\end{equation}

In practice, the diagonal terms reflect the signal variance, constant offset, and noise:
\begin{equation}
K_{EE}(\mathbf{r},\mathbf{r}) = \sigma_c^2 + \sigma_f^2 + \sigma_{n,E}^2,
\label{eq:diag-energy}
\end{equation}

while the force diagonal contains a metric-dependent second derivative and noise:
\begin{equation}
\big(\mathbf{K}_{FF}(\mathbf{r},\mathbf{r})\big)_{(i,d),(i,d)}
= -\frac{\sigma_f^2}{2}
\left.
\frac{\partial^2 \mathcal{D}^2}{\partial x_{i,d}\,\partial x'_{i,d}}
\right|_{\mathbf{x}'=\mathbf{x}=\mathbf{r}}
+ \sigma_{n,F}^2.
\label{eq:diag-force}
\end{equation}

When sampled configurations cluster too closely in geometry, the off-diagonal terms \(R_i\) grow large, and the Gershgorin bound can become small or negative—signalling poor conditioning and risk of numerical instability. Notably, the signal variance \(\sigma_f^2\) cancels out in the simple diagonal dominance test, indicating that configuration geometry and length scale govern stability, not the variance alone \cite{ababouConditionNumberCovariance1994}.
\subsection{Farthest point sampling}
\label{sec:otgpd:fps}
The upshot of the analysis in the previous section reveals that without sufficient geometric diversity, surrogate models risk numerical instability, manifesting as failed saddle searches or unphysical predictions. To mitigate this, we employ \gls{fps}, which systematically selects new configurations that maximize their separation from the existing set. \gls{fps} directly suppresses the magnitude of off-diagonal covariance terms, shrinking the Gershgorin radii and helps maintain the diagonal dominance needed for stable and physically meaningful surrogate surfaces.

\begin{figure}[htbp]
\centering
\includegraphics[width=1.1\textwidth]{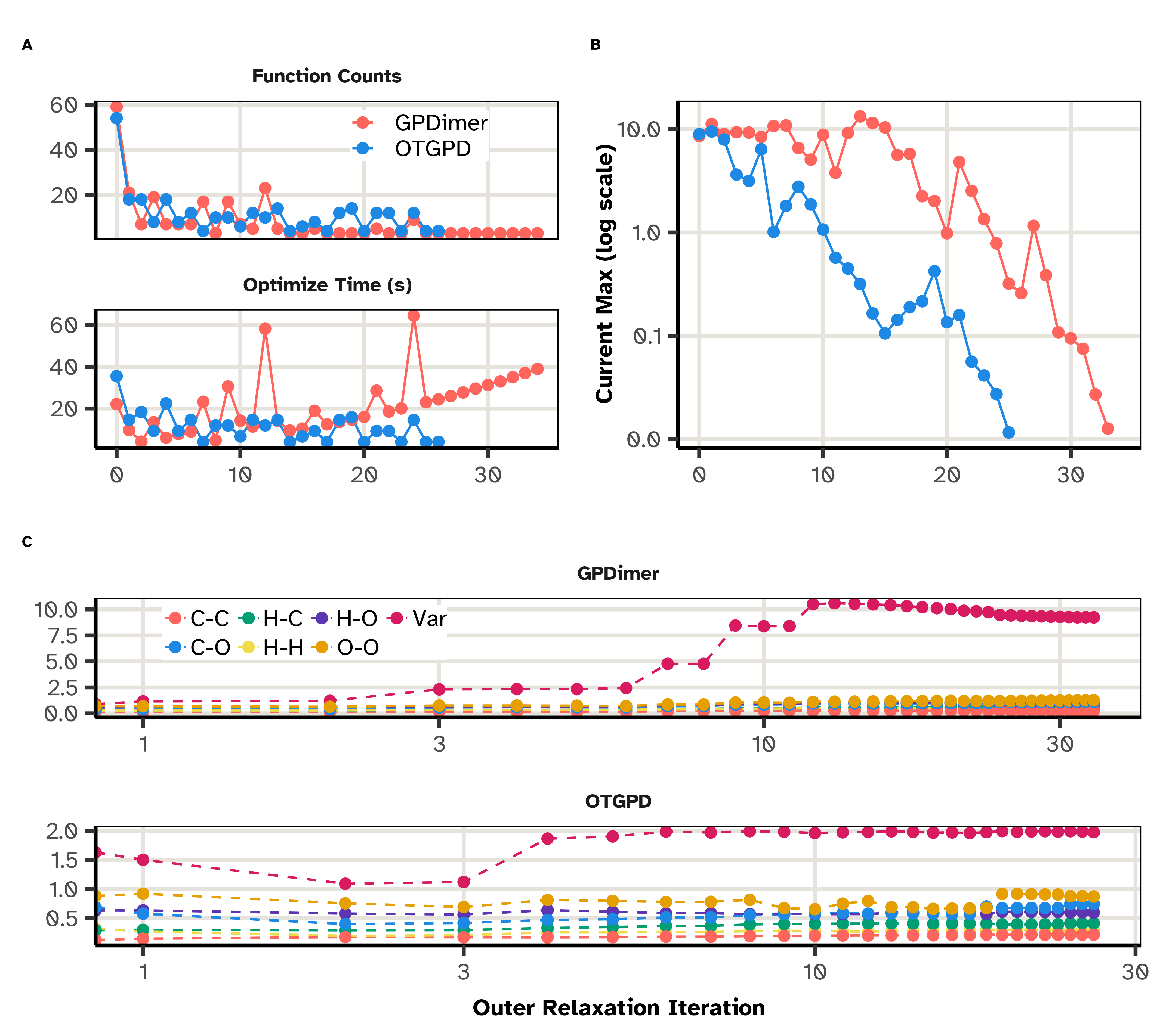}
\caption{\label{fig:s016_perf}Performance trace for the \texttt{singlet\_016} system (Figure fig:equiv:optgd), illustrating the comparative behavior of \gls{gpd} and \gls{otgpd} during saddle search optimization. (A) The per-iteration electronic structure function counts and wall time show that \gls{otgpd} (skyblue) consistently achieves lower and more stable computational cost per iteration compared to \gls{gpd} (coral), which exhibits pronounced spikes and variability. (B) Convergence profiles of the maximum force component (log scale) demonstrate smoother and more rapid relaxation for \gls{otgpd}, while \gls{gpd} progress stalls intermittently, reflecting underlying model instability. (C) Evolution of key hyperparameters over the course of the optimization, with the GP signal variance (magenta, Var) and interatomic distances (C-C, H-C, H-O, C-O, H-H, O-O) tracked for both methods. \gls{gpd} displays episodes of pathological variance growth, coinciding with force and runtime spikes, whereas \gls{otgpd} maintains stable and physically reasonable hyperparameter values throughout.}
\end{figure}

Figure \ref{fig:s016_perf} demonstrates the practical advantage of this approach in the \texttt{singlet\_016} system, where FPS and adaptive variance control enable robust and efficient saddle searches, in contrast to the instabilities observed in standard GPDimer runs, while also mitigating the effect of the size of the hyperparameter matrix optimization, which is the primary walltime bottleneck.

However, as demonstrated in the previous section, \gls{fps} is not merely an efficiency improvements; but an essential countermeasure to manage the numerical instabilities. By construction, \gls{fps} selects new data points that are maximally distant from the existing set. This strategy directly suppresses the magnitude of the off-diagonal covariance terms, systematically shrinking the Gershgorin radii (\(R_i\)). In doing so, \gls{fps} actively enforces the diagonal dominance (\(D_{ii} > R_i\)) required for a numerically stable and physically reliable surrogate model.
\subsection{Variance control and Hyperparameter stability}
\label{sec:otgpd:vconhypopt}
Figure \ref{fig:s016_perf} also shows a sudden jump in the hyperparameters, in particular, the variance. The effect of this hyperparameter is to basically make the model pathologically sensitive to the data points, which in turn will sample configurations which crash NWChem.  To counteract this instability and ensure robust performance, we implement two complementary mechanisms discussed in chapter \ref{sec:dataeff}: a direct control on the signal variance via an adaptive barrier, and a general heuristic for monitoring the stability of all hyperparameters. The adaptive behavior eliminates pathological variance growth while preserving the surrogate's ability to capture the true curvature of the \gls{pes}. Empirically, the combination of the adaptive barrier on \(\sigma_f^2\) and the general HOD mechanism provides robust control. The barrier acts as a targeted preventative measure against a known failure mode, while the HOD acts as a general safety net for the entire optimization process. Together, they reduce the number of failed saddle searches from roughly twelve percent to two percent, contributing significantly to the overall efficiency and reliability of the \gls{otgpd} method.
\subsection{Results}
\label{sec:otgpd:res}
We consider systems which have less than three fragments, and any calculation exceeding 240 minutes or which lead to NWChem failures or termination conditions other than success in EON are considered to be failed. 
\subsubsection{Reliability}
\label{sec:otgpd:res:reliable}
\begin{figure}[htbp]
\centering
\includegraphics[width=1.01\linewidth]{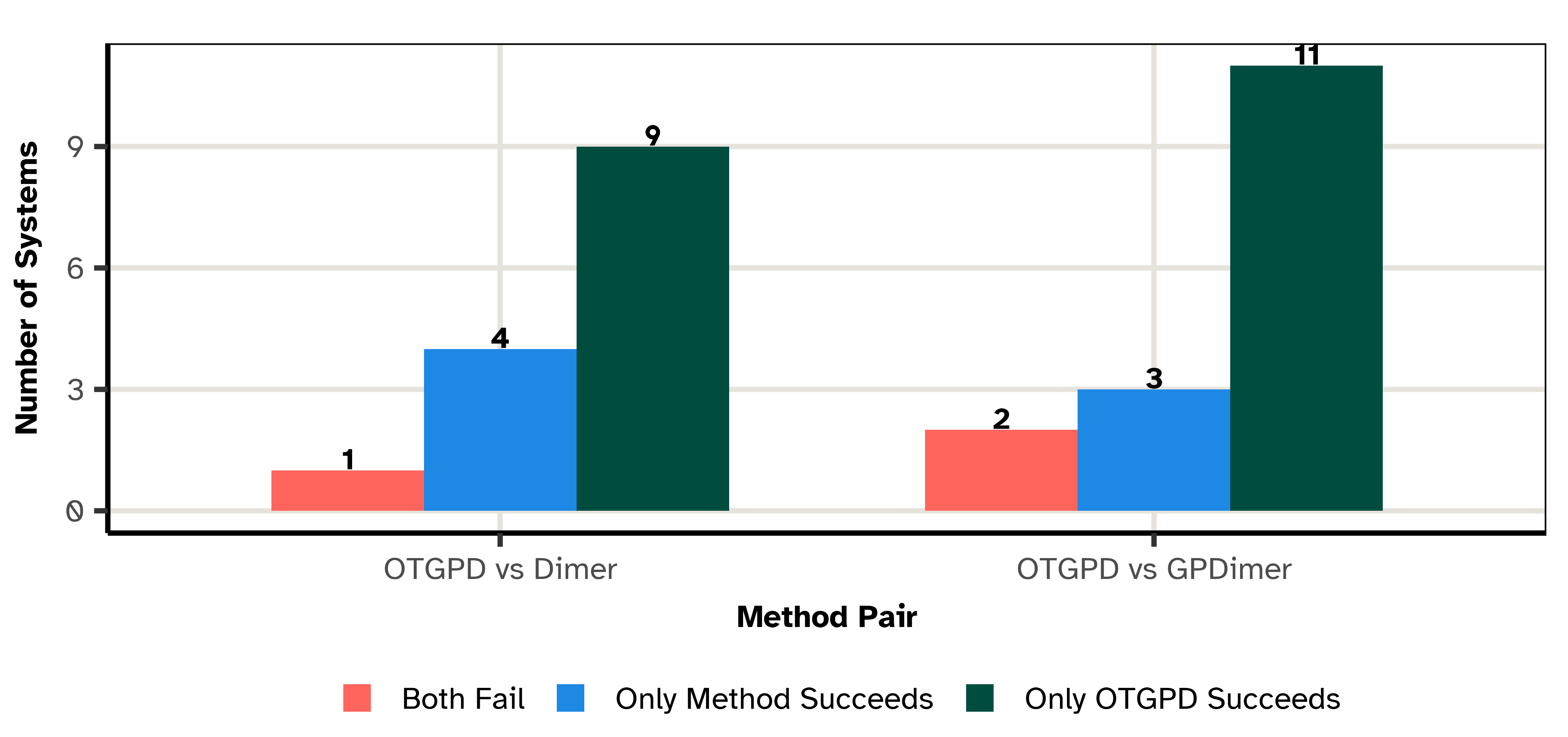}
\caption{\label{fig:otpgp_success}Reliability comparison of OTGPD against GPDimer and standard Dimer methods across 238 molecular systems. A calculation exceeding 240 minutes or raising an error in the electronic structure calculation counts as a failure. The bar chart shows the distribution of outcomes for each pairwise comparison: (red) systems where both methods fail, (blue) systems where only the alternative method succeeds, and (green) systems where only OTGPD succeeds. OTGPD uniquely finds the saddle point for 11 additional systems (4.6\%) compared to GPDimer and 9 additional systems (3.8\%) compared to standard Dimer, demonstrating measurable advantages in challenging cases.}
\end{figure}

\gls{otgpd} demonstrates superior robustness compared to existing dimer-based
methods, even considering the strange selection of data from Section
\ref{sec:datadrege}, as illustrated in Figure \ref{fig:otpgp_success}. While all
three approaches achieve comparable baseline success rates as shown in Table
\ref{tbl:succ:otgp}, the critical distinction emerges in systems where only one
method succeeds.

\begin{table}[htbp]
\caption{\label{tbl:succ:otgp}Success rates on systems of up-to two fragments}
\centering
\begin{tabular}{lrrr}
method & num\textsubscript{fragments} & n & success\\
\hline
Dimer & 1 & 26 & 92.3\\
Dimer & 2 & 212 & 96.7\\
GPDimer & 1 & 26 & 100\\
GPDimer & 2 & 212 & 96.2\\
OTGPD & 1 & 26 & 100\\
OTGPD & 2 & 212 & 97.6\\
\end{tabular}
\end{table}

The comparison reveals that OTGPD's advantage lies not in marginal improvements
to the baseline success rate, but in handling systems where conventional methods
fail. Against GPDimer, OTGPD uniquely succeeds on 11 systems while GPDimer
uniquely succeeds on only 3 (a \(3.7\times\) advantage). Against standard Dimer,
OTGPD uniquely succeeds on 9 systems compared to Dimer's 4 (a \(2.3\times\)
advantage). Conversely, cases where OTGPD alone fails are rare: only 2 systems
fail exclusively to OTGPD versus GPDimer, and only 1 system fails exclusively to
OTGPD versus Dimer.

The fact that the \gls{otgpd} captures difficult cases that other methods miss,
while rarely failing alone—demonstrates that \gls{gp} acceleration can provide
genuine robustness rather than simply shifting failure patterns. The method's
ability to navigate challenging optimization landscapes translates to practical
reliability improvements for automated saddle point discovery workflows.
\subsubsection{Identifying failure modes}
\label{sec:otgpd:res:failmodes}
The \gls{otgpd} framework successfully eliminates the signal variance instability
that plagued earlier iterations. However, our benchmarking revealed four
specific systems—D016, D084, D100, and S242—where the \gls{otgpd} failed to
converge, despite success in the \gls{gpd} baseline. A forensic analysis of these
trajectories indicates that these failures stem not from an intrinsic
algorithmic instability in the \gls{otgpd}, but rather from artifacts introduced
by the dimer initialization routine.

\begin{figure}[htbp]
\centering
\includegraphics[width=.9\linewidth]{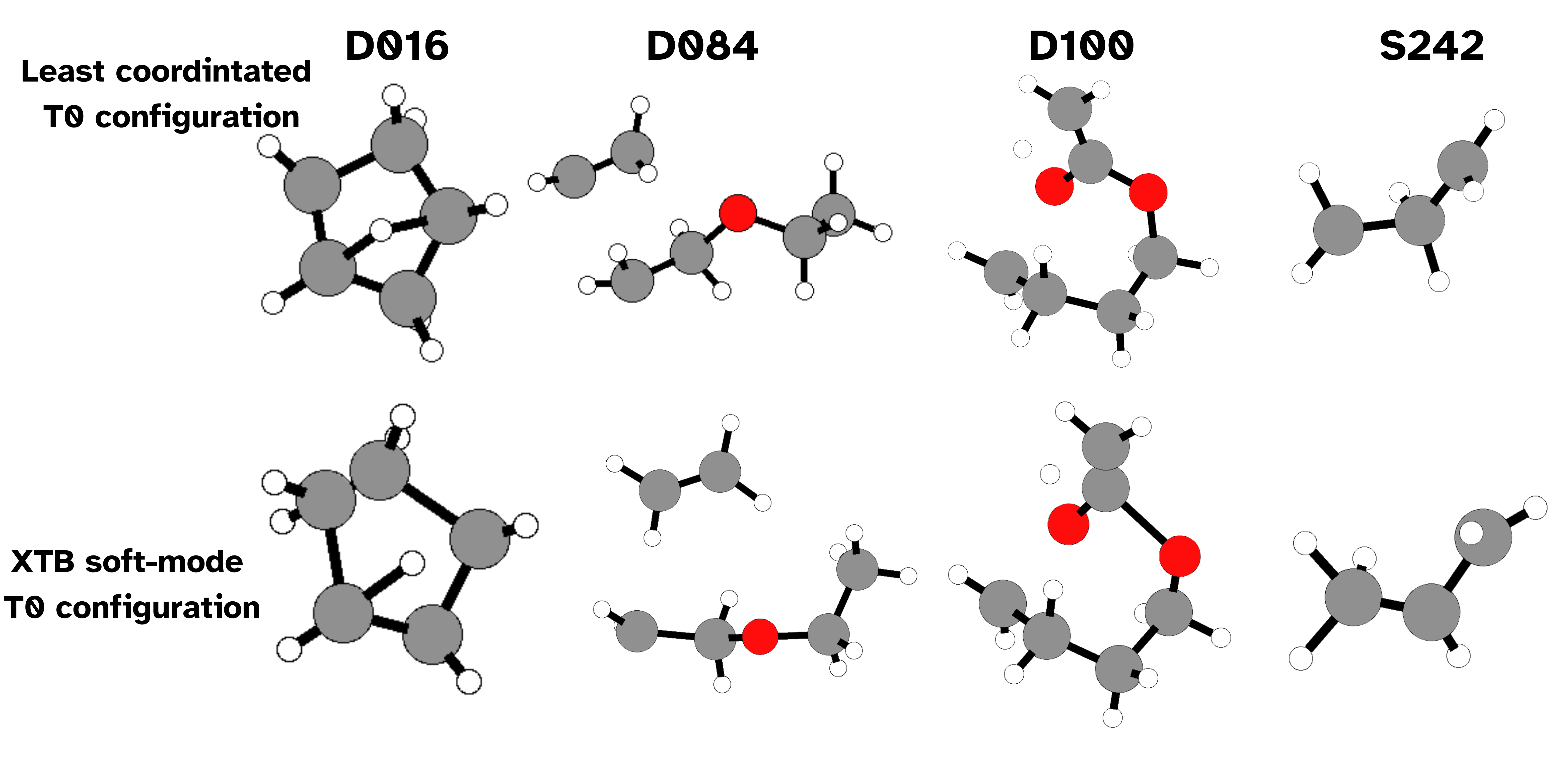}
\caption{\label{fig:suppl:initgpdotgpd}Initializations for the GPDimer (top) and OTGPD (bottom). The xTB initialization procedure in these four cases results in unphysical initial geometries, leading to failures in the optimization routine. Specifically, \texttt{D016} exhibits unphysically close carbon atoms, \texttt{D084} possesses a shortened carbon-oxygen bond, \texttt{D100} shows near-overlapping carbon atoms, and \texttt{S242} features a misplaced methyl group.}
\end{figure}

For the \gls{otgpd} benchmark, we generated initial dimer configurations by
displacing atoms along the softest mode identified by an inexpensive
semi-empirical (GFN-xTB) calculation. In these four specific cases, this
procedure produced pathological starting geometries characterized by excessively
high forces or unphysical atomic overlaps. Figure \ref{fig:suppl:initgpdotgpd}
contrasts these starting points. For example, system D016 initialized with
unphysically close carbon atoms, while S242 erroneously replaced a hydrogen on
the middle carbon with a methyl end-group.

When a GP-accelerated method receives such a ``poisoned'' baseline—a high-energy,
high-force configuration—the initial surrogate model incorporates this
unphysical data. The Gaussian Process effectively learns that these extreme
forces are characteristic of the landscape, leading to a warped posterior that
guides the search into unstable regions. The \gls{gpd} benchmark avoided this
specific pathology by utilizing a different initialization strategy based on
displacing the least coordinated atom.

To verify this diagnosis, we re-executed the \gls{otgpd} searches for these four systems using the valid, non-pathological initial configurations from the \gls{gpd} runs. Table \ref{tbl:fail_rerun} presents the results. In every case, the \gls{otgpd} successfully converged. Furthermore, it retained its performance advantage: for the challenging \texttt{D084} system, \gls{otgpd} required only 3.0 minutes (181 s) compared to 6.95 minutes (417 s) for \gls{gpd} and 4.25 minutes (255 s) for the standard Dimer. This confirms that the observed ``failures'' were initialization artifacts rather than deficiencies in the optimal transport framework.

\begin{table}[htbp]
\caption{\label{tbl:fail_rerun}Performance metrics for re-run systems (D016, D084, D100, S242) using identical initial configurations. All methods converge successfully. Times reported in minutes. The \gls{otgpd} maintains superior or competitive efficiency in both electronic structure calls and total wall time.}
\centering
\begin{tabular}{llrr}
System & Method & HF Calls & Time (min)\\
\hline
D016 & Dimer & 106 & 7.4\\
 & GPDimer & 20 & 3.7\\
 & OTGPD & 20 & 3.3\\
\hline
D084 & Dimer & 2666 & 255.4\\
 & GPDimer & 75 & 417\\
 & OTGPD & 65 & 181.4\\
\hline
D100 & Dimer & 214 & 24.3\\
 & GPDimer & 28 & 16.2\\
 & OTGPD & 28 & 18.52\\
\hline
S242 & Dimer & 249 & 14.6\\
 & GPDimer & 25 & 2.3\\
 & OTGPD & 33 & 1.85\\
\end{tabular}
\end{table}
\subsubsection{Linear bending angles and Sella}
\label{sec:otgpd:res:linbend}
\begin{figure}[htbp]
\centering
\includegraphics[width=.9\linewidth]{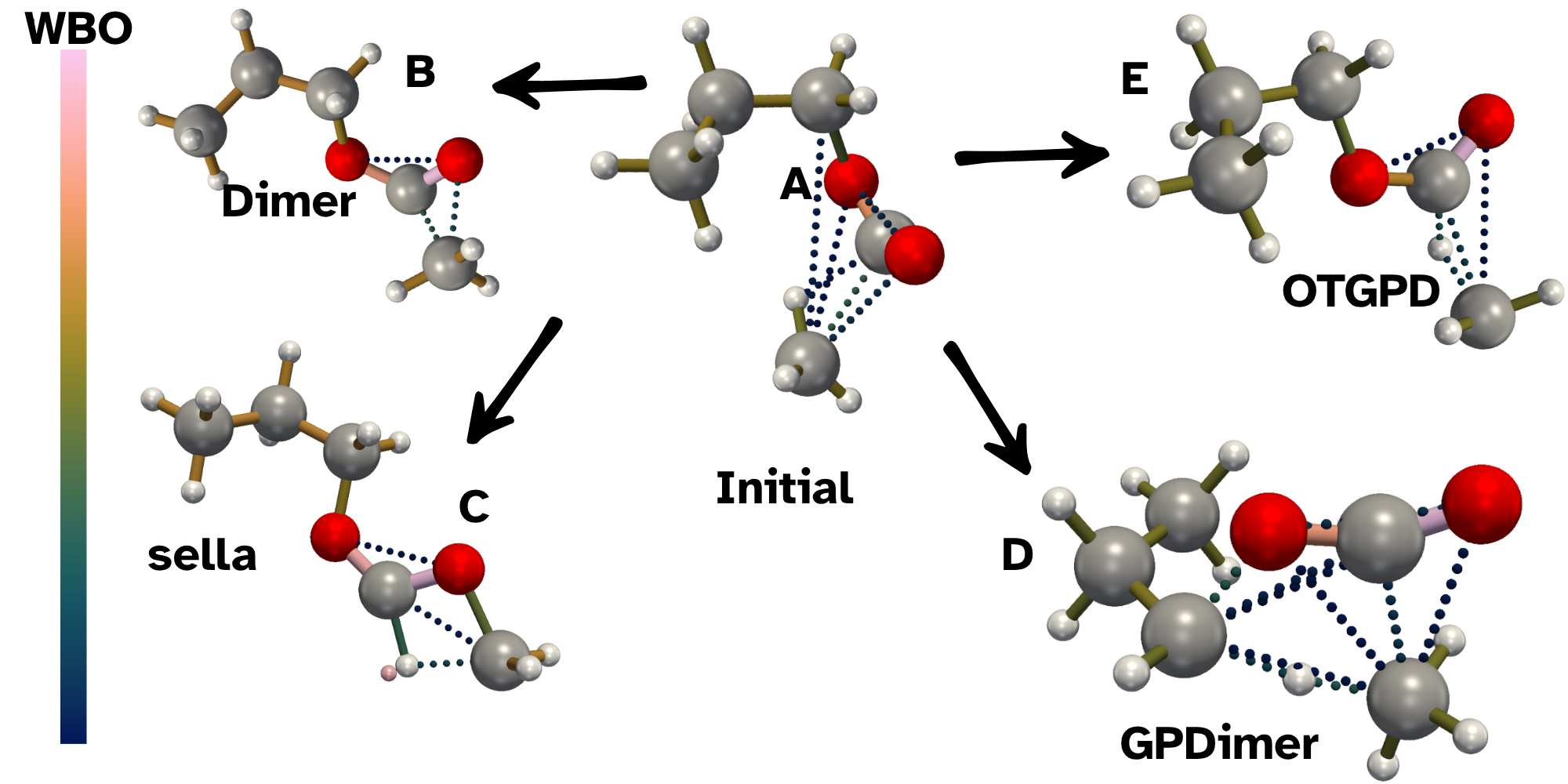}
\caption{\label{fig:equiv:optgd}Endpoints for saddle point search trajectories of the \texttt{singlet\_016} system starting from an initial configuration (A). The standard Dimer method (B) and the proposed OTGPD method (E) identifies the nearest transition state structure. The previous GPDimer method (D) and Sella (C) are guided towards a much more fractured state.}
\end{figure}

When the optimisation proceeds in Cartesian space we retain a clear mapping
between the optimisation variables and the molecular geometry, which permits a
post-hoc assessment of whether a reported saddle point corresponds to a
chemically meaningful transition state. Systems like \texttt{singlet\_016} in Figure
\ref{fig:equiv:optgd} clearly show a wide range of saddles connected to the same
initial state, one for each method, each of which are valid saddle points from a
mathematical perspective, but as shown in Section \ref{sec:gpjctc:sadcat} only
some correspond to physically relevant reaction pathways.

\begin{figure}[htbp]
\centering
\includegraphics[width=.9\linewidth]{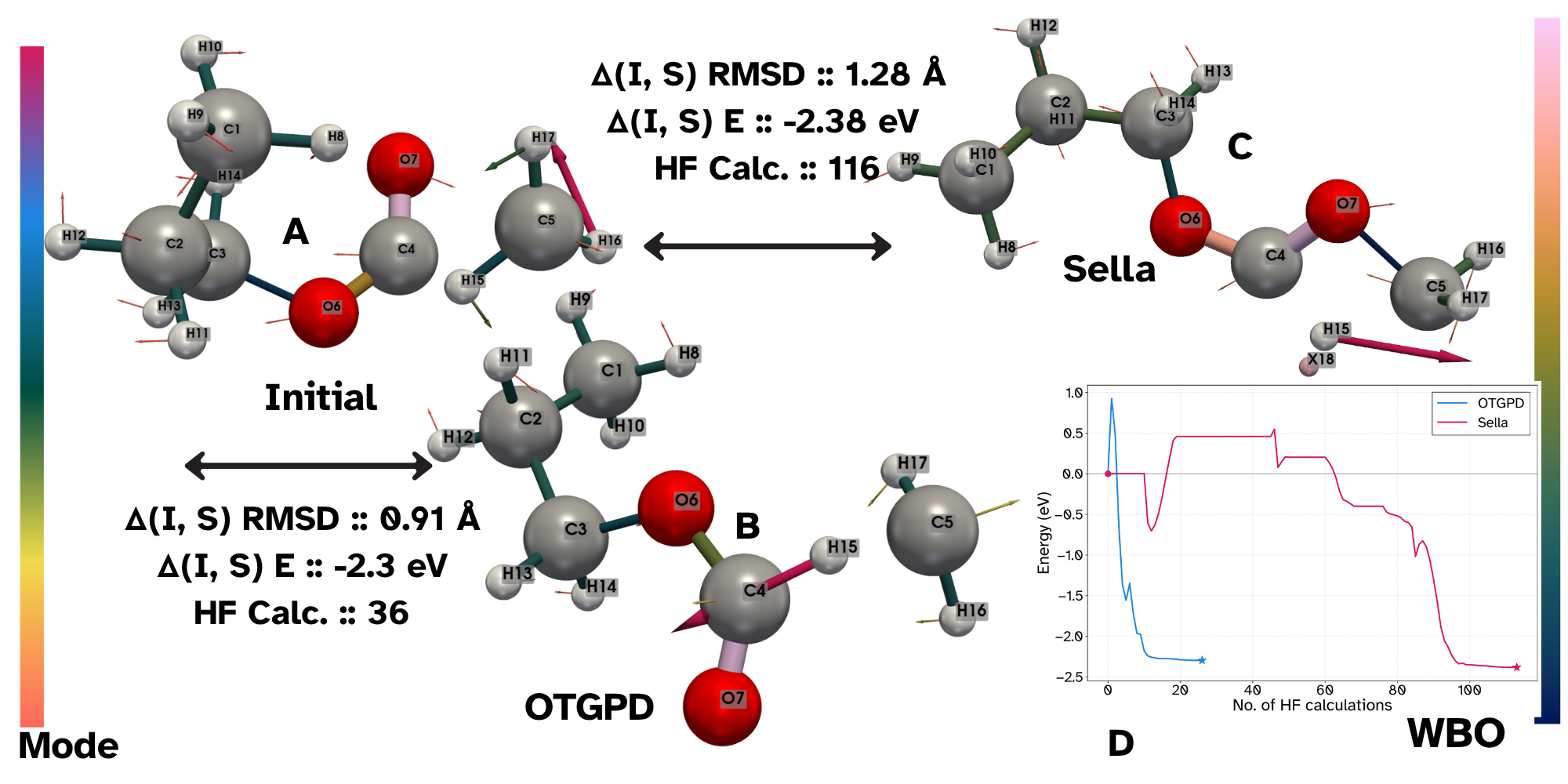}
\caption{\label{fig:otgpd_sella}Comparison of the OTGPD and Sella algorithms for a saddle point search on n-propyl acetate (\texttt{singlet\_016}), starting from the initial, non-equilibrium geometry. The OTGPD method efficiently locates the geometrically proximal saddle point corresponding to C–O bond cleavage in 36 steps. The Sella method follows a more computationally intensive path of 116 steps to find a more distant, nearly isoenergetic saddle corresponding to a 1,5-hydrogen atom transfer. The plot of the energy profiles for both searches highlights the significant difference in computational cost.}
\end{figure}

We compare the performance of the \gls{otgpd} and Sella algorithms on n-propyl
acetate (\(\mathrm{C_5H_{10}O_2}\)) in Figure \ref{fig:otgpd_sella} . Both search
methods commence from the same initial geometry, a structure that subsequent
optimization confirms does not represent a stable minimum on the potential
energy surface. The initial structure's softest vibrational mode corresponds to
a low-energy torsional motion of the hydrogens on the terminal methyl group
(C5).

The OTGPD method converges upon a proximal saddle point in just 36 PES
evaluations. This transition state corresponds to homolytic cleavage of the
central C4–O6 ester bond, with a barrier height of 2.3 eV and geometric
proximity to the initial structure (RMSD = 0.91 \AA{}). By identifying the
geometrically closest saddle, OTGPD successfully captures the most immediate
reaction pathway accessible from the initial geometry.

The Sella algorithm requires 116 HF calculations for convergence—a 3.2-fold
computational overhead. Rather than locating the nearby fragmentation pathway,
Sella explores a more complex trajectory and identifies a chemically distinct
saddle corresponding to a 1,5-hydrogen atom transfer (1,5-HAT) from C5 to the
carbonyl oxygen. This saddle possesses a nearly isoenergetic barrier height of
2.38 eV, yet lies significantly further from the initial structure (RMSD = 1.28
\AA{}). Sella has bypassed the more proximal and chemically direct C–O bond
cleavage saddle entirely.

\begin{figure}[htbp]
\centering
\includegraphics[width=0.8\textwidth]{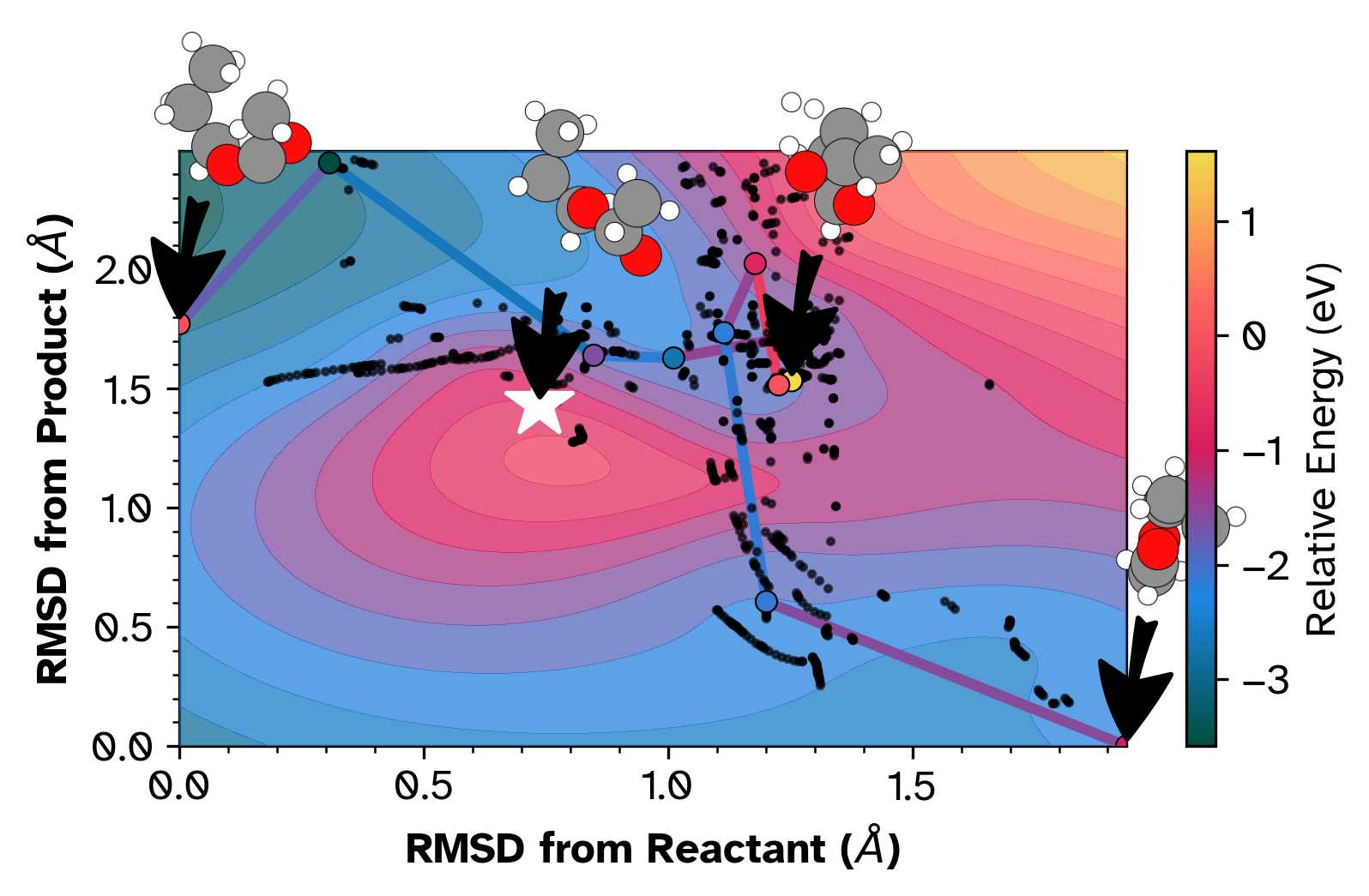}
\caption{\label{fig:s016_surface}A 2D landscape projection visualizing the potential energy surface of the n-propyl acetate system. This surface, described in Sec. \ref{sec:asd:pviz}, depicts the energy landscape as a function of observed paths during the optimization. The landscape clearly reveals several states. The proximal transition state (white star), corresponding to C-O cleavage, is the converged saddle point located by OTGPD and Dimer. The more distant saddle is the endpoint corresponding to the 1,5-HAT, located by Sella. This visualization strongly suggests that Sella's trajectory overshoots the first, more proximal saddle. We highlight the converged dimer saddle to illustrate that while the Sella trajectory passes near this configuration, it fails to localize the proximal transition state, instead proceeding to the distal 1,5-HAT saddle. The Sella trajectory passes near the dimer saddle configuration as reported earlier \cite{goswamiEfficientImplementationGaussian2025a}.}
\end{figure}

A two-dimensional landscape projection
\cite{goswamiTwodimensionalRMSDProjections2025} (Figure \ref{fig:s016_surface})
maps pathways from the initial, high-energy configuration to the deep
intermediate minimum and transition states. The white star marks the converged
\gls{otgpd} saddle, which is clearly proximal to the initial point, while the
surface shows the trajectory of paths explored during the optimization.

This behavior reveals a critical difference between the two approaches. While
both employ eigenvalue-following strategies, Gaussian process accelerated forms
of the dimer in cartesian coordinates efficiently converge to nearby saddles,
whereas Sella's search trajectory in internal coordinates may overshoot
geometrically proximal transition states in favor of lower-energy alternatives
located further from the starting configuration. For comprehensive reaction
exploration, this systematic overshoot would result in undercounting of the
local reaction network in automated discovery schemes. This would suggest that
the OTGPD's efficiency and geometric proximity make it superior for discovering
proximal transition states which is a prerequisite capability for
comprehensively cataloging accessible chemical transformations on the fly for
\gls{akmc}.
\subsubsection{Performance}
\label{sec:otgpd:res:perf}
\begin{figure}[htbp]
\centering
\includegraphics[width=1.1\textwidth]{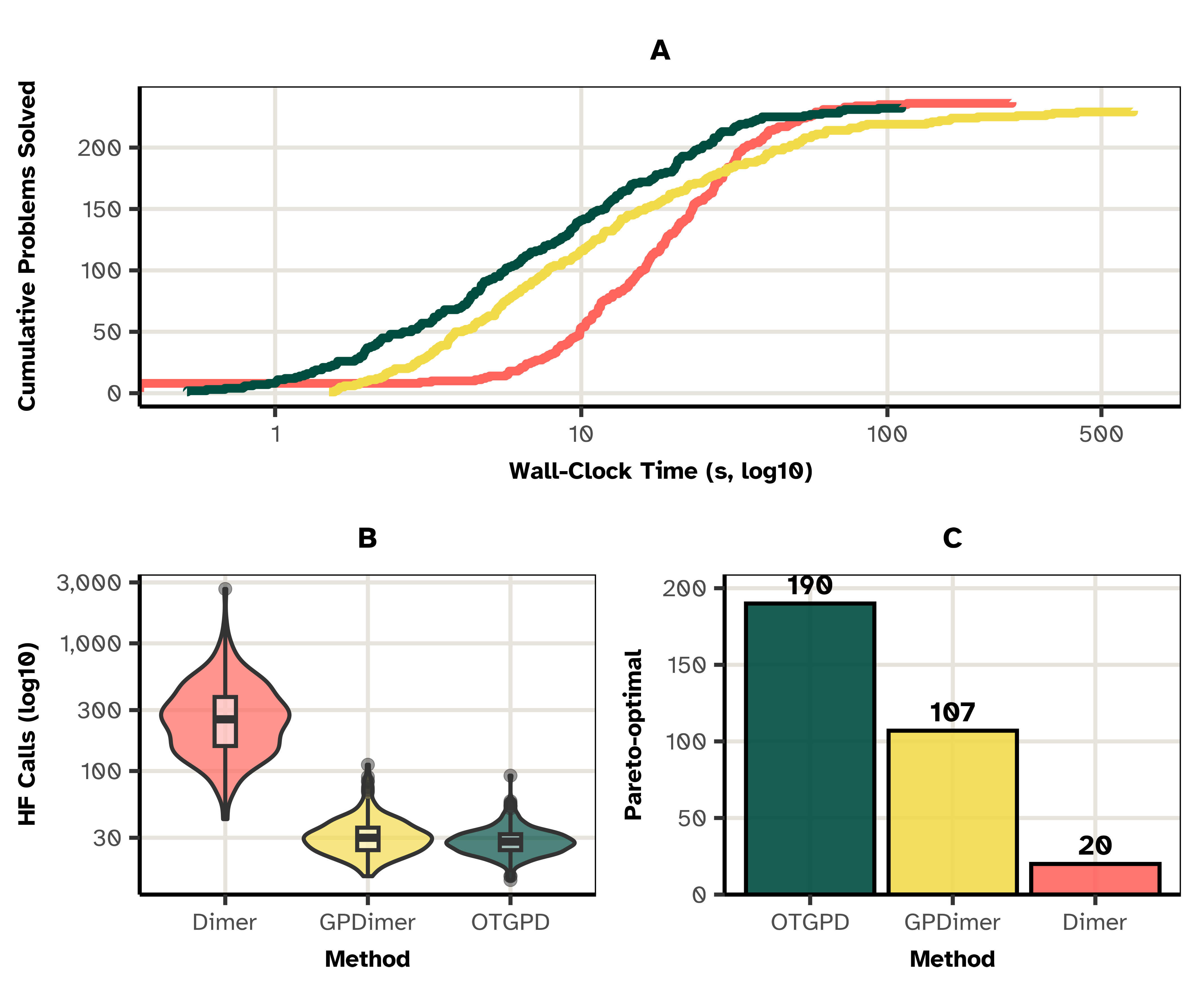}
\caption{\label{fig:otpgp_perf}Comparison of computational efficiency for the OTGPD, GPDimer, and standard Dimer methods. \textbf{\textbf{(A)}} A cactus plot shows the cumulative number of problems solved versus wall-clock time, demonstrating OTGPD's superior raw speed. \textbf{\textbf{(B)}} Violin plots of the number of Hartree-Fock (HF) calls show the order-of-magnitude improvement in data efficiency for the GP-accelerated methods. \textbf{\textbf{(C)}} A bar chart of the per-system Pareto-optimal count reveals that OTGPD most frequently provides the best trade-off between solution time and the number of HF calls, appearing on the frontier for 190 systems compared to 107 for GPDimer and 20 for the standard Dimer.}
\end{figure}

The raw solver throughput is visualized in the cactus plot (Figure
\ref{fig:otpgp_perf} A). This plot of cumulative problems solved versus wall-clock
time shows that \gls{otgpd} performance curve rises most steeply, indicating that
it solves a large number of problems in significantly less time than its
counterparts. \gls{gpd} follows as the next most efficient, while the standard
Dimer method exhibits a considerable lag, requiring more time to solve an
equivalent number of systems. Within the first 10 seconds of wall-clock time,
OTGPD successfully solves over 100 problems, while the standard Dimer has
converged on fewer than 50. While \gls{gpd} is a clear improvement over the Dimer,
it consistently lags OTGPD in the number of systems solved within any given time
budget.

This superior speed is partially accounted for by improved data efficiency, as
shown in Figure \ref{fig:otpgp_perf} B. The violin plots reveal that the number of
\gls{hf} calculations required by the standard Dimer is an order of magnitude
greater than that for the \gls{gp} -accelerated methods. We quantify this in the
median number of calculations: just 28 for OTGPD and 30 for GPDimer, compared to
254 for the standard Dimer. This drastic reduction in expensive electronic
structure calculations is the primary driver of the observed performance gains.

While both \gls{gp} methods exhibit similar data efficiency, the cactus plot shows
that \gls{otgpd} wall-clock performance is significantly better than \gls{gpd}. This
highlights the effect of the computational overhead reduction. By design, the
\gls{otgpd} minimizes this overhead, ensuring that the gains from reduced data
requirements translate more effectively into real-world speed.

The most sophisticated measure of performance comes from a per-system Pareto
optimality analysis (Figure \ref{fig:otpgp_perf} C. This moves beyond
single-metric comparisons to find the set of solutions that represent the best
possible trade-offs. For a set of solutions, a given solution is Pareto-optimal
if no other solution is superior in all objectives. Formally, a vector of
objectives \(F(x_A)\) dominates \(F(x_B)\), noted \(F(x_A) \prec F(x_B)\), if:

\begin{equation}
\label{eq:pareto}
F(x_A) \prec F(x_B) \iff \forall i, f_i(x_A) \le f_i(x_B) \land \exists j, f_j(x_A) < f_j(x_B)
\end{equation}

The analysis identifies which method resides on the Pareto frontier, which is
the set of non-dominated solutions for each system. The results show that
\gls{otgpd} is on the Pareto-optimal frontier for 190 systems, making it the
optimal choice nearly twice as often as \gls{gpd} (107) and almost ten times more
frequently than the standard Dimer (20). This confirms that the \gls{otgpd}
consistently finds the best balance of computational costs.
\subsection{Conclusions}
\label{sec:otgpd:conclusions}
This chapter transformed the Gaussian Process from a volatile statistical
interpolator into a robust engine for chemical discovery. We demonstrated that
naive application of machine learning to Cartesian landscapes frequently yields
pathological failures, such as the ``cold fusion'' artifacts observed in the
standard GPDimer. These failures stem not from a lack of data, but from a lack
of physical constraint.

To remedy this, we replaced the arbitrary Euclidean metric with the
permutation-invariant Intensive Earth Mover's Distance (EMD). This metric equips
the surrogate with a chemically meaningful sense of locality. It underpins the
adaptive trust radius, a mechanism that strictly enforces the validity of the
local approximation and prevents the optimizer from wandering into unphysical
regions of the potential energy surface. Furthermore, by linking data selection
to numerical stability via Farthest Point Sampling, we ensured that the
covariance matrices remain well-conditioned, satisfying Gershgorin bounds and
enabling reliable inversion even in high-throughput regimes.

Empirical validation confirms the efficacy of this approach. The \gls{otgpd} lies
on the Pareto-optimal frontier for the majority of test systems, offering the
most favorable trade-off between wall-clock time and electronic structure calls.
Crucially, the comparison with Sella reveals a distinct topological advantage:
while internal-coordinate methods often overshoot to find thermodynamically
stable but geometrically distant states, \gls{otgpd} reliably converges to the
proximal saddle point. This capability proves essential for kinetic Monte Carlo
simulations, where discovering the immediate connectivity of the reaction
network takes precedence over finding the global minimum.

Ultimately, the \gls{otgpd} represents the culmination of the efficiency
strategies developed throughout this thesis. It integrates the robust
saddle-finding logic of the Dimer method, the statistical power of Gaussian
Processes, and the rigorous stability controls derived from optimal transport
theory into a single, cohesive framework for autonomous reaction discovery.
\section{Summary}
\label{sec:summary}
\epigraph{The purpose of computing is insight, not numbers.}{Richard Hamming}
This dissertation presented a multi-faceted investigation into the role of
computational representation as a primary driver of progress in chemical
physics. We structured this inquiry around specific research objectives
targeting relativistic electronic structure, software architecture, statistical
benchmarking, and algorithmic efficiency. Here, we summarize the contributions
against those initial goals.

We began by questioning whether a squared Hamiltonian formulation within a
finite element framework could provide stability for all-electron relativistic
calculations. Chapter \ref{sec:fem} validated this hypothesis. The \texttt{featom}
solver, built upon a high-order \gls{fem} discretization of the squared Dirac
operator, successfully eliminated spectral pollution. It achieves sub-second
wall times for all-electron \gls{dft} calculations of heavy elements like Uranium,
outperforming state-of-the-art shooting methods in the relativistic regime while
avoiding the complexity of kinetic balance constraints.

Subsequently, we investigated whether modernizing legacy software architectures
could unlock novel scientific algorithms. Chapter \ref{sec:asp} demonstrated this
through the complete refactoring of the EON suite. By decoupling the potential
energy surface via a client-server architecture, we enabled the creation of the
hybrid \gls{roneb} method. This algorithm reduces the computational cost of
double-ended saddle searches by approximately 60\% compared to standard
approaches, a gain achievable only through the flexibility of the redesigned
state management system. In Chapter \ref{sec:gpjctc}, a C++ rewrite enabled the
Cartesian coordinate based \gls{gpd} to compete with state of the art internal
coordinate methods.

To ensure these gains were robust, we sought to move beyond average-cost
benchmarks and rigorously quantify algorithmic reliability. Chapter
\ref{sec:brmsgp} applied Bayesian generalized linear mixed models to the Dimer
method. This analysis overturned the conventional wisdom regarding rotational
minimization, providing robust statistical evidence that the Conjugate Gradient
optimizer offers superior reliability compared to \gls{lbfgs}. Furthermore, we
quantified the high cost of quaternion-based rotation removal within EON,
demonstrating that it incurs a significant computational penalty without a
credible improvement in convergence success.

Finally, we addressed the cubic scaling and instability bottlenecks of Gaussian
Process acceleration. Chapter \ref{sec:otgpd} introduced the \gls{otgpd} framework.
By implementing the Intensive Earth Mover's Distance for physically motivated
trust regions and Farthest Point Sampling for numerical conditioning, we
eliminated the ``cold fusion'' pathologies observed in previous methods. The
resulting algorithm places on the Pareto frontier for 190 out of 238 test
systems, effectively halving the time-to-solution compared to the standard
\gls{gpd} and demonstrating superior performance to internal-coordinate methods
for proximal saddle searches.

\section{Conclusions}
\label{sec:conclusions}
\epigraph{Dealing with failure is easy: Work hard to improve. Success is also easy to handle: You've solved the wrong problem. Work hard to improve.}{Alan Perlis}

As summarized in the preceding chapter, this thesis produced a suite of novel
methods and robust software tools, transforming raw computation into scientific
insight. Yet, as the doctrine of Alan Perlis suggests, every solution unmasks a
new layer of problems. This chapter moves beyond a summary of specific results
to synthesize their broader implications, confront the limitations that remain,
and outline a path forward for the future of computational science.
\subsection{The Science of Scientific Software}
\label{sec:org5fac158}
The development of powerful, specialized tools demands a significant investment
of expert effort, yet their long-term maintenance and the management of their
inherent technical debt represent a systemic challenge in academic research. The
future of the field depends not only on the creation of novel algorithms but on
the elevation of software construction to the status of a science itself. By
treating the software as a first-class research object, we ensure that the
insights gained from one generation of scientific inquiry provide a robust
foundation for the next.

Across the breadth of this thesis, we emphasized that efficient representations
emerge only through the effort expended in ``speaking binary.'' The era of
``formula translation''—blindly transcribing mathematical notation into code—has
ended. The constraints of modern hardware, particularly in the exascale era,
demand a deeper engagement with memory hierarchies, vectorization, and
concurrency. The success of the \gls{otgpd} serves as a testament to this
philosophy; it functions essentially as an exercise in rigorous profiling. We
identified that hyperparameter estimation constituted the wall-time bottleneck
and addressed it not through clearer mathematics, but through algorithmic
restructuring and data pruning. Thus, technical mastery over average-time
complexity must accompany theoretical innovation.
\subsection{The Ontological Gap: Statistics vs. Physics}
\label{sec:org6c66cbb}
A central tension persists throughout this work: the disconnect between the
statistical surrogate and the physical reality it models. A \gls{gp} defines a
\gls{mvn} over function values. We select the specific distribution that best
explains the observed data by maximizing the \gls{mll}. However, this statistical
procedure implies no guarantee that the model reproduces the true \gls{pes}
generated by electronic structure theory.

The true potential energy surface arises from the many-body expansion \cite{stoneTheoryIntermolecularForces2013}:
\begin{equation} V(x) = \sum_i V_1(i) + \sum_{i<j} V_2(i, j) + \sum_{i<j<k} V_3(i, j, k) + \dots + V_N(1, \dots, N) \label{eq:mbe} \end{equation}

This expansion possesses specific decay properties and symmetries not inherently
captured by the two-point covariances of a stationary Gaussian kernel. The GP
``anthropomorphizes'' the data; we often ascribe unjustified physical meaning to
hyperparameters like the signal variance or length scale. In reality, these
parameters merely maximize the likelihood of a specific dataset. When that
dataset consists of sparse, correlated points along a saddle-search
path—atypical of the global surface—the \gls{mll} surface often becomes shallow or
degenerate. This manifests visibly in the ``cold fusion'' pathologies observed in
the standard GPDimer, where the signal variance explodes to accommodate
high-force gradients, flattening the mean prediction and directing the search
into unphysical regions.

The \gls{otgpd} mitigates this not by forcing the GP to understand physics, but by
imposing physical constraints (trust regions, stability barriers) upon the
statistical engine. We acknowledge that the model operates as a mathematical
interpolator, not a physical simulator, and constrain it accordingly.
\subsection{Future Outlook: A BLAS for Chemical Kinetics}
\label{sec:org291380c}
The OT-GP framework, while successful, shares the worst-case cubic time
complexity of its predecessors. Pruning over dynamic data remains an open
challenge, particularly for systems exceeding the size regimes explored here. To
address this, the field requires a paradigm shift in software architecture.

We propose the vision of a ``BLAS for chemical kinetics.'' Just as the Basic
Linear Algebra Subprograms (BLAS) provide a standard, highly optimized
foundation for linear algebra, we require a standardized, decoupled layer for
kinetic primitives—saddle searches, rate evaluations, and traversing potential
landscapes. Such a framework would rely on polyglot libraries communicating
through zero-copy interfaces, allowing researchers to compose high-level
workflows (like \gls{akmc} or \gls{md}) from highly optimized, hardware-aware
components.

This architectural evolution paves the path for scaling the methods developed
herein. The \gls{otgpd} provides the foundational approach for the active learning
of high-energy transition state geometries. These datasets act as the critical
feedstock for training next-generation reactive machine-learned potentials. By
integrating these surrogates with symmetry-adapted representations or
reinforcement learning agents, we move closer to the ultimate goal: the
autonomous, reliable exploration of the vastness of chemical space.
\printbibliography[heading=bibintoc, title={References}] %
\end{document}